\documentclass[twoside,12pt]{article}
\usepackage{graphicx}

\def\Journal#1#2#3#4{{#1} {#2} (#4) #3 }

\newcommand{\be}{\begin{equation}}
\newcommand{\ee}{\end{equation}}
\newcommand{\bea}{\begin{eqnarray}}
\newcommand{\eea}{\end{eqnarray}}

\begin{document}

\title{ \vspace{1cm} Flavor Physics and CP Violation}
\author{Paoti Chang, Kai-Feng Chen, Wei-Shu Hou\footnote{Email: wshou@phys.ntu.edu.tw}
\\
\\
Department of Physics, National Taiwan University, Taipei 10617, Taiwan
}
\maketitle
\begin{abstract}
We currently live in the age of the CKM paradigm.
The $3\times 3$ matrix that links $(d,\ s,\ b)$ quarks to $(u,\ c,\ t)$
in the charged current weak interaction, being complex and nominally with 18 parameters,
can be accounted for by just 3 rotation angles and one $CP$ violating (CPV) phase,
with unitarity and the CKM phases triumphantly tested at the B factories.
But the CKM picture is unsatisfactory and has too many parameters.
The main aim of Flavor Physics and $CP$ violation (FPCP) studies is
the pursuit to uncover New Physics beyond the Standard Model (SM).
Two highlights of LHC Run~1 period are the CPV phase $\phi_s$ of $B_s$ mixing
and $B_s \to\mu^+\mu^-$ decay, which were found to be again consistent with SM,
though the saga is yet unfinished.
We also saw the emergence of the $P_5'$ angular variable
anomaly in $B^0 \to K^{*0}\mu^+\mu^-$ decay and $R_{K^{(*)}}$ anomaly
in $B \to K^{(*)}\mu^+\mu^-$ to $B \to K^{(*)}e^+e^-$ rate ratios,
and the BaBar anomaly in $B \to D^{(*)}\tau\nu$ decays,
which suggest possible New Physics in these flavor processes,
pointing to extra $Z'$, charged Higgs, or leptoquarks.
Charmless hadronic, semileptonic, purely leptonic and radiative $B$ decays
continue to offer various further windows on New Physics.
Away from $B$ physics,
the rare $K \to \pi\nu\nu$ decays and $\varepsilon'/\varepsilon$ in the kaon sector,
$\mu \to e$ transitions, muon $g-2$ and electric dipole moments of the neutron and electron,
$\tau \to \mu\gamma,\ \mu\mu\mu,\ eee$, and a few charm physics probes,
offer broadband frontier windows on New Physics.
Lastly, flavor changing neutral transitions involving
the top quark $t$ and the 125 GeV Higgs boson $h$,
such as $t\to ch$ and $h\to \mu\tau$, offer a new window into FPCP,
while a new $Z'$ related or inspired by the $P_5'$ anomaly,
could show up in analogous top quark processes,
perhaps even link with low energy phenomena such as muon $g-2$ or rare kaon processes.
In particular, we advocate the potential new SM, the two Higgs doublet model
without discrete symmetries to control flavor violation, as SM2.
As we are close to the alignment limit with $h$ rather SM-like,
flavor changing neutral Higgs couplings (FCNH) are suppressed
by a small mixing angle, but the exotic Higgs doublet possesses
FCNH couplings, which we are just starting to probe.
As LHC Run 2 runs its course, and with Belle II $B$ physics program to start soon,
there is much to look forward to in the flavor and CPV sector.
\end{abstract}
%\eject
\newpage

\tableofcontents

\section{Introduction %:\quad {\texttt{history \quad CKM \quad frontier}}
}

In the domain of Flavor Physics and CP Violation (FPCP),
we are currently under the Cabibbo--Kobayashi--Maskawa, or CKM, paradigm,
%which constitutes
one of the pillars of the Standard Model (SM) of particle physics.
Formulated roughly half a century ago and emergent since 40 years,
we have not seen any clear cracks in the past 20 years of B factory,
and now LHC(b), scrutiny.
In the current period of LHC Run~2, much more data would be collected,
while Belle II would finally start B physics data taking.
It is a good time to take stock and project what lies ahead,
towards the dawn of the 2020s.

Facing a ``strangely'' prolonged kaon lifetime,
young Cabibbo proposed~\cite{Cabibbo:1963yz} in 1963 a unitary symmetry approach:
in modern language of quarks, it is the linear combination of
$d' = d\cos\theta_C + s\sin\theta_C$ that enters the charged current.
With a small $\theta_C \simeq 14^\circ$, the Cabibbo angle,
one also understood neutron beta decay better.
As this was the time when the unified electroweak gauge theory was formulated,
Cabibbo's picture led to a new problem:
if the neutral current $Z^0$ boson couples\footnote{
Dropping Dirac matrices for simplicity.
}
to $\bar d'd'$,
it would lead to flavor changing neutral current (FCNC) $\bar dsZ^0$ coupling,
hence $K_L \to \mu^+\mu^-$ at a rate that was simply not observed at the time.
The electroweak theory had other more fundamental challenges to face, but in 1970,
Glashow, Iliopoulos and Maiani (GIM) proposed~\cite{Glashow:1970gm}
an elegant solution to the FCNC problem:
if $Z^0$ couples also to a second combination $s' = -d\sin\theta_C + s\cos\theta_C$
that is orthogonal to $d'$, then a new term $-\sin\theta_C\cos\theta_C$ would
cancel the offending $\sin\theta_C\cos\theta_C$ by unitarity.
The price, or prediction, for this to happen,
is the existence of a 4th quark ``charm'', $c$,
to pair with $s'$ in the charged current.
The existence of the charm quark was discovered~\cite{Aubert:1974js, Augustin:1974xw}
in the November Revolution of 1974, a watershed that precipitated the establishment of SM.
Both the electroweak gauge theory and the newly emerged SU(3)
color gauge theory of the strong interactions were by then calculable,
allowing one to even estimate the charm mass from rare kaon processes,
independent from its direct experimental discovery.

There is another great contribution from kaon physics:
the observation~\cite{Christenson:1964fg} of $CP$ violation (CPV) in 1964.
Though the treatment of quark phases was mentioned,
GIM did not~\cite{Glashow:1970gm} pursue CPV in their seminal work.
Instead, the issue was picked up in late 1972 by
two young physicists in the Far East.
Besides exploring several different models,
arguing that the two generation case possessed no phase in
the charged current coupling, Kobayashi and Maskawa showed~\cite{Kobayashi:1973fv} that
a single CPV phase remains when one has three generations of quarks.
A new doublet, $(t,\ b')$, couples to the charged current,
and $d'$, $s'$ and $b'$ are related to the mass eigenstates
$d$, $s$ and $b$ by a $3\times 3$ unitary matrix, $V$.
By argument of anomaly cancellation, a third doublet of leptons,
$(\nu_\tau,\ \tau)$ also had to exist, and remarkably,
in the same data that gave the $\psi$ and $\psi'$ peaks~\cite{Augustin:1974xw},
the $\tau$ lepton was discovered~\cite{Perl:1975bf} in 1975.
Repeating the $J$ discovery~\cite{Aubert:1974js}, the $b$ quark was
discovered~\cite{Herb:1977ek} through the $\Upsilon$ resonances in 1977
at hadron facilities.
But, although each and every new collider that followed pursued the $t$ quark,
it took almost 20 years for the Tevatron to eventually
discover~\cite{Abe:1995hr, Abachi:1995iq} the top in 1995,
which was \emph{unexpectedly heavy}.

\begin{figure}[ht]
\begin{center}
\begin{minipage}[t]{8 cm}
\vskip-0.3cm \hskip-1.4cm
\includegraphics[width=9.5cm]{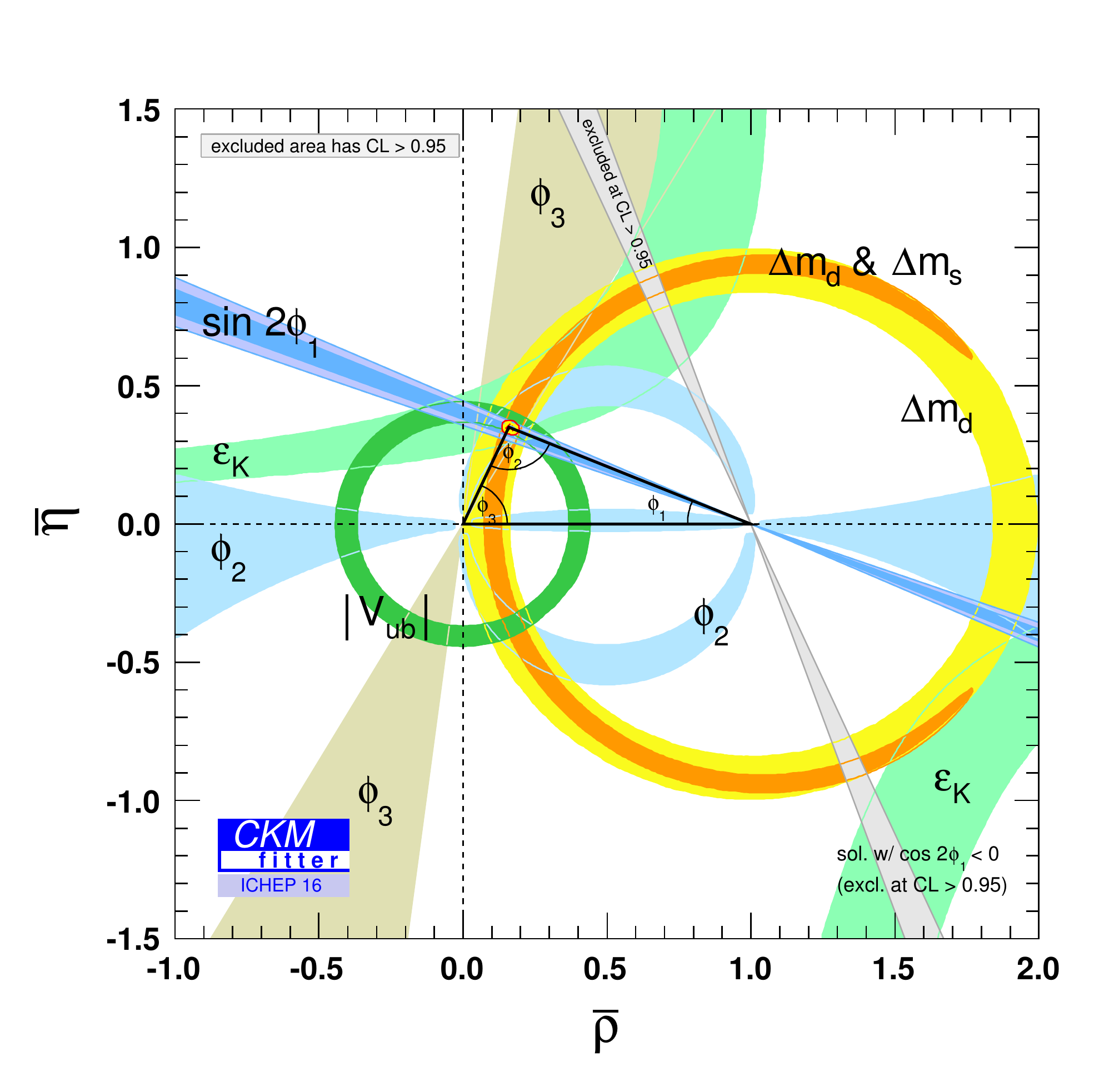}
\end{minipage}
\vskip-0.2cm
\begin{minipage}[t]{16.5 cm}
\caption{
CKM fit of Eq.~(\ref{eq:bdTri}) as of Summer 2016, with triangle
normalized to $|V_{cd}V_{cb}^*|$, or $-V_{cd}V_{cb}^*$ if one takes
the parametrization of Eq.~(\ref{eq:VCKM}).
 [Source: http://ckmfitter.in2p3.fr/]
 \label{fig:CKM2016}
}
\end{minipage}
\end{center}
\end{figure}

The heaviness of the top quark was in fact foretold by
the observation~\cite{Albrecht:1987ap} of large $B^0$--$\bar B^0$ mixing in 1987,
which illustrates the central role that B physics plays in FPCP.
This in fact traces back to the discovery~\cite{Lockyer:1983ev, Fernandez:1983az}
of prolonged bottom hadron lifetimes, and the equally astounding
discovery~\cite{Klopfenstein:1983nz, Chen:1984ci} that
$b \to u$ transitions are far more suppressed than $b \to c$ transitions.
A hierarchical structure
\be
|V_{ub}| \ll |V_{cb}| \ll |V_{us}| \simeq 0.22,
 \label{eq:hiera_V}
\ee
emerged for the CKM matrix $V$. This led Wolfenstein to write down
his namesake parametrization~\cite{Wolfenstein}.
In the phase convention~\cite{PDG} of keeping $V_{us}$ and
$V_{cb}$ real, the unique CPV phase is placed in $V_{ub}$,
hence in $V_{td}$ as well by unitarity of $V$.
The Wolfenstein form~\cite{Wolfenstein} of the CKM matrix is
\be
 V =
    \left( \begin{array}{ccc}
    V_{ud}  &   V_{us}  &  V_{ub}  \\
    V_{cd}  &   V_{cs}  &  V_{cb}  \\
    V_{td}  &   V_{ts}  &  V_{tb}  \end{array} \right)
 \simeq
    \left(  \begin{array}{ccc}
    1 - \lambda^2/2               &     \lambda     &  A\lambda^3(\rho - i\,\eta) \\
      - \lambda                   & 1 - \lambda^2/2 &  A\lambda^2 \\
    A\lambda^3(1 - \rho - i\, \eta) &   -A\lambda^2   &  1
   \end{array} \right),
 \label{eq:VCKM}
\ee
which any FPCP practitioner should memorize, together with\footnote{
We have kept the original Wolfenstein approximate definition
for ease of memorising, but as reflected in
the $\bar\rho$--$\bar\eta$ axes in Fig.~\ref{fig:CKM2016},
a more refined definition that holds to all orders in
$\lambda$ and is also rephasing invariant, is typically used:
$\lambda = |V_{us}| / \sqrt{|V_{ud}|^2 + |V_{us}|^2}$,
$A \lambda^2 = |V_{cb}| / \sqrt{|V_{ud}|^2 + |V_{us}|^2}$, and
$\bar\rho + i \bar\eta = V_{ud} V_{ub}^* / V_{cd} V_{cb}^*$.
}
\be
\lambda \equiv V_{us} \cong 0.22, \ \ \ \ A\lambda^2 \equiv V_{cb}
\simeq 0.041, \ \ \ \ A\lambda^3\sqrt{\rho^2+\eta^2} \equiv |V_{ub}| \sim
0.0036, \
 \label{eq:lambda}
\ee
which reflect and quantify the hierarchy pattern of Eq.~(\ref{eq:hiera_V}).
And the quest began for the CPV phase.

With the long B lifetime, rare B decays became interesting. On top of this,
the very heavy top, together with the associated nondecoupled effects of
large Yukawa coupling, made the B physics program very attractive.
So, when large $\Delta m_{B_d}\ (\simeq \Gamma_{B_d}/2$!) was observed,
on one hand it implied a heavy top quark, on the other hand,
``B factory'' planning started at SLAC and KEK,
towards the measurement of the CPV phase $\sin\phi_1/\beta$
by the elegant methods developed by Sanda and Bigi~\cite{Carter:1980tk, Bigi:1981qs}.
By the 1990's, construction started for the corresponding experiments and
\emph{asymmetric} $e^+e^-$ colliders, BaBar/PEP-II and Belle/KEKB.

%     - CKM fit plot (paradigm)

The CPV phases $\phi_3/\gamma \equiv \arg V_{ub}^*$
and $\phi_1/\beta \equiv \arg V_{td}$ appear as vertex angles
in the triangle or unitarity relation
\be
V_{ud}V_{ub}^* + V_{cd}V_{cb}^* + V_{td}V_{tb}^* = 0.
 \label{eq:bdTri}
\ee
The Wolfenstein form in Eq.~(\ref{eq:VCKM})
satisfies Eq.~(\ref{eq:bdTri}) to $\lambda^3$ order,
and can be further extended to $\lambda^5$ order.
The current status of our CKM paradigm can be summarized in a plot,
Fig.~\ref{fig:CKM2016},
using the KM notation of $\phi_1$, $\phi_2$, $\phi_3$
(equivalent to $\beta$, $\alpha$, $\gamma$ in the notation used by BaBar).
This is not only a \emph{beautiful} plot that incorporates Eq.~(\ref{eq:bdTri}),
it contains much information and measurements, details of which would be discussed in Sec.~2.
But the main point is, less than 15 years after discovery of large B mixing,
the world not only constructed two B factories,
but Belle and BaBar both observed~\cite{Aubert:2001nu, Abe:2001xe}
the CPV phase of $B^0$--$\bar B^0$ mixing in 2001,\footnote{
The year after, the two bi-annual ``Heavy Flavor'' and ``B Physics and CP Violation'' conferences
merged into the annual FPCP conference.
}
confirming the CKM paradigm, and the 2008 Nobel prize was awarded to Kobayashi and Maskawa
(together with Nambu for spontaneous symmetry breaking, SSB).

A CPV effect requires the interference between a CPV phase
and a $CP$-conserving phase~\cite{FlavTeV}.
The beauty of the Bigi--Sanda method is that the latter phase,
$e^{i\Delta mt}$, the time evolution of the $B^0$--$\bar B^0$ oscillation,
is directly measured ``\emph{in situ}'' by the experiments
(thanks to the development of silicon-based vertex detectors),
and is thus independent of so-called hadronic uncertainties.
The oscillation is a quantum phenomenon of entangled states.

The fact that the sides and angles of the unitarity triangle of Eq.~(\ref{eq:bdTri})
fit nicely together in Fig.~\ref{fig:CKM2016} and
holding up in the past 15 years of intense scrutiny,
means that we see no real cracks in the CKM paradigm,
even though we know it falls far short of the CPV strength needed for baryogenesis~\cite{Peskin:2008zz}.
While such precision measurements and ``CKM fits'' would certainly continue,
in general we seek  ``anomalies'' that suggest deviations from SM predictions,
i.e. hints for New Physics (NP). This is the frontier push of FPCP,
but here we are oftentimes at the mercies of ``hadronic uncertainties".
As an example, let us mention the well established~\cite{PDG} measurement of
difference of \emph{direct} CPV in $B\to K\pi$ decay,
\be
\Delta A_{K\pi} \equiv A_{K^+\pi^0} - A_{K^+\pi^-}
= 0.122 \pm 0.022, \quad {\rm (PDG)}
 \label{eq:DAKpi}
\ee
as first clearly demonstrated by Belle~\cite{Lin:2008zzaa}.

\begin{figure}[ht]
\begin{center}
\begin{minipage}[t]{9.5cm}
\hskip-0.2cm %\hskip-1.2cm
\includegraphics[width=9.5cm]{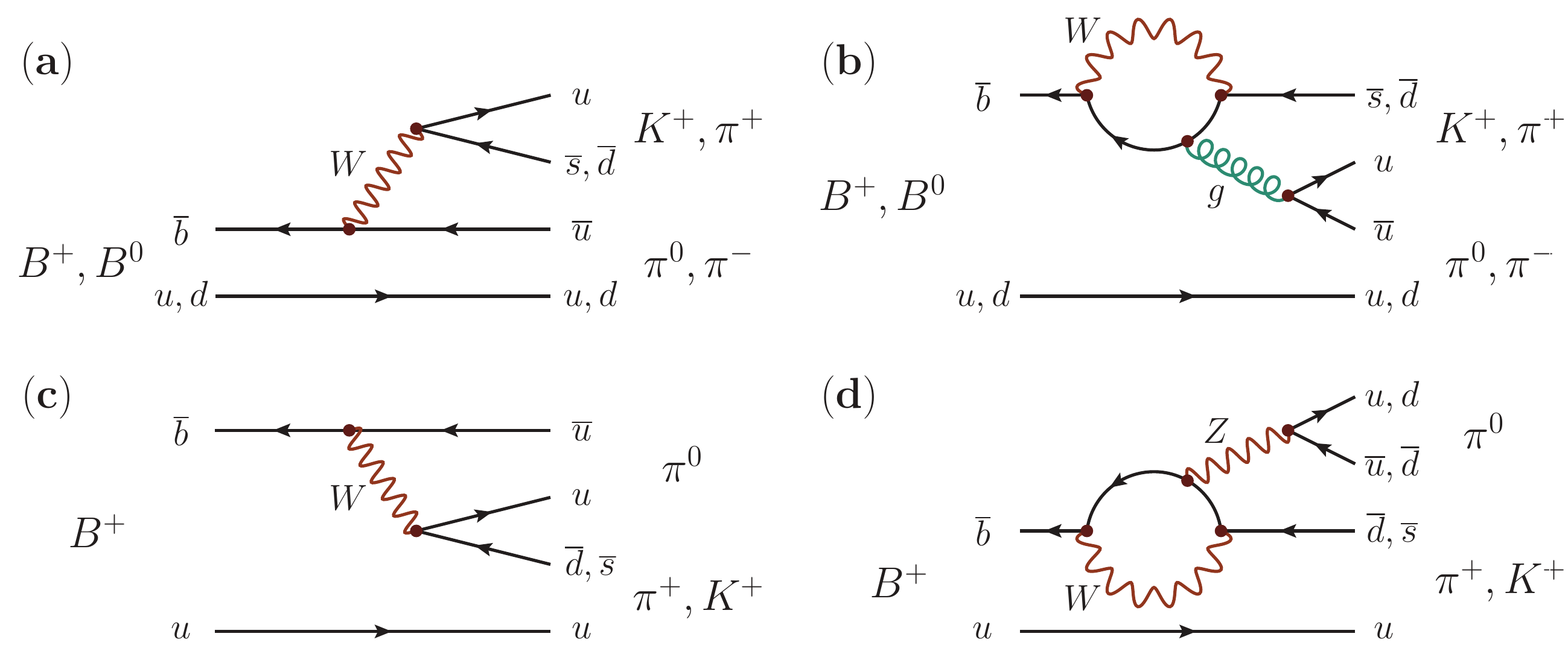}
\end{minipage}
\vskip-0.1cm
\begin{minipage}[t]{16.5 cm}
\caption{
Feynman diagrams of (a) tree, (b) strong penguin, (c) color-suppressed tree and (d) electroweak penguin
processes for $B^+ \to K^+\pi^0$ and $B^0 \to K^+\pi^-$ decays.
 \label{fig:BtoKpi}
}
\end{minipage}
\end{center}
\end{figure}

The well measured direct CPV (DCPV) in $B^0 \to K^+\pi^-$ mode,
\be
A_{K^+\pi^-} = -0.082 \pm 0.006, \quad {\rm (PDG)}
 \label{eq:AKpi}
\ee
arises from the interference between the subdominant tree amplitude $T$
with the dominant (strong) ``penguin'' amplitude $P$,
with $T$ carrying the CPV phase in $V_{ub}$
(see Fig.~\ref{fig:BtoKpi}).
The strength of $A_{K^+\pi^-}$ would depend on the relative
strong phase between $P$ and $T$, which may not be easy to predict.
But naively one would expect $A_{K^+\pi^0}$ to be similar in strength to $A_{K^+\pi^-}$,
since the two additional amplitudes,
electroweak (or $Z$) penguin $P_{\rm EW}$ and color-suppressed (tree) $C$,
are either suppressed by an extra power of $G_F$ or by color mismatch ($1/N_C$),
and are expected to be small.
The $P_{\rm EW}$ amplitude, however, is enhanced by nondecoupling of top~\cite{Hou:1986ug},
and could pick up NP phases by analogous enhancements.
But, as will be discussed in Sec.~3, we see no NP effect
in the CPV phase measurement of the companion $B_s^0$--$\bar B_s^0$ mixing amplitude,
$\phi_s$, hence NP effect through $b\to s$ $P_{\rm EW}$ amplitude seems unlikely.
The large shift in $A_{K^+\pi^0}$ value, even changing sign from $A_{K^+\pi^-}$,
would then have to arise from an unsuppressed $C$ amplitude that carries
a large strong phase difference with respect to (w.r.t.) $T$.
Such strong rescattering effects~\cite{Hou:1999st, Chua:2002wk} were originally
 (i.e. before the start of B factories)
thought to be perturbatively suppressed~\cite{Beneke:1999br, Keum:2000wi}
by the high $m_B$ scale, but they now turn around to haunt us.
The moral,\footnote{
 A second example is $\Delta A_{\rm CP} \equiv A_{KK} - A_{\pi\pi}$
 in $D^0 \to K^+K^-$ vs $\pi^+\pi^-$ decays,
 which arose~\cite{Aaij:2011in} then disappeared~\cite{Aaij:2014gsa} again
 with Run 1 data of the LHCb experiment.
 Had it stayed, it would still be even more likely than $\Delta A_{K\pi}$
 to be due to hadronic effects.
}
then, is one should avoid hadronic uncertainties if the interest is in NP.

In this review, we deemphasize processes that are susceptible to large hadronic corrections.
This includes lattice calculations of well-known hadronic parameters,
such as form factors or decay constants, even though these are promising directions in the long run.
Also, lattice calculations have not reached the stage of handling multiple dynamical scales,
which is often the case in rare B decays.
Another very important aspect of flavor physics we deemphasize is spectroscopy.
In fact, the discovery of new, unexpected hadrons such as~\cite{PDG}
$X(3872)$, $Y(4260)$ and $D_{sJ}(2320)$ are listed among the best cited
B factory papers, with a rather large theoretical and experimental following
purporting to four-quark or molecular states.
The LHCb experiment also claims to have discovered pentaquarks.
But while interesting, these have little to do with our quest for NP beyond SM,
and we refer to the companion article by Stone~\cite{Stone}.

%
%\hskip0.65cm - deemphasize hadronic (corr./latt./spec.):
%                 \ morals from $\Delta A_{K\pi}$ \& $\Delta A_{\rm CP}$

%
Instead, let us list the leading and recent flavor highlights (some of them ``anomalies''),
or otherwise probes of NP, at our current frontier:
\begin{itemize}
\item $\phi_s$ from $B_s \to J/\psi\,\phi$, $J/\psi\,f_0$;
\item $B_{s,\, d}^0 \to \mu^+\mu^-$;
\item $P_5'$ and $R_{K^{(*)}}$ anomalies in $B \to K^{(*)} \ell^+\ell^-$;
\item $B \to D^{(*)}\tau\nu$ (BaBar anomaly);
\item $B^+ \to \tau^+\nu$ and $b\to s\gamma$ ($H^+$ probes);
\item $K^+ \to \pi^+\nu\bar\nu$ and $K_L \to \pi^0\nu\bar\nu$;
\item muon $g-2$ anomaly and $\mu\to e$ processes;
\item $\tau\to\mu\gamma$ etc.;
\item $h(125) \to\mu\tau$ and $t\to ch$.
\end{itemize}

As a rule, however, ``anomalies'' come and go.
In fact, they mostly just go (away), as B factory workers could ruefully attest to.
Thus, which of the above anomalies are ``real''?
We do not know.
We have kept $g - 2$ only because it has been so long-standing,
and because of continued experimental (and lattice) interest.
The $P_5'$ ``anomaly'' is less long-standing while also
susceptible to hadronic effects.
It is kept because of its volatile experimental situation,
and because there might be a common trend with other
measurements, such as the $R_{K^{(*)}}$ anomaly.
Other hints, such as some discrepancy of direct measurements of
$\sin2\phi_1/\beta$ vs CKM fit result, or the long standing ``discrepancies''
of inclusive vs exclusive measurements of $V_{ub}$ and $V_{cb}$,
would be covered as traditional elements of CKM measurements in the following section.
Recent progress in lattice studies have reopened $\varepsilon'/\varepsilon$
as potentially harboring NP, which we briefly discuss with rare kaon decays,
while we do not really discuss $\varepsilon_K$
(which plays a supporting role in Fig.~\ref{fig:CKM2016}) as it has
well documented dependence on decay constant and ``bag'' parameter
which are being steadily improved by lattice.
Note that we did not list any charm process above.
Be it rare $D$ decays or $D^0$--$\bar D^0$ mixing, we have
no indications for NP yet in charm physics.
The reason is opposite to why B physics is prominent:
charm hadron lifetimes are unsuppressed by $V_{cs} \simeq 1$,
while loop-induced decays are suppressed by the smallness of
down type quark masses on the weak scale, hence suffer strong GIM cancellation.
New Physics in charm is hard to demonstrate, and we discuss it in a regular subsection.
In a similar vein, rare top decay is also a ``harder'' subject,
because the top quark decays faster than even the strong interaction time scale.
But its association with the Higgs particle, $h(125)$,
throws in some new light.

%     - summary of leading flavor anomalies

The Higgs boson $h$ was discovered at LHC Run 1,
but no New Physics appeared at 7 and 8 TeV.
For LHC Run 2 at 13 TeV, there is unfortunately a repeat:
again no New Physics has appeared so far,
with the 750 GeV ``diphoton'' disappearing with more data.
Although discovery is still possible with new particles
or effects with weaker coupling, there is no new high energy facility that is firmly in sight
and that would run after completion of the high luminosity LHC (HL-LHC) running around 2035.
This only enhances the importance of FPCP studies.
Although in general it probes NP only indirectly, it provides us
with hope that some indication or evidence for NP might
appear in the not so distant future.
One of course has New Physics (beyond SM) in the
neutrino sector, the fact that neutrinos mix hence have mass,
but that is covered by another review~\cite{SFKing}.

Bottom, charm, $\tau$ and top are traditionally viewed as heavy flavors,
but kaon and $\mu$ physics may still offer surprises.
In the following sections, we first discuss CKM measurements in Sec.~2,
then the highlight B anomalies (first 4 items above) in Sec.~3.
Some further discussion of $B$ decays are given in Sec.~4, including connection to Dark sector.
Rare K decays, mainly $K\to \pi\nu\nu$ (plus Dark sector connections),
are discussed in Sec.~5, with brief comments on $\varepsilon'/\varepsilon$.
This is followed by rare muon processes and $g-2$,
with a brief touch upon electric dipole moments.
Tau and charm physics are covered only briefly in Sec.~6,
but we introduce the new subject of flavor changing top and Higgs couplings in Sec.~7,
followed by our conclusion.
Our emphasis is on the physics and the experiment--theory interplay,
rather than on technical details,
with a view towards improvements and updates in the LHC Run 2 period.

\section{CKM Paradigm} \label{sec:CKM}

The CKM paradigm stands as the pillar of flavor physics and CP violation
of SM, but the CKM matrix elements are free parameters that have to be measured experimentally.
Our mission is to verify if the matrix is unitary, and whether
the matrix elements measured with various methods are consistent with each other.
If one can demonstrate violation of unitarity, or if convincing discrepancies
emerge somewhere, it may indicate New Physics.

The nine complex CKM matrix elements of $V$ can be expressed in
just four independent real parameters if $V$ is unitary, as given
in Eq.~(\ref{eq:VCKM}) in a convenient parametrization.
In this section, we first describe the methods to measure each matrix element,
then report the current experimental results.
%\emph
{The most recent fit results to extract the Wolfenstein parameters from a
global fit to available measurements will be discussed.}
Possible discrepancies will be highlighted together with the future prospect.

\subsection{Magnitudes %of Matrix Elements
} \label{sec:CKM-mag}

The magnitudes of CKM matrix elements are often extracted from
measurements of transition or decay rates and comparing with
theoretical calculations, where the magnitudes of the elements are the unknowns.
Since theoretical calculations typically involve nonperturbative parameters
such as decay constants and form factors, %which may not be precisely computed,
the extracted magnitudes have both experimental and theoretical uncertainties.

\subsubsection{%\boldmath
 The $2\times 2$ Submatrix} \label{sec:2x2}

The four elements of the $2\times 2$ submatrix in the upper left corner of
the CKM matrix are related to the Wolfenstein parameter $\lambda$ only,
with which the suggestion of unitarity first emerged~\cite{Cabibbo:1963yz, Glashow:1970gm}.

\

\noindent\underline{\it \bf\boldmath
$|V_{ud}|$ and $|V_{us}|$}\vskip0.1cm

The element $V_{ud}$ is the first to be determined, based on
$u\leftrightarrow d$ ($p \leftrightarrow n$) transitions, and is the most precisely known.
The magnitude of $V_{ud}$ is determined from the study of
$0^+\to 0^+$ nuclear beta decays, which is governed purely by the weak current.
The measured  half lifetimes, $t$, and Q values, which give the decay rate
factor ${\mathcal F}$, are used to determine $|V_{ud}|$ using the formula~\cite{vudfor},
\be
|V_{ud}|^2 = \frac{2984.45 \;\rm{sec}}{{\mathcal F} t (1+\Delta)}\,,
\ee
where $\Delta$ denotes the combined effect of electroweak radiative corrections,
nuclear structure and isospin violating nuclear effects.
The average over 14 precise results~\cite{vud} gives,
\be
   |V_{ud}| = 0.97417 \pm 0.00021,
 \label{eq:vud}
\ee
where the error is dominated by theoretical uncertainty.

Extractions of $|V_{ud}|$ are also performed with measurements of
neutron lifetime and the $\pi^+\to \pi^0 e^+\nu_e$ decay rate.
Both give results consistent with Eq.~(\ref{eq:vud}),
but with uncertainties larger by more than an order of magnitude.
Determination from neutron lifetime is limited by knowledge of
the ratio of axial-vector to vector couplings,
while semileptonic $\pi^+$ decay suffers from statistics.
The details can be found in the review by Blucher and Marciano in
PDG~\cite{PDG}.
%, ``$V_{ud}, V_{us},$ the Cabibbo Angle and CKM unitarity".

The magnitude of $V_{us}$ may be extracted from the decays of kaons, hyperons
and $\tau$ leptons. Two different approaches are employed:
extracting $|V_{us}|$ directly, and measuring $|V_{us}/V_{ud}|$.
The aforementioned review in PDG documents the efforts.

In direct extraction, $|V_{us}|$ is often determined from semileptonic
$K\to \pi \ell \nu$, or $K\ell 3$, decays.
The partial decay width is related to the product of $|V_{us}|$ and the form factor
at zero momentum transfer, $f_+(0)$, of the lepton--neutrino system.
Experimental inputs are decay branching fractions and kaon lifetimes,
and form factor measurements that enable theoretical computation of phase space integrals.
Theory is needed to compute short and long distance radiative corrections,
isospin breaking strength, and $f_+(0)$.
%Traditionally, only $K^0_L \to \pi e\nu$ decay is used to avoid
%possible isospin breaking corrections for $K^\pm$.
%Moreover, the decay to muon mode is also not used,
%due to the large uncertainty of phase space integral,
%which needs the measured form factors.
%
Recent high statistics and high quality measurements have facilitated
good constraints to justify the comparison between different decay modes.
PDG therefore follows the prescription of Ref.~\cite{flavia} to average results
from $K_L^0\to \pi e \nu$, $\pi \mu\nu$, $K^\pm \to \pi^0 e^\pm \nu$,
$\pi^0 \mu^\pm \nu$, and $K^0_S\to \pi e \nu$,
giving the product $|V_{us}|f_+(0) = 0.2165\pm 0.0004$.
The form factor average\footnote{
See the review
%``Leptonic Decays of Charged Pseudoscalar"
by Rosner, Stone, and Van de Water in PDG~\cite{PDG}.
}
$f_+(0) = 0.9677\pm 0.0037$ from the three-flavor lattice QCD calculations~\cite{three-flavor-lattice}
gives $|V_{us}| = 0.2237 \pm 0.0009$.

The second method aims at determining the ratio $|V_{us}/V_{ud}|$, and
$|V_{us}|$ can be obtained by use of the well determined $|V_{ud}|$.
This ratio can be extracted from the ratio of decay rates of
$K \to\mu\nu(\gamma)$ and $\pi\to\mu\nu(\gamma)$,
along with the ratio of decay constants from lattice QCD calculations.
The KLOE measurement of $K\to \mu \nu(\gamma)$~\cite{kloe} leads to
$|V_{us}| = 0.2254\pm 0.0008$, where the uncertainty is dominated by the ratio of decay constants.
Combining the two approaches to $|V_{us}|$ measurements, one gets
\be
   |V_{us}| = 0.2248\pm 0.0006.
 \label{eq:vus}
\ee

Other determinations of $|V_{us}|$ include extraction from
hyperon decays~\cite{vushyp}, and from semihadronic $\tau$ decays.
%Avoiding first order SU(3) breaking effects, the former finds $|V_{us}| = 0.2250\pm 0.0027$.
The former approach gives results consistent with Eq.~(\ref{eq:vus}) but with larger errors.
%A representative effort~\cite{mateu} considers the SU(3) breaking
%and gives $|V_{us}| = 0.226\pm 0.005$.
As for semihadronic $\tau$ decays, $\tau\to K\nu (+X)$, $\pi\nu (+X)$,
the average of inclusive and exclusive rates yields\footnote{
We skip detail discussions here, and refer to Ref.~\cite{hfag},
which is the HFAG update for the PDG 2015 update. Thus, PDG 2016 often uses this reference,
while in Sec.~3 we would more often quote the latest from HFAG.
}
$|V_{us}| = 0.2204\pm 0.0014$~\cite{hfag}.
The average $|V_{us}|$ value determined from $\tau$ decays
deviates from the kaon average by almost $3\sigma$,
and the discrepancy mainly comes from inclusive $\tau$ measurements,
which gives $0.2186\pm 0.0021$. Further investigation is needed.

We can check CKM unitarity using $|V_{ud}|, |V_{us}|$ and $|V_{ub}|$, where
%one can simply neglect the latter as it is known to be too small for any impact.
the latter is known to be too small for any impact.
Taking $|V_{ud}|$ and $|V_{us}|$ from Eqs.~(\ref{eq:vud}) and (\ref{eq:vus}),
respectively, the squared sum gives,
\be
   |V_{ud}|^2 + |V_{us}|^2\, (+\, |V_{ub}|^2) = 0.9995 \pm 0.0005.
 \label{eq:row1}
\ee
The good agreement with unitarity already confirms the radiative corrections
from Standard Model at better than 50$\sigma$ level~\cite{sirlin}.
Moreover, it also sets good constraints on New Physics effect~\cite{Cirigliano:2009wk}
that may contribute in nuclear beta decay, kaon decay and muon decay.

\
%\subsubsection

\noindent\underline{\it \bf\boldmath
$|V_{cd}|$ and $|V_{cs}|$} \vskip0.1cm

Early determinations of $|V_{cd}|$ were performed using neutrino scattering data.
The strategy is to compare the production rates of opposite sign dimuons
with single muons, from both neutrino and anti-neutrino beams on nucleus.
The nuclear scattering can be expressed as $\nu_\mu + N \to  \mu^- + c + X$,
where the $c$ quark hadronizes into a charm hadron and subsequently decays
semileptonically into a muon plus other particles.
The underlying process is a neutrino interacting with a $d$ or $s$ quark in a nucleus,
and similarly $\overline d$ or $\overline s$ quark for anti-neutrino beam.
The difference of the double-muon to single-muon production ratio
by neutrino and antineutrino beams is proportional to the charm cross section
of valence $d$ quarks, and therefore to $|V_{cd}|^2$ times
the average semimuonic branching fraction of charm hadrons, ${\cal B}_\mu$.
%\emph
{PDG 2016 uses the average over CDHS, CCFR and CHARM,}
${\cal B}_\mu |V_{cd}|^2 = 0.463\pm 0.034$, %~\cite{bvcd1},
and the average ${\cal B}_\mu =0.087\pm 0.005$ of two estimates~\cite{bvcd2,bvcd3},
to obtain $|V_{cd}| = 0.230\pm 0.011$.

Similar to the $|V_{us}|$ case, the magnitude of $V_{cd}$ can be extracted from
semileptonic and leptonic $D$ decays, where the form factor $f_+^{D\pi}(0)$ is
needed for the former, and the decay constant $f_D$ for the latter.
Using the form factor $f_+^{D\pi}(0) = 0.666\pm 0.029$~\cite{three-flavor-lattice}
from three-flavor lattice QCD calculations, and the $D\to \pi \ell \nu$
branching fractions measured from BaBar, Belle, BESIII and CLEO-c,
PDG obtains $|V_{cd}| = 0.214\pm 0.003\pm 0.009$,
where the first uncertainty is experimental
and the second is theoretical (form factor).
Averaging the $D^+\to \mu^+\nu$ branching fraction measurements
from BESIII and CLEO-c and using $f_D = 209.2 \pm 3.3$ MeV~\cite{three-flavor-lattice},
PDG gives  $|V_{cd}| = 0.219 \pm 0.005 \pm  0.003$.
The three different approaches give consistent values,
and the average by PDG yields
\be
   |V_{cd}| = 0.220\pm 0.005.
 \label{eq:vcd}
\ee

Analogous to $|V_{us}|$ and $|V_{cd}|$ extraction, the magnitude of $V_{cs}$
can be extracted from semileptonic $D$ decays and leptonic $D_s$ decays,
using the form factor and $D_s$ decay constant calculated from lattice QCD, respectively.
Using the form factor $f^{DK}_+(0) = 0.749\pm 0.019$~\cite{three-flavor-lattice}
and the average branching fraction of $D\to K\ell \nu$ decays,
one obtains $V_{cs} = 0.975\pm 0.007\pm 0.025$, where
the first uncertainty is experimental and the second
is mainly due to the form factor.
For leptonic decays, the average branching fractions $(5.56\pm 0.24)\times 10^{-3}$
and $(5.56\pm 0.22)\times 10^{-2}$ for $D^+_s\to \mu^+\nu$ and $\tau^+\nu$,
respectively, are documented in the same review in footnote 5.
 %``Leptonic Decays of Charged Pseudoscalar Mesons''.
The magnitude of $V_{cs}$ can be determined for each decay branching fraction
using the decay constant $f_{D_s} = (248.6\pm 2.7)$ MeV~\cite{three-flavor-lattice},
$D^+_s$ lifetime and mass, and the corresponding lepton masses.
The average of the two gives $|V_{cs}| = 1.008 \pm 0.021$,
where the uncertainty is dominated by the decay constant.
Combining with $D\to K\ell \nu$ decays, the average gives
\be
   |V_{cs}| = 0.995 \pm 0.016.
 \label{eq:vcs}
\ee

Another information on $|V_{cs}|$ can come from $W$ decay. The $W$
decay branching fraction to each lepton is inversely proportional to
$3[1+ \Sigma_{u,\, c,\, d,\, s,\, b}|V_{ij}|^2 (1+\alpha_s(m_W)/\pi) + ...]$.
Assuming lepton universality and using the branching fraction of
$W$ leptonic decay measured by LEP-2, one obtains
$\Sigma_{u,\, c,\, d,\, s,\, b}|V_{ij}|^2 = 2.002 \pm 0.027$~\cite{lep2w},
which is a good test of CKM unitarity.
The magnitude of $|V_{cs}|$ can also be determined from flavor tagged $W^+\to c\bar{s}$ decays.
However, The precision suffers from the low statistics and purity of flavor tagging.

\

\noindent\underline{\bf
Comments} \vskip0.1cm

As pointed out in the first section, the four magnitudes discussed here are
much larger than the others except $|V_{tb}|$. Numerous experimental
measurements together with progressively improved theoretical calculations
of form factors and decay constants yield  precise determination
of the four magnitudes. The obtained results show that
$|V_{ud}|$ is indeed close to $|V_{cs}|$, and $|V_{us}|$ close to $|V_{cd}|$.
The quadratic sum of $|V_{ud}|$ and $|V_{cd}|$ is very close to unity.  It is
quite impressive that various determinations using different decays or different
methods for the same amplitudes are  consistent with each other,
demonstrating the accuracy of the Standard Model.  The uncertainties of the
magnitudes except for $|V_{cd}|$ are dominated by theory uncertainties,
while experimental and theory uncertainties are comparable for $|V_{cd}|$.

So far the only notable discrepancy at $\sim 3\sigma$ level comes from
$|V_{us}|$ determination using exclusive and inclusive $\tau$ decays.
Owing to large data samples collected by a multitude of experiments
that enable the extraction in a systematic and coherent way,
PDG adopts the $|V_{us}|$ value determined from kaon decays as default.
The $|V_{us}|$ extraction using $\tau$ decays is facilitated by measuring the
decay rate with kaon (strange quark) in the final state.
The deviation in $|V_{us}|$ value between the kaon and $\tau$ methods
mainly arises from the inclusive method for $\tau$,
which is more challenging experimentally.
The result using the exclusive method is closer to the default $|V_{us}|$.
Therefore, a $|V_{us}|$ anomaly seems unlikely, not to mention New Physics effects.
It is the responsibility of the experiments to understand the deviation in
the inclusive mode and provide more precise measurements.

\subsubsection{%\boldmath
 $|V_{ib}|$ and $|V_{tj}|$}

%\

\noindent\underline{\it \bf\boldmath
$|V_{cb}|$ and $|V_{ub}|$} \vskip0.1cm

Semileptonic decays of $B^+$ and $B^0$ mesons proceed via leading order weak interactions.
Modes involving light leptons ($\ell = e$ or $\mu$)
are expected to be free from non-SM contributions.
Therefore, these decays serve the crucial role for determining $|V_{cb}|$ and $|V_{ub}|$,
which are extracted from measurements of exclusive and inclusive branching fractions,
and theoretical formulae based on the operator product expansion (OPE).
A dedicated PDG review documents the details of the extraction method and current results.
%``Semileptonic Bottom Hadron Decays and Determination of $V_{cb}$ and $V_{ub}$".

The differential branching fractions for exclusive $B\to D^{(*)} \ell \nu$ can be
expressed as the product of $|V_{cb}|^2$, form factor squared ${\cal F}^2(\omega)$,
where $\omega$ is the inner product of $B$ and $D^{(*)}$ four velocities,
and other factors that need to be computed.
For example, the differential decay rate of $B\to D^* \ell \nu$ is given by
\be
\frac{d\Gamma}{d\omega}(B\to D^* \ell\nu) = \frac{G^2_F m^5_B}{48\pi^3} |V_{cb}|^2 (\omega^2 -1)^{1/2} P(\omega)(\eta_{\rm ew}{\cal F(\omega )})^2, \label{eq:drate}
\ee
where $P(\omega)$ is a phase space factor,
$\eta_{\rm ew}$ accounts for leading electroweak corrections that include
an estimate of uncertainty from (still) missing long-distance QED radiative corrections.
Two unknowns in the equation are $|V_{cb}|$ and ${\cal F}(\omega)$.
The idea is to obtain $|V_{cb}|{\cal F}(\omega)$ from the measured differential rates and
extrapolate it to the end point of $\omega = 1$, which corresponds to
maximum momentum transfer for the leptons.
In the infinite mass limit, the form factor at the end point is unity.
For finite quark masses, ${\cal F}(1)$ needs to be computed
by lattice QCD~\cite{Bailey:2014tva, Lattice:2015rga, Na:2015kha}.
The same strategy can be applied to the $B\to D\ell \nu$ case.
Using all available data from LEP, CLEO, Belle and
BaBar, PDG 2016 gives
\be
|V_{cb}| = (39.2 \pm 0.7)\times 10^{-3}, \quad {\rm (exclusive)}
 \label{eq:vcb_excl}
\ee
where the dominant theoretical uncertainty comes from the form factor,
and the experimental uncertainty is due to the decay rate near $\omega = 1$.
Owing to possible New Physics in the data, the branching fraction measurements of
$B\to D^{(*)} \tau \nu$ were not included in the extraction of $|V_{cb}|$.
The discussion of $B \to D^{(*)}\tau\nu$ can be found in
the $B$ highlights given in Section 3.

Extracting $|V_{cb}|$ from inclusive decays requires the measurement of
total semileptonic decay branching fractions, moments of lepton energy,
and hadronic invariant mass spectra in $b\to c$ transitions.
Theoretical calculations for the decay rate can be performed reliably
in terms of nonperturbative parameters based on
$1/m_b$ expansion, where there are subtleties at $O(1/m_b^2)$~\cite{ope1,ope2}
 %in $\alpha_s$ and $b$ quark mass.
The nonperturbative parameters can be extracted from information of moments from experiments.
Measurements of inclusive processes have been performed using $B$ mesons
from $Z^0$ decays at LEP and $\Upsilon(4S)$ decays at $B$ factories.
Experimental information on the leptonic and hadronic invariant masses
in $B\to X_c \ell \nu$, and the photon moments from $B\to X_s \gamma$
all help in determining the nonperturbative parameters.
An average of experimental measurements leads to~\cite{PDG}
\be
|V_{cb}| = (42.2\pm 0.8)\times 10^{-3}. \quad {\rm (inclusive)}
 \label{eq:vcb_incl}
\ee

There is a slight tension ($2.9\sigma$) between the two extractions of $|V_{cb}|$.
Belle II will have two orders of magnitude more data in the near future,
which facilitates more precise measurements of the semileptonic decay rates.
However, one needs to reduce theoretical uncertainties to verify the
discrepancy between the two $|V_{cb}|$ extractions.
The PDG combination of the two, after scaling the error by $\sqrt{\chi^2}=2.9$, is
\be
|V_{cb}| =(40.5 \pm 1.5)\times 10^{-3}.
 \label{eq:vcb}
\ee

Similar to the inclusive $|V_{cb}|$ extraction,
the extraction of $|V_{ub}|$ from inclusive $B\to X_u \ell \nu$ decays
can also be pursued using the same effective theory.
However, the large inclusive $B\to X_c \ell \nu$ background poses
a challenge for measuring the branching fractions.
To calculate partial decay rates in regions of phase space
where the $B\to X_c \ell \nu$ decays are suppressed
requires a nonperturbative distribution function
called the shape function~\cite{shape1,shape2}.
Note that the nonperturbative physics for the $|V_{ub}|$ case is modelled by a few parameters.
At leading order, the shape function is universal, therefore
it can be extracted from the photon energy spectrum of the $B\to X_s \gamma$ decays.
An alternative approach is to measure the semileptonic decay rate more inclusively,
covering some parts of the $B\to X_c \ell \nu$ region.
Large data samples collected at the $B$ factories enable
the tagging of accompanying $B$ meson as decaying hadronically or semileptonically,
 %to purify the signal $B$ candidates.
hence a wider kinematic region for the signal can be allowed.
Averaging over the two approaches, PDG gives the inclusive average,
\be
|V_{ub}| = (4.49 \pm 0.16^{+0.16}_{-0.18})\times 10^{-3}, \quad {\rm (inclusive)}
 \label{eq:vub_incl}
\ee
where the first uncertainty is experimental and the second theoretical.
 %The detail

Exclusive charmless semileptonic $B$ decays provide another way to determine $|V_{ub}|$.
Although exclusive decays result in better signal-to-background ratios experimentally,
their branching fractions are rather small, leading to large statistical uncertainties.
Among all the $b\to u$ semileptonic decay modes,
the decay $B\to \pi\ell \nu$ is the most promising.
To extract $|V_{ub}|$, the form factor from theoretical calculations,
lattice QCD or light-cone sum rule, is needed.
With the available $B\to \pi \ell \nu$ form factor for
high leptonic momentum squared ($q^2$) region calculated by lattice QCD~\cite{lqcd1,lqcd2},
a fit to the experimental partial rates and lattice results
versus $q^2$ gives $|V_{ub}| = (3.72 \pm 0.16)\times 10^{-3}$~\cite{lqcd2}.
PDG rescales the uncertainty by $\sqrt{\chi^2/\rm{d.o.f.}} = 1.2$ and gives
\be
|V_{ub}| = (3.72\pm 0.19)\times 10^{-3}. \quad {\rm (exclusive)}
 \label{eq:vub_excl}
\ee

The inclusive and exclusive $|V_{ub}|$ results again show a $2.6\sigma$ tension.
Belle II will be able to reduce the experimental uncertainties in the near future,
but we need to pin down the theoretical uncertainties to verify the discrepancy.
The average over inclusive and exclusive results after scaling the uncertainty by 2.6 is
\be
|V_{ub}| = (4.09\pm 0.39)\times 10^{-3}.
 \label{eq:vub}
\ee

Information from other $B$ decays can be used to probe  $|V_{ub}|$.
For instance, the decay rate of $B^+\to \tau^+ \nu$ is directly proportional to $|V_{ub}|^2$.
Using the branching fraction of $B^+\to \tau^+\nu$,
the decay constant $f_B$ from lattice QCD calculations and $B^+$ lifetime,
PDG obtains $|V_{ub}| = (4.04\pm 0.38)\times 10^{-3}$, consistent with Eq.~(\ref{eq:vub}).
Since the decay $B^+\to\tau^+\nu$ is sensitive to the charged Higgs contribution at tree level,
this mode is often used to examine possible New Physics (see Sec. 3).
A large $\Lambda_b$ sample collected by LHCb facilitates the measurement of $|V_{ub}/V_{cb}|$.
Based on the lattice calculation of Ref.~\cite{Detmold:2015aaa},
LHCb reported $|V_{ub}/V_{cb}| = 0.083\pm 0.006$~\cite{ubcb_lhcb}
by measuring the ratio of branching fractions of
$\Lambda^0_b \to p^+ \mu^- \nu$ and $\Lambda^0_b\to \Lambda_c^+ \mu^-\nu$
with $q^2>15$ GeV$^2/c^2$ and $q^2 >7$ GeV$^2/c^2$, respectively.
The ratio of the two magnitudes can also be included
along with other determination methods, and therefore test the CKM paradigm.

%\subsubsection{

\

\noindent\underline{\it \bf\boldmath
$|V_{td}|$ and $|V_{ts}|$} \vskip0.1cm

Due to the difficulty of tagging $d$ or $s$ jets, it is unlikely
that $|V_{td}|$ and $|V_{ts}|$ can be determined from top quark decays.
Instead, these two CKM elements can be extracted from $B$--$\overline{B}$ oscillations
mediated by box diagrams, or from rare $K$ and $B$ decays via penguin loop diagrams.
In general, the ratio $|V_{td}/V_{ts}|$ is determined with better precision,
since the hadronic uncertainties in $B^0_d-\overline{B}^{0}_d$ and $B_s^0-\overline{B}_s^{0}$
mixings are equal in the flavor SU(3) limit, and therefore cancel in the ratio.

Experimental measurements of the mass difference for both $B^0_d$ and $B^0_s$
meson systems have reached rather good accuracy:
$\Delta m_d =(0.5064\pm 0.0019)\;\rm{ps}^{-1}$ and
$\Delta m_s =(17.757\pm 0.021)\;\rm{ps}^{-1}$~\cite{PDG}.
%(see the PDG review on ``$B^0-{\overline B}^0$ mixing").
Using the lattice QCD results
$f_{B_d} \sqrt{\hat{B}_{B_d}} =(216\pm 15)$ MeV and
$f_{B_s} \sqrt{\hat{B}_{B_s}} =(266\pm 18)$ MeV~\cite{three-flavor-lattice},
where $\hat{B}_{B_d}$ is the bag parameter, PDG obtains
\be
|V_{td}| = (8.2\pm 0.6)\times 10^{-3},\ \quad |V_{ts}| = (40.0\pm 2.7)\times 10^{-3},
 \label{eq:tdts}
\ee
where the uncertainties are dominated by lattice QCD calculations. As stated,
some of the uncertainties for the decay constants and bag parameters can be reduced by taking ratio.
Using $f_{B_d}\sqrt{\hat{B}_{B_d}}/ f_{B_s} \sqrt{\hat{B}_{B_s}}
 = 1.268\pm 0.064$~\cite{three-flavor-lattice}, the ratio of the two magnitudes is
extracted to be
\be
 |V_{td}/V_{ts}| = 0.215\pm 0.001\pm 0.011.
  \label{eq:rtdts}
\ee
The second uncertainty is from lattice QCD calculation, which is substantially reduced.

Other information related to $|V_{td}|$ or $|V_{ts}|$ come from rare $B$ decays. However, the
statistics of rare decays is typically limited, hence the extracted numbers may not be precise.
For instance, $|V_{td}/V_{ts}|$ can also be determined from the ratio of
$B \to \rho\gamma$ and $B\to K^*\gamma$ decay branching fractions.
To increase statistics, PDG combines both charged and neutral $B$ meson decay rates.
%\emph
{With the assumption of isospin symmetry
(e.g. $\Gamma(B^0\to \omega^0 \gamma) = \Gamma(B^0\to \rho^0 \gamma)$)
and taking the heavy quark limit, one can write}
$|V_{td}/V_{ts}|^2/\xi_\gamma^2 =
  [\Gamma(B^+\to \rho^+\gamma)+ 2\Gamma(B^0\to \rho^0 \gamma)]
 /[\Gamma(B^+\to K^{*+}\gamma)+\Gamma(B^0\to K^{*0}\gamma)]
 = (3.19\pm 0.46)\%$~\cite{hfag}.
The factor $\xi_\gamma$ contains information of hadronic physics.
Taking $\xi_\gamma$ from various theoretical computations
gives $|V_{td}/V_{ts}| = 0.214\pm 0.016\pm 0.036$, consistent with the value
from the mixing analysis but with larger uncertainties.

The decay rates of two very rare decays, $B_s\to \mu^+\mu^-$ and $K^+\to \pi^+ \nu \overline{\nu}$,
are proportional to $|V_{ts}V_{tb}|^2$ and $|V_{td}V_{ts}|^2$, respectively.
The current measured $B_s\to \mu^+\mu^-$ branching fraction
is consistent with the SM prediction,
while a few $K^+\to \pi^+ \nu \overline{\nu}$ events are observed,
suggesting a branching fraction that is slightly higher than SM expectations at $10^{-10}$ level.
Since possible New Physics may appear in these rare decays,
it is better to determine the corresponding CKM matrix elements precisely from other channels,
such that the SM branching fractions can be predicted with good accuracy.
More discussions for these two very rare decays are given in Sections 3 and 5, respectively.

%\subsubsection{

\

\noindent\underline{\it \bf\boldmath
$|V_{tb}|$} \vskip0.1cm

$|V_{tb}|$ can be extracted directly by measurement of
the single top production cross section at hadron colliders.
The combined CDF and D\O\ measurements give $|V_{tb}| = 1.02^{+0.06}_{-0.05}$~\cite{tev-vtb}.
ATLAS and CMS have measured single top production cross sections for $t$-channel,
$Wt$ and $s$-channel processes at 7, 8 and 13 TeV~\cite{lhctop}.
The average of the LHC measurements gives $|V_{tb}| = 1.005 \pm 0.036$,
assuming that systematic and theoretical uncertainties are correlated.
The average of the Tevatron and LHC measurements from PDG is
\be
|V_{tb}| = 1.009 \pm 0.031,
 \label{eq:vtb}
\ee
where error is dominated by experimental systematic uncertainties.

The second method to determine $|V_{tb}|$ uses the ratio of branching fractions,
\be
R = {\cal B}(t\to W b)/\Sigma_q {\cal B}(t\to W q) = |V_{tb}|^2,
\ee
where $q = d$, $s$ or $b$. Assuming unitarity, the ratio $R$ is $|V_{tb}|^2$.
Experiments at Tevatron and LHC have followed this method and
performed the extractions by selecting $t\bar{t}$ events with various number of $b$-tagged jets.
Since the measurements are not precise, experiments reported either an allowed range~\cite{vtb-d0}
or lower limits~\cite{vtb-cdf, vtb-cms} at 95\% C.L.
All experimental results are consistent with each other, and with Eq.~(\ref{eq:vtb}).

\

\noindent\underline{\bf
Comments} \vskip0.1cm

The current PDG values of $|V_{ib}|$ and $|V_{tj}|$ show, as expected, that
$|V_{tb}|$ is close to unity and the rest of  $|V_{ib}|$ and  $|V_{tj}|$ are
small. Based on single top production at LHC, the precision of $|V_{tb}|$
can be improved with more data, while the theory uncertainties need to be
reduced to extract $|V_{td}|$ and $|V_{ts}|$ precisely from $B^0-\overline{B}^0$
and $B_s^0-\overline{B_s}^0$ mixing analyses, respectively.
As for $|V_{cb}|$ and $|V_{ub}|$, the uncertainties of both magnitudes
will be reduced by 10 to 50 times more data at Belle II.
However, the discrepancies between the exclusive and inclusive methods
need to be understood. See section 4.2 for more discussion.

\subsection{%\boldmath
 Phases %of CKM elements
}

We turn to discuss the determination of CKM phases.
In SM, $CP$ violation involves the phases of CKM elements.
Therefore, CPV measurements in a host of $B$ decays
provide good constraints on the angles of the unitarity triangle
shown in the $\bar{\rho}-\bar{\eta}$ plane of Fig.~\ref{fig:CKM2016}.

%\subsubsection{Direct $CP$ violation}

We have already given a brief account of the pursuit of direct CPV,
and lamented that the DCPV difference in $B \to K^+\pi$ decay, Eq.~(\ref{eq:DAKpi}),
seems to be due to an ``unsuppressed color-suppressed tree diagram $C$'',
that carries also a large strong phase difference w.r.t. the tree amplitude $T$,
that makes the extraction of short distance information,
such as CKM elements, impossible, even at the $B$ decay scale.
We give a little more discussion in Sec.~\ref{sec:charmless}.
The precursor of realization of the ``hadronic menace''
was in fact in the pursuit of DCPV in kaon decay.

The measurement of the direct $CP$ violation parameter, Re($\varepsilon^\prime/\varepsilon$),
was viewed as a way to test CKM mechanism and constrain many New Physics scenarios.
%T
Theoretically, Re$(\varepsilon^\prime/\varepsilon)$ is proportional to Im$(V_{td}V^*_{ts})$.
But since the electroweak penguin contribution tends to cancel
the gluonic penguin contributions for large $m_t$~\cite{Flynn:1989iu},
this greatly amplifies the hadronic menace.
Therefore, it is rather difficult to use Re($\varepsilon^\prime/\varepsilon$)
to extract CKM parameters. Unlike $\varepsilon_K$, indirect CPV in neutral kaon mixing,
it does not appear in Fig.~\ref{fig:CKM2016} as a constraint.
%Measured predicted values for the parameter agrees with the experimental measurements,
%which suggests that $\overline{\eta}$ is positive.
%
Inspection of Fig.~\ref{fig:CKM2016}, however, shows that $\varepsilon_K$
as a constraint provides consistency support for the determination of
the $\phi_1/\beta$, $\phi_2/\alpha$ and $\phi_3/\gamma$ program at the $B$ factories,
therefore we do not go into any details.
Suffice to say that $\varepsilon_K$, the original discovery of CPV~\cite{Christenson:1964fg},
depends on the decay constant $f_K$ and bag parameter $\hat{B}_{K}$,
and carries its own hadronic uncertainties.
We note, however, that recent lattice developments suggest a major deficit for SM to
account for the measured Re($\varepsilon^\prime/\varepsilon$) result, suggesting
it could possibly be a renewed probe of New Physics.
More discussion is given in Sec.~\ref{sec:epsprime}.

\subsubsection{%\boldmath
 $\phi_1/\beta$ (and $\phi_2/\alpha$)
\label{sec:phi1}}

\noindent\underline{\it \bf\boldmath
$\phi_1/\beta$} \vskip0.1cm

The triumph of the B factories has been the clean measurement of the CPV phase angle $\phi_1/\beta$.
Thanks to the method developed by Bigi and Sanda~\cite{Carter:1980tk, Bigi:1981qs},
unlike our discussion so far about measurement of magnitudes of the CKM matrix,
the measurement of $\phi_1/\beta$ is especially free of hadronic uncertainties.
As the subject is relatively well known, let us be brief.

Precise determination of $\phi_1$ is done experimentally by studying
neutral $B$ meson decays into a $CP$ eigenstate $f$ via $b\to c\bar{c}s$ transitions,
which has no CPV phase in the decay amplitude, but makes these excellent probes
of CPV phase of the $B_d$ mixing amplitude. The time-dependent CPV asymmetry can be written as
\begin{equation}
  {\cal A}_f(t) = \frac{\Gamma(\overline{B}^0\to f) - \Gamma(B^0\to f)}{
                    \Gamma(\overline{B}^0\to f) + \Gamma(B^0\to f)} = S_f\sin(
  \Delta m_d\; t)-C_f\cos(\Delta m_d\; t),
 \label{eq:tcp1}
\end{equation}
where
\be
 S_f = \frac{2\, {\rm Im} \lambda_f}{1+|\lambda_f|^2}, \;\; C_f =\frac{
1-|\lambda_f|^2}{1 + |\lambda_f|^2}, \;\; \lambda_f = \frac{q}{p}\frac{\overline A_f}{A_f}.
  \label{eq:tcp2}
\ee
The mixing parameter $q/p$ can be approximated as
$V^*_{tb}V_{td}/V_{tb}V^*_{td} = e^{-2i\phi_1 +  O(\lambda^4)}$,
where $\lambda$ is defined in Eq.~(\ref{eq:lambda}).
$A_f\ ({\overline A}_f)$ is the amplitude of the decay $B^0\to f\;
({\overline B}^0\to f)$. Minus $C_f$ is the parameter of direct $CP$ violation.
If the amplitude is dominated by a single diagram such that the strong phase factorize,
$|\lambda_f| = 1$ and $S_f = \sin({\rm arg}\,\lambda_f) =\eta_f \sin2\phi_1$,
where $\eta_f$ is the $CP$ eigenvalue. Such is the case, e.g. for the golden $B \to J/\psi K_S$ mode.
If there are at least two diagrams with two CKM phases contributing to the amplitude,
$S_f$ may be sensitive to the difference of strong phases between the two diagrams,
and $C_f$ may not be zero. The $t$-depend study using Eq.~(\ref{eq:tcp1})
can therefore measure DCPV directly.

The extraction of $\phi_1$ with $B \to J/\psi K_S$ and other golden modes
was the main target for the $B$ factory experiments at the dawn of the new millennium.
Large mixing induced $CP$ violation of $O(1)$ was observed in such golden modes.
%$B^0\to {\rm charmonium} + K^0_{S,L}$.
The most recent average performed by HFAG~\cite{hfag} using the
precise measurements from BaBar and Belle, early LEP and CDF results
and recent LHCb measurements with 3 fb$^{-1}$, gives
\be
\sin2\phi_1 = 0.691 \pm 0.017.
 \label{eq_phi1}
\ee
The measured value of  $\sin\phi_1$ has a four-fold ambiguity, % in $\phi_1$.
which can be resolved by a global fit to all available measurements of CKM parameters,
such as the one displayed in Fig.~\ref{fig:CKM2016} from CKMfitter.
The four-fold ambiguity can be reduced to two, $\phi_1$  and $\pi +\phi_1$,
by a time dependent angular analysis of $B^0\to J/\psi K^{*0}$,
or a time dependent Dalitz plot analysis of $B^0\to {\overline D}^0 h^0$ ($h^0 =\pi^0,\ \eta,\ \omega$).
Results by BaBar and Belle on both type of modes show that negative $\cos 2\phi_1$ is unlikely,
as can be seen in Fig.~\ref{fig:CKM2016}.

The same time dependent analysis has been applied to $B^0$ decays via $b\to c\bar cd$ transitions.
However, $b\to d$ penguin pollution may cause $S_f$ and $C_f$ to deviate from
the values of $\sin 2\phi_1$ and $\cos 2\phi_1$, respectively.
Therefore, the experimental measurements of $S_f$ and $C_f$ for
$B^0 \to J/\psi \pi^0$, $B^0\to J/\psi \rho^0$ and $B^0\to D^{(*)+} D^{(*)-}$ are not
included in the global fit by the CKMfitter group,
although these results are consistent with that from the golden modes.
The $b\to s \bar qq$ penguin-dominated decays, on the other hand,
have the same CKM phase as the $b\to c\bar cs$ modes, and should yield the same $S_f$ and $C_f$.
But since New Physics phases may contribute in the penguin loop,
resulting in deviations for $S_f$ and $C_f$ from those measured in the golden modes
(which were indicated by early data), these time-dependent analyses provide
a clean probe for New Physics, which is discussed in Sec. 4.

%\subsubsection{

\

\noindent\underline{\it \bf\boldmath
$\phi_2/\alpha$} \vskip0.1cm

The second phase angle, $\phi_2$, of the triangle of Eq.~(\ref{eq:bdTri})
is the angle between $V_{td}V^*_{tb}$ and $V_{ud}V^*_{ub}$.
It can be extracted from time dependent analysis in $b\to u\bar ud$ transitions,
analogous to the $\phi_1$ case. But the $b\to d$ penguin
may have significant strength compared to the $b\to u\bar ud$ tree amplitude,
resulting in a phase shift from the CKM phase in $S_f$
and making the $\phi_2$ determination difficult.
Experimentally the angle $\phi_2$ has been extracted with
three decay channels: $B\to \pi\pi$, $\rho\rho$ and $\rho\pi$.

From the observation of direct $CP$ violation in $B\to \pi^+\pi^-$
we know that the $d$ penguin contribution is not small.
Therefore, $S_{\pi^+\pi^-}$ is not $\sin(2 \phi_2)$, but can be expressed as,
\be
 S_{\pi^+\pi^-} = \sqrt{1-C^2_{\pi^+\pi^-}} \sin(2\phi_2 + \Delta),
\ee
where $\Delta$ is the phase difference between the two amplitudes,
$e^{-2\pi\phi_3} {\overline A}_{\pi^+\pi^-}$ and $A_{\pi^+\pi^-}$.
This phase angle shift can be extracted with isospin analysis
using the branching fractions and direct $CP$ asymmetries for
the three $B\to \pi\pi$ decays~\cite{pipi-phi2}.
PDG uses the world average of BaBar, Belle and LHCb results~\cite{hfag},
$S_{\pi^+\pi^-}= -0.66\pm 0.06$, $C_{\pi^+\pi^-} = -0.31\pm 0.05$,
the branching fractions of all three modes, and $C_{\pi^0\pi^0} = -0.43^{+0.25}_{-0.24}$.
Since the analysis with the data used in 2015 leads to 16 mirror solutions,
while the branching fraction and $CP$ asymmetry of the $\pi^0\pi^0$ mode
have sizeable uncertainties, the constraint on $\phi_2$ is rather loose,
$0^\circ < \phi_2 < 3.8^\circ$, $86.2^\circ < \phi_2 <  102.9^\circ$,
$122.1^\circ<\phi_2 < 147.9^\circ$, and $167.1^\circ <\phi_2 < 180^\circ$ at 68\% confidence level.

Compared with $B\to \pi\pi$ , the $\rho$ is a vector meson, hence the
$\rho^+\rho^-$ final state from $B$ decays could be
a mixture of $CP$ even or $CP$ odd components.
This mixture state complicates the analysis, and the phase angle
cannot be extracted if the mixture is half and half.
Fortunately, experimental data show that the longitudinal polarization fractions
in $B^+\to \rho^+\rho^0$ and $B^0\to \rho^+\rho^-$ decays are close to unity,
indicating that final states are almost $CP$ even.
Moreover, the small branching fraction of $B^0\to \rho^0\rho^0$ is a factor of twenty less than
that of $B^0\to \rho^+\rho^-$ and   $B^+\to \rho^+\rho^0$, indicating the penguin contribution is small.
Isospin analysis is applied to the $\rho\rho$ modes using
the world average~\cite{hfag} of $S_{\rho^+\rho^-}$, $C_{\rho^+\rho^-}$,
the branching fractions of the three $B\to \rho\rho$ modes, together with
the $S_{\rho^0\rho^0}$ and $C_{\rho^0\rho^0}$ values from Ref.~\cite{rhorho-babar}.
The $\phi_2$ phase angle is obtained in six regions at 68\% confidence level:
$0^\circ<\phi_2<5.6^\circ$, $84.4^\circ <\phi_2<95.3^\circ$, $174.7^\circ<\phi_2 < 180^\circ$,
and the other three regions in mirror solutions $3\pi/2-\phi_2$.

Although the $B^0\to \rho^+\pi^-$ final state is not a $CP$ eigenstate,
both $B^0$ and ${\overline B}^0$ can decay to $\rho^\pm \pi^\mp$. Furthermore,
the decay  $B^0\to \rho^\pm\pi^\mp$ proceeds via the same diagrams as $B^0\to\pi^+\pi^-$,
making interference of four diagrams possible.
The time-dependent Dalitz plot analysis of $B^0\to \pi^+\pi^-\pi^0$ enables
the extraction of $\phi_2$ with a $\phi_2\to  \pi +\phi_2$ ambiguity
since the variation of strong phase in the interference regions of
$\rho^+\pi^-$, $\rho^-\pi^+$ and $\rho^0\pi^0$ can be known~\cite{dalitz}.
PDG averages the Belle~\cite{bellerhopi} and BaBar~\cite{babarrhopi} measurements
to give $\phi_2 = (54.1^{+7.7}_{-10.3})^\circ$ and $(141.8^{+4.7}_{-5.4})^\circ$.

Combining all the three $\phi_2$ measurements, $\phi_2$ is constrained to be rather close to $90^\circ$,
\be
 \phi_2 = (88.6^{+3.5}_{-3.3})^\circ.
 \label{eq_phi2}
\ee

\subsubsection{%\boldmath
 $\phi_3/\gamma$}

The third CKM angle $\phi_3$ is between $V_{ud}V^*_{ub}$ and $V_{cd}V^*_{cb}$,
and is practically the phase angle of $V_{ub}^*$ in the standard phase convention.
Since the top quark is not involved in the CKM elements for $\phi_3$,
it cannot be extracted using time-dependent analysis.
Three different methods are used to extract $\phi_3$ experimentally.
All employ the principle that the interference of
$B^-\to  D^{(*)0} K^{(*)-} (b\to c{\bar u} s)$ and
$B^-\to {\overline D}^{(*)0} K^{(*)-}$ ($b\to u{\bar c} s$, which carries the CPV phase)
can be studied by comparing some $D^{(*)0}$ and ${\overline D}^{(*)0}$ decays.
In principle, the interference permits the extraction of the $B$ and $D$ amplitudes,
the weak phase and the relative strong phases.
The crucial parameter is $r_B$, defined as the amplitude ratio,
\be
 r_B = \frac{A(B^-\to {\overline D}^{(*)0} K^{(*)-} )}{A(B^-\to D^{(*)0} K^{(*)-} )},
  \label{eq:rb}
\ee
which ranges from 0.1 to 0.2.
Note that a smaller $r_B$ causes the interference to be less effective.

The GLW method~\cite{glw1, glw2} considers $D$ decays to $CP$ eigenstates.
Therefore, both the Cabibbo-suppressed and favored decays
go into the same final state, facilitating the interference.
To enhance $r_B$ and make the amplitudes of $B$ and $D$ mesons
compatible for both decays, the ADS method~\cite{ads} considers
Cabibbo-allowed ${\overline D}$ decay and doubly-Cabibbo-suppressed $D$ decay.
Extensive measurements of $\phi_3$ are performed at the $B$ factories, CDF and LHCb,
which can be found in the HFAG web pages.

The most promising and effective method, if not limited by statistics,
comes from  the Dalitz plot method. The point is that some three-body $D^0$ decays,
such as $D^0 \to K^0_S\pi^+\pi^-$, have large branching fractions,
and the interference can be studied across the Dalitz plot
for such $D$ decays~\cite{dalitz1-phi3,dalitz2-phi3}.
Belle and BaBar have measured $\phi_3$ with good accuracy using the Dalitz plot method.
The obtained results are
$\phi_3 = (78^{+11}_{-12}\pm 4 \pm 9)^\circ$~\cite{belle-phi3} for Belle, and
$\phi_3 = (68\pm 14\pm 4 \pm 3)^\circ$~\cite{babar-phi3} for BaBar,
where the last uncertainty of these measurements is from $D$ decay modeling.
But the beauty of the Dalitz analysis is that,
by a bin-by-bin fit, one can fit for the variation of the strong phase
across the Dalitz plot, and the eventual measurement of $\phi_3$ would be model independent.
The LHCb result with this model independent approach yields $\phi_3 = (62^{+15}_{-14})^\circ$~\cite{lhcb-phi3}.

Combining the results from GLW, ADS and Dalitz analyses, PDG gives
\be
   \phi_3/\gamma = (73.2^{+6.3}_{-7.0})^\circ.
 \label{eq_phi3}
\ee
But with the large data sets expected by Belle II and LHCb in the future,
the error stands major improvement.
Combining Eqs.~(\ref{eq_phi1}), (\ref{eq_phi2}) and (\ref{eq_phi3}),
one can easily check that the three angles sum to $180^\circ$ within errors,
which can be further tested in the future.

\

\noindent\underline{\bf
Comments} \vskip0.1cm

When the $B$ factories were proposed, only the first phase angle $\phi_1$ was
assured to be measured with good accuracy. The angle $\phi_3$ was known to be
difficult at the beginning of the B factory era. With large accumulated data,
however, the data analysis is gradually improved,
and several new methods were brought out.
Now the uncertainty for $\phi_3$ is at the $7^\circ$ level.\footnote{
To show the volatility of the subject, as reported at EPS-HEP 2017,
the new combined LHCb result~\cite{lhcb-gamma-EPS}, 
$\gamma/\phi_3 = (76.8^{+5.1}_{-5.7})^\circ$ by a single experiment.
This leads to the HFLAV combined result of
$\gamma/\phi_3 = (76.2^{+4.7}_{-5.0})^\circ$ for EPS 2017,
which is at the $5^\circ$ level!
Belle II data is eagerly awaited.
}
When high quality data become available in the future,
it is likely that we can achieve better accuracy than originally conceived.
With Belle II data taking to start soon, and with the LHCb detector upgrade,
the angle $\phi_3$ is expected to reach an accuracy of $1^\circ$
for both Belle II and LHCb with full data samples
(with LHCb aiming to push below with a Phase 2 Upgrade).
It is worth reemphasizing that $\phi_3$ measurement is based on \emph{tree level} processes,
and accesses the phase of $V_{ub}^*$ directly.

\begin{table}[h]
\begin{center}
\caption{Summary of measured CKM matrix element and phase values.}\vskip0.2cm
\begin{tabular}{ccccc}
\hline\hline
                   & $|V_{ud}|$ & $|V_{us}|$ & $|V_{ub}|$ \\
 \hline
                   & $0.97417 \pm 0.00021$ & $0.2248 \pm 0.0006$ & $0.00409 \pm 0.00039$ \\
\hline\hline
                   & $|V_{cd}|$ & $|V_{cs}|$ & $|V_{cb}|$ \\
 \hline
                   & $0.220 \pm 0.005$ & $0.995 \pm 0.016$ & $0.0405\pm 0.0015$ \\
\hline\hline
 $|V_{td}/V_{ts}|$ & $|V_{td}|$ & $|V_{ts}|$ & $|V_{tb}|$ \\
 \hline
 $0.215 \pm 0.001 \pm 0.011$ & $0.0082 \pm 0.0006$ & $0.0400 \pm 0.0027$ & $1.009 \pm 0.031$ \\
\hline\hline
                   & $\sin2\phi_1$ & $\phi_2$ & $\phi_3$  \\
 \hline
                   & $0.691\pm 0.017$ & $(88.6^{+3.5}_{-3.3})^\circ$ & $(73.2^{+6.3}_{-7.0})^\circ$ \\
\hline\hline
\end{tabular}
\label{tab:sum}
\end{center}
\end{table}

\subsection{CKM Summary and Global Fit}

We have given a short review on the determination of CKM elements based on values given by PDG 2016.
The magnitudes of the matrix elements are extracted from various
semileptonic and leptonic decays of mesons, as well as muon and tau decays.
As for the phase angles, time-dependent $CP$ analysis is used for $\phi_1$ and $\phi_2$,
while $\phi_3$ can be extracted by using $B^-\to D^{(*)0} K^{(*)-}$ decays with various methods.
Footnote 7 below makes clear the impact of LHCb, especially on $\gamma/\phi_3$.

The PDG 2016 averages of the measured magnitudes of CKM matrix elements,
as well as the three phase angles, are summarized in Table~\ref{tab:sum}.
In most cases, extractions using different channels for the same
elements or phase angles show consistent results, strongly
supporting the CKM paradigm. Among these, many already
reach good accuracy with errors dominated by theory uncertainties.
However, there are also noticeable tensions in some of the measurements.
The discrepancies between the inclusive and exclusive determinations of
$|V_{cb}|$ and $|V_{ub}|$ need to be resolved with efforts from both experiment and theory.
An almost $3\sigma$ tension on $|V_{us}|$ coming from inclusive $\tau$ decays
also need to be understood.

The CKM paradigm stands tall, but there is still work to be done.

%\subsubsection{

\

\noindent\underline{\it \bf\boldmath
CKM Global Fit} \vskip0.1cm

With the nine magnitudes of the CKM matrix elements and three angles
summarized in Table~\ref{tab:sum}, the unitarity of the matrix and
sum of the three angles can be readily checked.
%As shown in Sec.~\ref{sec:vud-vus},
The squared sum of the first row of the CKM matrix gives
$|V_{ud}|^2 + |V_{us}|^2 + |V_{ub}|^2 = 0.9996\pm 0.0005$,
which can be compared with Eq.~(\ref{eq:row1}),
where adding $|V_{ub}|^2$ has moved the fourth decimal place.
The squared sums for 1st column, 2nd row and 2nd column are,
$|V_{ud}|^2 + |V_{cd}|^2 + |V_{td}|^2 = 0.9975 \pm 0.0022$,
$|V_{cd}|^2+|V_{cs}|^2+|V_{cb}|^2 = 1.040\pm 0.032$ and
$|V_{us}|^2 + |V_{cs}|^2 + |V_{ts}|^2 = 1.042\pm 0.032$, respectively,
where uncertainties of the latter two are dominated by $|V_{cs}|$.
These squared sums are all consistent with unity.
The angle sum of $\phi_1, \phi_2$ and $\phi_3$ is $\big(183^{+7}_{-8}\big)^\circ$,
consistent with a triangle.

Since the matrix elements can be determined more precisely using a global fit
to all available measurements under the constraint of three generation unitarity,
PDG conducts a fit using two approaches:
the frequentist statistics~\cite{hocker} used by CKMfitter~\cite{ckmfitter},
as well as a Bayesian approach from UTfit~\cite{utfit1, utfit2}.
Fit results from both approaches are similar, where
the Wolfenstein parameters obtained from the frequentist approach are
\be
 \lambda = 0.22506 \pm 0.00050, \quad  A = 0.811\pm 0.026,  \quad
 \bar\rho = 0.124^{+ 0.019}_{-0.018}, \quad \bar \eta = 0.356 \pm 0.011,
\ee
which can be compared with the Bayesian approach:
 $\lambda = 0.22496\pm 0.00048, \; A = 0.823\pm 0.013, \; {\bar \rho} = 0.141\pm 0.019$ and
${\bar \eta} = 0.349\pm 0.012$.
The fitted magnitudes of the CKM matrix elements are
\be
   (|V_{ij}|) =
    \left( \begin{array}{ccc}

     0.97434^{+0.00011}_{-0.00012} & 0.22506 \pm 0.00050 & 0.00357 \pm 0.00015 \\
     0.22492 \pm 0.00050           & 0.97351 \pm 0.00013 & 0.0411  \pm 0.0013  \\
     0.00875^{+0.00032}_{-0.00033} & 0.0403  \pm 0.0013  & 0.99915 \pm 0.00005
           \end{array} \right),
 \label{eq:Vij_fit}
\ee
and the Jarlskog invariant~\cite{jarlskog},
defined as ${\rm Im}\, (V_{us}V_{cb}V^*_{ub}V^*_{cs})$, is
\be
J = \left(3.04^{+0.21}_{-0.20}\right)\times 10^{-5}.
\ee
Eq.~(\ref{eq:Vij_fit}) is consistent with Table~\ref{tab:sum}, but with higher precision.

\

\noindent\underline{\it \bf\boldmath
Comments} \vskip0.1cm

So far we have briefly described the determination of the magnitudes and
phase angles of the CKM matrix elements. A total of 12 parameters are
extracted from hundreds of experimental measurements,
from the decays of light and heavy mesons, baryons, tau leptons, W bosons and top quarks,
together with perturbative and nonperturbative theoretical calculations based on SM.
Data analyses were performed to measure the decay rates, mixing parameters,
time-dependent or direct $CP$ violating asymmetries, etc.
No apparent anomaly is found in the CKM sector,
and all the obtained parameters give a consistent picture.
This is rather remarkable, indicating that
the Higgs mechanism with Yukawa coupling and the quark mixing matrix
constitute the fundamental theory that governs the particle physics realm.
Consistent experimental measurements also imply that New Physics
contribution, if any, is rather small.
Therefore, it maybe more fruitful to search for New Physics
in the channels where the SM contribution is forbidden or highly suppressed.

%\hskip0.65cm  Sides \quad Angles \quad \& Measurements

\section{Recent B Highlights \label{sec:B-high}}

% ?s; Bq ? ?+??; B ? K(?)l+l?;  B ? D(?)??, ??; (3-body CPV)
% Need one representative figure for each topic?
% phi_s => HFAG combine
% BMM => Nature combine contour? <=
% K*ll => LHCb p5' <=
% D(*)tau nu => HFAG R_D(*) plot
% tau nu => just wording?
% 3-body CPV => just wording? <= single large Acp plot?

The B factories were very successful, measuring $\sin2\phi_1/\beta$ precisely
and free of hadronic uncertainties. They confirmed the Kobayashi--Maskawa picture,
and further sealed the CKM paradigm with a plethora of measurements.
It is quite amazing that the tiny CPV phase in $B_s$ mixing, $\phi_s = -2\beta_s$,
is now also measured to be consistent with SM expectation at $ -0.04$.
It is not ``anomalous'', but there were excitements along the path of measurement,
which is still partially reflected in the latest HFAG~\cite{Amhis:2016xyh} combination plot,
Fig.~\ref{fig:dGammas_vs_phis}, as one moved from the Tevatron to the LHC.
The very rare decay $B_s \to \mu^+\mu^-$ was toted as a great vehicle for discovering
the effect of SUSY in the flavor sector, but it again turned out to be consistent with SM.
These two measurements are triumphs of experimental physics,
and epitomize the success of the LHC(b).

While the B factories saw many anomalies, they have mostly gone away.
There are, however, two anomalies that emerged during the LHC Run 2 period,
and not just from the LHC.
One is the so-called ``$P_5'$ anomaly'' uncovered by LHCb in the
angular analysis of $B \to K^*\mu^+\mu^-$ decay.
It is reminiscent of the $A_{FB}$ anomaly (which got eliminated by LHCb)
from the B factories, but not the same.
It may call for an extra $Z'$ boson.
Another is called the ``BaBar anomaly'' in the $B \to D\tau\nu$ and $D^*\tau\nu$
final states, which cannot be understood in the usual (SUSY) type of
two Higgs doublet models (called 2HDM-II).
In this section, we would go through these recent $B$ highlights,
along with a few lesser effects.
As an anticlimax, we mention the spectacular DCPV asymmetries observed over
the Dalitz plot of three body $B$ decays, to reemphasize that hadronic effects remain a menace.

\subsection{%\boldmath
 CPV Phase $\phi_s$}

Recall that it was the discovery~\cite{Albrecht:1987ap} of large $B_d$ mixing,
i.e. $\Delta m_d$ is comparable to $\Gamma_d$,
that stimulated the construction of the B factories.
From that perspective, it was the magnificent observation~\cite{Abulencia:2006ze}
of $B_s$ mixing at the Tevatron that was the precursor to the measurement of $\phi_s$
at the LHC, which we now discuss.
The new ``B factory'', called LHCb, a hybrid of a fixed-target and collider detector,
was already under construction at that time.
A new era of $B$ physics at hadronic colliders was dawning.

\vskip0.3cm
\begin{figure}[ht]
\begin{center}
\includegraphics[width=10cm]{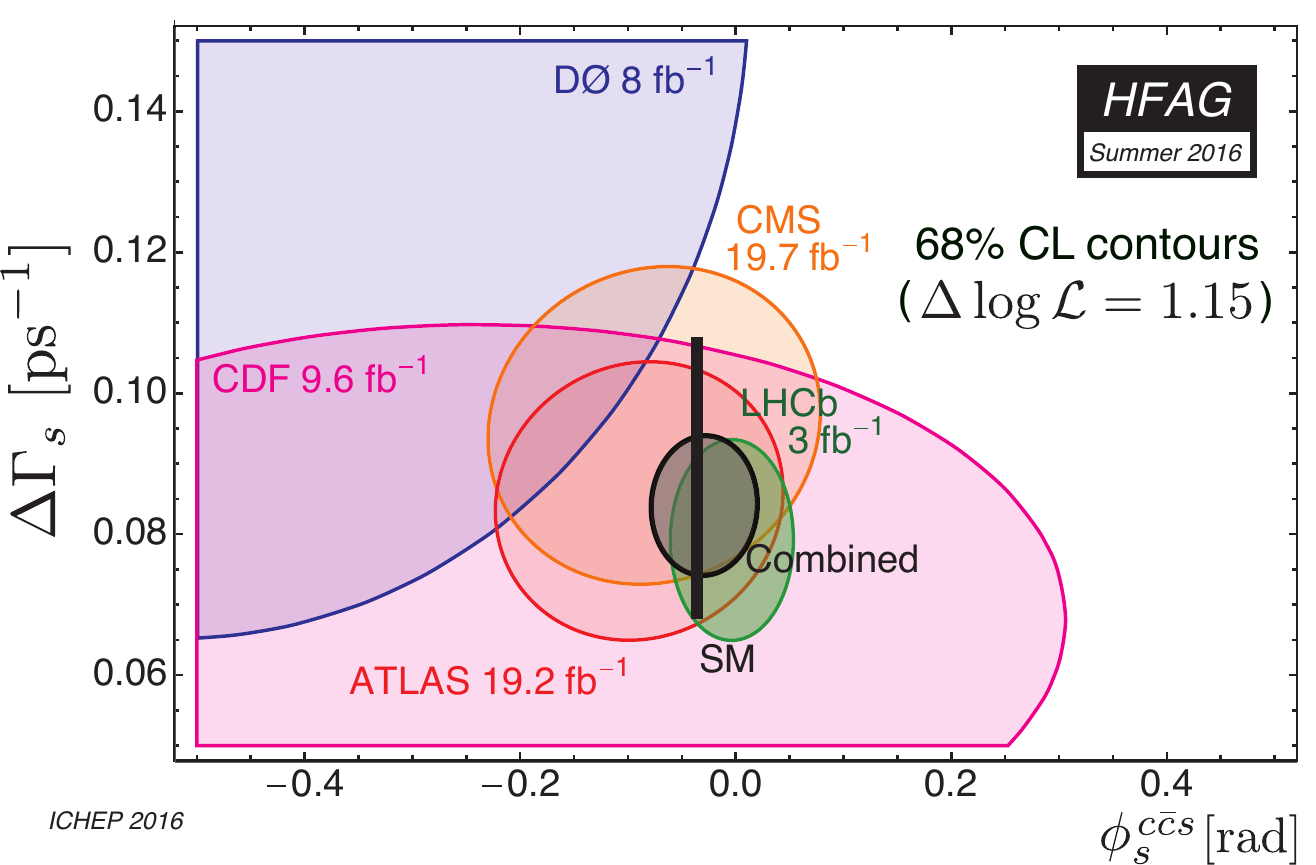}
\vskip-0.1cm
\caption{
Measured 68\% C.L. contours in the $\phi_s$--$\Delta\Gamma_s$ plane
from all relevant experiments. %ATLAS, CDF, CMS, D\O\ and LHCb.
The combined contour (black ellipse) is consistent with SM expectation (thick black line).
%The combined result is consistent with the SM prediction.
[Source: http://www.slac.stanford.edu/xorg/hfag/]
}
 \label{fig:dGammas_vs_phis}
\end{center}
\end{figure}

Inspection of Fig.~\ref{fig:dGammas_vs_phis} shows that $\phi_s$ is measured
together with $\Delta\Gamma_s$, which is different from the $B_d$ case.
As discussed already with $\sin2\phi_1/\beta$ measurement,
for neutral $B_q$ meson decaying into a $CP$ eigenstate that is common to
both $B_q$ and $\overline B_q$ mesons, the interference between the amplitudes
of the direct decay and decay after mixing generates a time-dependent CPV asymmetry.
For the $B_s$ and $\overline B_s$ meson system, however, $b \to c\bar cs$ decay is not Cabibbo suppressed
like $b \to c\bar cd$ decay, which can link between $\bar bs$ and $b\bar s$ mesonic states.
That is, the mixing of $B_s$ and $\overline B_s$ mesons has a non-negligible absorptive part,
or a width difference $\Delta\Gamma_s$, in addition to $\Delta m_s$.
Thus, the CPV asymmetry study fits for both the weak phase $\phi_s$ and
the decay width difference $\Delta \Gamma_s$.
The expected decay width difference in SM is determined~\cite{Artuso:2015swg} to be
\begin{equation}
\Delta \Gamma_s^{\rm (SM)} = 0.088 \pm 0.020~{\rm ps}^{-1},
 \label{eq:dGs_SM}
\end{equation}
where $\Delta\Gamma_s$ is defined as the difference between decay widths
$\Gamma_{L}$ and $\Gamma_H$  of the light and heavy $B_s$ states.
The weak phase $\phi_s$ is predicted to be rather small but with high precision
($\beta_s$ is used by CDF, while $\phi_s$ is used by D\O\ and LHCb),
\begin{equation}
\phi_s^{\rm (SM)} = -2\beta_s^{\rm (SM)}
 = -2 \arg(-V_{ts}V_{tb}^* / V_{cs}V_{cb}^*) = -0.0376^{+0.0008}_{-0.0007},
 \label{eq:phis_SM}
\end{equation}
where the numerical value is from indirect determination by
a global fit to experimental data~\cite{Charles:2011va}.
Since the prediction is rather precise, any deviation between the measured value
and Eq.~(\ref{eq:phis_SM}) would be strong hint for New Physics,
and it was stressed that many New Physics scenarios could
greatly enhance $\phi_s$~\cite{Dunietz:2000cr}. %~\cite{Buras:2009if, Chiang:2009ev}.

From the relation to the CKM matrix elements as displayed in Eq.~(\ref{eq:phis_SM}),
one can see that the unitarity triangle\footnote{
One can verify easily by use of Eq.~(\ref{eq:VCKM}),
but expanding to $O(\lambda^5)$ terms.
}
\be
V_{us}V_{ub}^* + V_{cs}V_{cb}^* + V_{ts}V_{tb}^* = 0
 \label{eq:bsTri}
\ee
is very squashed in SM compared with Eq.~(\ref{eq:bdTri}), where
the measurement of the latter is displayed in Fig.~\ref{fig:CKM2016}.
This enhances one's appreciation of the current world average,
largely based on LHC Run 1 data,
\begin{equation}
\phi_s = -0.030 \pm 0.033,
 \label{eq:phis}
\end{equation}
which is quite consistent with the SM prediction, as one can see pictorially
from Fig.~\ref{fig:dGammas_vs_phis}.
But the game is far from over, and one needs much improvement in data size,
as well as analysis methods, to scrutinize whether there is New Physics in $\phi_s$,
or it is fully consistent with SM.
However, at the dawn of the LHC era, there was much anticipation
for a potentially large, New Physics CPV phase in $\phi_s$. Though this wish or hope
was dashed by the measured value of Eq.~(\ref{eq:phis}),
it was truly a recent highlight of $B$ physics.
So let us give some account.

The process $B_s \to J/\psi\, \phi$ is the main decay used to
measure $\phi_s$ as well as $\Delta \Gamma_s$.
The $B_s \to J/\psi\,\phi$ decay final state is a mixture of $CP$ eigenstates,
and an angular analysis is required to disentangle the $CP$-odd and $CP$-even components.
A time-dependent angular analysis has to be carried out by measuring the decay angles of
the final state particles and the proper decay length of the $B_s$ candidate.
The differential decay rate can be expressed as,
\begin{equation}
{d^4\Gamma[B_s(t)] \over d\Theta dt}
= \sum^{10}_{i=1} \mathcal{O}_i(\alpha, t)\,
g_i(\Theta),
\end{equation}
where $\mathcal{O}_i$ are time-dependent functions, $g_i$ the angular functions,
$\alpha$ represents the physics parameters of interest ($\Delta \Gamma_s$ and $\phi_s$),
and $\Theta$ and $t$ represent the measured angles and proper decay time of $B_s$ decay.
Because of the finite width difference, the functions $\mathcal{O}_i$ can be expanded as,
\begin{equation}
\mathcal{O}_i(\alpha,t) \propto
e^{-\Gamma_s t} \Big[
a_i \cosh\left(\Delta\Gamma_s t/2\right) +
b_i \sinh\left(\Delta\Gamma_s t/2\right) +
c_i \cos\left(\Delta m_s t\right) +
d_i \sin\left(\Delta m_s t\right)
\Big],
\end{equation}
where the non-oscillatory $\Delta\Gamma_s$ effects also enter.
The extraction of $\phi_s$ mostly comes through the parameters $b_i$ and $d_i$.
The angular parameters symbolized by $\Theta$ contain the polar and azimuthal angles
$\theta_T$ and $\phi_T$ of the $\mu^+$ in the $J/\psi$ rest frame,
and the helicity angle $\psi_T$ of the $\phi$ meson w.r.t. the direction of $K^+$.
The details of the description of the time-dependent decay rate
can be found in Ref.~\cite{Dighe:1995pd} (see also Ref.~\cite{Artuso:2015swg} for a review).

From the angular distribution alluded to above, one can already tell the complexity of the analysis.
The required ingredients include a well modeled detection efficiency along the decay angles,
decay proper time, and the flavor of the $B_s$ meson.
In particular, flavor tagging is a critical element in the analysis,
where the tagging power in recent studies is still at the order of 1--3\%.
Possible interference between the $J/\psi\,\phi$ and $J/\psi\, K^+K^-$ components is
one of the dominant systematic issues, and a detailed study of the $K^+K^-$
invariant mass distribution is needed to disentangle the interference pattern.
There are several similar decays that can be further used to determine $\phi_s$,
such as $B_s \to \psi(2S)\, \phi$ and $J/\psi f_0$ (i.e. $J/\psi\, \pi^+\pi^-$),
which require similar or simpler modeling of the differential decay rate.
The $B_s \to J/\psi\, \phi$  mode dominated early studies,
but for the measurement by LHCb, the $J/\psi f_0\; (\to \pi^+\pi^-)$
mode was critical as well (see e.g. Ref~\cite{Artuso:2015swg}).

To aim at measuring $|\phi_s| < 0.1$ appeared rather daunting at the Tevatron.
However, with the discovery~\cite{Lin:2008zzaa} of large DCPV difference in $B \to K\pi$,
it was argued~\cite{Hou:2005hd} that a New Physics phase in the electroweak penguin
that could potentially explain the effect of Eq.~(\ref{eq:DAKpi}),
could also show up as large CPV in $B_s$ mixing, and that CDF and D\O\
had the chance to get the first glimpse of an $O(1)$ value for $|\sin2\beta_s|$.
Intriguingly, the Tevatron results indeed suggested~\cite{Bona:2008jn} $\phi_s$
that is away from the predicted value of $-0.04$.
The situation is somewhat still preserved in the latest HFAG combination
plot shown in Fig.~\ref{fig:dGammas_vs_phis}.
The final CDF and D\O\ results tend towards rather sizable negative values but with large error bars
(while one can only tell that $\Delta\Gamma_s > 0$).
There was thus much anticipation at the turn-on of the LHC, where the earliest LHCb result was
not inconsistent~\cite{Golutvin11} with that from the Tevatron.
Alas, the hope was dashed by the LHCb announcement at Lepton--Photon 2011 held in Mumbai,
that $\phi_s$ also turned out to be consistent with SM.
Further results from ATLAS and CMS,
and in particular the LHCb result with full Run 1 data using both $B_s \to J/\psi\, \phi$
and $J/\psi\, f_0\; (\to \pi^+\pi^-)$ modes are much closer to the SM prediction,
with the current world average given in Eq.~(\ref{eq:phis}).
The uncertainty is still dominated by statistics, so further improvement is
expected with more LHC data and improvement of analysis techniques.
More refined algorithms (as well as statistics) should also improve systematic errors.
But the game has definitely switched from large deviation search to precision test of SM.

The reverse argument of Ref.~\cite{Hou:2005hd}, that no sign of New Physics in $\phi_s$
makes a $Z$-penguin explanation of Eq.~(\ref{eq:DAKpi}) implausible,
points to the color-suppressed amplitude $C$ as culprit,
causing us to lament in the Introduction that hadronic effects
are still a menace in $B$ decays.
Theorists never predicted $|C| \sim |T|$.

As seen from Fig.~\ref{fig:dGammas_vs_phis}, the measurement of 
$\Delta \Gamma_s$ is now also in agreement with SM.
To a lesser extent compared with $\phi_s$ measurement,
due to its connection with the like-sign dimuon charge asymmetry $A^b_{sl}$,
$\Delta \Gamma_s$ had its own saga.
Based on the final data set from D\O~\cite{Abazov:2013uma},
the measured $A^b_{sl}$ value differs from the SM prediction by 3.6$\sigma$,
consistent with what was reported at the dawn of the LHC.
Although it has been difficult for other experiments to confirm or rule out this asymmetry,
it can be expressed in terms of the wrong-sign asymmetries for $B_d$ and $B_s$ mesons,
$a^d_{sl}$ and $a^s_{sl}$, respectively.
The wrong-sign asymmetry $a^d_{sl}$ for $B_d$ is measured by B factory experiments
and shows no deviation from SM.
Attributing the ``D\O\ anomaly'' to the wrong-sign $B_s$ asymmetry
implies a value for $a^s_{sl}$ that is much larger than in SM,
which could probe the product of $\Delta \Gamma_s$ and $\tan\phi_s$.
This was quite interesting before summer 2011, when $\phi_s$ might turn out sizable.
%it has direct impact on $\Delta \Gamma_s$.
Since the prediction of $\Delta \Gamma_s$ based on OPE is quite robust,
and long distance effects have been demonstrated to have little impact~\cite{Chua:2011er},
the issue has gone away with $\phi_s$ being consistent with the small value expected in SM
(but ``D\O\ anomaly'' remains).
%any deviation from the SM prediction can be interpreted
%as New Physics in $B_s$ mixing.

The current $\Delta \Gamma_s$ measurement has comparable precision as SM expectation.
With the upcoming improved measurement of $B_s \to J/\psi\phi$ and similar decays, together with
the improvements of SM-based calculations, the situation might be different in the future.
This also echoes the recent experimental development of the
$B_s$ effective lifetime measurement using $B_s \to \mu^+\mu^-$ events.
As discussed in the next subsection, only the heavier state of $B_s$ is allowed
in this super-rare decay process.

\vskip0.3cm
\begin{figure}[t]
\begin{center}
\includegraphics[width=10cm]{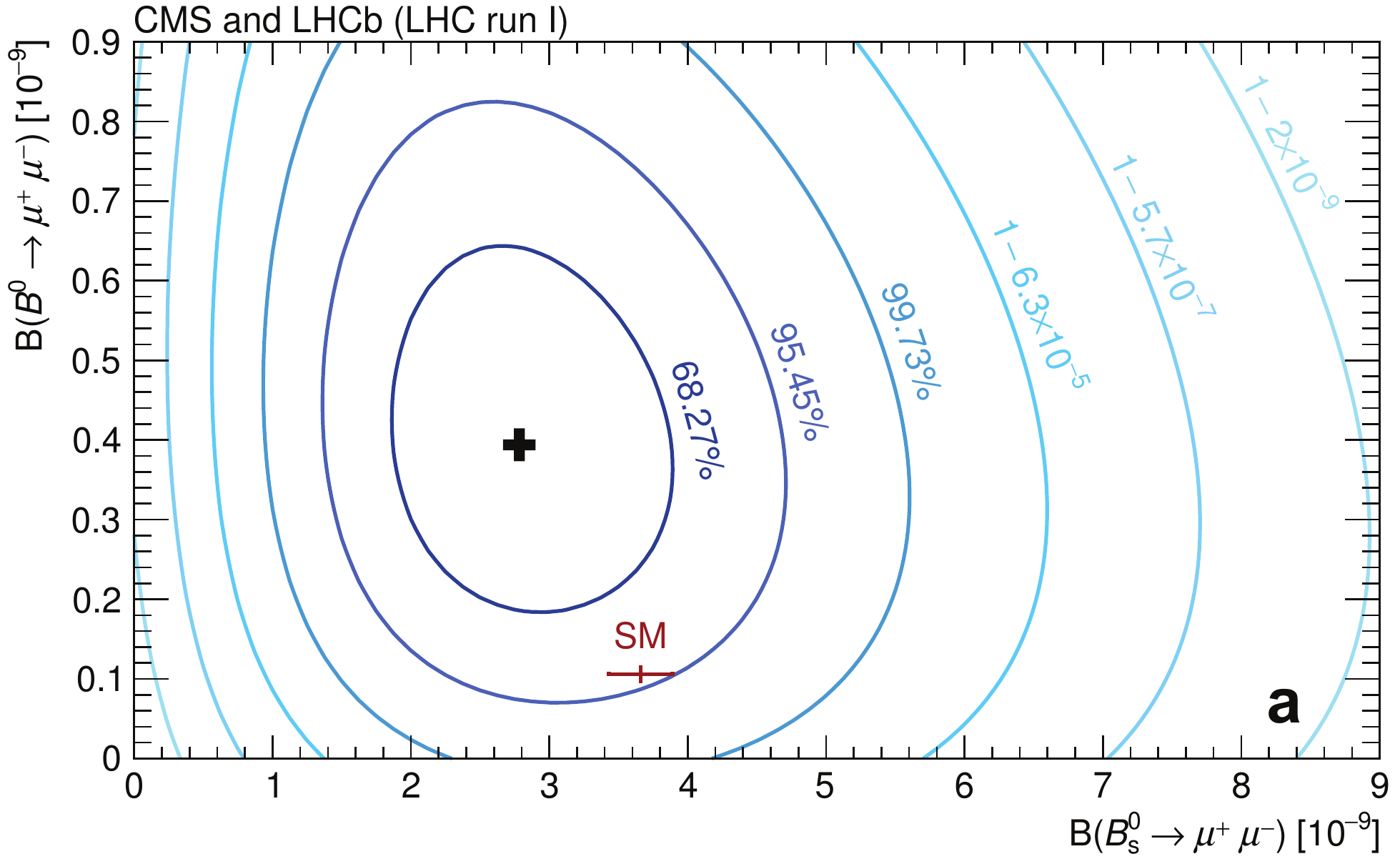}
\caption{
Likelihood contours in the $\mathcal{B}(B_s \to \mu^+\mu^-)$ vs $\mathcal{B}(B^0 \to \mu^+\mu^-)$ plane.
The black cross indicates the best-fit value, while the SM prediction (and its uncertainty) is also marked.
 [Source: courtesy CMS and LHCb collaborations, from Ref.~\cite{CMS:2014xfa}]
 }
\label{fig:bd_versus_bs}
\end{center}
\end{figure}

\subsection{%\boldmath
 $B_s \to \mu^+\mu^-$ and $B^0 \to \mu^+\mu^-$
 \label{sec:Bqmumu}}

Fig.~\ref{fig:bd_versus_bs} is a highlight achievement of $B$ physics at LHC Run 1:
the joint measurement~\cite{CMS:2014xfa} of $B_s \to \mu^+\mu^-$ by CMS and LHCb
that is consistent with the SM expectation,
and a mild hint for the $B^0 \to \mu^+\mu^-$ process that is further suppressed by $|V_{td}/V_{ts}|^2$.
But this statement does not really capture how much a highlight it actually is.

Just like the absence of $K_L \to \mu^+\mu^-$ and other FCNC effects
led to the invention of the GIM mechanism~\cite{Glashow:1970gm},
the $B_s\to\mu^+\mu^-$ and ${B^0}\to\mu^+\mu^-$ processes are
forbidden at tree level and suppressed by GIM cancellation in SM.
It suffers from further suppression from quark annihilation within the $B_q$ meson,
hence suppressed by $f_{B_q}^2/m_{B_q}^2$, and by helicity suppression,
hence another factor of $m_{\mu}^2/m_{{B_q}}^2$.
%, where $m_{\mu}$ and $m_{{B}}$ are the masses of the muon and the ${B}$ meson, respectively.
%
The branching fractions of these two decays are predicted~\cite{Bobeth:2013uxa} reliably in SM as
\bea
\mathcal{B}^{\rm SM}({B}_{s}\to\mu^+\mu^-) &=& (3.66\pm0.23)\times10^{-9},
 \label{eq:Bsmumu_SM} \\
\mathcal{B}^{\rm SM}({B^0}\to\mu^+\mu^-) &=& (1.06\pm0.09)\times10^{-10}.
 \label{eq:Bdmumu_SM}
\eea
These rates are indeed tiny, but not totally out of reach experimentally,
considering the huge $b$-hadron production cross section at hadron colliders
and the simple dimuon tracks from a well-defined vertex,
and as evidenced in the measurement displayed in Fig.~\ref{fig:bd_versus_bs}.
Because of their extreme rarity, these modes are great probes of New Physics
--- any deviation from the SM branching fractions could provide a smoking gun signal for New Physics.
Alternatively, measurements that confirm the SM predictions would be
a triumph of experimental physics, and provide stringent constraint of New Physics.
%The two charged-track nature of the signal makes the search in principle straightforward.

Introducing new particles or non-SM couplings in the loop processes
can modify these two decays significantly.
In SUSY models, $\mathcal{B}({B_s}\to\mu^+\mu^-)$ can be proportional to
$\tan^6\beta$, which can easily increase the branching fractions
by orders of magnitude~\cite{Huang:1998vb, Hamzaoui:1998nu,
 Choudhury:1998ze, Babu:1999hn, Bobeth:2001sq},
an effect largely arising from additional Higgs bosons,
and because of the aforementioned multiple suppression of $B_q \to \mu^+\mu^-$ rates.
%Consider several SM extensions, such as a two-Higgs doublet model~\cite{Ellis:2006jy}, or leptoquarks~\cite{Davidson:2010uu}, may enhance the decay branching fractions. By introducing some other alternative models, the decay rates might be reduced~\cite{Ellis:2007kb}.
%
The potential for enhancement by three orders of magnitude over SM expectation
galvanized the great pursuit at the Tevatron, with the hope to
discover SUSY indirectly through the flavor sector.
This nicely illustrates the complementarity between
direct search at colliders, compared with
indirect search inferred by rare decay studies.

As the Tevatron pursuit pushed down the orders of magnitude and reaching the endpoint of sensitivities,
but just before early LHC Run-1 data was fully analyzed,
the CDF experiment reported some excess in $B_{s} \to \mu^+\mu^-$,
eventually publishing the large decay rate of $\left( 1.3^{+0.9}_{-0.7}\right) \times 10^{-8}$
with the full CDF data set~\cite{Aaltonen:2013as}.
This is still close to four times larger than Eq.~(\ref{eq:Bsmumu_SM}),
and would constitute good evidence for New Physics, such as SUSY,
if the effect was true.

However, the CDF excess was refuted already by CMS and LHCb right at the time of announcement,
with CDF data already exhausted and unable to extend further.\footnote{
Recall that the Tevatron collider was shut down in 2011.}
As LHC data accumulated, and after LHCb and CMS each announced
their evidence for $B_s \to \mu^+\mu^-$ with full Run 1 data~\cite{PDG},
it took the combination of CMS and LHCb data to announce the 6.2$\sigma$
discovery~\cite{CMS:2014xfa} of $B_{s} \to \mu^+\mu^-$ decay,
and a hint of $B^0 \to \mu^+\mu^-$ decay was also found.
The best fit branching fractions are
\bea
\mathcal{B}({B}_{s}\to\mu^+\mu^-) &=& (2.8^{+0.7}_{-0.6})\times10^{-9},
 \label{eq:Bsmumu} \\
\mathcal{B}({B^0}\to\mu^+\mu^-) &=& (3.9^{+1.6}_{-1.4})\times10^{-10},
 %\quad {\rm (CMS} + {\rm LHCb, 2015)}
 \label{eq:Bdmumu}
\eea
where the two dimensional likelihood contours have already been
shown above in Fig.~\ref{fig:bd_versus_bs}.
%the $\mathcal{B}(B_s \to \mu^+\mu^-)$--$\mathcal{B}(B^0 \to \mu^+\mu^-)$ plane
The branching fraction of $B_s \to \mu^+\mu^-$ is in agreement with (if not slightly lower than)
the prediction of Eq.~(\ref{eq:Bsmumu_SM}),
while the $B^0 \to \mu^+\mu^-$ rate is higher than the expectation,
but still within 2$\sigma$ of  Eq.~(\ref{eq:Bdmumu_SM}).
Note that the $\pm2\sigma$ (or 95.45\%) confidence interval for $\mathcal{B}({B^0}\to\mu^+\mu^-)$
is evaluated to be $[1.4, 7.4]\times10^{-10}$, which is a two-sided bound,
and both CMS and LHCb have some hint of events.
But because of possibility of peaking backgrounds and issues of muon quality (see below),
the experiments have not emphasized this so far.\footnote{
ATLAS has released a corresponding measurement~\cite{Aaboud:2016ire},
but with no real excess found. The negative central value of fit has caused concern for some.
}
A new study more optimized towards $B^0 \to \mu^+\mu^-$ is also called for.

%But again, SM reigns, as LHC Run 1 data again demonstrate, which we describe below.
%
Note that the ratio of branching fractions between the two decays also provides
a different test to the New Physics~\cite{Buras:2003td}.
However, theories with the property of minimal flavor violation (MFV),
i.e. flavor violation occurring only in CKM sector, preserve the same value as in SM.
With the measured branching fractions consistent with SM,
these decays can be used as a powerful constraint on New Physics scenarios,
e.g. the three orders of magnitude~\cite{Huang:1998vb, Hamzaoui:1998nu,
 Choudhury:1998ze, Babu:1999hn, Bobeth:2001sq}
``lever arm'' is now fully explored.
The large parameter space of interest can go beyond the reach of LHC collision energy.
This is nicely illustrated by Fig.~2 from Ref.~\cite{Straub:2010ih}.
The ``Straub plot'', which we do not show here, exhibits the correlations between
$B_s\to\mu^+\mu^-$ and ${B^0}\to\mu^+\mu^-$ branching fractions in several different SM extensions,
mostly different SUSY-based flavor models (which are practically outdated),
and the progress of LHC measurements compared with the Tevatron era.
Much parameter space has now been eliminated by direct searches at the LHC experiments,
but the measurements of $B_s\to\mu^+\mu^-$ and ${B^0}\to\mu^+\mu^-$ also
contribute as one of the strongest constraints to date,
and concurs with direct search. For example: no sign of SUSY.

Although the experiments are not yet paying much attention, and
in part because of empirical evidence in support of MFV,\footnote{
Or, MFV is based on such evidence: absence of flavor violation other than from CKM paradigm.
}
it is not easy for an enhancement of $B_d \to \mu^+\mu^-$ as indicated by Eq.~(\ref{eq:Bdmumu}).
A fourth generation~\cite{Hou:2013btm} of quarks can achieve it,
but it is disfavored by the observation of the 125 GeV Higgs boson.
Or one can tune parameters in exotic Higgs representations in SUSY-GUT
framework~\cite{Dutta:2015dla}, but we have no evidence for SUSY.
Very few models can comfortably enhance $B_d \to \mu^-\mu^-$,
where the $B_d$ vs $B_s$ ratio test~\cite{Buras:2003td} provides another avenue.
We note that a fourth generation violates MFV3 (MFV with 3 generations, i.e. SM-like)
but not the principle of MFV.

Let us turn to the actual experimental measurement, which is still a work in progress.
We note that the situation is different from the Tevatron era,
where one was shooting for sizable enhancement above SM,
and the situation for signal versus background is quite different.
Just to illustrate: had the CDF central value of Ref.~\cite{Aaltonen:2013as}
been true in Nature, CMS and LHCb would have easily verified it in 2011 already.
The challenge now is of course the tiny signal from SM rate,
and background reduction is the essential piece of the analysis.
The $B_{s,\,d} \to \mu^+\mu^-$ signal consists of
a pair of opposite sign muons from a single displaced vertex,
with invariant mass in agreement with the nominal $B_{s,\,d}$ mass within resolution.
The combinatorial background, which arises mostly from muons originating from
two different semi-leptonic $B$ meson decays, can be reduced by requiring a good reconstructed vertex,
plus low hadronic activity associated with the reconstructed candidate.
When one or both of the two muon candidates arise from misidentified hadrons,
the situation can be more challenging. In particular,
consider $B$ meson decay to two charged hadrons ($B_{s,d} \to K^+K^-$, $\pi^+\pi^-$ or $K^+\pi^-$),
if both hadrons are misidentified as muons,
a bump very close to the\ signal can be produced (peaking background),
which is difficult to suppress except by requiring more stringent muon reconstruction quality.

Another important aspect of the analysis is how to normalize the resulting decay rate.
A typical choice is the $B^+ \to J/\psi (\to \mu^+\mu^-) K^+$ decay,
which has a large, well-measured, branching fraction,
\begin{equation}
\mathcal{B}(B_{s,d} \to \mu^+\mu^-) = {N_{\rm sig} \over N(B^+ \to J/\psi K^+)} \cdot
\mathcal{B}(B^+ \to J/\psi K^+) \cdot {\epsilon(B^+) \over \epsilon(B_{s,d})}
\cdot {f_u \over f_{s,d}},
\end{equation}
where $N_{\rm sig}$ is the signal yield for $B_{s,d} \to \mu^+\mu^-$,
and $f_u$ and $f_{s,d}$ are respective hadronization fractions.
The detection acceptance times reconstruction efficiency ($\epsilon$)
for $B^+$ and $B_{s,d}$ should be corrected accordingly.
The two muons from $J/\psi$ decays can also be used to reduce
the systematic uncertainties related to muons.
The ratio of $B_s$ and $B^+$ hadronization composition needs to be determined and additional systematic uncertainty must be introduced for $B_{s} \to \mu^+\mu^-$ decay.
If the $B_s \to J/\psi K^+K^-$ absolute branching fraction can be measured better in the near future
(in particular by Belle-II), it can be a good alternative choice of normalization.

A recent update by LHCb~\cite{Aaij:2017vad} that includes both Run-1 and
the newly collected 2016 data confirms the observation of $B_{s} \to \mu^+\mu^-$ decay,
\begin{equation}
\mathcal{B}({B}_{s}\to\mu^+\mu^-) = (3.0\pm0.6^{+0.3}_{-0.2})\times10^{-9}, \quad
\mathcal{B}({B^0}\to\mu^+\mu^-) < 3.4\times10^{-10}, \quad {\rm (LHCb,\ up\ to\ 2016)}
\end{equation}
while the excess of $B^0 \to \mu^+\mu^-$ is reduced with respect to the Run 1 only analysis.
In the same presentation, LHCb also announced the first effective lifetime measurement for ${B}_{s}\to\mu^+\mu^-$, with initial state consisting of only the heavy $B_s$ in SM.

Further investigations are definitely required.
The upcoming LHC data would clearly provide better sensitivity,
but collisions at higher pile-up (number of $pp$ collisions per beam crossing) rate also
degrade the performance, especially for the general purpose CMS experiment.
Furthermore, as the measurement of $B_{s} \to \mu^+\mu^-$ decay is already well established,
the new focus would be to have the $B^0 \to \mu^+\mu^-$ rate measured for the first time.
The $B^0 \to \mu^+\mu^-$ decay suffers more background contamination,
where even $B_{s} \to \mu^+\mu^-$ constitutes a critical background.
An improved dimuon mass resolution is an essential requirement for
separating $B^0$ from $B_{s}$ events at the upcoming analysis updates.
Both CMS and LHCb experiments expect to discover $B^0 \to \mu^+\mu^-$
with HL-LHC operations, if the SM branching fraction is assumed.
However, if the actual rate is larger than SM expectation,
the first evidence for $B^0 \to \mu^+\mu^-$ decay could be found
within the next few years, which makes this program very interesting.
Given the historic interest in $B_s \to \mu^+\mu^-$,
theorists are encouraged to come up with models that
could considerably enhance $B^0 \to \mu^+\mu^-$,
as the experimental situation is not settled yet.

\subsection{%\boldmath
 $P_5'$ and $R_{K^{(*)}}$ anomalies in $B \to K^{(*)} \ell^+\ell^-$
\label{sec:P5'}}

FCNC decays such as $B \to K^{*} \ell^+\ell^-$ are of particular interest since
New Physics phenomena can be observed indirectly through their influence in the loop diagrams.
Many observables can be probed as a function of $q^2 = m^2_{\ell^+\ell^-}$ with a single decay channel,
for example, the differential branching fraction ($d\mathcal{B}/dq^2$),
forward--backward asymmetry of the leptons ($A_{FB}$),
and longitudinal polarization fraction of the $K^{*0}$ ($F_L$).
Theoretical calculations for these observables are suitably robust,
and any difference between experimental measurements and SM predictions
can be interpreted as a hint of New Physics beyond SM
(See, e.g. Refs.~\cite{Ali:1999mm, Buchalla:2000sk, Altmannshofer:2008dz}).
%~\cite{Yan:2000dc,Altmannshofer:2008dz,Ali:1999mm,Buchalla:2000sk,Feldmann:2002iw}.

Take, for example, the forward--backward asymmetry, $A_{FB}$.
With the observation~\cite{Hou:1986ug} that the $Z$-penguin would
actually dominate the $b \to s\ell^+\ell^-$ processes with large $m_t$,
$A_{FB}$ was subsequently pointed out~\cite{Ali:1991is} as an interesting observable.
While one is familiar with $A_{FB}$ measurements at the $Z$ peak,
$A_{FB}$ in $B \to K^{*} \ell^+\ell^-$ is in fact closer to the classic measurement
of Ref.~\cite{Prescott:1978tm} that confirmed electroweak theory, only better:
the enhanced virtual $Z$ effect is brought down to below $m_B^2$
to interfere with the virtual photon effect, also loop-induced,
and the resulting $A_{FB}$ can be measured within $0 < q^2 < m_B^2$.

Measurement of $A_{FB}$ was first done at the B factories~\cite{PDG},
but when Belle announced~\cite{Wei:2009zv} a 2.7$\sigma$ deviation from SM expectation
in 2009, it caused some sensation.
The deviation was mostly in the low $q^2$ region,
where the SM prediction of $A_{FB}$ crosses from positive to negative values around $q^2 = 4$~GeV$^2$.
The $A_{FB}$ measured by Belle and BaBar tend to be positive for all $q^2$.
This ``anomaly'' was very difficult to accommodate within SM, since
the predicted $q^2$ region of zero-crossing is quite robust.
However, the anomaly was \emph{not} confirmed by new LHCb measurement with much larger statistics,
and neither by subsequent CMS analysis.
Current measurements of $A_{FB}$ are consistent with SM expectations.

\begin{figure}[ht]
\begin{center}
\includegraphics[width=10cm]{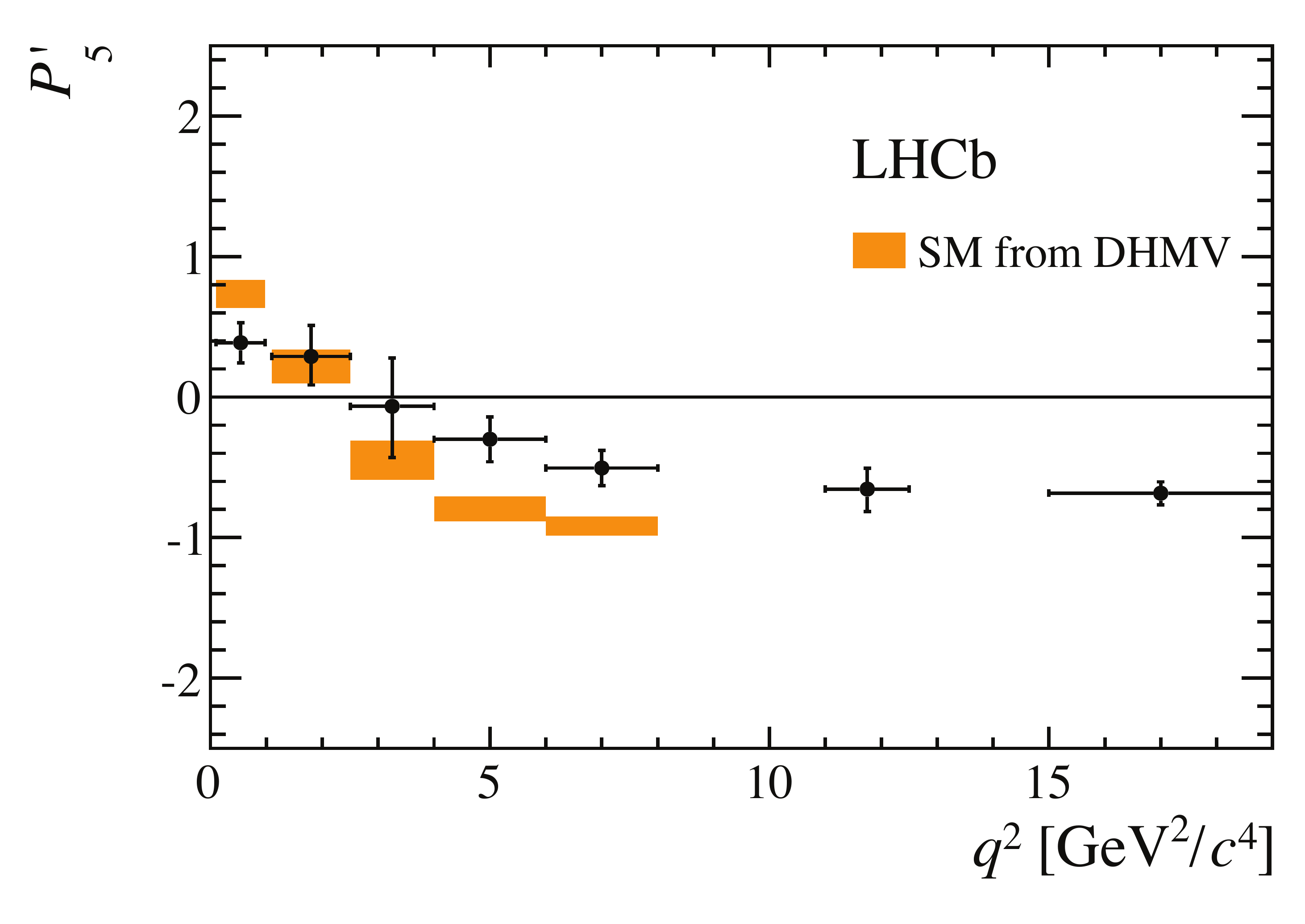}
\caption{
The $P^\prime_5$ angular observable in bins of $q^2$ from LHCb Run 1 data,
where some deviation is seen from SM-based predictions (shaded boxes) for 4 GeV$^2 < q^2 < 8$ GeV$^2$.
See text for discussion.
[Source: courtesy LHCb experiment, from Ref.~\cite{Aaij:2015oid}]
}
\label{fig:lhcb_p5prime}
\end{center}
\end{figure}

But this is definitely not the end of the story, as the $B \to K^*\ell^+\ell^-$
process is rich with observables.
To reduce uncertainties from form factors, several new observables were proposed,
such as those denoted by $P^\prime_{n}$~\cite{Descotes-Genon:2013vna}
 %, $P^\prime_5$, $P^\prime_6$ and $P^\prime_8$
which are claimed to be largely free from form factor uncertainties in the low-$q^2$ region.
Using these observables in the analysis, LHCb found~\cite{Aaij:2015oid} some
new discrepancies between measurements and expectations
in some $q^2$ bins in the $P^\prime_5$ observable,
as can be seen from Fig.~\ref{fig:lhcb_p5prime}.

%\vskip0.3cm

The decay has four charged particles in the final state, $B^0 \to K^{*0} \mu^+\mu^- \to K^+\pi^-\mu^+\mu^-$. A full description of the decay requires three angular observables:
the helicity angle $\theta_K$ of the $K^{*0}$ candidate,
the helicity angle $\theta_\ell$ of the dimuon system, and
the angle $\phi$ between the $K^{*0} \to K^+\pi^-$ and dimuon planes.
The CP-averaged differential angular distribution can be expressed as,
\begin{eqnarray}
{1 \over d\Gamma/dq^2}
{d^4\Gamma \over dq^2 d\Theta}&=& {9\over 32\pi}\bigg[
   {3\over 4}(1-F_L) \sin^2\theta_K + F_L \cos^2\theta_K
  +{1\over 4}(1-F_L)\sin^2\theta_K\cos2\theta_\ell \nonumber \\
&&- F_L\cos^2\theta_K\cos2\theta_\ell + S_3 \sin^2\theta_K \sin^2\theta_\ell\cos2\phi
  + S_4 \sin2\theta_K \sin2\phi_\ell\cos\phi \nonumber \\
&&+ S_5 \sin2\theta_K \sin\theta_\ell\cos\phi +{4\over 3} A_{FB} \sin^2\theta_K \cos\theta_\ell
  + S_7 \sin2\theta_K \sin\theta_\ell\sin\phi \nonumber \\
&&+ S_8 \sin2\theta_K \sin2\theta_\ell\sin\phi + S_9\sin^2\theta_K\sin^2\theta_\ell\sin2\phi\bigg],
\end{eqnarray}
where $S_n$ ($n=3, 4, 5, 7, 8, 9$) are angular observables from the %bilinear combinations of the
$B^0 \to K^{*0} \mu^+\mu^-$ decay amplitude,
which are generally functions of the Wilson coefficients and form factors,
hence functions of $q^2$ as well.
The $P^\prime_n$ observables are defined by combining $F_L$ and $S_n$,
with reduced form-factor uncertainties:
\begin{equation}
P^\prime_{n} = {S_{n} \over \sqrt{F_L(1-F_L)}}.
 %{\rm and}~~~P^\prime_{6} = {S_{7} \over \sqrt{F_L(1-F_L)}}~.
\end{equation}

A full angular analysis of $B^0 \to K^{*0} \mu^+\mu^-$ decay is required
to determine these observables of interest.
Contamination from a possible S-wave component and
the associated interference should be taken into account.
LHCb carried out the measurements with two different statistical methods,
likelihood fit and principle moment analysis, and reported~\cite{Aaij:2015oid}
the $q^2$-dependent angular observables for full Run 1 data, where
the measured $P^\prime_5$ as a function of $q^2$ is already shown in Fig.~\ref{fig:lhcb_p5prime}.
There is some deviation between measured values and the SM prediction,
denoted as DHMV~\cite{Descotes-Genon:2014uoa}, in the region of $4 < q^2 < 8$~GeV$^2$.
The significance is at 3.4$\sigma$ level,
although the significance of deviation was actually slightly higher in
an earlier LHCb publication with 1 fb$^{-1}$ data, which is a cautionary note.
A further caution is that, although this deviation could well be due to New Physics,
it could still plausibly be explained by introducing large hadronic effects.
For example, subsequent to the LHCb result,
it was pointed out~\cite{Lyon:2014hpa, Jager:2014rwa, Ciuchini:2015qxb}
that hadronic effects are very difficult to estimate around the charm threshold,
$q^2 \cong 4m_c^2 \simeq 6.8$ GeV$^2$, precisely the region of deviation indicated by LHCb.

\vskip0.3cm
\begin{figure}[ht]
\begin{center}
\includegraphics[width=7cm]{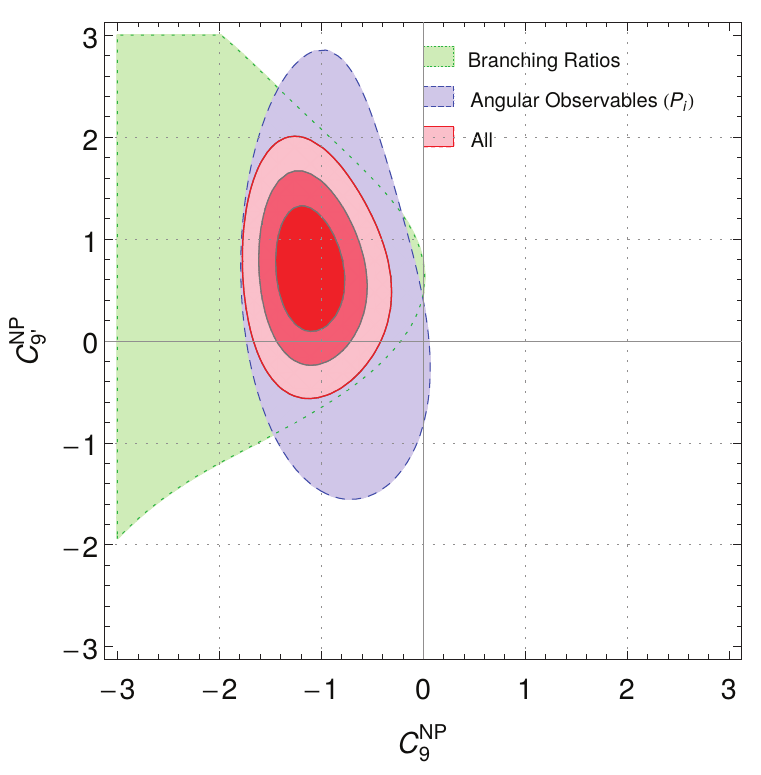}
\caption{
Combined fit to $B^0 \to K^{*0}\mu^+\mu^-$ angular and rate observables illustrates
a possible New Physics shift in the $C_9$ coefficient by $-1$,
and some shift in other Wilson coefficients.
[Source: courtesy S. Descotes-Genon and J. Matias, from Ref.~\cite{Descotes-Genon:2015uva}]
}
\label{fig:C9_contour}
\end{center}
\end{figure}

While we prefer to wait for Run 2 data to pan out, the $P_5'$ anomaly is,
however, well known and popular.
A representative global fit~\cite{Descotes-Genon:2015uva} to New Physics,
taking into account many measurements is given in Fig.~\ref{fig:C9_contour}.
An order $-1$ shift in the $C_9$ Wilson coefficient of one of
the $Z$-penguin operators
\be
   O_9 = (\bar s\gamma_\mu Lb)(\bar \ell\gamma_\mu\ell),
 \label{eq:O9}
\ee
seems commonly needed, suggesting the presence of an extra $Z'$ boson.
The point was first made in Refs.~\cite{Descotes-Genon:2013wba} and \cite{Altmannshofer:2013foa},
and subsequently by many authors.
A more recent fit can be found in Ref.~\cite{Altmannshofer:2017fio}.

Belle has updated~\cite{Wehle:2016yoi} the analysis for both $B^0 \to K^{*0} \mu^+\mu^-$
and $K^{*0} e^+e^-$ modes with full data, using the same approach as LHCb.
The result is consistent with LHCb, and a mild discrepancy of
2.6$\sigma$ is found in a similar $q^2$ region for the muon channel.
However, the electron channel shows better agreement with SM expectations.
Very recently, ATLAS and CMS have also measured the $P^\prime_5$ observable.
Interestingly, CMS finds~\cite{CMS:2017ivg} better agreement with DHMV predictions
with no real deviation in the target $q^2$ region, hence is at some variance
with LHCb (and Belle).
%, DHMV of Ref.~\cite{Descotes-Genon:2014uoa}, and what was called HEPfit,
%which is a fit in Ref.~\cite{Ciuchini:2015qxb}) to LHCb results with hadronic effects.
% while understandable LHCb agrees with HEPfit, but so does Belle data.
The ATLAS result~\cite{ATLASp5prime} is consistent with LHCb
in the $4 < q^2 < 6$~GeV$^2$ bin and away from DHMV, but does not give results for higher $q^2$.
%also agrees with DHMV better, and is slightly lower than the LHCb results.
Currently, LHCb has the best statistical uncertainty,
but given the scatter of measured values, and that statistical errors are still quite large,
together with the issue of hadronic uncertainties, it seems
premature to make a conclusion at this moment.
The upcoming LHC and Belle-II data, together with improved SM predictions,
should shed further light on this intriguing search.

On a slightly different front, as a test of lepton universality, LHCb measured
the ratio of branching fractions of $B^+ \to K^+ \mu^+\mu^-$ vs $K^+ e^+e^-$ decays~\cite{Aaij:2014ora},
in the range of $1 < q^2 < 6$ GeV$^2$. The ratio, denoted by $R_K$, is determined to be
\begin{equation}
R_K = 0.745^{+0.090}_{-0.074} \pm0.036. \quad {\rm (LHCb\ Run\ 1},\ 1 < q^2 < 6\ {\rm GeV}^2)
 \label{eq:RK}
\end{equation}
which is slightly below the SM expectation of unity, but still compatible within 2.6$\sigma$.
This is echoed by the mild deviation of angular distributions between
$B^0 \to K^{*0} \mu^+\mu^-$ and $K^{*0} e^+e^-$ modes found by Belle~\cite{Wehle:2016yoi}.
What has excited many theorists is the recent announcement by LHCb the
measurement of $B^+ \to K^{*0} \mu^+\mu^-$ vs $K^{*0} e^+e^-$ decays~\cite{Aaij:2017vbb},
\begin{equation}
R_{K^{*0}} = 0.69^{+0.11}_{-0.07} \pm 0.05, \quad {\rm (LHCb\ Run\ 1},\ 1.1 < q^2 < 6\ {\rm GeV}^2)
 \label{eq:RKst}
\end{equation}
which is in good agreement with Eq.~(\ref{eq:RK}), and is by itself
of order 2.5$\sigma$ from SM. Together they offer strong support for
the aforementioned shift in $C_9$.
LHCb, however, also gave the number
\begin{equation}
R_{K^{*0}} = 0.66^{+0.11}_{-0.07} \pm 0.03, \quad {\rm (LHCb\ Run\ 1},\ 0.045 < q^2 < 1.1\ {\rm GeV}^2)
 \label{eq:RKst-1}
\end{equation}
which is slightly lower but rather similar to Eq.~(\ref{eq:RK}),
and lower than SM by $\sim 2.2\sigma$.
But this poses some puzzle~\cite{Altmannshofer:2017yso}
because the lower bin is dominated by $B\to K^{*0}\gamma^*$
that peaks strongly towards small $q^2$,
with a well measured and much larger $B\to K^{*0}\gamma$ parent
that should not (QED) distinguish between $e^+e^-$ and $\mu^+\mu^-$.
LHCb probably aimed at using this lower bin\footnote{
The electromagnetic penguin does not contribute at low $q^2$
to $B \to K\ell^+\ell^-$.
}
as ``validation'', but instead itself showed some disagreement
with SM.

Further investigations are clearly needed, and improved measurements from LHCb
and Belle II are expected in the near future.
Given that several anomalies are dominated by LHCb,
a second experiment like Belle II is especially called for
to provide cross check.

\subsection{%\boldmath
 BaBar anomaly: $B \to D^{(*)}\tau\nu$ %(and $B \to \tau\nu$)}
 \label{sec:babar-A}}

\begin{figure}[ht]
\begin{center}
\vskip0.3cm
\includegraphics[width=10cm]{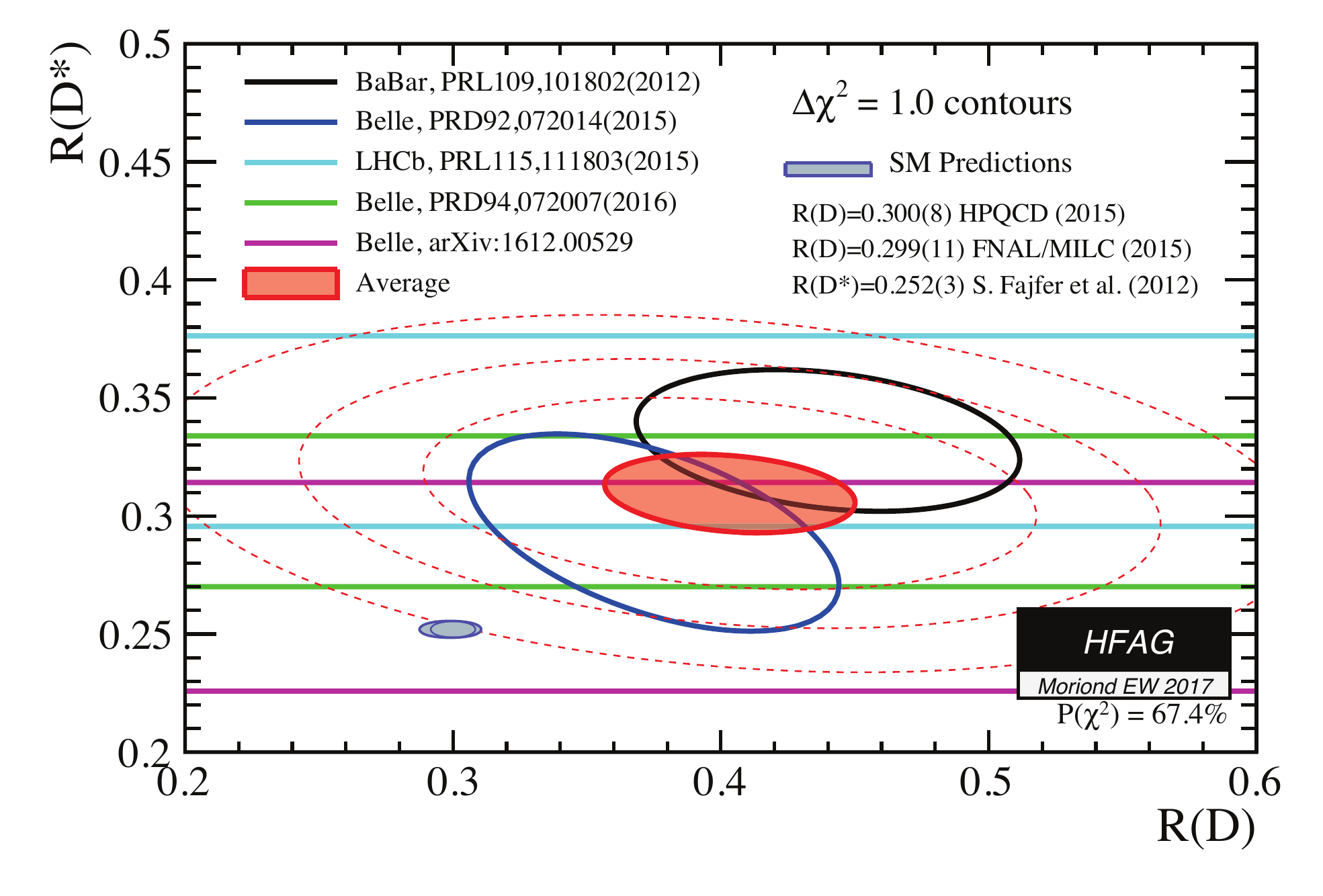}
\vskip-0.1cm
\caption{
Current status of $R_{D}$ and $R_{D^*}$ measurements, illustrating the ``BaBar anomaly''
in $B \to D^{(*)}\tau\nu$ decays.
The average value is shown as a filled ellipse, together with associated uncertainty contours
(dashed), which is apparently $4\sigma$ away from SM expectations.
[Source: HFAG Winter 2017]
}
\label{fig:rdrds}
\end{center}
\end{figure}

The measurement of $B \to \tau\nu$ decay is a highlight of the B factories,
but by the time of LHC running, it had quieted down, and we defer the
discussion to the next section.
However, application of the techniques developed for studying $B \to \tau\nu$
lead to a new anomaly~\cite{Lees:2012xj} uncovered by BaBar in 2012,
that the rates and differential distributions
in $B \to D^{(*)}\tau\nu$ decays seem to deviate not only from SM,
but cannot be understood by charged Higgs boson ($H^+$) effects from
the usual 2HDM-II that arises under SUSY.
This has caused a lot of interest, both experimental and theoretical,
and the current measurement status is summarized in Fig.~\ref{fig:rdrds}.

The $\tau$ leptons appearing in $B$ decay final states were seldom addressed
before the B factory era because of complexity of detection and reconstruction,
since there are multiple invisible neutrinos\footnote{
In the ARGUS and CLEO era, techniques were developed for detecting
a single missing neutrino: missing energy/momentum but vanishing missing mass.
These techniques were used to measure semileptonic $B$ decays
such as $B \to D^{(*)}e\nu$ and $D^{(*)}\mu\nu$.
}
in the final state.
%
%Recent studies on the leptonic $B \to \tau\nu$ and the semi-leptonic $B \to D^{(*)}\tau\nu$ decays
%show some interesting hints of deviation from the SM and induced many activities in the field.
%The predictions of these decays are rather robust in the SM and any deviation from the calculation
%can be interpreted as the contributions from new, not-yet-discovered, physics phenomena.
%In particular, if an enhancement found in the decay branching fractions of $B \to \tau\nu$ or
%$D^{(*)}\tau\nu$, but not in the decay involving other two lepton flavors, $B \to (e,\mu)\nu$ or $D^{(*)}(e,\mu)\nu$, a charged Higgs boson is an excellent candidate process which can contribute.
%On the other hand, the measurements of $B \to \tau\nu$ and $B \to D^{(*)}\tau\nu$ can be used to
%constrain the charged Higgs model itself, which is complementary to the direct searches at the
%high energy colliders.
%
But as described in Sec.~4, $B \to \tau\nu$ emerged as an excellent probe of $H^+$,
and with the large statistics expected at the B factories, Belle and BaBar developed tools
to probe such decays.
The B factories rely on the coherent production of $B$ meson pairs from
the mother $\Upsilon(4S)$ state.
``Tagging" one of the $B$ mesons by full reconstruction with either hadronic or semileptonic decays,
the kinematics of the other $B$ meson becomes known,
which makes it possible to search for signals such as $B \to \tau\nu$.
Due to the good coverage of B factory detectors, few missing particles are expected,
and the signal can be selected and identified with little extra measured energy,
after subtracting off the tagged $B$ meson candidate and the signal $\tau$ lepton.
A hadronic tag measures full final state four momentum of the tagged $B$ meson
and gives better identification, but suffers from low reconstruction/tagging efficiency.
On the other hand, the semileptonic decay tag requires a $D^{(*)}$ meson
and another charged lepton within an event hence has much higher efficiency,
but due to the presence of an additional neutrino on tag side,
the kinematic constraints are slightly weaker, resulting in a higher background level.
The measurement of $B \to \tau\nu$ (as well as $B \to \mu\nu$) 
would be discussed in the next section.

The semileptonic $B \to D\tau\nu$, $D^{*}\tau\nu$ decays proceed though $b \to c$ transitions,
and with branching fractions expected at the couple percent level,
they are technically not really rare decays.
But the difficulty of detection and the lack of anticipation for any New Physics
delayed the study of these decays until late in the B factory era.
One can measure the differential partial width in terms of $q^2$,
the squared invariant mass of the $\tau$ lepton and neutrino.
The lower bound of $q^2$ depends on the $\tau$ mass, while the upper bound
is decided by the mass difference between the $B$ and $D^{(*)}$ mesons.
By use of ratios of branching fractions, the hadronic uncertainties
as well as experimental errors are reduced.
The dependence on the CKM element $V_{cb}$ and other common factors also cancel out.
The most up-to-date predictions~\cite{Fajfer:2012vx, Bigi:2016mdz}
of the ratios, $R_D$ and $R_{D^*}$, are quite accurate,\footnote{
The more recent Ref.~\cite{Bernlochner:2017jka}
gives $R_{D^*} = 0.257 \pm 0.003$, which is slightly higher.
}
\bea
R_D^{\rm SM} &=& {\mathcal{B}(B \to D \tau \nu) \over \mathcal{B}(B \to D \ell \nu)}
  = 0.299 \pm 0.003,
 \label{eq:RD-SM} \\
R_{D^*}^{\rm SM} &=& {\mathcal{B}(B \to D^* \tau \nu) \over \mathcal{B}(B \to D^* \ell \nu)}
  = 0.252 \pm 0.003,
 \label{eq:RDst-SM}
\eea
where $\ell = e,\ \mu$.
The experimental measurements of $R_D$ and $R_{D^*}$ rely on studies of
the signal $B \to D^{(*)}\tau\nu$ and the normalization channels $B \to D^{(*)}\ell\nu$.

The analyses carried out by BaBar~\cite{Lees:2012xj, Lees:2013uzd}
and Belle~\cite{Huschle:2015rga, Sato:2016svk} are similar to the studies for $B \to \tau\nu$
described above, with one $B$ meson tagged with hadronic or semileptonic decays,
but the signal $B$ meson is reconstructed with a charged electron or muon, plus a $D^{(*)}$ meson.
The signal $\tau$ decay $\tau \to \ell \overline{\nu}_{\ell} \nu_\tau$
gives exactly the same visible final state particles as the normalization channel.
The two additional neutrinos from $\tau$ decays lead to a wider missing mass distribution
calculated by all of the visible particles seen in the event,
while the normalization channel has nearly zero missing mass since there is only one neutrino.
A very recent Belle measurement~\cite{Hirose:2016wfn} of $R_{D^*}$
also looks into hadronic $\tau$ decays.
Interestingly, LHCb is able to measure $\overline B \to D^{*+}\tau\nu$  as well.
Although the forward hadron collider environment is not as clean as at the B factories,
LHCb reconstructs the $B$ meson with a $D^{*+}\ (\to D^0 \pi^+)$ meson plus a muon candidate.
The feature of a highly boosted $B$ meson is utilized by requiring a significant
distance between the $B$ candidate decay vertex and the proton-proton collision point.
Given the missing neutrinos, the momentum of $B$ candidate is in fact unknown,
but can be calculated in an approximate way~\cite{Aaij:2015yra}.

The HFAG average of all available experimental measurements of $R_D$ and $R_{D^*}$
is already given in Fig.~\ref{fig:rdrds}. Open ellipses indicate BaBar and
Belle analyses with hadronic tag, while other measurements are given by horizontal lines.
The average value is shown as a filled ellipse,
together with the associated uncertainty contours, which clearly exceeds the SM prediction.
The numerical values together with their uncertainties are given by
\bea
R_D &=& 0.403 \pm 0.040 \pm 0.024,
 \label{eq:RD} \\
R_{D^*} &=& 0.310 \pm 0.015 \pm 0.008.
 \label{eq:RDst}
\eea
Deviations of $2.2\sigma$ and $3.4\sigma$ from SM predictions
(Eqs.~(\ref{eq:RD-SM}) and (\ref{eq:RDst-SM})) are found for $R_D$ and $R_{D^*}$, respectively.
Considered together, however, a deviation of $3.9\sigma$ is estimated
with correlations taken into account.

Could the deviation be due to a statistical fluctuation? More data is needed.
But the deviations of $R_D$ and $R_{D^*}$ are enthusiastically embraced by theorists,
and the leading candidate New Physics are two Higgs doublet models
beyond~\cite{Crivellin:2012ye, Fajfer:2012jt} 2HDM-II,
and ``flavored'' leptoquarks (see e.g. Ref.~\cite{Sakaki:2013bfa}).
Note that leptoquarks are also popular for explaining the $P_5'$ and $R_K$ anomalies
(see e.g. Refs.~\cite{Hiller:2014ula} and \cite{Gripaios:2014tna}).
Both leptoquark and charged Higgs explanations are challenged~\cite{Faroughy:2016osc}
by direct search for $\tau$ pairs at high $p_T$,
while charged Higgs effects to explain the BaBar anomaly
tend to cause problems~\cite{Li:2016vvp, Alonso:2016oyd} with $B_c$ lifetime,
or the $B_c \to \tau\nu$ rate.
We cannot do justice to the theory activity, but our view is that
more data, especially from Belle II and LHCb,\footnote{
A result from LHCb~\cite{LHCb-had-tau} for FPCP 2017 using
$\tau^\pm \to \pi^\pm\pi^+\pi^-(\pi^0)\nu$
gives $R_{D^*} = 0.285 \pm 0.019 \pm 0.029$,
which is lower in value than using the muonic decay of $\tau$,
and the HFLAV (previously HFAG) combination is now
$R_{D^*} = 0.304 \pm 0.013 \pm 0.007$, compared with Eq.~(\ref{eq:RDst}).
The ``BaBar anomaly'' is still quite volatile on experimental side.
}
is needed to clarify the situation, and that flavored leptoquarks
might be more exotic than $H^+$ beyond 2HDM-II
(we will come back to this in Sec.~\ref{sec:S2HDM}).

If the enhancement on the $B \to D^{(*)}\tau\nu$ decay rate is really due to some
new particle (or effective operators or operator coefficients that differ from SM),
it should modify also the kinematics of the decay itself.
Hence a further detailed study of the momentum distribution for the final state particles
may provide additional discriminant power.
A $q^2$ dependent measurement of the ratios can indicate how the model should be modified.
These extended studies will require much more statistics, which fortunately
would become available in the not so distant future.
We note in passing that $Z$ and $\tau$ decays also lead to
strong constraints on models that try to explain $R_D$ and $R_{D^*}$.
See e.g. Ref.~\cite{Feruglio:2016gvd}.

\subsection{CPV in Three-body B Decays}

We have emphasized in Sec.~2 the importance of 3-body Dalitz analysis as part of the ``$DK$'' program
for extracting the CPV phase angle $\phi_3/\gamma$. In this Highlight section,
as an anticlimax, let us discuss CPV in the 3-body phase space of charmless $B$ decays
as a highlight of sorts, but to illustrate the depressing nature of the ``hadronic menace''
over our goal of uncovering New Physics.

CPV rate difference arises when two interfering amplitudes have
difference in both CP conserving and CP violating phase.
The former usually arise from hadronic or strong interactions,
while in SM the latter arise from the CKM matrix.
We have mentioned in the Introduction that the large DCPV difference (more than 10\%)
between $B^+ \to K^+\pi^0$ and $B^0 \to K^+\pi^-$, Eq.~(\ref{eq:DAKpi}),
could arise from a New Physics CPV phase in the $Z$ penguin.
But this hope was dashed by no indication of such phase in the
accompanying $B_s$ mixing CPV phase, $\phi_s$ (Eq.~(\ref{eq:phis}), compared with Eq.~(\ref{eq:phis_SM})),
and thus one would have to accept that the $B \to K\pi$ DCPV difference is
due to the interference of a hadronically enhanced (hence hadronic phase)
color-suppressed amplitude $C$ with the dominant strong penguin amplitude $P$.

\vskip0.3cm
\begin{figure}[ht]
\begin{center}
\includegraphics[width=9.5cm]{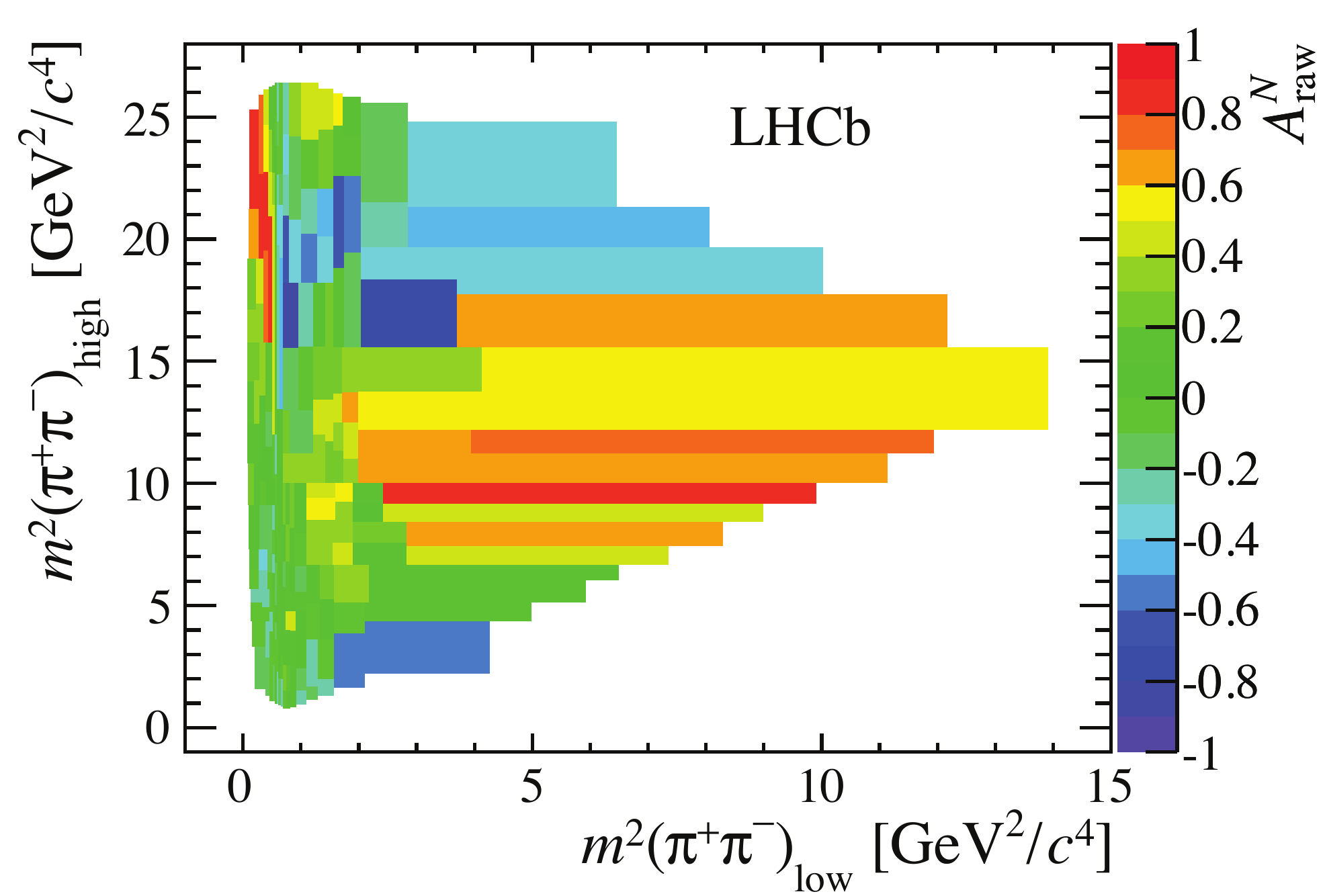}
\caption{
The measured raw asymmetry in $B^+ \to \pi^+\pi^-\pi^+$ decay Dalitz space,
%(left) and the example invariant mass distribution of the $B^+ \to \pi^+\pi^-\pi^+$
%candidates restricted to a local sector (right).
where DCPV asymmetries can reach $O(1)$.
[Source: courtesy LHCb collaboration, from Ref.~\cite{Aaij:2014iva}]
}
\label{fig:3body_cpv}
\end{center}
\end{figure}

LHCb has performed precision analysis of three body charmless
$B^+ \to K^+\pi^-\pi^+$, $K^+K^-K^+$, $\pi^+\pi^-\pi^+$ and $\pi^+ K^+K^-$ decays
with Run 1 data and found significant direct CP asymmetries~\cite{Aaij:2014iva}.
For example, the integrated asymmetry is about 6\% for $B^+ \to \pi^+\pi^-\pi^+$
with 4.3$\sigma$ significance.
But the CPV asymmetries can be measured in the Dalitz plot,
and huge local asymmetries are uncovered.
As shown in Fig.~\ref{fig:3body_cpv}, the local asymmetries in $B^+ \to \pi^+\pi^-\pi^+$ decays
can be over $80\%$, and change rapidly across the Dalitz plot.
As data accumulate with time, these colorful (when seen offline) plots could
turn into an artist's palette,
but unfortunately such large direct CPV asymmetries can be readily accommodated within SM.
The tree level $b \to u\bar ud$ process carries the CPV phase,
but the strong penguin can rescatter the CKM favored $b \to c\bar cd$ process to charmless final states
and bring in an amplitude of comparable strength across Dalitz space, and carry varying strong phase.
Furthermore, presumably there are resonance interference and final state rescattering effects
between all kinds of charmless final states.
Thus, even if there is an effect from some New Physics,
it will be fully enveloped by these hadronic effects,
and there is no way to disentangle them. Beauty may not tell the truth.

%\hskip0.65cm  $\phi_s$; \quad  $B_q \to \mu^+\mu^-$;
%        \quad $B \to K^{(*)}\ell^+\ell^-$; \quad $B \to D^{(*)}\tau\nu$, $\tau\nu$; \quad  (3-body CPV)

\section{Beauty and Dark}

In this section we start with some charmless hadronic rare $B$ decays that
remain of interest, then turn to various processes that provide probes to
New Physics beyond the SM, some of which provide stringent constraints.
We also briefly cover the search for dark particles at the B factories.

\subsection{%\boldmath
 Charmless Hadronic %$B$ Decays
\label{sec:charmless}}

One curiosity before the B factory era was the absence of
$B$ decays to charmless baryonic final states, compared with e.g. $B \to K\pi$, $\pi\pi$ decays.
This has to do with the effective four-fermion interaction structure of weak decay,
while the composition of baryons is more complicated than mesons.
It was suggested~\cite{Hou:2000bz} that, because of extra suppression of two body modes,
perhaps charmless baryonic decays would first emerge in 3-body final states.
Sure enough, the first observation of $B$ meson decay to charmless baryonic final states
was the $p\bar pK^+$ mode discovered by Belle~\cite{Abe:2002ds} in 2002,
which opened a new subfield of $B$ physics with its interesting topics of study.
Despite search efforts at the B factories~\cite{PDG}, it was not until the LHC era
and with the huge statistics of LHCb, that first evidence for two-body
$B \to p\bar p$~\cite{Aaij:2013fta} and $p\bar \Lambda$~\cite{Aaij:2016xfa} decays
start to emerge at the $10^{-8}$ and $10^{-7}$ level, respectively.
The $B$ meson is the only meson that can have baryons in its decay final state.
This notion led to the suggestion~\cite{Hou:2005iu} to search for baryon number violation (BNV)
in heavy meson or fermion decays, but it was realized that these were
already highly constrained by the extremely strong bounds on proton decay.\footnote{
This did not stop, for example, the search for BNV in top decays by CMS~\cite{Chatrchyan:2013bba}.
}

Hadronic uncertainties make it difficult to examine hadronic $B$ decays
for physics beyond SM, as we have seen for the $B \to K\pi$ direct CPV difference.
%Therefore, observables suitable for new physics search are often ratios,
%where some of the hadronic uncertainties may cancel.
There are two CPV variables that meet the purpose better.
The first is $S_f$ in time dependent $CP$ violation analysis as discussed in Sec.~\ref{sec:phi1}.
The $S_f$ values in the tree-dominated $b\to c{\overline c} s$ processes
and the penguin dominated $b\to s q{\overline q}$ processes are expected to be very similar.
Any significant deviation ($\Delta S$) from the mean of $b\to c{\overline c} s$ processes
may indicate New Physics phase in the loop.

%\vskip-0.3cm
\begin{figure}[t]
\begin{center}
\includegraphics[width=9.5cm]{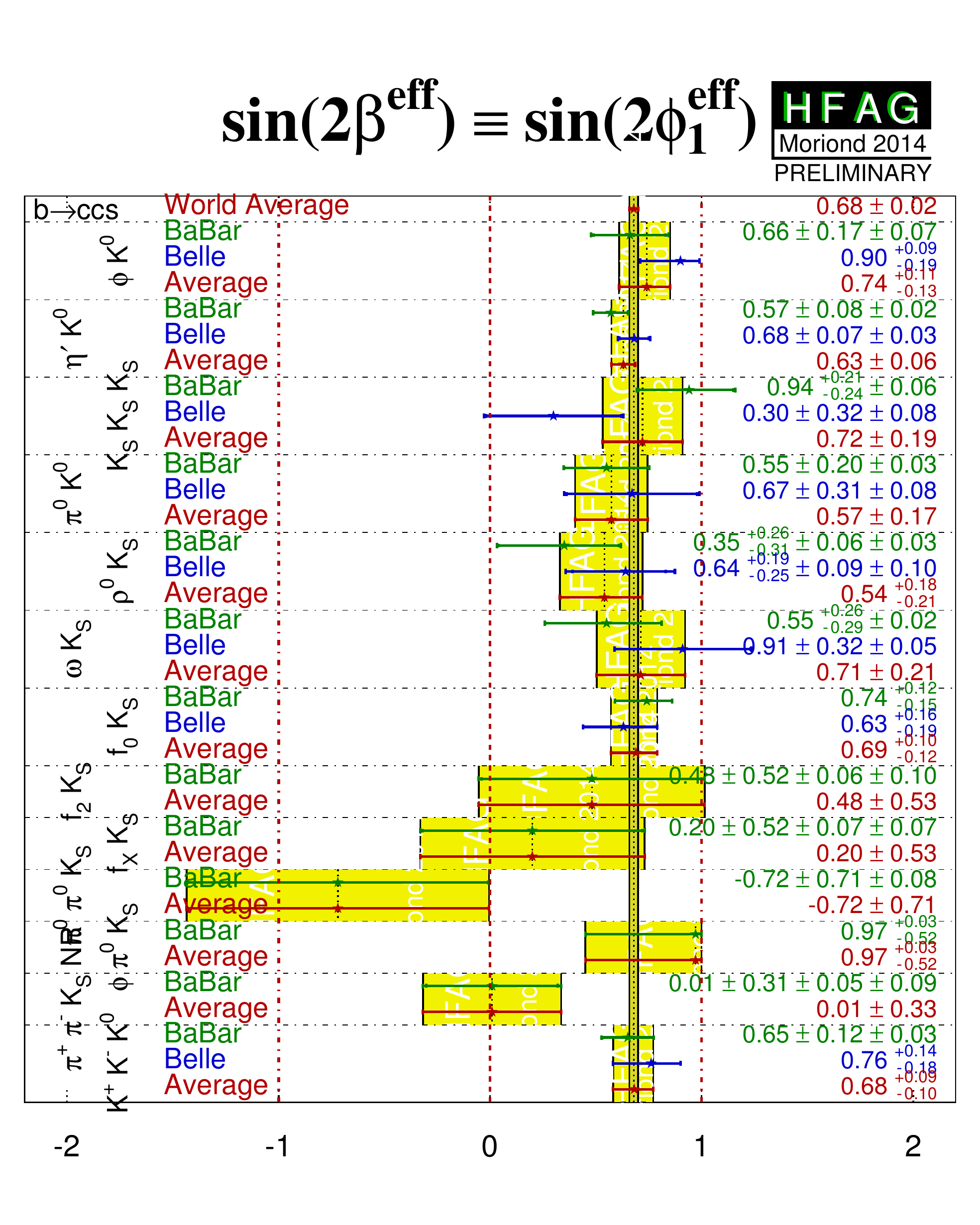}
\end{center}
\vskip-0.5cm
\caption{
The $S$ values for decay modes related to $b\to s$ penguin,
superimposed with the well measured $\sin2\phi_1$ from $b \to c\bar cs$ modes
(narrow vertical band). [Source: HFAG Winter 2014]
\label{fig:dspeng}
}
\end{figure}

Among the various $b\to s$ penguin decay modes, two channels are the most promising:
$B^0\to \phi K^0$ and $B^0\to \eta^\prime K^0$.
The former is from  pure $b \to s\bar ss$ penguin transition,
while the latter receives tree contributions but provides the largest rate.
The current experimental averages from BaBar and Belle are
$S_{\phi K^0} = 0.74^{+11}_{-13}$ and $S_{\eta^\prime K^0} = 0.65\pm 0.06$~\cite{hfag}, respectively,
consistent with $\sin2\phi_1 = 0.691\pm 0.017$ determined in charmonium plus $K^0$ modes.
The $C_f$ values in the three cases are all consistent with zero,
showing no direct $CP$ violation. %, hence reduced hadronic uncertainties.
The study of such $\Delta S$ values became one of the major analysis subjects
at the B factories since 2003, and especially when Belle
experienced a sign flip of $S_{\phi K^0}$ in the summer update of 2004.
Fig.~\ref{fig:dspeng} shows the $S$ values measured from various decays
that receive contributions from $b\to s$ penguins.
The naive average from all $b \to sq\bar q$ decay modes gives~\cite{hfag}
\be
S(b\to sq\bar q) = 0.68^{+0.09}_{-0.10},
\ee
which is in good agreement with Eq.~(\ref{eq_phi1}).
It remains one of the clean channels to probe for New Physics in the future.
One would need at least an order of magnitude more data to tell if
$\Delta S$ really deviates from zero.
The verdict will be reached by Belle~II, with the help of LHCb, in the 2020s.

The second possible approach to probe New physics in charmless hadronic $B$
decays is still through the DCPV asymmetries such as in $B\to K\pi$ modes.
As we have seen in Eq.~(\ref{eq:DAKpi}), large DCPV difference between
the neutral and charged $B$ decays is experimentally well established,
which is the difference between
$A_{CP}(K^+\pi^-) = -0.082\pm 0.006$ for $B^0\to K^+\pi^-$ (Eq.~(\ref{eq:AKpi})) and
$A_{CP}(K^+\pi^0) = 0.040\pm 0.021$ for $B^+\to K^+\pi^0$.
The decay $B^+\to K^+\pi^-$ is expected to proceed through
external $W$-emission tree diagram $T$ and gluon mediated penguin diagram $P$,
as illustrated in Figs.~\ref{fig:BtoKpi}(a) and (b).
These diagrams also govern $B^+\to K^+\pi^0$ decay, but extra diagrams called
color-suppressed tree $C$ and electroweak penguin $P_{\rm EW}$,
Figs.~\ref{fig:BtoKpi}(c) and (d), contribute in addition.
The electroweak penguin effect is rather small in SM,
and furthermore, it does not bring in any CPV phase.
At the start of the millennium, it was generally thought that
the magnitude of $C$ is small compared to $T$ (recall the $1/N_C$ expansion of QCD),
and carry the same CPV phase, so it was expected that the $CP$ asymmetries for $
B^+ \to K^+\pi^0$ and $B^0 \to K^+\pi^-$ decays should be rather similar,
i.e. the naive expectation was $\Delta A_{K\pi} \simeq 0$.
With the observation of a distinctively large difference, Eq.~(\ref{eq:DAKpi}),
the explanation can come from either large color-suppressed tree through hadronic effects,
or from large electroweak penguin effect with New Physics CPV phase, which would be more exciting.
although the electroweak, or $Z$, penguin and the box diagrams for $B_s$--$\overline B_s$ mixing
should probe similar physics. But this reasoning is not foolproof.

One way to test which explanation is correct is to
check a CPV asymmetry sum rule~\cite{sumrule_gronau, sumrule_soni},
\be
A_{K^+\pi^-} + A_{K^0\pi^+}\frac{{\cal B}(K^0\pi^+)}{{\cal B}(K^+\pi^-)}
\frac{\tau_0}{\tau_+} = A_{K^+\pi^0} \frac{2{\cal B}(K^+\pi^0)}{{\cal B}(K^+\pi^-)}\frac{\tau_0}{\tau_+} +A_{K^0\pi^0}\frac{2{\cal B}(K^0\pi^0)}{{\cal B}(K^+\pi^-)},
 \label{eq:acprule}
\ee
where $\tau_0$ and $\tau_+$ are $B^0$ and $B^+$ lifetimes, and
${\cal B}(K\pi)$ is the branching fraction of $B\to K\pi$.
What is missing is a precision measurement of $A_{K^0\pi^0}$.

The branching fractions of all $K\pi$ decays will become precision measurements in the near future.
The asymmetry $A_{K^0\pi^+}$ is expected to be zero due to its pure penguin nature,
and the current combined result, $A_{K^0\pi^+}= -0.017\pm 0.016$~\cite{hfag} supports the SM  expectation.
The crucial information for checking the sum rule of Eq.~(\ref{eq:acprule}) is $A_{K^0\pi^0}$,
where time-dependent $CP$ asymmetry analysis is needed.
The average asymmetry from BaBar and Belle is
\be
A_{K^0\pi^0}=0.01\pm 0.10,
 \label{eq:AK0pi0}
\ee
where the statistical uncertainty is around 0.13 for each experiment.
If we assume Eq.~(\ref{eq:acprule}) holds and $A_{K^0\pi^+}= 0$,
the expected $CP$ violating asymmetry of $B^0\to K^0\pi^0$ is
\be
A_{K^0\pi^0} = -0.129\pm 0.027, \quad {\rm (sum\ rule\ projection)}
 \label{eq:sr_projection}
\ee
which is sizeable and negative. The dominant uncertainty comes from $A_{K^+\pi^0}$,
which will be reduced in the future.
Since there is a $\pi^0$ in the final state for $B^0\to K^0\pi^0$,
only Belle~II would be able to provide the update.
%If the central value of $A_{K^0\pi^0}$ is indeed around $-0.12$, the sum rule
%can be verified from an experimental measurement with an uncertainty around 0.03.
%
To test the sum rule at the error of Eq.~(\ref{eq:sr_projection}),
assuming systematic errors will improve accordingly, one would need a data
sample $\sim 20$ times larger than at the B factories,
which seems within reach at Belle II.

We have mentioned the rather large local asymmetries measured~\cite{Aaij:2014iva} by LHCb in three body
charmless $B^+ \to K^+\pi^-\pi^+$, $K^+K^-K^+$, $\pi^+\pi^-\pi^+$ and $\pi^+K^-K^+$ decays.
Given the complex rescatterings of the final states involved,
a three body sum rule analogous to Eq.~(\ref{eq:acprule}) may seem impractical.
But the observed asymmetries seem paired in sign, i.e.
$A_{CP}(B^+ \to K^+\pi^-\pi^+) \simeq +0.025$ vs $A_{CP}(B^+ \to K^+K^-K^+) \simeq -0.036$, and
$A_{CP}(B^+ \to \pi^+\pi^-\pi^+) \simeq +0.058$ vs $A_{CP}(B^+ \to \pi^+K^-K^+) \simeq -0.12$.
Assuming that these four modes reflect inclusive
$b \to s\bar qq$, $s\bar ss$, and $b \to d\bar qq$, $d\bar ss$,
where $q = u,\ d$, one can check a rather old perturbative two-loop sum rule~\cite{Gerard:1990ni},
that inclusive asymmetries for charmless $b\to s$ and $b\to d$ each largely
cancel between tree dominant and penguin dominant modes.
A naive check using the LHCb result, without taking into account
momentum-dependent $K^+$ vs $\pi^+$ detection efficiencies,
we find the cancellation to work well for inclusive charmless $b\to s$ decays,
but slightly less well with inclusive charmless $b\to d$ decays.
But this check needs to be done in more detail across the Dalitz plot, while
the $V_{ub}$ and $V_{td}$ values used in Ref.~\cite{Gerard:1990ni} should be updated.

In this vein, we recall that inclusive charmless $b\to s\bar qq$
(where $q$ now stands for $u$, $d$ and $s$) decay is predicted
robustly at the 1\% level, with $b\to d\bar qq$ slightly smaller.
The rate for charmless $b\to s\bar qq$ is dominated by the timelike~\cite{Hou:1988wt} penguin
(Fig.~\ref{fig:BtoKpi}(b) with the spectator quark line removed),
while charmless $b\to d\bar qq$ penguin is tree-dominant.
Direct measurement of these inclusive rates, which are at the same footing
as $b\to s\ell^+\ell^-$ and $b \to u\ell\nu$, should be contemplated.
Furthermore, as we will discuss $b\to s\gamma$ in a following subsection,
we mention that the analogous lightlike penguin, $b\to sg$, where the emitted gluon
is ``on-shell'' ($m_{g^*} < 1$ GeV), is predicted at the 0.2\% level.\footnote{
Given the agreement of $b\to s\gamma$ with SM, there is little room for large enhancement.
But enhancement up to 10\% level~\cite{Grzadkowski:1991kb} had been entertained
in the CLEO era, though the original reasons have by now evaporated.
}
This is an important rare $B$ decay that we do not have a good handle on,
but hopefully can be addressed in the super B factory era to come.
Just like the BaBar anomaly in the long neglected $B \to D^{(*)}\tau\nu$ mode,
the difficulty to measure may be hiding some New Physics.
Do we know it is not 1\%, rather than 0.2\%?

\subsection{%\boldmath
 Semileptonic %$B$ Decays
}

As mentioned in Sections 2 and 3, there are unresolved tensions and even new
physics hints in semileptonic $B$ decays. The $2.9\sigma$ and $2.6\sigma$ discrepancies
between exclusive and inclusive measurements of $|V_{cb}|$ and $|V_{ub}|$, respectively,
need to be resolved by combined experimental and theoretical efforts,
as currently experimental and theoretical uncertainties are on similar footing.
Although it is unlikely to find New Physics by measuring
semileptonic $b\to c\ell\nu$, $u\ell\nu$ decays, and that the discrepancies are
likely due to our lack of knowledge about nonperturbative hadronic effects,
$V_{cb}$ and $V_{ub}$ are fundamental parameters,
which need to be precisely determined to test the CKM paradigm.
Measuring $|V_{cb}|$ and $|V_{ub}|$ is one of the topmost missions in flavor physics,
and certainly a subject of high priority at Belle~II,
with many workshops and private meetings dedicated to their determination.

The combined $3.9\sigma$ discrepancy from SM in $R_D$ and $R_{D^*}$
in $B \to D\tau\nu$, $D^*\tau\nu$ decays is very enticing,
and needs to be further addressed and confirmed with better significance.
The analysis method is already proven to work and
measuring $\tau$ polarization~\cite{Hirose:2016wfn} in $B\to D^*\tau \nu$ is another triumph.
Besides requiring more data, the systematic uncertainties that arise from the
unmeasured backgrounds, such as $B\to D^{**} \ell \nu$ and  hadronic $B$ decays,
need to be reduced by measuring these branching fractions.
With both Belle II and LHCb data, the $R_D$, $R_{D^*}$ anomaly should be confirmed
or disproved by early 2020s.

Two possible hints of New Physics from $B\to K^{(*)}\ell^+\ell^-$ decays are
discussed in Sec~\ref{sec:P5'}. The $P_5^\prime$ and $R_K^{(*)}$ issues should
reach a conclusion with LHCb, CMS, ATLAS and Belle II data by early 2020s.
A similar decay that sheds light on New Physics is $B\to K^{(*)} \nu \bar{\nu}$,
where the branching fractions in SM are estimated with smaller theoretical uncertainty
due to absence of long-distance effects from electromagnetic penguin contributions.
Besides revealing the same possible New Physics that contributes to
the decay $B\to K^{(*)}\ell^+\ell^-$, the requirement of missing four-momentum with missing mass,
owing to the unobserved $\nu\bar{\nu}$ pair, can imitate
the signature of other weakly interacting particles,
such as low mass dark matter particles, right-handed neutrinos, or SUSY particles.
Therefore, The decay rates or other observables should be
measured as a function of the dineutrino momentum squared ($q^2$).
Technically, background contaminations differ for each $q^2$ bin,
while binning also depends on available statistics.
Therefore, it is better to optimize the event selections bin by bin,
rather than throwing a common requirement for all events.

Experimentally, much like $B\to \tau\nu$ decay, the $B\to K^{(*)}\nu \bar{\nu}$ decays
are identified by fully reconstructing the accompanying $B$ meson, and requiring a
reconstructed $K^{(*)}$ meson without any other particle in the event.
Although tagging the other $B$ reduces the efficiency,
the kinematic variable $q^2$ can be resolved as
$q^2 = (P_{\rm beam} - P_{ \rm Btag} - P_{K^{(*)}})^2$.
BaBar has performed a study with 10 bins in $q^2$~\cite{Lees:2013kla},
and the result shows a small excess (6 events with 2.9 background)
in the low $q^2$ bins, $0< q^2/m^2_B < 0.3$ for the $B^+ \to K^+\nu\bar{\nu}$,
although the lowest bin for both $K^+\nu\bar{\nu}$ and  $K^{*0}\nu\bar{\nu}$ indicate excess.
Since the significance for four $B\to K^{(*)} \nu\overline{\nu}$ decay channels are low,
the branching fraction upper limits at 90\% C.L. are given from $3.2\times 10^{-5}$ to
$ 11.5\times 10^{-5}$~\cite{Lees:2013kla}, even a two-sided bound, and potentially
an order of magnitude higher than SM expectations.
Although the significance of the excess is small,
one should make diligent updates in the future.
Belle should perform a binned analysis with existing data and compare with BaBar,
while the order of a hundred events is expected by the end of Belle II,
and the situation can be better clarified.
A model discussion of the BaBar hint is given in Sec.~\ref{sec:KOTO}.

\subsection{Leptonic and Radiative}

\noindent \underline{\it\bf Leptonic Decays} \vskip0.1cm

Unlike charmless hadronic $B$ decays, purely leptonic $B$ decays are largely free from
hadronic uncertainties, and are therefore ideal for New Physics search.
For neutral $B$ mesons, the decay $B_s^0\to\mu^+\mu^-$ is finally observed by
CMS and LHCb, as discussed in the highlight section, Sec.~\ref{sec:Bqmumu}.
As already commented there, the LHC experiments can eventually measure
the even rarer $B^0\to \mu^+\mu^-$ with the Phase~II upgrade, if the value is at SM expectation.
But discovery could come earlier if the rate is enhanced, and background issues understood.
Belle~II is expected to accumulate around 5 to $10 \times 10^{10} B{\overline B}$ pairs,
which is at the threshold of evidence according to the SM expectation at $10^{-10}$ level,
Eq.~(\ref{eq:Bdmumu_SM}).
But again, if the rate is enhanced, Belle~II can also be part of the game,
in a much cleaner environment.
So, the LHC experiments should strive to overcome background issues for $B^0\to \mu^+\mu^-$.

Searches for $B_{(s)}^0\to \tau^+\tau^-$ have been conducted by BaBar and LHCb using
$2.32\times 10^8$ $B{\overline B}$ pairs and 3 fb$^{-1}$, respectively.
The obtained upper limits are all at $10^{-3}$ level~\cite{tautau-babar, tautau-lhcb}.
These are around four to five orders higher than SM expectations,
 ${\cal B}(B^0 \to \tau^+\tau^-) = (2.22\pm 0.19)  \times 10^{-8}$ and
 ${\cal B}(B_s^0 \to \tau^+\tau^-) = (7.73\pm 0.49) \times 10^{-7}$~\cite{Bobeth:2013uxa}.
Unless there is a breakthrough in analysis technique, the experimental measurements
for $B^0_{(s)} \to \tau^+\tau^-$ cannot reach the SM value in the next 10 years.
Searching for $B^0_{(s)}\to e^+e^-$ is more challenging since it is
helicity suppressed and expected to be extremely small.
Lepton flavor violating neutral $B$ decays
such as $B_{(s)}^0\to \mu\tau$ can also be entertained.

The situation for the charged $B$ case would have a neutrino in the final state,
with $B^-\to \tau^-\bar\nu$ already observed, which was a highlight at the B factories.
The decay of $B \to \tau\nu$ proceeds through $b \bar u$ weak annihilation,
and the decay branching fraction is estimated rather well,
\be
\mathcal{B}^{\rm SM}({B}^- \to \tau^- \overline{\nu}_\tau)
 = (0.75^{+0.10}_{-0.05})\times 10^{-4},
 \label{eq:Btaunu-SM}
\ee
based on inputs from lattice calculations and
unitarity-constrained $|V_{ub}|$ value from other measurements~\cite{Charles:2011va}.
Both BaBar and Belle studied $B \to \tau\nu$ decays as data accumulated,
and earlier measurements~\cite{Aubert:2008ac, Lees:2012ju, Hara:2010dk} hinted
at higher central values compared with Eq.~(\ref{eq:Btaunu-SM}),
which drew a lot of attention from both theory and experiment communities.
With improvements in reconstruction and tagging techniques, analyses based on
the final Belle data set~\cite{Adachi:2012mm, Kronenbitter:2015kls} show a weaker tension with SM.
The combination of the best available measurements is compatible with
Eq.~(\ref{eq:Btaunu-SM}) within 1.4$\sigma$,
\be
\mathcal{B}({B}^- \to \tau^- \bar{\nu}_\tau)
 = (1.06\pm0.19)\times 10^{-4}.
 \label{eq:Btaunu}
\ee
Instead of showing an anomaly in the decay of $B \to \tau\nu$ itself, the result
provides a strong constraint on the extended Higgs sector,
using the simple multiplicative factor~\cite{Hou:1992sy} 
in the (SUSY type) two Higgs doublet model, 2HDM-II,
\be
\mathcal{B}^{\rm 2HDM-II} = r_H\,\mathcal{B}^{\rm SM}({B}^- \to \tau^- \bar{\nu}_\tau),
\quad \ \ r_H = \left(1 - \tan^2\beta\,\frac{m_{B^+}^2}{m_{H^+}^2}\right)^2.
\ee
where $\tan\beta$ is the ratio of vacuum expectation values of the two Higgs doublets,
which, like $m_B$ and $m_{H^+}$, is also a physical parameter in the model.
A large swath in the $m_{H^+}$ vs $\tan\beta$ plane is excluded by Eq.~(\ref{eq:Btaunu}).
The measurement is expected to improve with the upcoming Belle~II data.
If the current central value stays, it might hint at a deviation from SM.
Otherwise, it would put an even stronger constraint on the charged Higgs sector,
or on any similar New Physics process.
This is an unusual tree level effect.
But the situation is a little volatile, because the BaBar anomaly
seems to rule out 2HDM-II, and perhaps the extended Higgs sector is more intricate,
which we would return to comment in Sec.~\ref{sec:S2HDM}.
Regardless, measurement of $B \to \tau\nu$ would provide a useful probe in the future.

We remark that, assuming SM, the $|V_{ub}|$ value
determined from the $B \to \tau \nu$ branching fraction is similar to
the combined inclusive and exclusive $|V_{ub}|$ determination, both in value and uncertainty.
The next in line is $B^-\to \mu^-\bar\nu$, which is simpler to measure,
except the branching fraction is more suppressed.
The expected ratio of branching fractions for the $\mu\nu_\mu$ and $\tau\nu_\tau$ modes
depends simply on the $B$ meson and lepton masses
for SM or the popular 2HDM-II,
\be
{\cal B}(B\to \mu\nu_\mu)/{\cal B}(B\to \tau\nu_\tau) = {m^2_\mu(m^2_B-m^2_\mu)}/{m^2_\tau(m_B^2-m^2_\tau)}.\quad ({\rm SM\ \&\ 2HDM\;II})
 \label{eq:ratio_munu-taunu}
\ee
Using the current averaged branching fraction of Eq.~(\ref{eq:Btaunu}),
one gets the expected branching fraction
${\cal B}(B^-\to \mu^-\nu_\mu) = (4.87\pm 0.87)\times 10^{-7}$.

Since the muon momentum in the $B$ rest frame is monoenergetic
with vanishing recoil mass,
%the corresponding momentum lies in a small region in the $e^+-e^-$ rest frame
%due to a small boost. Therefore,
one can utilize this information with other variables to identify the signal.
Both BaBar~\cite{munu-babar} and Belle~\cite{munu-belle} searched for $B \to \mu \nu_\mu$ decay
and the obtained branching fraction upper limits are at the $10^{-6}$ level.
On closer scrutiny, however, Belle's limit of
\be
\mathcal{B}({B} \to \mu \bar{\nu}) < 1.7 \times 10^{-6}, \quad ({\rm Belle,\ 277M}\ B\overline B)
 \label{eq:Bmunu-Belle}
\ee
at 90\% C.L. is based on only a fraction of the full data,
while the BaBar limit of
\be
\mathcal{B}({B} \to \mu \bar{\nu}) < 1.0 \times 10^{-6}, \quad ({\rm BaBar,\ 468M}\ B\overline B)
 \label{eq:Bmunu-BaBar}
\ee
at 90\% C.L. uses their entire data set.
Belle has not made a general full update with the 772M $B\overline B$ at hand,
but instead analyzed full data using~\cite{Yook:2014kga}
hadronic $B$ tag (full reconstruction) on the other $B$ meson.
This is, however, counterproductive, since the efficiency of $B$ tag is at $10^{-3}$ level,
and the reason it was developed was for events like $B \to \tau\nu$ and $B\to D^{(*)}\tau\nu$
where there is missing mass in addition to missing energy.
Taking the tagging efficiency hit on such a rare decay,
no wonder the limit obtained with 2.9 times the data used in Ref.~\cite{munu-belle},
is poorer, i.e. worse than Eq.~(\ref{eq:Bmunu-Belle}).
The method used by BaBar, and the 2007 paper by Belle, dates to CLEO,
which reconstructs all the energy and momentum of the other $B$ more loosely,
demanding no extra missing energy, hence total missing mass is zero.
In this way, the low efficiency of full reconstruction is avoided.

Judging from the BaBar limit, Eq.~(\ref{eq:Bmunu-BaBar}),
is only a factor of two above SM,
assuming Belle can do similarly with a larger data set,
and given the tendency of enhancement in $B\to \tau\nu$ (Eq.~(\ref{eq:Btaunu})),
a hint for $B \to \mu\nu$ may well emerge with B factory data,
especially if one could combine the Belle and BaBar data sets.
Even with the few $\times 10^{10}$ $B\overline B$ events in the future
of Belle II, one should avoid using full
reconstruction of the accompanying $B$ meson for this channel.
Such a requirement will simply kill most of the signal events.
The point has been made in the FPCP2012 theory summary~\cite{Hou:2012hh},
but unfortunately it still has not been followed through by Belle.
%A better strategy is to have loose requirement on the tag side and utilize the
%$\mu^+$ momentum, kinematic and shape variables to identify signals.
%
We could further emphasize that measurement of $B \to \mu\bar\nu$
is a pursuit in itself. After all, the BaBar anomaly points
beyond the usual 2HDM-II, and a deviation of
${\cal B}(B\to \mu\nu_\mu)/{\cal B}(B\to \tau\nu_\tau)$
from Eq.~(\ref{eq:ratio_munu-taunu}) would be rather interesting.
\vskip0.2cm

\noindent {\it\bf Note Added}: As announced at EPS 2017,
Belle has finally performed~\cite{Belle-munu} a regular (untagged) 
search for $B \to \mu\nu$ with full data set, finding the value 
$\mathcal{B}({B} \to \mu \bar{\nu}) = (6.46 \pm 2.22 \pm 1.55) \times 10^{-7}$,
which has a significance of 2.4$\sigma$.
Belle and BaBar should now follow through with a B factory combination
as suggested in Ref.~\cite{Hou:2012hh}.
It seems that, before long, $B \to \mu\nu$ search would be 
an end in itself at Belle II.

\

\noindent \underline{\it\bf Radiative Decays} \vskip0.1cm

Radiative $B$ decays provide important information on flavor physics in SM,
but are also sensitive to New Physics in the penguin loop.
$B\to K^*\gamma$ was the first electromagnetic penguin
$b\to s \gamma$ transition to be observed, in 1993 by CLEO~\cite{K*gamma}.
%Moreover, the decay rate
%of $B\to X_s\gamma$ is proportional to the top quark mass. The large $K^*\gamma%$ rate measured in 1993 suggest that top quark is heavy.
The inclusive $B\to X_s \gamma$ branching fraction were subsequently
measured by CLEO, then BaBar and Belle, providing a star
constraint on $H^+$ bosons, \emph{from flavor physics}.
Technically, the $B\to X_s \gamma$ signal is identified by requiring
a minimum energy cut on photon candidates in the $e^+e^-$ center of mass frame,
and then either reconstruct $X_s$ with a kaon plus one to four pions,
or estimate the number of background events from off-resonance data
for $e^+ e^-\to q{\bar q}$ $(q= u, d, s, c$ quarks),
and background photons from $\pi^0$ as well as $\eta$ decays in $B$ events.
The measured branching fractions are then extrapolated to $E_\gamma >1.6$ GeV
using the method given in Ref.~\cite{Buchmuller}.
The averaging is then performed by HFAG, which currently gives
\be
{\cal B}(B\to X_s\gamma)|_{E_\gamma >1.6} = (3.49\pm 0.19)\times 10^{-4},
 \label{eq:bsgamma1.6}
\ee
where the systematic uncertainty dominates. This inclusive rate provides
a rather stringent bound on charged Higgs mass that is
independent of $\tan\beta$ in 2HDM-II.
Using the current NNLO calculations~\cite{misiak}, Eq.~(\ref{eq:bsgamma1.6})
gives the limit of
\be
M_{H^+} > 480\ {\rm GeV/c}^2, \quad (b \to s\gamma,\ {\rm 2HDM\ II})
\ee
at 95\% C.L. for 2HDM-II. While sophisticated NNLO calculations,
requiring the participation of a large number of theorists (18 for Ref.~\cite{misiak}),
is needed for proper numerical extraction, the reason that the above limit is powerful is
similar to why $b \to s\gamma$ receives large enhancement~\cite{Bertolini:1986th, Deshpande:1987nr}
from QCD corrections: suppression of the leading contribution by power GIM cancellations,
where QCD alleviates at $\alpha_s$ order by bringing in logarithms,
or $H^+$ effects which can therefore have large impact.
The independence of $\tan\beta$ is even more subtle.
It is because the dipole transition is of $\sigma_{\mu\nu}m_b R$ form,
where $R$ is the right-handed projection. To extract an $m_b$,
it could be by Yukawa coupling of the $H^+$ to top at one side of the loop,
while to bottom at the other side to give $m_b$ factor.
Such terms would bring, in 2HDM-II,
a $\cot\beta$ factor to counterbalance a $\tan\beta$ factor,
hence become $\tan\beta$ independent.\footnote{
This effect therefore does not~\cite{Hou:1987kf} happen for 2HDM-I.
}
It so happens that the effect carries the same sign as short distance SM effect,
so always enhances the amplitude~\cite{Grinstein:1987pu, Hou:1987kf},
hence sensitively probes $m_{H^+}$, independent of $\tan\beta$.

In fact,
a recent Belle update~\cite{Belle:2016ufb} of $B \to X_s\gamma$ is
slightly lower than SM expectation,
which results in a more stringent bound~\cite{Misiak:2017bgg} of $m_{H^+} > 570$ GeV
in 2HDM-II, raising issues also of theory--experiment correspondence
(such as photon energy $E_\gamma$ cut).
Such high masses are not yet sufficiently probed by direct search.
Together with challenges to 2HDM-II such as from the BaBar anomaly,
perhaps one should broaden scope in considering
$b\to s\gamma$ bound on charged Higgs mass.

Let us refrain from discussing inclusive and exclusive $B\to X_d\, \gamma$ modes,
as it is quite challenging for the inclusive case.
Many exclusive channels of $b\to s\gamma$ decays have been measured,
including mesons or baryons from $s$-quark fragmentation,
and measuring the $M_{X_s}$ spectrum for inclusive process.
Inclusive and exclusive $B \to X_{s,\, d}\,\gamma$ decays are a mainstay for
the B factories and Belle~II.
%An alternative way for $M_{X_s}$ is to fully reconstruct the accompanying $B$ meson
%and look for an energetical photon in the signal side. Since the four momentum of the beam is
%known,the mass of the $X_s$ can be computed using the four momenta of the
%accompanying $B$ meson and the photon. To reduce the $d\gamma$ contamination,
%a kaon tag is required. With a few times $10^{10}$ $B$ mesons expected in
%Belle II, the $M_{X_s}$ distribution can be precisely determined.
%
%Similar to the exclusive $B\to X_s\gamma$ case, exclusive $B\to X_d\gamma$
%decays will be explored with Belle II data. It's expected that several $X_d$
%fragmentation channels will be detected and the $M_{X_d}$ distribution may be
%measured with various combinations of pions in the final state.
%
What we wish to stress is that, during the B factory era, the experimental
effort that measured $B \to \tau\nu$, and the joint experiment--theory
co-development for studying inclusive $b\to s\gamma$,
were two highlights that together provide much more stringent constraints
on the charged Higgs boson $H^+$ than is currently achieved at the LHC
by direct search. This will continue for a while into the Belle II era.
It is a domain that LHCb has less to say.
But we caution again that, with the BaBar anomaly in $B\to D^{(*)}\tau\nu$,
perhaps the enlargement of the Higgs sector is different
from 2HDM-II that is championed by SUSY, and we look forward to
further developments in the coming few years.

\subsection{Dark Connection}

The overwhelming evidence for dark matter and the long standing muon $g-2$ anomaly
may suggest particle physics beyond SM.
The extension of SUSY is a natural possibility that offers to explain both effects.
However, lack of evidence from LHC direct search may imply that it might be some other extension.
Recently, models of the dark sector introduce a new hidden U(1) interaction,
under which the dark matter particles are charged~\cite{darkt1, darkt2, darkt3}.
An Abelian gauge field, the dark photon $A^\prime$, couples the dark sector to SM
particles through kinetic mixing~\cite{Holdom:1985ag} with hypercharge of SM.
Since the dark sector models do not by themselves constrain the dark photon mass,
dark sector particles with masses ranging from a few MeV/c$^2$ to few GeV/c$^2$
can be probed at B factories.
%(or kaons, Sec.~\ref{sec:kaon}).

LHC, B factories, beam dump and kaon experiments have all
conducted  searches for dark sector particles.
The high luminosity B factory with a clean $e^+e^-$ environment is
an ideal place for dark particle search.
The searches for dark particles by BaBar and Belle can be divided into four different categories:
\begin{enumerate}
\item Search for dark photon in $e^+ + e^- \to \gamma + A^\prime$  with $A^\prime$
 decaying into two charged leptons or invisible final states~\cite{darkpho1, darkpho2}.
\item Search for dark Higgs ($h^\prime$) in Higgsstrahlung process,
 $e^+ + e^- \to A^\prime + h^\prime$, $h^\prime \to A^\prime A^\prime$~\cite{darkhigg1, darkhigg2}.
\item Search for muonic dark force, $e^+ + e^- \to \mu^+\mu^- Z^\prime$,
 $Z^\prime \to \mu^+\mu^-$, where $Z^\prime$ is a dark boson coupling only
 to 2nd and 3rd generation leptons~\cite{darkforce}.
\item Search for dark vector gauge boson in $U^\prime \to \pi^+\pi^-$ in
 $D^0\to K^0_S\eta$, $\eta\to U^\prime \gamma$~\cite{darkbos}.
\end{enumerate}
Although no signals were observed in any search, much parameter space has been
either excluded or given upper limits.
We will give some explicit discussion of dark boson search with kaons in Sec.~\ref{sec:KOTO},
which also relates back to $B$ physics as probe.
There, we would also give some further discussion on the Dark connection
of flavor physics.

The goal of B factory experiments is to study Flavor Physics and $CP$ Violation,
aiming at finding New Physics.
The large $e^+ + e^- \to q + {\bar q}$ sample provides opportunities to do other physics.
Some may be within Standard Model, such as QCD fragmentation, but some may provide a
clean probe for New Physics. Dark sector search certainly belongs to the latter case.
The extension to the Standard Model in the Dark direction may have impact on flavor physics.
If some dark sector model is confirmed, it is important to understand flavor
physics and the dark sector model together in a coherent way.

%\

%\hskip0.65cm  (hadronic,) Charmless; Semileptonic; \quad Leptonic / Radiative
      % Dark Connection

\section{Strange \label{sec:kaon}}

As mentioned in the Introduction, kaon physics is the
granddaddy and progenitor of much of FPCP.
Yet it has not run out of its bag of tricks,
namely $K^+ \to \pi^+\nu\bar\nu$ and $K_L \to \pi^0\nu\bar\nu$,
the pursuit of which is quite at the forefront of NP probes~\cite{Buras:2016egb},
and constitutes a precision test of SM.
These two processes are induced by the $Z$-penguin (and associated box diagram) in SM and theoretically clean,
i.e. precisely what is left from nondecoupled top after GIM cancellations,
and does not suffer much from hadronic uncertainties~\cite{Buras:2015qea},
\bea
{\cal B}^{\rm SM}(K^+ \to \pi^+\nu\bar\nu) &=& (8.4 \pm 1.0) \times 10^{-11},
 \label{eq:K+SM} \\
{\cal B}^{\rm SM}(K_L \to \pi^0\nu\bar\nu) &=& (3.4 \pm 0.6) \times 10^{-11}.
 \label{eq:KLSM}
\eea
These predicted SM rates do depend on CKM angles and CPV phase $\phi_3$,
the dependence of which we do not exhibit.
We focus on the experimental pursuit, plus a twist that was
uncovered recently that augments the prospects for the $K_L$ process.
Besides these and other topics, it is interesting that
$\varepsilon'/\varepsilon$ has resurfaced again to be on the forefront.
This is due to decades of improved lattice calculations,
which we mention only briefly.

\subsection{%\boldmath
 $K^+\to \pi^+\nu\bar\nu$ and NA62 \label{sec:NA62}}

\begin{figure}[ht]
\begin{center}
\begin{minipage}[t]{9.5 cm}
\hskip1cm
 \includegraphics[width=7.5cm]{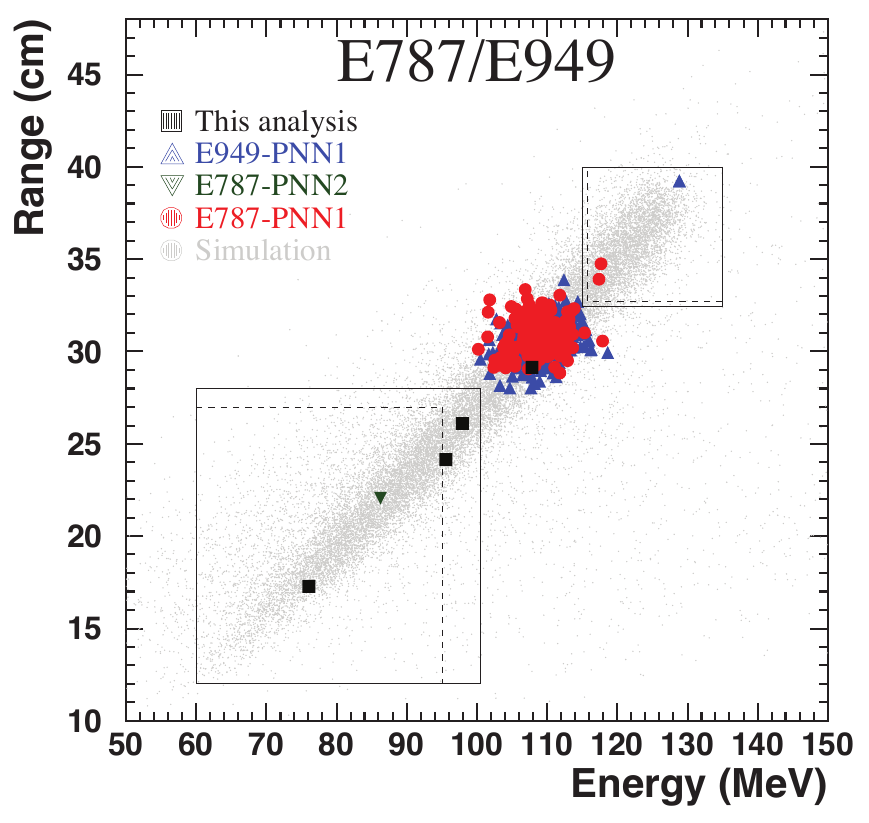}
\end{minipage}
\vskip-0.2cm
\begin{minipage}[t]{16.5 cm}
\caption{
Seven $K^+\to \pi^+\nu\bar\nu$ events observed by E787/E949
on either side of $K^+\to \pi^+\pi^0$ backgrounds.
% that have passed selection cuts.
[Source: courtesy E949 Collaboration, from Ref.~\cite{Artamonov:2009sz}, copyright APS]
 \label{E949}
}
\end{minipage}
\end{center}
\end{figure}

In the long experimental pursuit at Brookhaven, E787 and E949
observed altogether seven $K^+\to \pi^+\nu\bar\nu$ decay events,
giving the value~\cite{Artamonov:2008qb, Artamonov:2009sz}
\be
{\cal B}(K^+ \to \pi^+\nu\bar\nu)= \bigl(17.3^{+11.5}_{-10.5}\bigr)
 \times 10^{-11}, \quad \textrm{(E787/E949)}
 \label{eq:K+949}
\ee
which is on the high side of, but not inconsistent with, Eq.~(\ref{eq:K+SM}).
As shown in Fig.~\ref{E949}, the seven signal events are in two signal boxes
straddling a rather ``bright'' region that reflects $K^+ \to \pi^+\pi^0$ background events.
 %that remain after selection cuts.
The mild excess of Eq.~(\ref{eq:K+949}) motivated the NA62 experiment at CERN SPS,
with the main goal of making 10\% (assuming Eq.~(\ref{eq:K+SM}))
measurement of ${\cal B}(K^+ \to \pi^+\nu\bar\nu)$,
or a total of $\sim 10^{13}$ $K^+$ decays.

NA62 opts for $K^+$ decay in flight, compared with
decay at rest for E787/E949.
The signal boxes are quite similar to E787/E949, however,
and require precision kinematic reconstruction
of $m_{\rm miss}^2 = (P_{K^+} - P_{\pi^+})^2$.
Upstream there is kaon ID for the $K^+$ beam, with precision measurement
of the 75 GeV beam momentum using Si pixel beam tracker.
Timing is of essence, as the $K^+$ then enters a 65 m long decay region,
and downstream one has charged particle tracking with excellent timing,
$e/\mu/\pi$ PID, and hermetic photon rejection.
One selects a solo downstream track that matches the beam track and calorimeter energy,
with $p_{\pi^+} < 35$ GeV to optimize $K^+ \to \mu^+\nu$ rejection,
as well as photon rejection since $p_{\pi^0} > 40$ GeV
from $K^+ \to \pi^+\pi^0$ decay.
After detector commissioning in 2014,
low intensity runs were conducted in 2015 for detector quality verification.
The upshot is that timing and kinematic resolutions,
as well as $\mu/\pi$ separation (by RICH detector) are all close to design,
while calorimeter studies are still ongoing.
With the low statistics of 2015 run, photon rejection already gives
$\pi^0$ rejection at $O(10^6)$, where target is $O(10^8)$.
With commissioning over, the 2016 data should be enough to address the latter target.

NA62 has an approved run program until end of 2018, i.e. until LS2 of LHC.\footnote{
Long Shutdown 2, scheduled for 2019 and 2020.
}
The 2016 data (which is up to 30\% nominal beam intensity) is expected to reach SM sensitivity,
the 2017 data should be able to surpass the precision reached
at Brookhaven considerably, while by end of 2018,
the data should allow NA62 to reach 10\% precision of SM (i.e. $O(100)$ events).
Thus, one can expect that, by time of LS2,
our knowledge of $K^+\to \pi^+\nu\bar\nu$ should improve dramatically.
NA62 could confirm the enhancement suggested by Eq.~(\ref{eq:K+949})
 --- New Physics,
indicate a significant deviation from Eq.~(\ref{eq:K+SM}),
or confirm this SM prediction. There is much to look forward to.

With active charged particle and photon detection plus PID capabilities,
NA62 can do a lot more than $K^+\to \pi^+\nu\bar\nu$ search,
and with a total of $10^{13}\ K^+$ decays,
the topics studied depends much on trigger bandwidth hence strategy.
Let us discuss a few examples.
What first comes to mind is 3-track trigger with dileptons,
namely $K^+ \to \pi^+\ell^\pm\ell^{\prime\mp}$ and $K^+ \to \pi^-\ell^+\ell^{\prime+}$.
While LFV search with $\ell = \mu$ and $\ell^\prime = e$ is obvious,
let us discuss the dimuon cases.

NA48/2, the predecessor to NA62, has recently published~\cite{CERNNA48/2:2016tdo}
their search for the LNV process $K^+ \to \pi^-\mu^+\mu^+$,
\be
{\cal B}(K^+ \to \pi^-\mu^+\mu^+) < 8.6 \times 10^{-11}, \quad \textrm{(NA48/2)}
 \label{eq:LFV-NA48-2}
\ee
at 90\% C.L., probing for Majorana neutrino via $K^+ \to \mu^+N_4$
with $N_4 \to \pi^\mp\mu^\pm$~\cite{Atre:2009rg}.
Dimuon resonances have also been searched for in
LNC $K^+ \to \pi^+\mu^+\mu^-$ decay~\cite{CERNNA48/2:2016tdo}.
Nothing was found, but the point is that, with 50 times the expected number of $K^+$
than NA48/2 and with better detector resolution,
NA62 can vastly improve the above bound, as well as exotic particle search.
We further note that $K^+ \to \pi^+\mu^+\mu^-$ and $\pi^+e^+e^-$
may show correlations~\cite{Crivellin:2016vjc} with the NP hint in
$P_5^\prime$ and $R_K$ observables in $B\to K^*\mu^+\mu^-$ and $K\ell^+\ell^-$,
and are worthy of pursuit. These kaon modes, however, have long distance effects
to disentangle, which demand a lot of studies.

\begin{figure}[ht]
\begin{center}
\begin{minipage}[t]{9 cm}
\hskip1cm
\includegraphics[width=7.5cm]{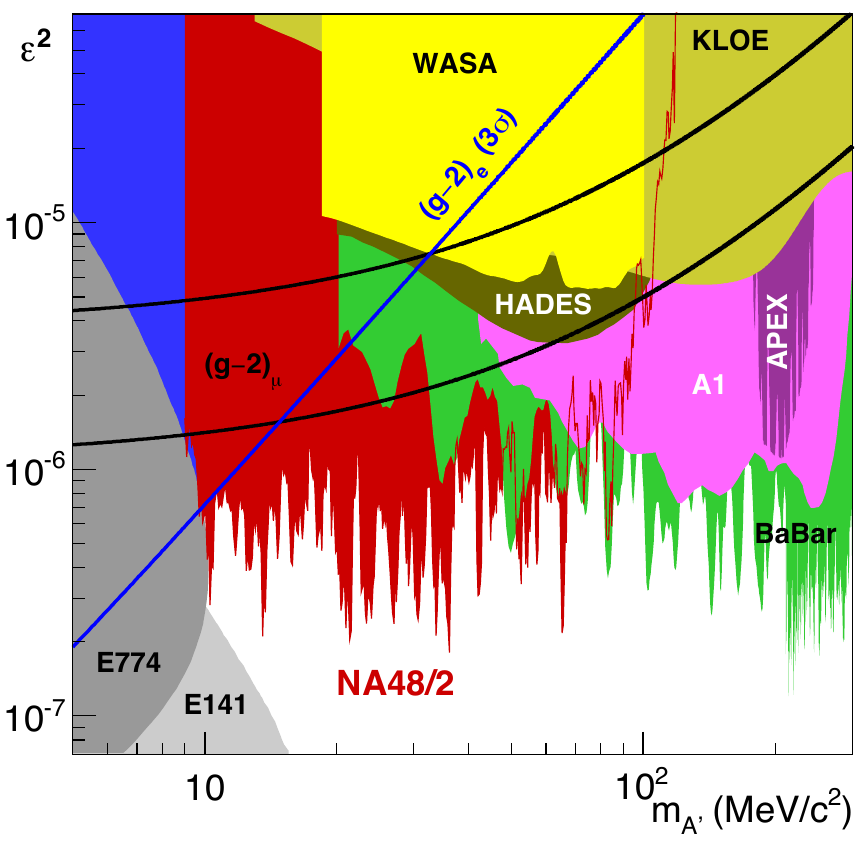}
\end{minipage}
\vskip-0.2cm
\begin{minipage}[t]{16.5 cm}
\caption{
Combined experimental bounds rule out the dark photon $A'$
as explanation for muon $g-2$ (curved band), with NA48/2~\cite{Batley:2015lha}
closing the last window in $m_{A'}$-$\varepsilon^2$ plane.
These results assume $A'$ couple to quarks and decay
dominantly to SM fermions.
 [Source: courtesy NA48/2 Collaboration, from Ref.~\cite{Batley:2015lha}]
 \label{fig:DP-NA48-2}
}
\end{minipage}
\end{center}
\end{figure}

Another example we give concerns rare/exotic $\pi^0$ decays.
With $10^{13}\ K^+$ decays and with $K^+ \to \pi^+\pi^0$ at $\sim 21\%$,
NA62 is a ``$\pi^0$ factory''.
Again, its predecessor NA48/2 has searched for the dark photon $A'$ via
$\pi^0 \to \gamma A'$, followed by prompt $A' \to e^+e^-$ decay.
Such a dark photon with $\gamma$--$A'$ mixing parameter $\varepsilon \sim 10^{-3}$
and mass below 100 MeV could possibly explain the muon $g-2$ anomaly.
Altogether, 17 million $\pi^0 \to \gamma e^+e^-$ decays were fully reconstructed
by NA48/2, with no signal for $A'$ observed.
Combined with other experiments, the NA48/2 result~\cite{Batley:2015lha}
practically rules out the dark photon as an explanation of muon $g-2$
(assuming $A' \to e^+e^-$ decay is predominant),
as illustrated in Fig.~\ref{fig:DP-NA48-2}.
The NA48/2 bound, however, is limited by irreducible $\pi^0 \to \gamma A'$
Dalitz decay background, so only modest improvement is expected with the
larger NA62 data. Furthermore, an $e^+e^-$ trigger may need to be scaled down.
There is also $\pi^0 \to \nu\nu$, or invisible $\pi^0$ decay
that in principle can be probed with $\pi^+$ tagging.
We defer the discussion to the next subsection.

Beyond LS2 of the SPS/LHC complex, NA62 can further improve the studies mentioned above.
There is further thought for a year-long run in ``beam dump'' mode
to probe for hidden sector candidates in MeV--GeV range,
which we do not discuss any further here.

\subsection{%\boldmath
 $K_L\to \pi^0\nu\bar\nu$ and KOTO \label{sec:KOTO}}

If NA62 evolved from NA48, then the KOTO experiment currently running
at KEK's J-PARC facility descended from KTeV, with E391A as intermediate step.
The current best direct search limit~\cite{Ahn:2009gb},
\be
{\cal B}(K_L\to \pi^0\nu\bar\nu) < 2.6 \times 10^{-8}, \quad \textrm{(E391A)}
 \label{eq:E391A}
\ee
at 90\% C.L. is still the one from E391A,
the first dedicated $K_L\to \pi^0\nu\bar\nu$ search experiment.
Eq.~(\ref{eq:E391A}) is, however, 3 orders of magnitude
above the SM expectation, Eq.~(\ref{eq:KLSM}),
and a factor of 20 weaker than an indirect bound extracted from
the E949 result on $K^+\to \pi^+\nu\bar\nu$, Eq.~(\ref{eq:K+949}),
assuming isospin and lifetime ratio factors~\cite{Grossman:1997sk},
\be
{\cal B}(K_L\to \pi^0\nu\bar\nu) < 1.4 \times 10^{-9}, \quad \textrm{(E949/GN)}
 \label{eq:E949/GN}
\ee
which illustrates how difficult things are for KOTO.

Compared with $K^+\to \pi^+\nu\bar\nu$ decay, $K_L\to \pi^0\nu\bar\nu$ is
indeed a really challenging mode to search for.
There is no way to detect the incoming $K_L$ (without risking conversion
to $K_S$), which decays in flight.
One knows only the beam direction and energy profile,
and the appearance of just a single $\pi^0$ in the detector, plus nothing else.
More specifically, the experimental signature is
$2\gamma$, plus ``nothing'', plus missing $p_T$.
One actually \textit{assumes} the detected $2\gamma$ are from $\pi^0$ decay,
and from $\pi^0$ direction one calculates the decay vertex $z$
along beam direction, and hence the $p_T$ of the $\pi^0$.
A signal box is defined in $z$ and $p_T$, and the remaining job,
in addition to getting as many primary beam protons on target as possible,
is to understand and reject background.
This is quite the poor-man's $K \to \pi\nu\bar\nu$ process,
with no luxury of ``kinematic control''.
Besides moving from KEK PS to J-PARC, the main changes from E391A to KOTO
are the reuse of CsI crystals from KTeV for signal photon detection,
more hermetic photon veto for $K_L \to \pi^0\pi^0$ suppression,
and waveform digitization.

\begin{figure}[ht]
\begin{center}
\begin{minipage}[t]{9 cm}
\hskip0.6cm
 \includegraphics[width=8.5cm]{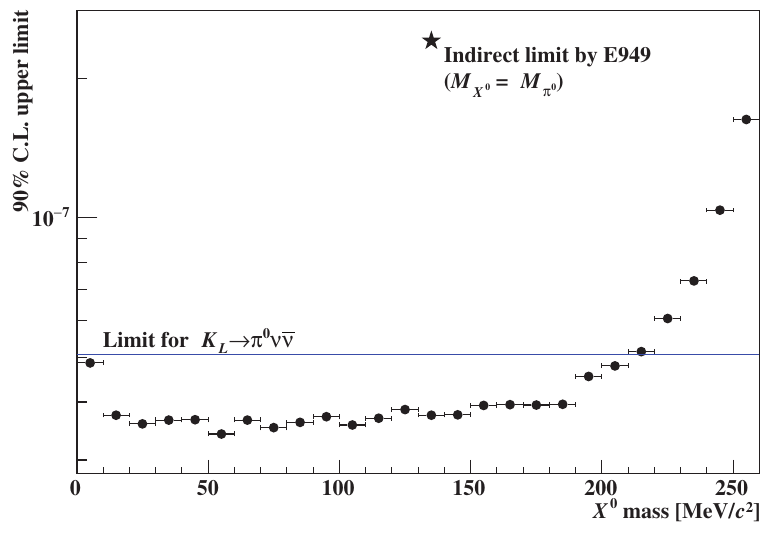}
\end{minipage}
\vskip-0.2cm
\begin{minipage}[t]{16.5 cm}
\caption{
Upper limit on ${\cal B}(K_L \to \pi^0X^0)$ vs $m_{X^0}$ at the 90\% C.L.,
compared with limit (horizontal line) on ${\cal B}(K_L \to \pi^0\nu\bar\nu)$
and the indirect limit (star) by E949.
 [Source: courtesy KOTO Collaboration, from Ref.~\cite{Ahn:2016kja}]
 \label{KOTO13}
}
\end{minipage}
\end{center}
\end{figure}

If the path for long-lived neutral kaon rare decay search is long and tortuous,
it suffered further some bad luck.
After just 100 hours into data taking in 2013,
the beam was stopped by an unrelated accident at J-PARC,
and data taking did not resume until two years later.
The hiatus of two years, however, was perhaps a mixed blessing.
Analyzing the meager 2013 data, one event was found in the signal box
when $0.34 \pm 0.16$ events were expected.
The main backgrounds were beam halo neutrons that hit the CsI calorimeter,
and physics backgrounds, e.g. from $K_L \to \pi^+\pi^-\pi^0$ events
where the charged pions go down the beampipe.
Steps were taken, both in added hardware
and in software improvement, to mitigate these problems
(the only approach since E391A period),
and these were tested with the first equivalent amount of data taken in 2015,
leading to improved understanding hence further control of background.
Only by the KAON2016 conference did KOTO announce their final result~\cite{Ahn:2016kja}
of $5.1 \times 10^{-8}$ at 90\% C.L. based on 2013 data,
which is weaker than Eq.~(\ref{eq:E391A}) by E391A.
The paper, however, announced also a first direct search limit,
\be
{\cal B}(K_L\to \pi^0X^0) < 3.7 \times 10^{-8}, \quad \textrm{(KOTO, $m_{X^0} \simeq m_{\pi^0}$)}
 \label{eq:KOTO_X0}
\ee
which is more than a factor of 6 improvement~\cite{Ahn:2016kja} of
an indirect result from $K^+ \to \pi^+ X^0$ search by E949~\cite{Artamonov:2009sz},
where $X^0$ is an invisible object of $\pi^0$ mass.
The three bounds are shown in Fig.~\ref{KOTO13}.

The new interest in the process of Eq.~(\ref{eq:KOTO_X0}) needs some explanation.
It reflects some longstanding oversight in the arduous planning of a $K_L$ experiment,
self-hypnotized by the ``Grossman--Nir bound''.
The main target for NA62 and KOTO are to measure the SM processes
$K^+\to \pi^+\nu\bar\nu$ and $K_L\to \pi^0\nu\bar\nu$, respectively.
The two branching ratios are related by the Grossman--Nir bound~\cite{Grossman:1997sk} of
\be
{\cal B}(K_L\to \pi^0\nu\bar\nu) < 4.3 \times
 {\cal B}(K^+\to \pi^+\nu\bar\nu), \quad \textrm{(Grossman--Nir)}
 \label{eq:GrossmanNir}
\ee
which is rather solid as it is based on isospin symmetry
and the fact that the $K_L$ process is CP violating while $K^+$ process is a bulk measure,
with the numerical factor dominated by the lifetime ratio.
It is by inserting the E787/E949 upper bound implied by Eq.~(\ref{eq:K+949})
into this relation that one arrives at Eq.~(\ref{eq:E949/GN}),
which we shall refer to as the ``GN bound''.
While the Grossman--Nir bound of Eq.~(\ref{eq:GrossmanNir}) is robust,
it is the numerical ``GN bound'' of Eq.~(\ref{eq:E949/GN})
that everyone seems to be conditioned by,
though it is a thorn-in-flesh for KOTO people in their heroic efforts,
as if their ``business'' would not really start until the
magical number of Eq.~(\ref{eq:E949/GN}) is breached.

Inspecting the detection methods of $K_L\to \pi^0\nu\bar\nu$ and $K^+\to \pi^+\nu\bar\nu$,
with the fact that the $\nu\bar\nu$ pair is not actively detected
but known rather just as missing energy-momentum,
a loophole was recently pointed out~\cite{Fuyuto:2014cya} that allows KOTO to have
NP discovery potential above the ``GN bound'' of Eq.~(\ref{eq:E949/GN}).
Recall that the $K^+$ experiments use kinematic control to
define signal boxes around $K^+ \to \pi^+\pi^0$ decay,
the reason being that the 21\% branching fraction makes the latter too ``bright'' to behold.
Even then, photon veto is critical for $\pi^0$ rejection.
Thus, E787/E949 and NA62 all avoid the $m_{\pi^0}$ window,
which KOTO, without the luxury of kinematic control,
can turn disadvantage into opportunity.
In fact, and as already mentioned in Sec.~\ref{sec:NA62},
E949 had tried putting a bound on
$K^+ \to \pi^+X^0$ with $X^0$ in the $\pi^0$ mass window,
by effectively searching for $\pi^0 \to nothing$ with $\pi^+$ tagging.
Unfortunately, the bound~\cite{Artamonov:2005cu}
\be
{\cal B}(\pi^0 \to \nu\bar\nu) < 2.7 \times 10^{-7}, \quad \textrm{(E949)}
 \label{eq:E949 pi0nunu}
\ee
at 90\% C.L. results in
${\cal B}(K^+ \to \pi^+X^0) < 5.6 \times 10^{-8}$~\cite{Artamonov:2009sz}
that is $\sim 200$ times worse than the one on $K^+ \to \pi^+\nu\bar\nu$.
This is the indirect bound on $K_L \to \pi^0X^0$ (deduced by reversing Grossman--Nir argument)
that the recent direct search~\cite{Ahn:2016kja} by KOTO has surpassed
with small 2013 data set, Eq.~(\ref{eq:KOTO_X0}).

\begin{figure}[ht]
\begin{center}
{\hspace{0.3cm}
\begin{minipage}{2in}
 \includegraphics[width=2in]{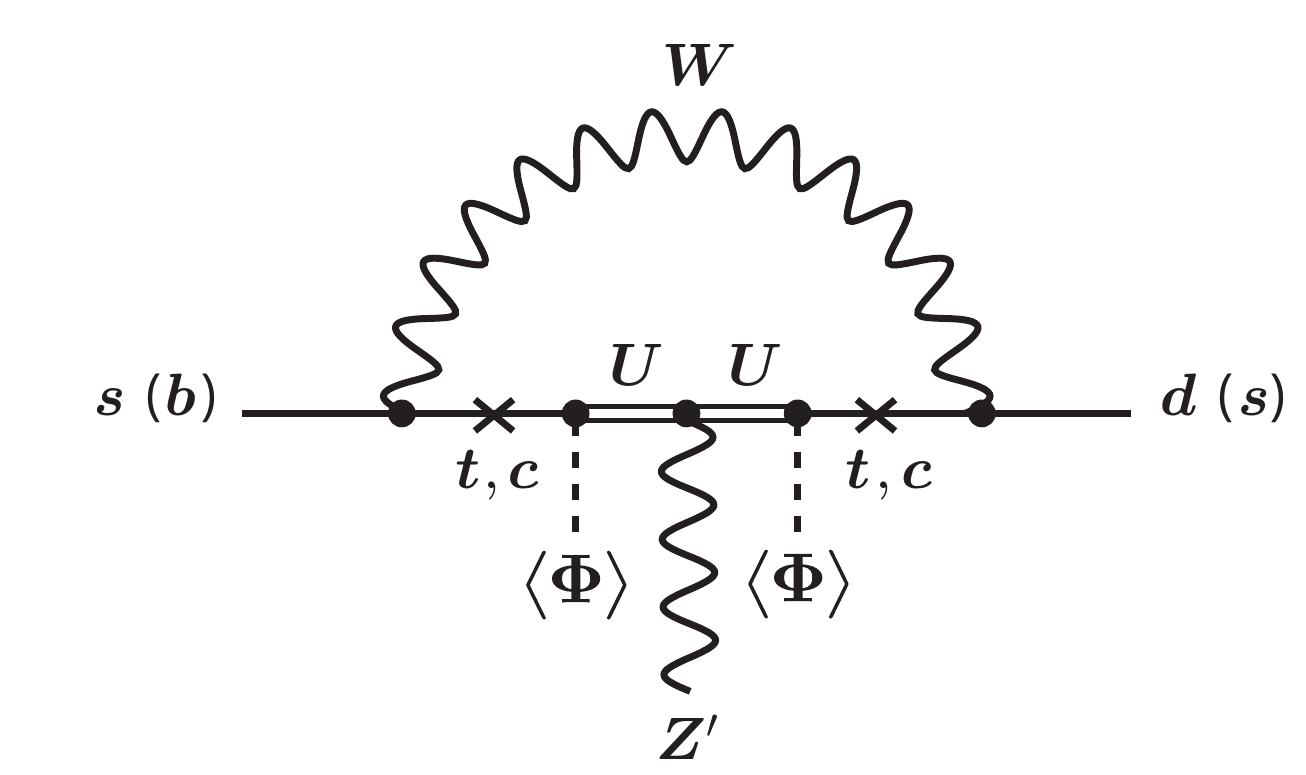}
\end{minipage}\hspace{1cm}%
\begin{minipage}{3.5in}
 \includegraphics[width=3.5in]{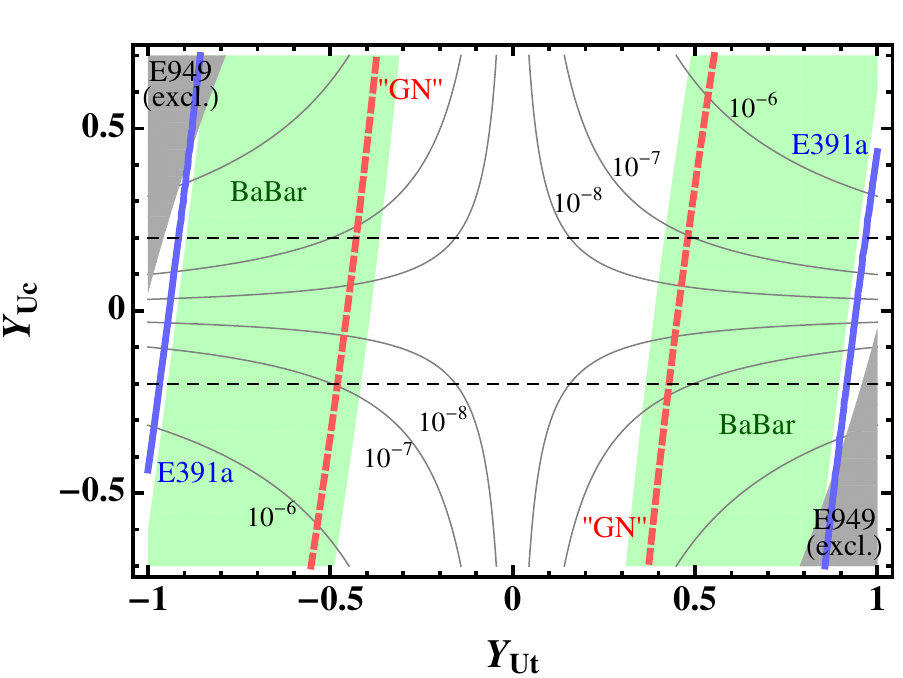}
\end{minipage}
}
%\vskip-0.1cm
\begin{minipage}[t]{16.5cm}
\caption{
[left] Induced $s\to dZ'$ and $b\to sZ'$ transitions with
$L_\mu - L_\tau$ charged vector-like $U$ quark; %~\cite{Altmannshofer:2014cfa};
[right] for $m_{Z'} = 135$ MeV ($Z' \to \nu\bar\nu$ 100\%),
the BaBar allowed 2$\sigma$ range of ${\cal B}(B \to K^{(*)}Z')$ span
the E391A bound (solid) of Eq.~(\ref{eq:E391A})
and ``GN bound'' (thick-dashed) of Eq.~(\ref{eq:E949/GN}),
while a weaker indirect bound by E949 %~\cite{Artamonov:2009sz}
on $K^+ \to \pi^+Z'$ is at the corner of the plot
in the exotic $Y_{Uc}$--$Y_{Ut}$ Yukawa plane,
where horizontal lines indicate Cabibbo angle strength,
and ${\cal B}(t \to cZ')$ contours are in the backdrop.
 [Source: Ref.~\cite{Fuyuto:2014cya}, copyright APS]
 \label{beyonGN}
}
\end{minipage}
\end{center}
\end{figure}

What type of NP could this be? In general, any ``dark'' $X^0$
with mass $m_{X^0} \sim m_{\pi^0}$ emitted
from $K\to \pi$ transition that goes through the detector undetected,
would slip through NA62 but can be caught by KOTO.
As existence proof, Ref.~\cite{Fuyuto:2014cya} gave an explicit model
based on gauged $L_\mu - L_\tau$ symmetry that was originally~\cite{Altmannshofer:2014cfa}
motivated by the $P_5^\prime$ anomaly (Sec. 3), but redirected~\cite{Altmannshofer:2014pba}
to account for muon $g-2$.
The latter demanded the $Z'$ to be lighter than 400 MeV.\footnote{
This light $Z'$ cannot be behind $P_5^\prime$,
but it is just right for kaon physics.
}
To couple to quarks, one introduces a vector-like quark $U$ that carries U(1)$'$ charge,
which mixes with normal $u$-type quarks. A weak boson loop then converts~\cite{Fuyuto:2015gmk}
the effective $Z'$ couplings of $u$-type quarks into $sdZ'$ and $bsZ'$ couplings
 (see Fig.~\ref{beyonGN}[left]).

Stringent constraints from narrow dimuon bump search by LHCb~\cite{Aaij:2015tna}
in $B\to K^*\mu^+\mu^-$ disfavors the $Z'$ above dimuon threshold
 ($Z'$ decay should be rather prompt~\cite{Fuyuto:2015gmk}).
Below $\mu^+\mu^-$ threshold, only $Z' \to\nu\bar\nu$ is allowed,
and interestingly, there is a mild hint~\cite{Lees:2013kla} from BaBar
in $B \to K^{(*)}\nu\bar\nu$ search for a $K^+$ or $K^{*0}$ recoiling against
an unobserved low mass system.\footnote{
Comparing with kaon decays, note that there are
no worries of $B\to K\pi^0$ background, as it is also a penguin loop process,
and the detection environment is very different.}
A similar analysis has not been
carried out yet by Belle. Within the model, the BaBar result does favor
the range between Eqs.~(\ref{eq:E391A}) and (\ref{eq:E949/GN}),
as can be seen from Fig.~\ref{beyonGN}[right],
which has specific parameters $m_{Z'} = 135$ MeV, $g' = 10^{-3}$
for the $Z'$ gauge coupling, and $m_U = 2$ TeV.
This example illustrates the genuine discovery potential for the $K_L \to \pi^0X^0$ process
with 2015 data, which is 20 times larger than 2013 data, and should allow KOTO
to explore $K_L \to \pi^0\nu\bar\nu$ down to the GN bound of Eq.~(\ref{eq:E949/GN})
with improvements stated in Ref.~\cite{Ahn:2016kja}.

We note that, as KOTO continues to take data,
and with new measurements of $K^+ \to \pi^+\nu\bar\nu$ by NA62 coming soon,
the suggestion of Ref.~\cite{Fuyuto:2014cya} still holds,
that the $K_L \to \pi^0X^0$ process could still be discovered above
the $K_L \to \pi^0\nu\bar\nu$ rate implied by
a new $K^+ \to \pi^+\nu\bar\nu$ upper bound through Eq.~(\ref{eq:GrossmanNir}).

The above loophole originates from the ``brightness'' of $K^+\to \pi^+\pi^0$ decay,
such that $K^+$ experiments use kinematic control to exclude events with missing mass
in the $m_{\pi^0}$ region. While this means extra business for KOTO to pursue,
is there anything that can be done by NA62? The answer is yes, by perfecting
their measurement of $\pi^0 \to \nu\bar\nu$ (or, $nothing$), which is indeed
on NA62's non-$\pi\nu\bar\nu$ agenda.
Recall the bound of Eq.~(\ref{eq:E949 pi0nunu}).
The limit is not so good because of photon detection inefficiency~\cite{Artamonov:2005cu},
such that a small fraction of $\pi^0\to\gamma\gamma$ events could mimic $\pi^0 \to nothing$.
With non-$K_{\pi2}$ background expected at $\sim 3$ events,
E949 saw 99 events~\cite{Artamonov:2005cu}, and
attributed the excess to photon detection inefficiencies.
These could come from calorimeter sampling fluctuations for lower energy photons,
or photonuclear interactions (e.g. with neutrons) for higher energy photons.
For NA62 to compete in $K^+\to\pi^+X^0$ detection,
investment should be made in understanding the photon detection inefficiency,
to see how much they can effectively improve the bound of Eq.~(\ref{eq:E949 pi0nunu})
on $\pi^0 \to nothing$. Roughly speaking: how well can NA62 stare into the brightness?

Back on the main pursuit of $K_L \to \pi^0\nu\bar\nu$ search,
the 2013 proton beam power for KOTO was 24 kW, but 42 kW has already been attained since.
The plan is to increase to 100 kW by 2019.
Together with further photon/neutron discrimination by adding
photon sensors in front of CsI crystals to measure shower depth,
KOTO should be able to push below $10^{-10}$~\cite{Yamanaka:2016wmx}.
To attain a genuine measurement of SM prediction of Eq.~(\ref{eq:KLSM}),
i.e. $O(100)$ events, KOTO had an ambitious machine and detector
upgrade plan called Step 2 in its proposal.
But judging from the timeline so far, this probably would not occur until
HL-LHC era, i.e. 2025 and beyond.
In part because of this, and in part planning for their own HL-LHC era run plan,
preliminary design studies have started in NA62~\cite{NA62-KLEVER}
for a 5 year run at the SPS targeting 2026 start,
to observe $O(60)$ $K_L \to \pi^0\nu\bar\nu$ events (assuming Eq.~(\ref{eq:KLSM})).

It seems that rare kaon search would continue beyond another decade,
but major progress is expected within the next few years,
either by discovery of NP in $K^+ \to \pi^+\nu\bar\nu$, $K_L \to \pi^0\nu\bar\nu$
or $K_L \to \pi^0X^0$ processes,
or confirming SM in the $K^+$ mode.

\

\noindent \underline{\it\bf Dark Addendum}: Not a Dark Photon? \vskip0.1cm

We have seen that KOTO has a unique probe into a dark boson $X^0$ with mass around the $\pi^0$,
and an example given was gauged $L_\mu - L_\tau$ symmetry with heavy vector like quarks.
A dark object at specific mass, and a built-up model to hide it, all sounds  ``fantastic''.
We wish to make a little detour to illustrate some points,
even though some of it is not about FPCP, but perhaps ``the flavor of Dark Matter''.

Dark Matter (DM), that the bulk of matter of the Universe
(let alone the ``Dark Energy'' that predominates in the current era)
are not seen, but ``felt'' through their gravitation effect,
is one of the main motivations for New Physics beyond SM.
The premium example is the so-called WIMP (weakly interacting massive particle),
with the lightest SUSY particle offering an ideal candidate.
Alas, both direct search for SUSY at LHC, and DM search with a plethora of means,
turn out empty handed so far. We cannot go deep into this Dark subject,
but wish to extend our discussion of $B$ and $K$ physics,
in particular the $K\to\pi + nothing$ searches that we have just discussed.

The NA48/2 experiment claims to have closed the window for a
dark photon $A^\prime$ explanation of the muon $g-2$ anomaly, see Fig.~\ref{fig:DP-NA48-2}.
But there is a catch: $A'$ has to decay via $e^+e^-$ dominantly.
This may seem plausible, as $m_{A'} < m_{\pi^0}$ in the remnant window
that NA48/2 covered, which corresponds to $\varepsilon \sim 10^{-3}$.
The $\varepsilon$ parameter induces ``kinetic mixing''~\cite{Holdom:1985ag}
between $\gamma$ and $A'$, and $10^{-3}$ is typically viewed as some
loop-induced value.
It was then pointed out~\cite{Davoudiasl:2014kua} that $A'$ could decay invisibly,
e.g. $A' \to \chi\chi$ where $\chi$ is some light dark matter particle,
which would then evade the bounds from NA48/2 and other experiments,
therefore still provide an explanation of muon $g-2$.
This motivated a fixed target experiment, NA64 at SPS,
that rose to the challenge of directly searching for such an $A'$.
Utilizing 100 GeV electron beam in $e^-Z \to e^-ZA'$,
where $Z$ denotes a high mass nuclear and $A'$ is produced
via kinetic mixing with bremsstrahlung photon,
but decays invisibly~\cite{Davoudiasl:2014kua} hence would lead to large missing energy.
An early summer 2016 run with $2.75 \times 10^9$ electrons-on-target (EOT),\footnote{
A relative low number compared to low energy muons discussed in the next section.
}
saw no such events~\cite{Banerjee:2016tad},
excluding the invisible $A'$ with mass below 100 MeV
as an explanation of the muon $g - 2$ anomaly.
A small region above 100 MeV but below the dimuon threshold~\cite{Davoudiasl:2014kua} remains.
NA64 completed a run in Fall 2016 that accumulated an order of magnitude more EOT,
and should be able to cover this region. %Would it be discovery, or exclusion?
Subsequent to Ref.~\cite{Banerjee:2016tad}, however,
the BaBar experiment has ruled out~\cite{darkpho2} the
invisibly decaying dark photon explanation of muon $g - 2$.

But there is another catch: what if $A'$ is not a dark ``photon'',
that it does not couple to electrons, but only to muons to induce the $g-2$ anomaly?
The $L_\mu - L_\tau$ model provides an excellent and motivated example.
Because of the built-in cancellation, the $\mu$ and $\tau$ loop induced
$\gamma$--$Z'$ kinetic mixing is quite suppressed, $\varepsilon \sim 10^{-5}$,
which is way out of reach for NA64.
Stimulated by models such as $L_\mu - L_\tau$,
there is a proposed concept to use $\mu$-beam to conduct the analogous
$\mu Z \to \mu ZZ'$~\cite{Gninenko:2014pea} with missing energy from $Z' \to nothing$.
However, it does not seem feasible that it can be realized before LS2 of SPS/LHC,
i.e. to run before 2021.
But the KOTO experiment can probe, and is already probing this relatively exquisite
$Z'$ scenario, albeit depending on vector-like quark assumptions~\cite{Fuyuto:2015gmk}.
On the other hand, while the muon beam can test the generic $L_\mu - L_\tau$ model,
KOTO has its own advantage, too:
any Dark $X^0$ from $K_L \to \pi^0 + X^0$ can be probed,
so long $X^0$ is not far from $\pi^0$ mass, and decays invisibly.
To be a bit whimsical, since the devil hides in the details,
this seems like a long standing little detail that ...

Our point is to illustrate that dark photon and dark boson search
is ongoing, and flavor physics offers its own angle.
Let's hope $K \to \pi + nothing$, and its sister studies in
$B \to K^{(*)} + nothing$, could shed light on \emph{Dark} physics beyond SM.

\subsection{%\boldmath
 $\varepsilon^\prime/\varepsilon$ \label{sec:epsprime}}

The measurement of direct CPV in kaon decay,
namely $\varepsilon^\prime/\varepsilon$, was a true \emph{tour de force}
of experimental physics, culminating in the measured values of
$(28.0 \pm 3.0\ {\rm (stat)}\ \pm 2.8\ {\rm (syst))} \times 10^{-4}$
 by KTeV~\cite{AlaviHarati:1999xp} and
$(18.5 \pm 4.5\ {\rm (stat)}\ \pm 5.8\ {\rm (syst))} \times 10^{-4}$
 by NA48~\cite{Fanti:1999nm} in 1999,
using subsets of data taken in the late 1990s.
This is 35 years after the original observation of indirect CPV~\cite{Christenson:1964fg}.
The final result, combining~\cite{PDG} the two experiments with full data,
gives
\be
{\rm Re}(\varepsilon^\prime/\varepsilon)\vert_{\rm exp} = (16.6 \pm 2.3) \times 10^{-4},
 \label{eq:epsprime}
\ee
which is 3 orders of magnitude smaller than $|\varepsilon_K| \simeq 2\times 10^{-3}$ itself.
DCPV in kaon decays is really small.
This is in contrast with DCPV in $B\to K\pi$ decays,
measured~\cite{Aubert:2004qm, Chao:2004mn} at strength of 10\%
and only 5 years after $\varepsilon^\prime/\varepsilon$,
and just 3 years after the measurement~\cite{Aubert:2001nu, Abe:2001xe}
of indirect CPV in $B^0 \to J/\psi K_S$, which is order one.
CPV, including DCPV, effects in $B$ system are large.

Interest in the measurement of $\varepsilon^\prime/\varepsilon$
rose after the discovery of the $b$ quark~\cite{Herb:1977ek},
allowing theoretical estimates with three generations of quarks,
where early predictions were at the $10^{-2}$ level.
As DCPV involves the interference between $K^0$ and $\overline{K^0}$ amplitudes
to different final state isospin~\cite{Wu:1964qx}, the trick is to measure
the double ratio,
\be
R = \frac{\Gamma(K_L \to \pi^0\pi^0)/\Gamma(K_S \to \pi^0\pi^0)}
         {\Gamma(K_L \to \pi^+\pi^-)/\Gamma(K_S \to \pi^+\pi^-)}
 \simeq 1 - 6\,{\rm Re}(\varepsilon^\prime/\varepsilon),
 \label{eq:doubleratio}
\ee
plus other ingenious ways to reduce error.
However, by the time early measurements pushed below the percent level,
the heaviness of the top quark, first hinted by the sizable
$B^0$--$\bar B^0$ mixing~\cite{Albrecht:1987ap},
implied an enhanced electroweak penguin~\cite{Hou:1986ug} that could
cancel against~\cite{Flynn:1989iu}  the earlier estimates based on
the strong penguin, resulting in a much smaller $\varepsilon^\prime/\varepsilon$ value.
It was this trick of Nature that caused Winstein to lament~\cite{Winstein:1991gz}
that perhaps we will not be able to distinguish between
the three generation SM and the ``superweak'' model of Wolfenstein~\cite{Wolfenstein:1964ks},
which predicted $\varepsilon^\prime/\varepsilon = 0$.
The experimental effort, of course, went on regardless of this.
The result of Eq.~(\ref{eq:epsprime}) therefore deals the deathblow
to the superweak model, and is a triumph of kaon physics.

The case would be settled as that,
because although the top mass became precisely known,
the strong penguin amplitudes involve various hadronic matrix elements
that are hard to estimate precisely.
Thus, $\varepsilon^\prime/\varepsilon$ does not seem like a
good observable to probe for NP, despite its suppressed value due to
accidental cancellations, which affirms our point of
avoiding probes that are susceptible to large hadronic effects.
However, motivated in part by the latter, and the ability to
put the QCD Lagrangian on the spacetime ``lattice" and use
the computer to ``measure'' the involved path integrals,
a recent result from lattice QCD motivates us to
put $\varepsilon^\prime/\varepsilon$ back as a measurable to watch,
and the comparison is between lattice predictions versus experimental measurement.

Lattice QCD started in the 1970s to approach the fundamental problems
of QCD itself. But practitioners started to tackle the measurements of,
e.g. $K\to \pi\pi$ hadronic matrix elements since the 1980s.
In an effort that parallels the experimental measurement of $\varepsilon^\prime/\varepsilon$,
after 30 years of development, the RBC+UKQCD collaboration
announced recently~\cite{Bai:2015nea} the lattice QCD measurement of
\be
{\rm Re}(\varepsilon^\prime/\varepsilon)\vert_{\rm latt.}
 = {\rm Re}\left\{\frac{i\omega e^{i(\delta_2 - \delta_0)}}{\sqrt2 \varepsilon}
                  \left[\frac{{\rm Im}A_2}{{\rm Re}A_2}
                      - \frac{{\rm Im}A_0}{{\rm Re}A_0}\right]
           \right\}
 = (1.38 \pm 5.15\ {\rm (stat)}\;\pm 4.59\;{\rm (syst)}) \times 10^{-4},
 \label{eq:RBC-UKQCD}
\ee
where $A_0$ and $A_2$ are the $I = 0$ and $2$ decay amplitudes to $\pi\pi$ final states,
respectively, $\delta_I$ are the strong phase shifts,
and $1/\omega = {\rm Re}A_0/{\rm Re}A_2 \simeq 22.5$ is from experiment.
A slightly earlier result~\cite{Blum:2015ywa} had calculated $A_2$ to good
($\sim 10\%$) precision, supporting the $1/\omega$ value.\footnote{
Suggesting the cancellation of two major contributions to $A_2$,
but no cancellation for $A_0$, as the source of the $\Delta I = 1/2$ rule,
that $1/\omega \simeq 22.5$ is a large number.
}
The main result in Eq.~(\ref{eq:RBC-UKQCD}) calculated ${\rm Re}A_0$ on one hand
that is in agreement with experiment, and on the other hand,
the computation of ${\rm Im}A_0$ predicts $\varepsilon^\prime/\varepsilon$.
The numerical result, consistent with zero, is in striking contrast with experiment,
Eq.~(\ref{eq:epsprime}), but the disagreement is only at 2.1$\sigma$ level,
in good part because of lattice errors.
Although some theorists have taken this ``anomaly'' very seriously~\cite{Buras:2016egb},
the RBC+UKQCD collaboration itself projects that improvements in
statistics, larger lattice volume and more lattice spacing values,
would take a $\sim$ 5 year or so effort to pan out.
If the discrepancy stays, it would turn into a true anomaly in the kaon sector.
Would this motivate a renewed effort for experimental measurement?
Maybe, but we caution further that, on the lattice side, a second,
independent crosscheck would also be desirable.

\section{Muon and EDM}

The muon was the harbinger of fermion generation repetition,
and at $\sim 200$ times the electron mass but with a rather long lifetime (compared with $\tau$),
it became a handy, versatile tool that probes across subfield boundaries.
For instance, the proton size problem~\cite{Hill:2017wzi} was recently
revealed by precision study of muonic hydrogen~\cite{Antognini:1900ns}.
We cannot do justice to ``muon physics", but would focus on
rare decays or (charged) lepton flavor violation,
anomalous magnetic moment (muon $g-2$), then crossover to comment on
electric dipole moment searches. That is, we focus on potential probes for NP.
For a more comprehensive discussion on precision muon physics, see Ref.~\cite{Gorringe:2015cma}.

\subsection{%\boldmath
 LFV $\mu\to e$ Processes \label{sec:cLFV}}

We now know that lepton flavor is violated (LFV) in neutral lepton sector
in the form of neutrino mixing, or that neutrinos have mass (true sign of NP).
But so far we have not observed any LFV in charged lepton sector.

The $\mu \to e\gamma$ transition is analogous to $b\to s\gamma$
(and certainly preceded it conceptually), but the ``penguin'' loop
is extremely suppressed by neutrino mass (at the eV level, compared with $m_t \cong 173$ GeV)
and is practically negligible.
It is thus a very good probe of NP, and has been pursued
ever since the muon became known to mankind.
The current bound~\cite{MEG:2016wtm} of
\be
{\cal B}(\mu \to e\gamma) < 4.2 \times 10^{-13}, \quad {\rm (MEG)}
 \label{eq:MEG}
\ee
at 90\% C.L. is the final result of the MEG experiment, and improves
the result of the previous MEGA experiment~\cite{Brooks:1999pu}
by a factor of 30. Earlier MEG results are documented in PDG~\cite{PDG}.

The signal is rather distinct: with $\mu^+$ decay at rest,
one has back to back $e^+$ and photon in coincidence and with equal energy.
As an $n$th generation experiment in the long quest for $\mu \to e\gamma$,
the MEG experiment has a large Liquid Xenon photon detector, opposite
an ultra-light drift chamber accompanied by fast timing counters
all placed in an innovative gradient field COBRA magnet.
Energy and angular resolution plus timing work together
to suppress the main background of accidental coincidence
between a regular $\mu \to e\nu\bar\nu$ decay and another that
emits a photon, or from $e^+e^- \to \gamma\gamma$.
For this reason, one uses a DC rather than pulsed muon beam.
The result of Eq.~(\ref{eq:MEG}) corresponds to
$7.5 \times 10^{14}$ stopped $\mu^+$ accumulated during 2009--2013.

With null observation at the LHC, traditional MSSM and SUSY-GUT
model expectations for $\mu \to e\gamma$ are now somewhat mute.
But given that pockets (cracks) of parameter space remain,
it would be interesting to watch the correlation~\cite{Kersten:2014xaa}
between $\mu \to e\gamma$ and muon $g-2$ in SUSY context.
Motivated by large neutrino mixing~\cite{PDG} and still in context of SUSY,
linking with flavor symmetries suggest~\cite{Blankenburg:2012nx}
${\cal B}(\mu \to e\gamma)$ at the $10^{-13}$ level,
while linking~\cite{Antusch:2006vw} with seesaw and baryogenesis-through-leptogenesis
still leaves some parameter space to be probed.
From the latter perspective, however,
$\tau\to \mu\gamma$ may not be observable at Belle II.
In any case, the MEG-II upgrade is currently under construction,
and is expected to enter physics run for 3 years starting 2017.
The aim is to improve the MEG limit by another order of magnitude,
to the $4 \times 10^{-14}$ level.
Besides increased beam current,
there is a factor of two improvement in resolution in all aspects,
e.g. changing from 2$^{\,\prime\prime}$ PMT to SiPM for the LXe photon detector.

If the $\mu\to e\gamma$ dipole transition process probes loop effects,
a different class of experiments such as $\mu N \to eN$ and $\mu\to e\bar ee$
probe contact 4-fermion interactions.
The current limit on $\mu\to e$ conversion is held by SINDRUM II~\cite{Bertl:2006up},
\be
R_{\mu e} = \frac{\Gamma(\mu + (A,\;Z) \to e + (A,\;Z))}
                 {\Gamma(\mu + (A,\;Z) \to \nu_\mu + (A,\;Z-1))}
 < 7 \times 10^{-13}, \quad {\rm (SINDRUM\ II)}
 \label{eq:SINDRUM-II}
\ee
at 90\% C.L., with muonic atoms formed on gold nucleus.
The conversion electron has energy basically the same as the muon mass.
The limit of Eq.~(\ref{eq:SINDRUM-II}) is a bit dated,
and improvement with same method appears stiff,
since the SINDRUM II experiment already used $\sim 1$ MW proton beam power.
The next generation experiments are based on the idea~\cite{Dzhilkibaev:1989zb}
of using a strong solenoidal B field to confine soft pions
 then collect the decay muons,
which drastically reduces the requirement on proton beam power.
Thus, two new experiments, aiming for a staggering \emph{4 orders of magnitude}
improvement, invest on superconducting solenoids.
Background from beam pion capture, followed by nuclear $\gamma$ decay
with $\gamma$ conversion resulting in $e^+$,
is mitigated by using pulsed proton beam and waiting out the prompt decay.
Muon decay in orbit, where the tail of energy distribution
above $m_\mu/2$ can enter the signal region, constitutes the main background,
and detector resolution is key.

The Mu2e experiment is under construction at Fermilab.
It aims for commissioning the solenoids in 2020, and physics run starting 2021
for 3 years to accumulate a total of $10^{18}$ stopped muons,
with the goal of reaching $R_{\mu e} < 10^{-16}$~\cite{Bartoszek:2014mya}.
The COMET experiment~\cite{Kuno:2013mha} at J-PARC takes
a staged approach aiming for fast start.
For COMET Phase I, detector and facility preparation is underway, aiming for
2018 run start that could reach below $10^{-14}$.
Phase II would use a more sophisticated S-shaped muon transport solenoid,
higher beam power and improved detector, and aims for run start in 2022,
also with a goal to reach below $10^{-16}$.
Both experiments would use muonic atoms formed on Al, as well as 8 GeV protons.
There is another experiment, DeeMe~\cite{DeeMe} at J-PARC, that is based on a different approach.
It would use 3 GeV proton beam at $\sim 1$ MW beam power
on a thick target for pion production, decay \emph{and} muon stopping,
then collect the electrons from $\mu \to e$ conversion
and use a second beamline as part of the ``spectrometer''.
DeeMe aims for physics run in 2017, and could reach below $10^{-13}$,
with improvement possible by optimizing target and running longer.

\begin{figure}[ht]
\begin{center}
\begin{minipage}[t]{9.5cm}
%\hskip0.5cm
 %\epsfig{file=mec_new.eps,scale=0.48}
 \includegraphics[width=9cm]{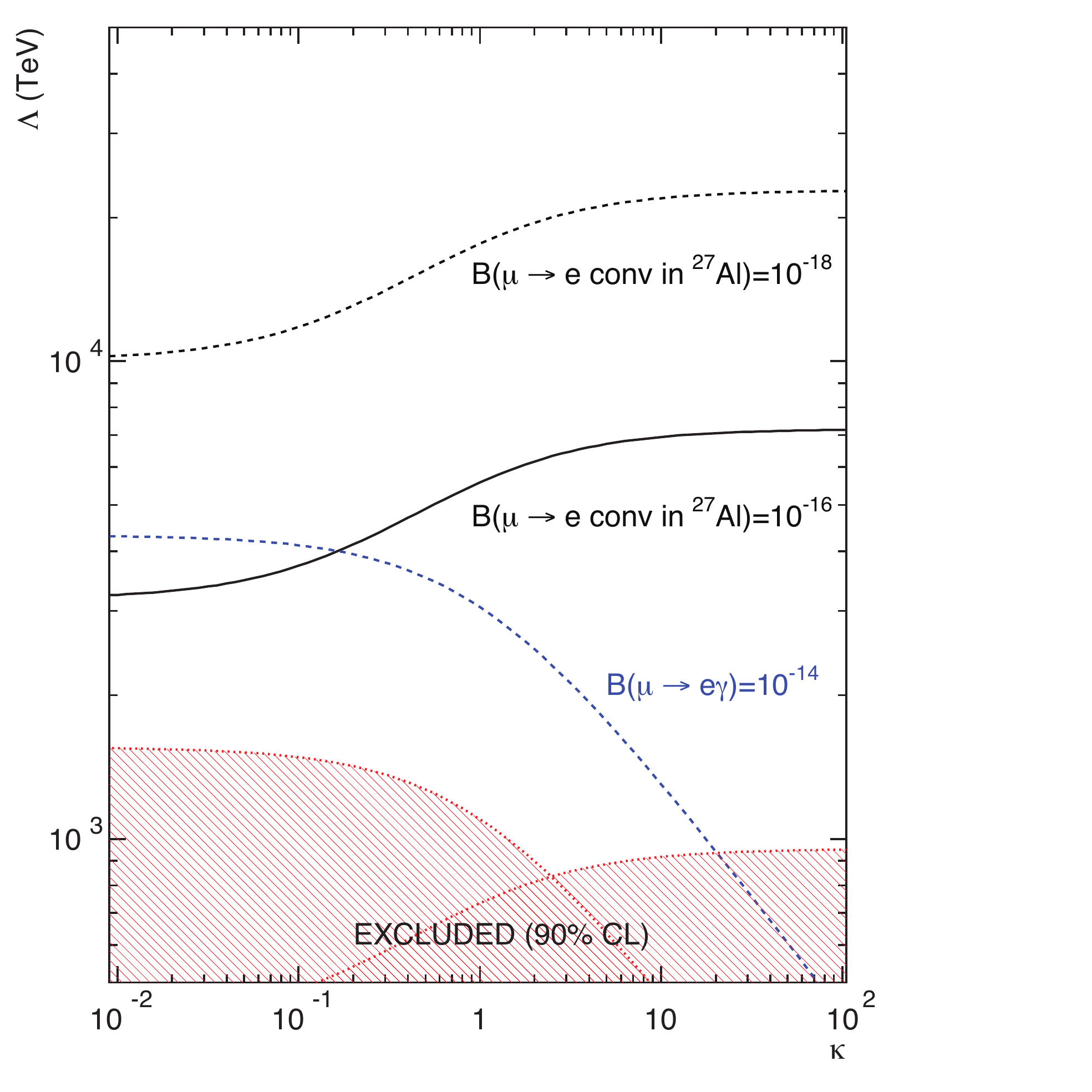}
\end{minipage}
\vskip-0.4cm
\begin{minipage}[t]{16.5 cm}
\caption{
Mass scales probed by $\mu \to e\gamma$ and $\mu \to e$ conversion,
where the $\kappa$ parameter ``dials'' between dipole-like (small $\kappa$)
to contact-like (large $\kappa$) interactions.
MEG-II aims to push the $\mu \to e\gamma$ excluded region
close to the lower dotted curve, while
%The lower right excluded region is the SUNDRUM II result of
%Eq.~(\ref{eq:SINDRUM-II}).
COMET and Mu2e aim to push the $\mu\to e$ conversion excluded region
beyond the solid curve.
 [Source: courtesy A. de Gouv\^ea, updated from Ref.~\cite{deGouvea:2013zba}]
 \label{deGouvea}
}
\end{minipage}
\end{center}
\end{figure}

Thus, there is a robust ongoing program for $\mu \to e\gamma$
and $\mu\to e$ conversion search. If the former is related to
SUSY and neutrino mixing, the latter has even broader coverage of possible NP,
as it probes new contact interactions including $Z'$, extra Higgs,
heavy neutrinos, leptoquark, compositeness, etc.,
i.e. similar to the LHC, with higher reach in NP scale, probing up to $10^4$ TeV.
We cannot do justice to the potential physics contact,
but refer the reader to Ref.~\cite{deGouvea:2013zba} for a review.
We just use a figure (Fig.~\ref{deGouvea}) from this reference to illustrate the
NP scale probed by these experiments.

Finally, analogous to $\mu eqq$ contact terms probed by $\mu\to e$ conversion,
$\mu^+ \to e^+e^+e^-$ could probe $\mu eee$ contact terms.
The old but current limit~\cite{PDG} at $10^{-12}$ is from SINDRUM almost 30 years ago.
The Mu3e experiment~\cite{Mu3e} at PSI aims to push the limit down to $10^{-16}$
by ultimately having $2\times 10^9$ muon decays {\it per second}.
Phase I running with $10^8$ muon decays per second could start in 2017 or 2018.
Note that, analogous to Fig.~\ref{deGouvea} for $\mu \to e\gamma$ vs $\mu\to e$ conversion,
in the dipole-like region, $\mu^+ \to e^+e^-e^+$ should be scaled by $\alpha$
when compared with $\mu \to e\gamma$ in rate.
However, for contact-like interactions, $\mu^+ \to e^+e^+e^-$
wins over $\mu \to e\gamma$ in energy scales probed.
As an explicit model that illustrates the probing power of $\mu\to e$ transitions,
let us take a recent study~\cite{Beneke:2015lba} of warped extra dimensions,
where KK excitations of multi-TeV now seems beyond the reach of LHC.
After applying $\mu \to e\gamma$ and $\mu \to e$ conversion
as the most stringent constraints, the combination of
MEG-II, Mu3e, and certainly Mu3e and COMET as well,
could explore the effects of extra dimensions on LFV
beyond $m_{\rm KK} > 20$ TeV.

\subsection{Muon g\ --\ 2 \label{sec:g-2}}

It can be said that particle physics started in 1947 in part
with Schwinger's seminal calculation of
 $a_e \equiv g_e/2 - 1 = \alpha/2\pi$ at one loop,
the ``anomalous'' gyromagnetic ratio (i.e. deviation from $g = 2$).
In flavor language, one could have called this the first ``penguin'' diagram.
It is pretty amazing that, 70 years later, we still have a ``muon $g-2$ anomaly",
now meaning deviation between experiment and SM prediction,
\be
\Delta a_{\mu} = a_{\mu}({\rm Expt}) - a_{\mu}({\rm SM})
 = (274 \pm 76) \times 10^{-11},
 \label{eq:Davier}
\ee
where the experimental value is measured by BNL-E821~\cite{Bennett:2006fi},
$(11659208.0 \pm 5.4 \pm 3.3) \times 10^{-10}$,
and the SM expectation is from the recent update~\cite{Davier:2016iru} of
hadronic vacuum polarization (HVP) contribution.
The more than $3\sigma$ deviation has persisted for more than a dozen years,
and could be handily explained~\cite{Stockinger:2006zn} by MSSM.
Although SUSY has not been sighted yet at the LHC, the persistent discrepancy
has motivated serious efforts to remeasure $a_\mu$,
as well as to refine the theoretical calculation.

After relocating and refurbishing the BNL muon storage ring,
Fermilab-E989 (also called Muon g-2) experiment~\cite{Grange:2015fou}
has meticulously shimmed the magnetic field to 3 times better uniformity
than at BNL, and fully around the 44 m storage ring circumference.
With improved detectors and other technologies,
and with over 20 times more muons, the aim is to
improve the experimental error by a factor of 4,
from 540 ppb down to 140 ppb.
Initial physics run should commence in 2017 and continue until 2019.
But one also has to improve the theory with a concerted~\cite{Benayoun:2014tra} effort.
The QED (calculated to $O(\alpha^5)$!) and electroweak uncertainties are under control,
but improvement of hadronic uncertainties are critical.
For HVP contribution, which enters at $O(\alpha^2)$ level,
we have already mentioned the recent update~\cite{Davier:2016iru}
that utilize $e^+e^- \to$ hadrons data.
For hadronic light-by-light (HLbL) scattering contribution
that enters at $O(\alpha^2)$ level,
recent lattice progress~\cite{Blum:2016lnc} seems to halve
both the numerical value and (statistical) error
compared with models~\cite{Prades:2009tw}.
If so, it would heighten the ``anomaly".
However, systematic errors from finite volume and lattice spacing
are still under investigation, which could increase~\cite{Blum:2016lnc} the value.
But in any case the two approaches are independent,
with unrelated systematic errors.
Note that lattice is also catching up~\cite{Blum:2015you} with
the numerical HVP contribution.\footnote{
For those interested in following lattice developments,
including $K$, $D$ and $B$ meson physics,
we refer to the Flavor Lattice Averaging Group (FLAG)~\cite{FLAG}.
}

We cannot do full justice to this important subject that is somewhat outside of our scope,
but look forward to major progress on the lingering muon $g-2$ anomaly
in the coming few years.

\subsection{Electric Dipole Moments (EDM) \label{sec:EDM}}

We have given only a cursory account of the long-standing muon $g-2$ anomaly,
in part because of its flavor conserving nature.
Unlike magnetic moments, however, electric dipole moments of
nondegenerate particles violate $CP$,\footnote{
It is $T$-violating, hence $CP$-violating by $CPT$ invariance.
}
and is thus close to our core subject.
But again we give only a cursory account,
as the experimental studies are both diverse and rather specialized.
For instance, measurements of neutron EDM, $d_n$,
typically utilize trapped ultracold neutrons (UCN),
while the extraction of electron EDM, $d_e$, involves
molecular and atomic (and even nuclear, e.g. mercury EDM) physics.
On the particle physics side, SM effects from CPV phase in CKM matrix
contribute only at rather high loop order, and current experiments are many
orders of magnitude away from SM expectations,
i.e. $d_n^{\rm SM} \sim 10^{-32}\ e\,$cm, $d_e^{\rm SM} \sim 10^{-40}\ e\,$cm.
Thus, discovery of EDM would definitely imply NP,
albeit in a rather indirect way.
We will briefly discuss the two cases of $d_n$ and $d_e$ as examples,
referring to more specialized reviews, such as Ref.~\cite{Engel:2013lsa},
for more detail.

If new CPV phases in SUSY were of ${\cal O}(1)$, %were of $O(1)$,
and SUSY particles have mass around the weak scale,
then neutron EDM should have appeared above the $10^{-24}\ e\,$cm level.
The current experimental bound~\cite{Afach:2015sja}, from ILL in France,
gives (final update from 2006 result~\cite{PDG})
\be
d_n^{\rm Expt} < 3.0 \times 10^{-26}\ e\,{\rm cm}, \quad {\rm (ILL)}
 \label{eq:nEDM}
\ee
at 90\% C.L., which means either CPV phases are small in MSSM,
or SUSY breaking scale is considerably higher than 1 TeV.
Judging from the lack of evidence for SUSY at the LHC so far
 (at neither 8 nor 13 TeV),
the latter is becoming more and more likely.
But it also means that \emph{$d_n$ could appear at any time}.
Of course, the smallness of $d_n$ already gave the puzzle
of an extremely small $\theta_{\rm QCD}$,
and the possibility of axion as explanation.
However, neutron EDM also probes quark EDMs $d_u$, $d_d$ as well as
the corresponding chromo-dipole moments, hence fascinating for theorists.

On the experimental side, there is a world-wide race now to reach below
$10^{-27}\ e\,$cm using UCN sources, to improve Eq.~(\ref{eq:nEDM})
by two orders of magnitude.
For example, the ILL setup was moved to PSI, where a dedicated UCN source was built.
The latter moderates spallation neutrons through heavy water, then solid D$_2$ crystals.
The PSI nEDM experiment~\cite{nEDM-PSI} began data taking in 2015 and
will last until 2017. The sensitivity should reach below $10^{-26}\ e\,$cm.
The n2EDM experiment to follow targets reaching below $10^{-27}\ e\,$cm,
where data taking could start in 2020.
Across the Atlantic, the SNS nEDM experiment~\cite{nEDM-SNS}
at the Oak Ridge SNS (Spallation Neutron Source) adopts a different
and novel~\cite{Golub:1994cg} approach, using superfluid $^4$He as
both the UCN moderator, as well as the high voltage insulator to sustain high electric field.
It further uses $^3$He as co-magnetometer and superconducting shield,
to control and measure magnetic field systematics.
With demonstration phase close to completion,
large scale integration and commissioning should converge at ORNL by 2019.
The sensitivity, assuming 3 years of running,
aims at $2\times 10^{-28}\ e\,$cm by the early 2020s.

We note that there is ongoing R\&D to pursue proton EDM, $d_p$,
measurement using a storage ring. This bears some analogy with
muon storage ring study of $g-2$ (which would measure $d_\mu$ parasitically),
but would be an ``all electric'' ring with no B field.
The target is to reach sensitivity of $10^{-29}\ e\,$cm, but schedule is not yet clear.

The current leading edge of charged particle EDM search is that of the electron,
$d_e$, where the limit from the ACME experiment~\cite{Baron:2013eja} gives,
\be
d_e^{\rm Expt} < 8.7 \times 10^{-29}\ e\,{\rm cm}, \quad {\rm (ACME)}
 \label{eq:eEDM}
\ee
at 90\% C.L. ACME utilizes polar thorium monoxide (ThO) molecule,
which has internal effective electric field $E_{\rm eff}$ of
order 84 GV/cm. We cannot describe the methodology here,
which uses molecular beams and lasers,
but since this is a first generation experiment of the type,
and there are other approaches (e.g. $^{199}$Hg, $^{224}$Ra, etc.) as well,
the result of Eq.~(\ref{eq:eEDM}) stands further improvement.
As an example of the theories probed, we quote the study~\cite{Inoue:2014nva}
of 2HDM-II with CPV in Higgs potential, which induces
mixing between $CP$-even and $CP$-odd neutral Higgs bosons.
It is found that, at present, the ThO result on $d_e$ poses the
most stringent constraint, while neutron and mercury constraints
are less stringent, and furthermore suffer from hadronic and
nuclear matrix element uncertainties.
However, given the expected progress, this indirect
probe of NP scales would be
complementary to the direct search at the LHC.
Note that 2HDM-II naturally follows from, but does
not necessarily imply, SUSY. Together, these two types of
NP provide sufficient motivation for the continued quest of EDMs,
and the NP-CPV phases carried by scalar particles are
being probed by current EDM searches.
If a discovery is made, one would need a lot of improvement in
hadronic and nuclear matrix element estimates to disentangle the
underlying NP~\cite{Engel:2013lsa} from the multiple probes.

\section{Tau/Charm}

In this section we put the discussion of tau and charm physics
together, in part because each discussion is short,
and also because ``tau/charm factory'' are often discussed
together due to similar production energy for an $e^+e^-$ collider.

The third generation $\tau$ lepton provides a unique and clean probe of lepton universality and
charged lepton flavor violation, which extends from the previous section.
%These studies are very clean, given that the dilution from SM is tiny.
The $\tau$ lepton itself is also a powerful tool for searches in high energy collisions.
For example, the lepton-flavor violating $h \to \tau\mu$ decays (next section) can arise from
several possible extensions to SM, including additional Higgs doublets at tree level,
or new vector bosons at loop level.
Studying $\tau$ lepton decays can also probe fundamental SM parameters,
such as measuring the CKM element $|V_{us}|$ by comparing $\tau$ decay widths
to hadrons with and without strange quark.
From detailed comparisons with perturbative QCD corrections,
parameters such as the strong coupling constant $\alpha_s$
at $\tau$ mass can be extracted.

Many recent contributions come from lepton collider experiments,
in particular the B factory experiments Belle and Babar, which each have several hundred million
$\tau$-pair events produced in a relatively clean environment.
New contributions from LHC experiments are expected as well.

The second generation up-type quark, charm, is the third heaviest known quark,
with mass slightly below the $\tau$ lepton.
%Physics of the charm quark is very rich:
%charm mesons or baryons, which contain one or more charm quarks,
%or the charmonium bound states which are formed by a pair of charm-anticharm quarks,
% play an important role in the studies of weak or strong interactions.
A highlight of charm physics is that four quark or ``molecular'' states, as well as pentaquark states,
emerged from discoveries of exotic charmonia.
But we shall refrain from discussing this rich field of exotic charmed meson
and charmonium spectroscopy, as these do not pertain to the physics beyond SM
that we are interested in, and refer the reader to the companion review~\cite{Stone}.
%Charm quarks mostly decay weakly into a strange quark with a $W^\pm$ boson as the mediator,
%or via annihilation with an anticharm quark in charmonium states.
%Thus, charm particles provide a unique platform to examine the dynamics
%associated with up-type quarks in a bound state, including $D^0$--$\overline D^0$ oscillations.

The charm system is in principle an excellent place to look for New Physics contributions,
in the sense that the decay branching fractions of FCNC processes as well as CPV asymmetries
are very small in SM.
Any nonzero value from experimental measurement can be a strong hint of physics beyond SM.
However, compared with B physics where leading decays are suppressed by $|V_{ub}| \ll |V_{cb}| \sim 0.04$,
the potential NP effects are hampered by fast and unhindered decay rates
as the leading weak decay is Cabibbo allowed, $|V_{cs}| \cong 1$.
Thus, charm as a probe for NP is not particularly promising.
On the other hand, knowledge of charm quark decays are useful for many studies in the
B physics sector, for example the CKM angle $\gamma$ ($\phi_3$) measurements
via the various ``$DK$'' methods rely on $D$ decays as part of the program.

Reflecting the Cabibbo allowed leading decay but Cabibbo suppression of loop processes,
unlike the bottom or strange systems, the unitary triangle for charm is very squashed,
resulting in rather tiny $CP$ violation effects.
CPV in the charm sector is restricted to Cabibbo-suppressed (CS) decay modes
(or through $D^0$--$\overline D^0$ mixing, discussed in Sec.~\ref{sec:Dmix-CPV}).
The expected DCPV asymmetries in SM are of order $10^{-3}$,
but the estimations are not very robust due to large uncertainties from
penguin amplitudes and other hadronic effects.
New Physics can enhance the DCPV asymmetries, and it had been commonly assumed that
finding a DCPV asymmetry at $10^{-2}$ level could be a hint for new CPV sources.
However, recent estimates~\cite{Brod:2011re,Bhattacharya:2012ah} suggest that
DCPV asymmetries of ${\cal O}(10^{-2})$ can be accommodated within SM.
Thus, finding DCPV asymmetries larger than ${\cal O}(10^{-2})$ is
needed for establishing NP in the charm sector.
Nevertheless, one should probe every possible corner experimentally.

Since the new millennium, measurements of CPV in charm sector have been dominated by Belle and BaBar,
and more recently joined by the LHCb experiment.
%contributes significantly to many of the CPV measurements,
%thanks to the huge production cross section at the LHC.
The experimental precision %of measured CPV asymmetries
has been pushed to the level of $O(10^{-2})$ with uncertainties of a few per mille.
Finding a truly sizable CPV asymmetry in charm sector would be difficult to accommodate by SM,
but the current results are still within the expectations from hadronic uncertainties.

In this section, we first discuss briefly recent tests for lepton universality and
lepton flavor violation in the $\tau$ sector (deferring $h \to \mu\tau$ to Sec.~\ref{sec:S2HDM}).
We then discuss the most recent results
associated with the charm quark, focusing mainly on $CP$ violation.

\subsection{%\boldmath
 New Physics in $\tau$ Decay}

In SM, the charged current weak interaction couples with the same strength to all lepton flavors.
Tests of lepton universality can probe this assumption
and provide a strong constraint to the model extensions, for example to some
lepton-specific two Higgs doublet models proposed to solve the muon $g-2$ anomaly.
Although the direct tests of lepton universality with $\tau$ decays
do not show any hint of deviation from SM expectation,
there are some unresolved tensions in semileptonic $B$ meson decays,
such as $B \to D^{(*)}\tau\nu$ and $B \to K\ell\ell$,
as introduced in Sec.~\ref{sec:B-high}.

%Consider a $\tau$ lepton decaying into a lighter lepton $e$ or $\mu$,
%accompanied by two neutrinos $\nu_\tau \overline{\nu}_e$ or $\nu_\tau \overline{\nu}_\mu$.
%The process is exactly governed by t
The charge current  induced leptonic
%though a $W$ boson, and the
decay width is
%can be expressed as
%
\be
\Gamma(\tau\to\ell\nu\overline{\nu}_\ell)
= {B(\tau\to\ell\nu\overline{\nu}_\ell) \over \tau_\tau}
= {G_\tau G_\ell \, m^5_{\tau} \over 192\pi^3}
f\left( {m_\ell^2 \over m_\tau^2} \right)
%R_W^\tau R_\gamma^\tau\,,
\left(1 + \frac{3}{5}\frac{m_\tau^2}{M_W^2}\right)
\left[1 + \frac{\alpha(m_\tau)}{2\pi}\left(\frac{25}{4} - \pi^2\right)\right],
\ee
where $G_\ell = g_\ell^2/4\sqrt{2}M_W^2$ for $\ell = e,\ \mu$ are the Fermi coupling constants,
%and can be expressed by the weak interaction coupling constant and mass of $W$ boson,
$f$ is a phase space factor, %$R_V^\tau$ are
and we have made explicit the radiative corrections~\cite{MarSir}.
By inserting the ratios of partial widths and the measured $\tau$ lifetime,
the latter significantly improved by a Belle measurement~\cite{tau_tau-Belle},
the following ratios of coupling constants are obtained by HFAG 2016~\cite{Amhis:2016xyh}
using purely leptonic processes,
\begin{equation}
\left( {g_\tau \over g_\mu} \right) = 1.0010\pm0.0015,~~
\left( {g_\tau \over g_e} \right) = 1.0029\pm0.0015,~~
\left( {g_\mu \over g_e} \right) = 1.0019\pm0.0014. \quad {\rm (HFAG)}
\end{equation}
If semi-hadronic processes such as $\tau\to (K,\pi)\nu_\tau$ or
$(K,\pi)\to \mu\overline{\nu}_\mu$ are also considered,
the ratio ${g_\tau / g_\mu}$ can be further combined to $1.0000 \pm 0.0014$,
which is in remarkable agreement with lepton universality,
providing a very strong constraint to New Physics models in the lepton sector.

Just like the pursuit of $\mu \to e$ transitions, observing lepton flavor violation
(LFV) would be an unambiguous signal of New Physics.
LFV processes are absent at tree level in SM,
and can occur only through tiny neutrino masses at loop level,
highly suppressed by the equivalent of the GIM mechanism.
%and the resulting decay rate is not accessible by any feasible experimental effort.
The decay rate for $\tau \to \mu\gamma$ is negligible in SM, %at O($10^{-40}$),
while the stringent $\mu\to e\gamma$ bound of Eq.~(\ref{eq:MEG}) dampens one's hope.
But several NP scenarios can in principle increase the rate for this mode,
or some other decay channels, to more accessible values.

The sensitivity to a particular $\tau$ decay channel is very model dependent.
Let us mention a few examples.
Consider Higgs-mediated decays in SUSY seesaw models~\cite{Dedes:2002rh},
where LFV can arise through the renormalization of soft SUSY breaking terms.
%early results suggested that $\tau\to \mu\mu\mu, e\mu\mu$ could reach up to $4\times 10^{-4}$.
But %this may no longer be relevant,
as we have no evidence for SUSY so far,
LHC Run1 results seem to push expectations for $\tau\to \mu\mu\mu$, $e\mu\mu$
towards $10^{-9}$~\cite{Goto:2014vga},
which are not quite within experimental reach in the near future.
By introducing additional heavy right-handed Majorana neutrinos,
or additional left-handed and right-handed neutral singlets~\cite{Cvetic:2002jy},
the branching fractions of $\tau\to \mu\gamma,\ e\gamma$ and
$\tau\to \mu\mu\mu,\ eee$ can be raised to approximately
$O(10^{-10})$--$O(10^{-8})$, which is still pertinent.
Adding a non-universal gauge boson $Z^\prime$ in topcolor-assisted technicolor models,
the branching fraction of $\tau \to eee$ or $\mu\mu\mu$
can be as large as $10^{-8}$ within a range of parameter space~\cite{Yue:2002ja}.
But it is not clear whether these models still stand with LHC data,
and correlation with the stringent MEG limit on $\mu \to e\gamma$ is a concern.
Finally, in the 2HDM-III that we would discuss in Sec.~\ref{sec:S2HDM},
because of flavor changing neutral Higgs couplings and other new Yukawa couplings,
$\tau\to\mu\gamma$ might be close~\cite{Chiang:2016vgf} to the current experimental limit.
Suffice it to say that many proposals can raise the LFV branching fractions to
the level of $O(10^{-10})$--$O(10^{-8})$, %(even $O(10^{-7})$?),
and also generate other possible LFV B meson or Higgs decays.
While NP models are now more and more constrained by LHC search and $\mu\to e\gamma$,
the experimental study of rare $\tau$ decays provide complementary information,
and should continue in any case.

\begin{figure}[ht]
\begin{center}\hskip-0.3cm
\includegraphics[width=14cm]{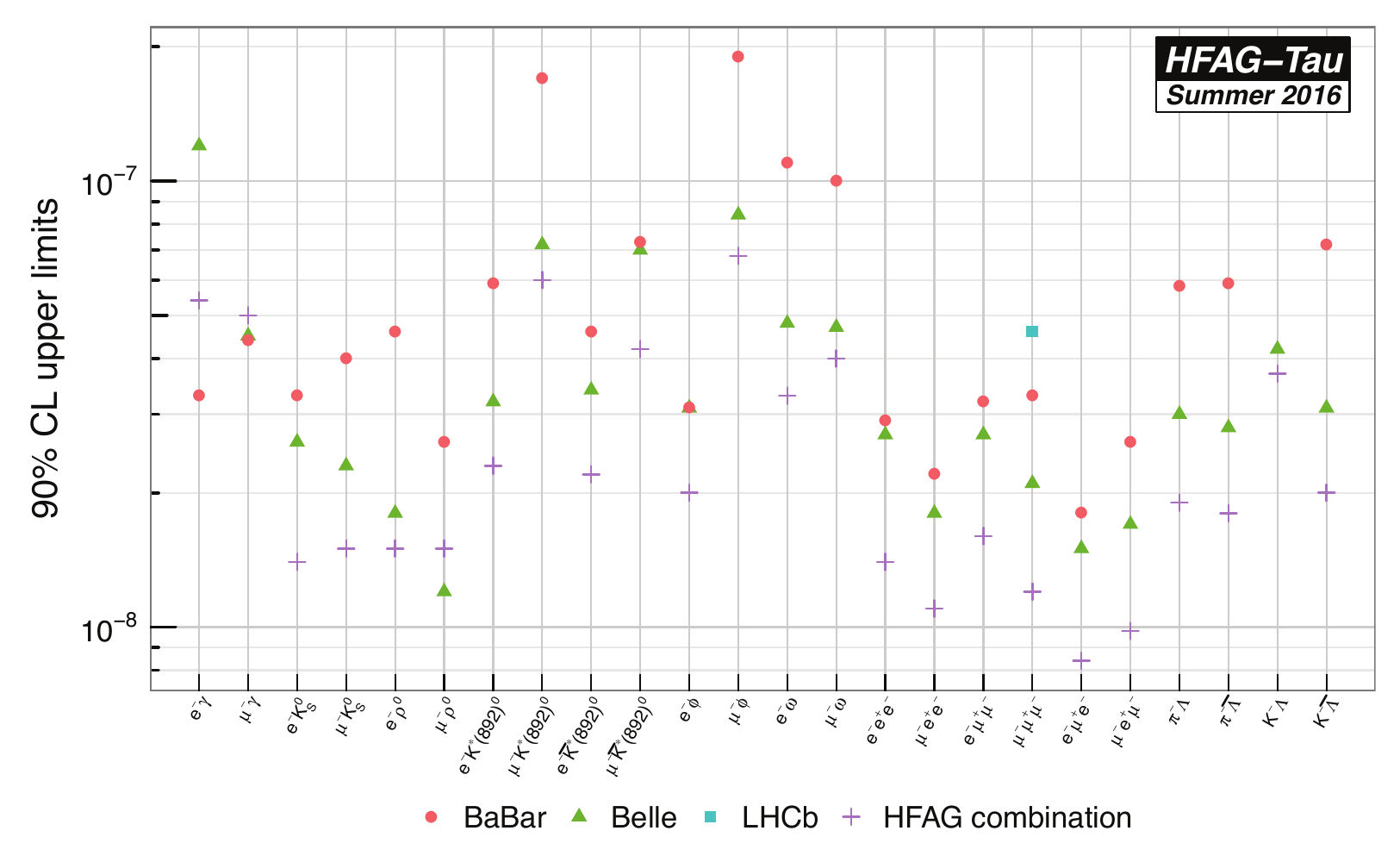}
\caption{
Upper limits on LFV $\tau$ decays,
including published individual limits and combined values
 (marked by $+$).
Due to statistical fluctuations of single results,
the combined limits are not necessarily tighter. %than individual limits.
 [Source: HFAG 2016] %http://www.slac.stanford.edu/xorg/hfag]
}
\label{fig:tau-lfv-combs}
\end{center}
\end{figure}

The summary of experimental upper limits from HFAG~\cite{Amhis:2016xyh}
is given in Fig.~\ref{fig:tau-lfv-combs},
which shows a plethora of search modes, especially as compared with rare muon decays
of the previous section.
Most of the existing measurements are from Belle and BaBar.
The most stringent limits are for three light charged lepton final states,
i.e. combinations of $e$ or $\mu$, with an entry of $\tau \to \mu\mu\mu$ from LHCb.
The best single limits are from Belle with a data set consisting of
7.2 million $\tau$ pairs~\cite{Hayasaka:2010np}.
The analysis was carried out in a very clean environment with
expected background level of $<0.2$ events.
No candidate events were observed in any combination of lepton flavors.
Given that all measurement errors are statistics dominant,
combinations of limits have been carried out by HFAG based on the standard $CL_{s}$ method.
Due to possible statistical fluctuations of single results,
the combined limit could be weaker than from individual experiments.
Most of the combined limits are in the range of $10^{-8}$--$10^{-7}$,
which are still 1--2 orders of magnitude higher than estimates by
various NP models, as illustrated in the previous paragraph.
%, which do not concern us until a discovery emerges.
The exception is $\tau\to \mu\gamma$, in conjunction with
$h \to \mu\tau$ search at the LHC, as discussed further in Sec.~\ref{sec:S2HDM}.

The upcoming Belle~II program, together with advances from LHCb,
should provide substantial update to these tests of LFV,
and more theoretical attention is encouraged.

\subsection{%\boldmath
 Direct CPV in Charm}

Non-zero CPV asymmetry requires at least two interfering amplitudes,
and with both weak and strong phase differences.
In SM, DCPV in charm sector can appear in single Cabibbo suppressed decays,
such as interference between $c \to d$ tree and the highly suppressed $c \to u$ penguin processes.
%Direct CPV (DCPV) asymmetries in charm mesons occur when the amplitude for $D$ meson decay,
%differs from the amplitude for,
%
Consider a typical $D$ meson decay with amplitude $\mathcal{A}(D \to f)$,
and $\mathcal{A}(\overline{D} \to \bar{f})$ for anti-$D$ meson decay.
The DCPV asymmetry can be represented by a time-integrated asymmetry in the form of
\begin{equation}
A_{CP} = {\Gamma(D \to f) - \Gamma(\overline{D} \to \bar{f})
 \over \Gamma(D \to f) + \Gamma(\overline{D} \to \bar{f})}.
\end{equation}

For the $CP$ eigenstate two-body final states $f = K^+K^-$ and $\pi^+\pi^-$,
$A_{CP}$ can be expressed in two terms:
 the component associated with DCPV in the decay amplitudes and
 the component associated with indirect CPV in the mixing,
  or from interference between mixing and decay.
The indirect CPV term is in general independent of the decay channels,
hence the difference in CPV asymmetries between $D^0 \to K^+K^-$ and $D^0 \to \pi^+\pi^-$ decays,
$\Delta A_{CP} = A_{CP}(D^0 \to K^+K^-) - A_{CP}(D^0 \to \pi^+\pi^-)$, is
a good observable of DCPV.
The LHCb experiment caused a splash when it announced evidence~\cite{Aaij:2011in}
for nonzero $\Delta A_{CP}$, which subsequently faded with more data~\cite{Aaij:2014gsa}.
The most recent measurement from LHCb~\cite{Aaij:2016cfh} tags the flavor of charm
by the charge of the pion from $D^{*+} \to D^0 \pi^+$ decays,
and gives
\be
\Delta A_{CP} = -0.10\pm0.08\pm0.03\%,
\ee
which unfortunately is still consistent with zero, and in any case at the per mille level.

\begin{figure}[ht]
\begin{center}
\includegraphics[width=10cm]{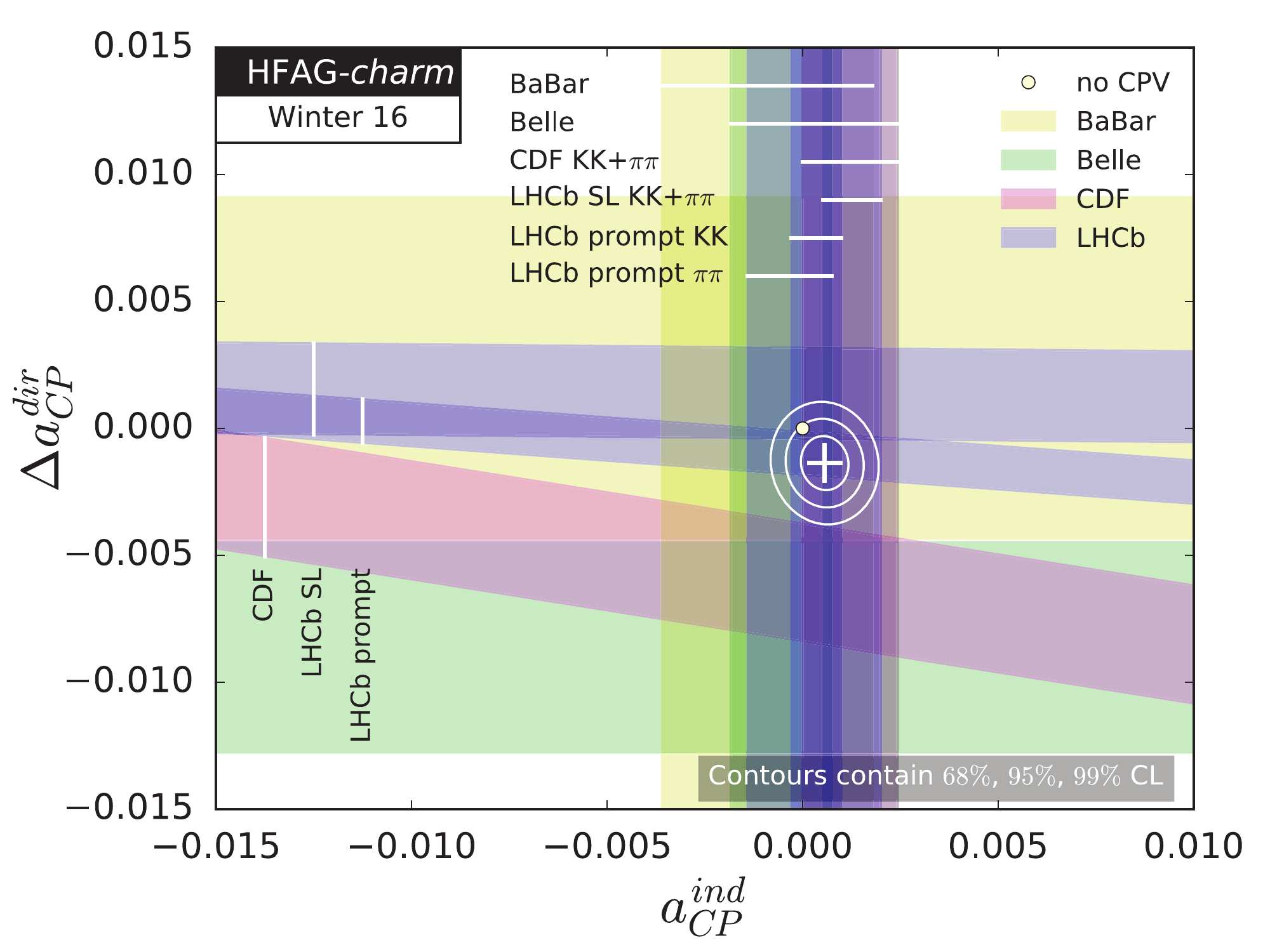}
\caption{
Combination plot for $\Delta a_{CP}^{\rm dir}$ and $a_{CP}^{\rm ind}$ from HFAG,
with 68\%, 95\%, and 99.7\% C.L. contours around the best fit value marked by $+$.
%, and is consistent with null CPV
 [Source: http://www.slac.stanford.edu/xorg/hfag]
}
 \label{fig:deltaACP_AGamma_fit}
\end{center}
\end{figure}

As mentioned, the interest is in finding CPV asymmetries above the percent level.
Following the formulation in Ref.~\cite{Gersabeck:2011xj},
the time-dependent observable $\Delta A_{CP}$ can be expressed by
\begin{equation}
\Delta A_{CP} = \Delta a_{CP}^{\rm dir} \left(1 + y_{CP}\, \overline{\langle t \rangle}/\tau\right)
 + a_{CP}^{\rm ind}\, {\Delta\langle t \rangle/\tau},
\end{equation}
where $\Delta\langle t \rangle$ is the difference of
mean decay time between $KK$ and $\pi\pi$ final state
and $\overline{\langle t \rangle}$ is their average,
$\tau$ is the true lifetime of the $D^0$ meson,
and $y_{CP} = (0.866 \pm 0.155)\%$~\cite{hfag} is a CPV observable related to
the mixing parameter $y$ (see the next subsection for a more detailed discussion).
$\Delta a_{CP}^{\rm dir}$ and $a_{CP}^{\rm ind}$ measure direct and indirect CPV,
and the current world best average from HFAG~\cite{Amhis:2016xyh} is
\begin{equation}
\Delta a_{CP}^{\rm dir} = (-0.137 \pm 0.070)\%, \quad
a_{CP}^{\rm ind} = (0.056 \pm 0.040)\%. \quad {\rm (HFAG16)}
 \label{eq:Del-aCP-t}
\end{equation}
The precision of these measurements has reached $O(10^{-3})$ already,
but the central values are within\ $2\sigma$ from the null CPV hypothesis,
%of $\Delta a_{CP}^{\rm dir} = 0$, $a_{CP}^{\rm ind} = 0$,
as can be seen in Fig.~\ref{fig:deltaACP_AGamma_fit}.

There are many other potential measurements in hadronic two-body $D$ decays.
For example, the $D^0 \to K^0_S K^0_S$ decay can have an enhanced CP violation even within the CKM picture.
An estimate~\cite{Nierste:2015zra} of penguin annihilation and exchange amplitudes
gives an upper bound on DCPV at 1.1\%, while the most recent measurement from Belle
has attained an uncertainty of 1.5\%~\cite{Abdesselam:2016gqq},
but with no indication for DCPV.
Many other measurements have uncertainties of a few percent,
so there is still plenty of room to look for CPV and New Physics.

Looking for DCPV with three-body or multi-body charm decays may
have better potential, since CPV can occur in the full decay phase space,
and local asymmetries can be larger than the integrated ones
(e.g. in effective two-body decays as discussed above).
In addition to a direct comparison between positively and negatively tagged distributions,
a more generic amplitude analysis can also be applied and has
direct access to the phase information of the decay amplitudes.
The triple-product correlations evaluated from the independent vectors
from a multi-body decay can also be studied and provide a different probe of $T$-odd asymmetries.
Note that multi-body charm decays are also essential for
the measurement of CKM angle $\phi_3$.

From an experimental point of view, the multitude of multi-body decays provide
a measurement bonanza.
The analyses require advanced techniques such as Dalitz analysis,
and modeling of decay amplitudes with kinematic variables of the decay daughters.
In a recent study of three-body $D^0 \to K_S K^\pm \pi^\mp$ decay by LHCb~\cite{Aaij:2015lsa},
the magnitude and phase information is studied in the Dalitz plot,
and a search for time-integrated CPV shows no clear evidence,
within an uncertainty of few \%.
Given the complexity of multibody decays, several new experimental methods are being developed.
For example, a model-independent technique, called ``energy test'', is introduced by
LHCb to study the Cabibbo-suppressed decay $D^0 \to \pi^+\pi^-\pi^+\pi^-$~\cite{Aaij:2016nki}.
The method provides an unbinned description of the decay,
and can be used to look for local excess of CPV asymmetries.
The technique is used to examine both $P$-even and $P$-odd CPV effects.
The $P$-even test separates the events according to flavor,
$D^0$ vs $\overline{D}^0$, for comparison.
The $P$-odd test distinguishes the events according to both flavor and sign of the triple products.
The core method is built on a test statistic $T$, which is
used to compare the average distances of events in phase space.
If the events in both samples are identical, the value of $T$
should be randomly distributed around zero.
Based on the resulting $T$ values and the simulated test statistic distributions,
a $p$-value of $(0.6 \pm 0.2)$\% (or 2.7 standard deviations) is
found in the $P$-odd CPV test, while the $P$-even test shows no hint.

All of these results on DCPV tests can be substantially improved in
the near future with a much larger data set from (HL-)LHC and Belle-II.
In particular, the experimental sensitivities are approaching SM-allowed upper bounds,
and discovery of CPV effects in the charm system could still be just around the corner.
But as stressed in the Introduction, we do not consider DCPV,
be it in $B$ or $D$ decays, as promising probes for NP.

\subsection{%\boldmath
 Indirect CPV in Charm
 \label{sec:Dmix-CPV}}

The mixing of neutral charm mesons is well-established.
The two mass eigenstates $|D_{1}\rangle$ and $|D_{2}\rangle$ of a
neutral charm and anticharm meson system can be expressed as
linear combinations of flavor eigenstates $|D^0\rangle$ and $|\overline{D}^0\rangle$,
\begin{equation}
|D_1\rangle = p\, |D^0\rangle + q |\overline{D}^0\rangle, \quad
|D_2\rangle = p\, |D^0\rangle - q |\overline{D}^0\rangle,
\end{equation}
with complex coefficients satisfying $|p|^2+|q|^2 = 1$.
The mixing parameters are defined as
\begin{equation}
x = 2(m_2 - m_1)/(\Gamma_1 + \Gamma_2), \quad
y = (\Gamma_2 - \Gamma_1)/(\Gamma_1 + \Gamma_2),
\end{equation}
where $m_{1,2}$ and $\Gamma_{1,2}$ are the masses and decay widths of the two mass eigenstates, respectively.
CP violation is established if a non-unity value is found for the parameter $\lambda_f$,
\begin{equation}
\lambda_f = - \left|{q \over p}\right| \left| {\overline{A}_{\bar{f}}\over A_f} \right| e^{-i\phi},
\end{equation}
where $A_f$ and $\overline{A}_{\bar{f}}$ are the amplitudes
for $D^0$ and $\overline{D}^0$ decaying into final state $f$ and $\bar{f}$,
and $\phi = \arg{(q/p)}$ is the CPV weak phase.
If $|q/p|$ deviates from unity, or if phase $\phi$ is nontrivial,
one has indirect CPV.

The decay $D^0 \to K^-\pi^+$  proceeds through a Cabibbo-favored (CF) tree diagram.
Decay to the charge-conjugate final state, $D^0 \to K^+\pi^-$, proceeds through
a Doubly Cabibbo-suppressed (DCS) process, or a mixing process together with a CF decay,
$D^0 \to \overline{D}^0 \to K^+\pi^-$.
The relative rates of DCS to CF decays is expressed as
\begin{equation}
R(t)^\pm \approx R_D^\pm + \sqrt{R_D^\pm} y^\prime\, {t\over\tau}
+ {(x^{\prime\pm)^2} + (y^{\prime\pm})^2 \over 4} \left({t\over\tau} \right)^2,
\end{equation}
where the $+$ and $-$ signs in the exponent denote the decays of $D^0$ or $\overline{D}^0$, respectively.
The primed parameters, $x^\prime$ and $y^\prime$, are rotated from the $x$ and $y$ parameters
by a strong phase difference between CF and DCS amplitudes.
The most recent measurement from LHCb~\cite{Aaij:2016roz} using
double-tagged $\overline{B} \to D^{*+} \mu^-X, D^{*+} \to D^0 \pi^+$ events, gives
\begin{eqnarray}
 \nonumber
& R^+_D = (3.474 \pm 0.081) \times 10^{-3}, & R^-_D = (3.591 \pm 0.081) \times 10^{-3}, \\
 \nonumber
& (x^{\prime+})^2 = (0.11 \pm 0.65) \times 10^{-4}, & (x^{\prime-})^2 = (0.61 \pm 0.61) \times 10^{-4}, \\
& y^{\prime+} = (5.97 \pm 1.25) \times 10^{-3}, & y^{\prime-} = (4.50 \pm 1.21) \times 10^{-3}.
  \quad \ \ {\rm (LHCb, Run 1)}
\end{eqnarray}
These results, together with many other recent measurements,
provide strong constraints on $|q/p|$ and $\phi$,
in particular for the region near $\phi = 0$.
A recent measurement using four-body $D^0 \to K^+\pi^-\pi^+\pi^-$ decay
has also been carried out~\cite{Aaij:2016rhq}.
The analysis measures the ratio between DCS and CF in bins of decay time,
\begin{equation}
R(t) \approx (r^{K3\pi}_D)^2 - r^{K3\pi}_D R^{K3\pi}_D
y^\prime_{K3\pi}\, {t\over\tau}  + {x^2+y^2\over 4} \left({t\over\tau}\right)^2~,
\end{equation}
where the measured $r^{K3\pi}_D$ is the phase space averaged ratio of DCS to CF amplitudes,
the coherence factor $R^{K3\pi}_D$ covers the phase difference between
DCS and CF amplitudes, $y^\prime_{K3\pi}$ is rotated from the mixing parameters $x$ and $y$
with the average strong phase difference $\delta_D^{K3\pi}$.
These parameters have been most precisely determined to date as
 $r^{K3\pi}_D = (5.67\pm0.12)\times 10^{-2}$ and
 $R_D^{K3\pi} y^\prime_{K3\pi} = (0.3 \pm 1.8) \times 10^{-3}$,
and are essential for the CKM $\phi_3$ angle measurement.

\begin{figure}[ht]
\begin{center}
\includegraphics[width=7cm]{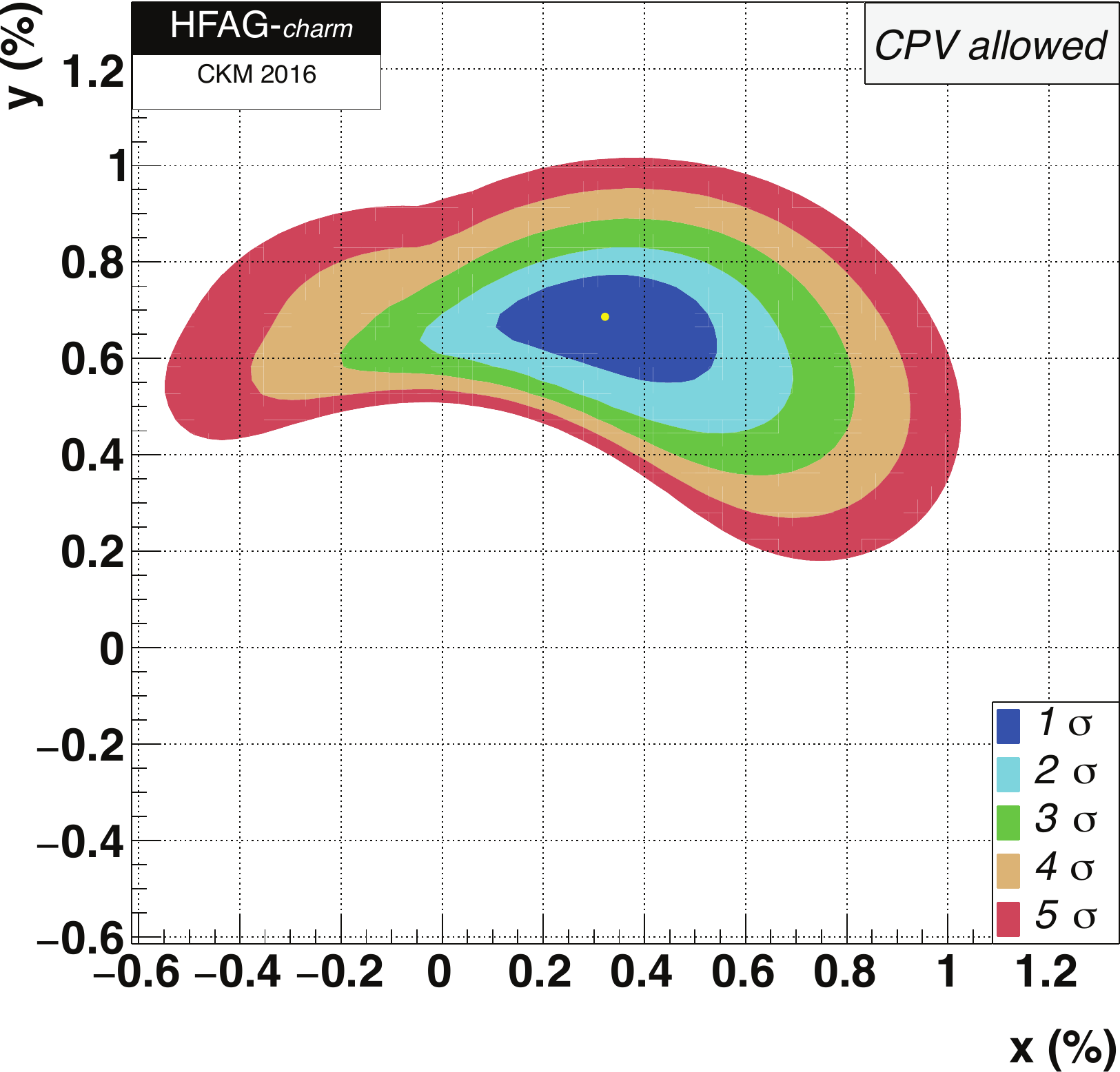} \hskip0.8cm
\includegraphics[width=7cm]{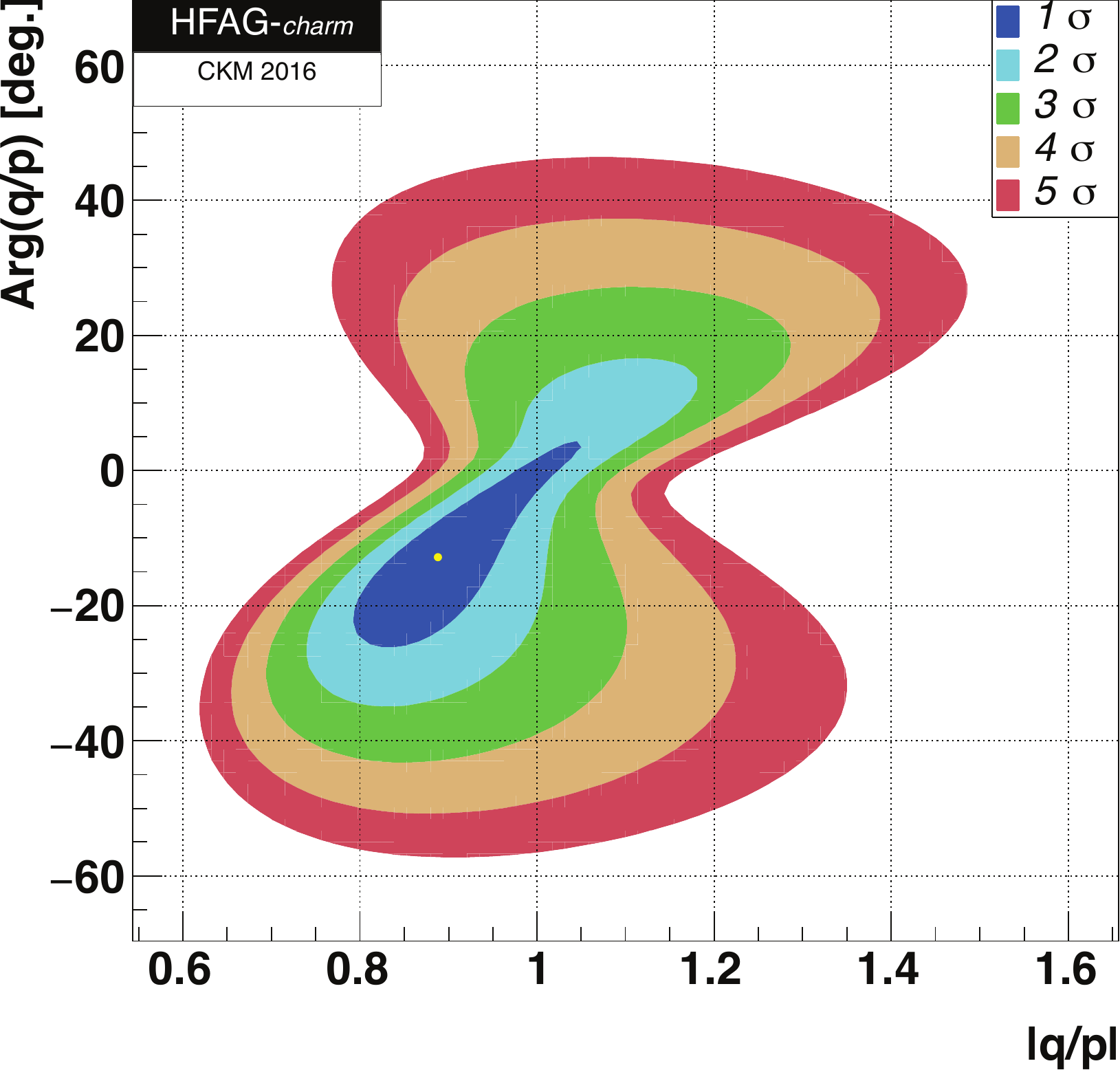}
\caption{
Two dimensional contours for [left]charm mixing parameters $x$ and $y$,
and [right] the CPV parameters $|q/p|$ and $\phi$.
The best fit point is indicated as a white dot,
and the $1\sigma$--$5\sigma$ contours are shown.
 [Source: HFAG 2016]
}
\label{fig:xy_qp2d_fit}
\end{center}
\end{figure}

Figure~\ref{fig:xy_qp2d_fit} shows the two dimensional contours from HFAG~\cite{Amhis:2016xyh}
for the mixing parameters $x$ and $y$, and the $CP$ parameters $|q/p|$ and $\phi$.
Charm mixing is firmly established
 ($>11.5\sigma$ away from no mixing hypothesis) from the combined fit.
However, the fit also shows no indication of $CP$ violation, i.e. within $1\sigma$
from the $CP$-conserving reference point $|q/p|=1$ and $\phi=0$.

%Apart from the discussions above,
A complementary study for CPV in the charm system
can be achieved by measuring the two observables $A_\Gamma$ and $y_{CP}$.
The observable $A_\Gamma$ measures the decay time asymmetry of $D^0$ and $\overline{D}^0$
decaying into the same $CP$ eigenstate $f$, including $K^+K^-$ and $\pi^+\pi^-$.
The explicit definition is given by
\begin{equation}
A_\Gamma
= {\Gamma({D}^0 \to f) - \Gamma(\overline{D}^0 \to {f}) \over \Gamma({D}^0 \to f) + \Gamma(\overline{D}^0 \to {f})}
= {1\over 2}\left[
\left( \left| {q\over p} \right| - \left| {p\over q} \right| \right) y \cos\phi -
\left( \left| {q\over p} \right| + \left| {p\over q} \right| \right) x \sin\phi
\right],
\end{equation}
which contains the CPV effects in mixing and in interference. Given there
is no hint for DCPV asymmetry in charm measurements, $A_\Gamma$
is mainly a pure probe of indirect CP violation.
It relates to the time-dependent asymmetry
$A_{CP}(t) \approx a^{\rm dir}_{CP} + a^{\rm ind}_{CP}\,{t/\tau}$,
 where $a^{\rm ind}_{CP}$ is just $-A_\Gamma$ in the limit of no DCPV.
The observable $y_{CP}$ measures the difference of effective lifetime
for decay into CP eigenstate and
decay into a flavor-specific state, such as $D^0 \to K^-\pi^+$,
\begin{equation}
y_{CP}
= {\tau({D}^0 \to K^-\pi^+) \over \tau({D}^0 \to K^+K^-)} - 1
= {1\over 2}\left[
\left( \left| {q\over p} \right| + \left| {p\over q} \right| \right) y \cos\phi -
\left( \left| {q\over p} \right| - \left| {p\over q} \right| \right) x \sin\phi
\right]~,
\end{equation}
in absence of DCPV. If $CP$ is conserved,
the $y_{CP}$ observable is equal to the mixing parameter $y$.

With a simultaneous fit to $D^0 \to K^+K^-$, $\pi^+\pi^-$, and $K^-\pi^+$ candidates
from the final Belle data set~\cite{Staric:2015sta}, the result is
\begin{equation}
y_{CP}= (+1.11 \pm 0.22 \pm 0.09)\%, \quad
A_{\Gamma} = (-0.03 \pm 0.20 \pm 0.07)\%. \quad {\rm (Belle)}
 \label{eq:AGa-Belle}
\end{equation}
The measured values are consistent with world averages\footnote{
The current HFAG fit is $y_{CP} = (0.835 \pm 0.155)\%$~\cite{Amhis:2016xyh},
consistent with the input used for arriving at Eq.~(\ref{eq:Del-aCP-t}),
and consistent with the Belle measurement of Eq.~(\ref{eq:AGa-Belle}) as well.
}
and show no indication of CP violation.
LHCb has recently updated the $A_\Gamma$ measurements with two different methods.
Unbinned maximum likelihood fits have been carried out to measure effective lifetimes with
$D^0 \to K^+K^-$ and $\pi^+\pi^-$ events from 2012 data~\cite{LHCb:2016ifp}.
The observable $A_\Gamma$ is measured separately in $K^+K^-$ and $\pi^+\pi^-$ channels,
and then combined with 2011 results:
\begin{equation}
A_{\Gamma} (K^+K^-)= (-0.14 \pm 0.37 \pm 0.10)\times 10^{-3}, \ \;
A_{\Gamma} (\pi^+\pi^-) = (0.14 \pm 0.63 \pm 0.15)\times 10^{-3}. \quad {\rm (LHCb)}
\end{equation}
Another analysis is carried out with $D^0 \to K^+K^-$ and $\pi^+\pi^-$ events
with full LHCb Run 1 sample~\cite{LHCb:2016lbh}, measuring the asymmetries in
bins of proper decay time, $A_{CP}(t) \approx a^{\rm dir}_{CP} + a^{\rm ind}_{CP}\, {t/\tau}$.
The detection biases are corrected using control samples of $D^0 \to K^-\pi^+$ events.
The analysis finds consistent results,
\begin{equation}
A_{\Gamma} (K^+K^-)= (-0.30 \pm 0.32 \pm 0.14)\times 10^{-3}, \ \;
A_{\Gamma} (\pi^+\pi^-) = (0.46 \pm 0.58 \pm 0.16)\times 10^{-3}, \quad {\rm (LHCb)}
\end{equation}
which is the most precise measurement of CPV in charm to date.
By combining with other existing measurements, the average from HFAG
\be
A_\Gamma = (-0.032 \pm 0.026)\%, \quad ({\rm HFAG})
\ee
is not so different from Eq.~(\ref{eq:AGa-Belle}),
and consistent with no $CP$ violation at few per mille level.
We do not have any hint for NP so far in charm sector.

It is expected that some of these studies can be further improved both in statistics,
by including a much larger data set from (HL-)LHC, and in systematics,
by introducing more robust modeling and methodology.
In any case, these studies are also essential for the $\phi_3$
CKM phase angle measurement, and further investigations are necessary.

%\hskip0.65cm Rare tau; \quad rare D; \quad D mixing

%
%\bea
%T \vert \Phi > &= &V \left\{ \vert \Phi > + \frac{1}{\omega  - H_0 +
%i\epsilon } V \vert \Psi >\right\}\nonumber \\
%& = &  \left\{ V + V \frac{1 }{\omega  - H_0 +i\epsilon } T\right\} \vert
%\Phi >.
% \label{eq:lipschw}
%\eea
%

\section{Top/Higgs}

We now discuss a subfield that was not traditionally regarded
as part of flavor physics (and associated CPV): top and Higgs decay.
These involve the two heaviest particles we know, and
the champion top quark receives mass from the runner-up Higgs field,
with a Yukawa coupling $\lambda_t \cong 1$ that
generates most FPCP effects within SM.
But the top quark itself decays before it can hadronize into
a ``top meson'', and the extremely short lifetime makes
rare top decays a somewhat depressed field.
Nevertheless, FCNC $t \to cZ$ decays have been searched
for since~\cite{PDG} the Tevatron discovery of the top.
Furthermore, the $P_5'$ anomaly suggests there could
be analogous top FCNC $tcZ'$ couplings of a new $Z'$ boson.

With the discovery of the Higgs boson $h(125)$, however,
it is the top in conjunction with the Higgs that is
drawing attention:  $t \to ch$ decay, which is possible because $m_h < m_t$.
As these are the newest particles we have uncovered,
from a purely experimental point of view,
the flavor-changing neutral Higgs (FCNH) $tch$ coupling
is of fundamental interest, \emph{because it can be probed directly}.
There is no such tree level coupling within SM after mass diagonalization,
and the loop-induced effect can be ignored for all practical purposes.
In 2HDM extensions, prejudice towards FCNH couplings had been
to ``remove before birth'', by the so-called
\emph{Natural Flavor Conservation} (NFC) condition of
Glashow and Weinberg~\cite{Glashow:1976nt},
where one invokes a discrete symmetry to forbid FCNH
 (thereby 2HDM-I and 2HDM-II).
But it was pointed out long ago~\cite{Fritzsch:1977vd, Cheng:1987rs} that,
given the ``trickling off'' flavor pattern observed in quark masses and mixings,
Nature does have her schemes regarding flavor,
while NFC itself turns out ``anthropological'' (human, not ``Natural'').
FCNH involving 3rd generation fermions in a 2HDM extension is \emph{natural},
which was thereby called 2HDM-III~\cite{Hou:1991un}.
It should in fact be called the Standard 2HDM, or SM2 for short,
as it just follows SM without \emph{ad hoc} assumptions.
Now that ATLAS and CMS experiments search directly for
$t \to ch$ (Sec.~\ref{sec:tcH}) and $h \to \mu\tau$ (Sec.~\ref{sec:S2HDM}) decays,
these are in fact true, current frontiers, where discovery could emerge at anytime.

\subsection{%\boldmath
 Top FCNC: $t \to cZ^{(\prime)}$ \label{sec:tcZ}}

By the GIM mechanism, there is no $tcZ$ coupling at tree level in SM,
with loop effects further suppressed by GIM cancellation.
The case is worse than charm:
whatever makes the corresponding $B$ decays favorable,
the situation is turned upsidedown for top.
Unlike the prolonged $B$ lifetime, the top is the shortest-lived ever
of all known particles,
while if the nondecoupling of the top quark makes
rare $B$ decays interesting, the near degeneracy of
the $d$, $s$ and $b$ quark masses at the $M_W$ and $m_t$ scales
means GIM cancellation is very effective.
This is what we mentioned earlier that $t \to cZ$ decay
is negligible for all practical purposes within SM.
New Physics might have saved the day if there were
new particles at the weak scale. Alas, we have not found any
at the LHC so far, neither at 8 TeV nor at 13 TeV collision energies,
with scales pushed to a few TeV, i.e. considerably higher than the
weak scale.

The current world best limit is from CMS Run 1 data~\cite{Chatrchyan:2013nwa},
\be
{\cal B}(t \to cZ) < 0.05\%, \quad {\rm (CMS)}
 \label{eq:tcZ-CMS1}
\ee
at 95\% C.L., with slightly weaker but comparable limit from ATLAS~\cite{Aad:2015uza}.
This result translates into ``a constraint on the KK gluon to be
heavier than 1.1 TeV''~\cite{Chatrchyan:2013nwa}, which is actually quite
good for a bound coming from an indirect rare top decay probe.
However, it should be clear that if there were KK gluons at 1 TeV scale,
they would have been discovered by direct search at the LHC already.
Indeed, compared with pre-LHC estimates,
warped extra dimension benchmarks for $t \to cZ$ have
moved down~\cite{Azatov:2014lha} to $10^{-5}$ level with LHC Run 1 data,
and suppressed by top compositeness scale $M_*^{-4}$,
which means the number would shrink further with 13 TeV bounds on $M_*$.
Other models, even in most favorable considerations
in say $R$-parity violating SUSY, the situation is no better~\cite{Bardhan:2016txk}.
The projections from theory seem to be beyond reach
even at the HL-LHC with 3000 fb$^{-1}$, where
estimates by CMS~\cite{CMS:2013xfa} and ATLAS~\cite{ATLAS:2013hta}
are at $0.01\%$ order.

We note that limits on $t \to cg$ and $c\gamma$ are better,
and in particular, single top production processes can effectively probe
smaller branching fractions than $t \to cZ$,
e.g. ${\cal B}(t\to ug) < 4 \times 10^{-5}$~\cite{Aad:2015gea} from ATLAS,
with $t\to cg$ limit five times weaker.
In our view, however, modeling for FCNC involving gluons or photons
would be in general much more contrived,
while $t\to u$ transitions do not seem favorable compared to $t\to c$ transitions
from the known flavor pattern. Search, of course, should continue.
What may be more interesting is $tZ$ associated production
through the process,
\be
g + c \to t + Z,
 \label{eq:tZ}
\ee
from $tcZ$ coupling. A recent study by CMS~\cite{Sirunyan:2017kkr} using 8 TeV data,
sets comparable limit on ${\cal B}(t \to cZ)$ as Eq.~(\ref{eq:tcZ-CMS1}).
In this connection, we remark that there may well exist
a weaker coupled $Z'$ boson associated with $t\to c$ transitions.
While the left-handed $tcZ'$ coupling suggested by the $P_5'$ anomaly
is too weak to be directly probed at the LHC,
the anomaly has inspired the suggestion of a right-handed $tcZ'$ coupling
that is not much constrained by $B$ decays~\cite{Fuyuto:2015gmk}.
A recent study which could be, but not necessarily,
related to the gauged $L_\mu - L_\tau$ model,
 %(which suggests large ${\cal B}(Z' \to \mu\mu)$),
suggests~\cite{Hou:2017ozb} that $tZ'$ associated production
analogous to Eq.~(\ref{eq:tZ}) might lead to discovery
with even 100--300 fb$^{-1}$ at 13--14 TeV collision energies.

Our point is to emphasize that, despite some pessimism for the $t \to cZ$ probe,
the top quark might still be a key to potential New Physics discoveries in the future.

%The top may not yet have exhausted its potential of tricks for NP at the LHC.

\subsection{%\boldmath
 Top FCNH: $t \to ch$ \label{sec:tcH}}

Tree level $tch$ coupling is certainly absent in SM,
and GIM cancellation is effective at the loop level,
such that $t \to ch$ decay is again negligible in SM for all practical purposes.
The usual 2HDM-I and 2HDM-II follow the NFC condition~\cite{Glashow:1976nt} of
Glashow and Weinberg, such that FCNH is again absent at tree level:
each type of fermion charge receive mass from one and only one Higgs doublet,
so the Yukawa and mass matrices are always simultaneously diagonalized.
With 2HDM-II automatic in MSSM, it became the most familiar two Higgs doublet model.
Cheng and Sher, however, offered some dissent~\cite{Cheng:1987rs} a decade after the NFC condition.
Generalizing the Fritzsch ansatz~\cite{Fritzsch:1977vd}, they noted that if
the generic Yukawa matrices for multi-Higgs models follow
an $\sqrt{m_im_j}\,$-like pattern, then low energy FCNC can be naturally suppressed.
It was then pointed out~\cite{Hou:1991un} that the $tch$ coupling would be
the most interesting, with $t \to ch$ the one to watch if $m_h < m_t$.

This is now indeed the case, and we advocated~\cite{Chen:2013qta} $t\to ch$ search at the LHC.
Sure enough, efforts within ATLAS~\cite{TheATLAScollaboration:2013nia}
on $t\to ch(\to \gamma\gamma)$ search,
or in association with CMS~\cite{Craig:2012vj} and
in multi-lepton final states, were already ongoing,
and within two years of Higgs boson discovery, the limits already reached
the sub-percent level~\cite{Aad:2014dya, Khachatryan:2014jya}, which is quite remarkable.
Both ATLAS and CMS embarked on $t\to ch$ search in $t\bar t$ events
with $h \to \gamma\gamma,\; WW^*,\; \tau^+\tau^-$, as well as $b\bar b$,
and the Run 1 combined limits for 8 TeV collisions are,
\be
{\cal B}(t \to ch) < 0.46\%, \ \ 0.40\% \quad {\rm (LHC\ Run\ 1)}
 \label{eq:tch}
\ee
at 95\% C.L. for ATLAS~\cite{Aad:2015pja} and CMS~\cite{Khachatryan:2016atv},
respectively.\footnote{
Based on 36.1$^{-1}$ data at 13 TeV, ATLAS has recently reported~\cite{Aaboud:2017mfd} 
the new limit of ${\cal B}(t \to ch) < 0.22\%$ at 95\% C.L. 
}
On the one hand, these sub-percent limits are rather impressive,
as they could have already resulted in discoveries, and could still in principle emerge tomorrow,
and all this on the backdrop of the prevailing NFC prejudice beforehand.
On the other hand, judging from the time it took the experiments to conduct their study,
the dominant $h \to b\bar b$ final state~\cite{Kao:2011aa}
seems to suffer from serious background, and it remains to be seen
how improvements would unfold with Run 2 data at 13 TeV.
Based on the cleaner $h\to \gamma\gamma$ mode, however,
ATLAS projects~\cite{ATL-tch-gaga} a final reach at HL-LHC of
${\cal B}(t \to ch) < 1.5 \times 10^{-4}$ at 95\% C.L.

It should be noted that, with $h^0$ denoting the 125 GeV boson,
all data point towards the decoupling or alignment limit,
that $h^0$ is very close to the SM Higgs boson.
In the context of 2HDM, it means that its non-standard couplings,
such as $tch$, are modulated by a small mixing angle
with the exotic $CP$-even Higgs boson $H^0$,
\be
\rho_{ct}\,\cos(\beta-\alpha)\,\bar cth^0,
 \label{eq:rho-ct}
\ee
where we have kept the 2HDM-II notation of $\cos(\beta-\alpha) \simeq 0$
as the $h^0$--$H^0$ mixing angle, and $\rho_{ct}$ is
the FCNH Yukawa coupling of heavy Higgs $H^0$.\footnote{
The physical $ctH^0$ coupling would be modulated by $\sin(\beta-\alpha) \simeq \pm 1$.
}
Thus, FCNH for $h$ is ``protected'' to be small, which is
a generalization from the Cheng--Sher argument.
Ref.~\cite{Chen:2013qta} stressed that $\rho_{ct} \sim 1$ is possible,
while the analogous $\rho_{cc}$ and $\rho_{tt}$ could also be ${\cal O}(1)$,
and are not well constrained. Given our poor handle on $h \to c\bar c$ measurement,
2HDM-III highlights the importance of precision measurement of
$h\to \tau\tau,\; c\bar c$, and $gg$ with more data.
The latter could be enhanced by $\rho_{tt}$, but
$h \to WW^*,\; ZZ^*,\; \gamma\gamma$ must be close to SM expectation
because $\cos(\beta-\alpha) \sim 0$.
As for the Yukawa coupling $\rho_{bb}$ of heavy Higgs $H^0$,
it enters $b\to s\gamma$ loop via the charged Higgs,
modulated by CKM matrix elements, and also receives a
chiral enhancement factor $m_t/m_b$,
and has been shown~\cite{Chen:2013qta} to be $\sim 0.01$
if $\rho_{ct}$ is sizable. Hence $h \to b\bar b$ is also SM-like.
The Yukawa pattern discussed here can be checked at LHC Run 2.
Note that for large $\rho_{ct} \sim 1$, the heavy neutral Higgs bosons
$H^0$ and $A^0$ could be searched for in $t\bar c$
final states~\cite{Altunkaynak:2015twa}, opening up a new search program.

\subsection{%\boldmath
 Standard 2HDM: $h \to \mu\tau$ (and CPV) \label{sec:S2HDM}}

In the previous subsection, we have given an almost experimental account
on $t \to ch$ search at the LHC, to make clear that ATLAS and CMS would
simply do it, regardless of ``doctrines'' such as NFC, or prejudices from MSSM
(i.e. 2HDM-II). In this subsection, we broaden the view
to advocate that 2HDM-III~\cite{Hou:1991un} with FCNH should in fact be called the
``Standard 2HDM'', or SM2 for short, in the same way we treat SM:
 let Nature have her say.

Motivated in part by the $tch$ discussion,
as well as the difficulty of Higgs search at the Tevatron,
$h \to \mu\tau$ search was suggested~\cite{DiazCruz:1999xe, Han:2000jz}
 at the turn of the millennium.\footnote{
Another motivation is the observation of large $\nu_\mu$--$\nu_\tau$ mixing
in the late 1990s. But this does not really translate to large $h \to\mu\tau$,
because of the extreme lightness of neutrinos and the large mixing angles of
the PMNS matrix (CKM counterpart in lepton sector). We are not sure that
neutrino masses are generated by the same Higgs mechanism.
But again, the experiments can simply do it once $h(125)$ is discovered,
while Nature can have her say.
}
Fast forward to LHC Run~1, it is remarkable that, reconstructing $\tau$ leptons
in the electronic and hadronic decay channels, CMS uncovered~\cite{Khachatryan:2015kon}
a 2.4$\sigma$ effect with 8 TeV data,
\be
{\cal B}(h \to \mu\tau) = \bigl(0.84^{+0.39}_{-0.37}\bigr)\%, \quad {\rm (CMS\ 8\ TeV)}
 \label{eq:hmutau-CMS}
\ee
or ${\cal B}(h \to \mu\tau) < 1.51\%$ at 95\% C.L., which aroused quite some interest.
The corresponding bound~\cite{Aad:2015gha} from ATLAS with 8 TeV data
is $1.84\%$ at 95\% C.L., which is not inconsistent.
Alas, an early peek by CMS at 13 TeV with 2.3 fb$^{-1}$ data~\cite{CMS:2016qvi}
did not quite support Eq.~(\ref{eq:hmutau-CMS}),
finding ${\cal B}(h \to \mu\tau) < 1.20\%$ when $1.62\%$ was expected.\footnote{
Corresponding to ${\cal B}(h \to \mu\tau) = \bigl(-0.76^{+0.81}_{-0.84}\bigr)\%$,
a negative central value, which tends to affect theorists.
}
It could be that the 8 TeV excess was an upward fluctuation,
but it could also be that the early 13 TeV result had a downward fluctuation.
Suffice it to say that ${\cal B}(h \to \mu\tau) \sim 0.5\%$ is perfectly possible.
Unfortunately, neither CMS nor ATLAS updated this Higgs decay mode at ICHEP 2016
held in Chicago, nor at 2017 Winter conferences.
But clearly the issue can be clarified with the large amount
of 2016 data at 13 TeV, and with full Run 2 data.

\begin{figure}[ht]
\begin{center}
%\hskip-0.5cm
\includegraphics[width=7.3cm]{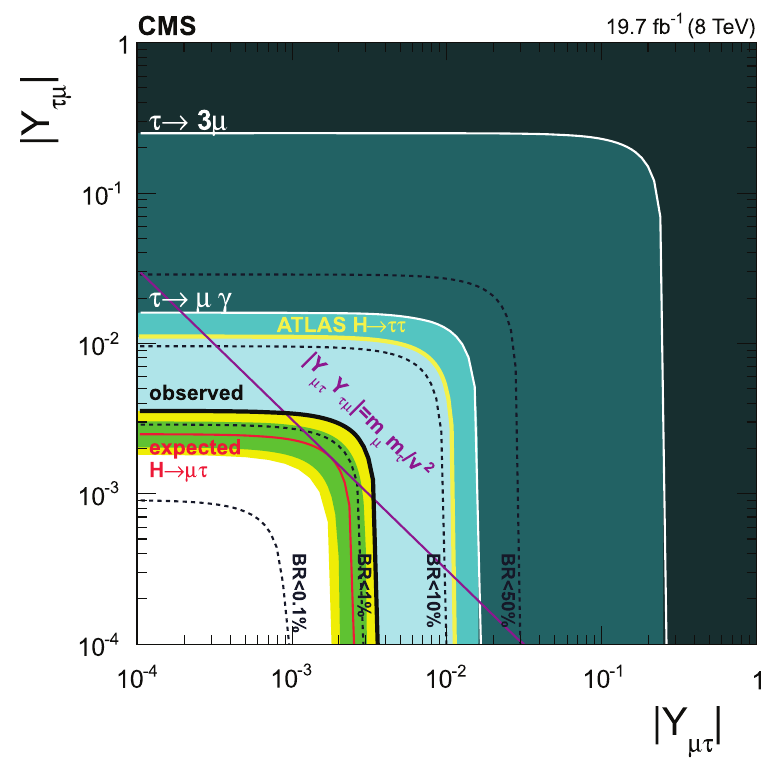} \hskip0.01cm
\includegraphics[width=7.35cm]{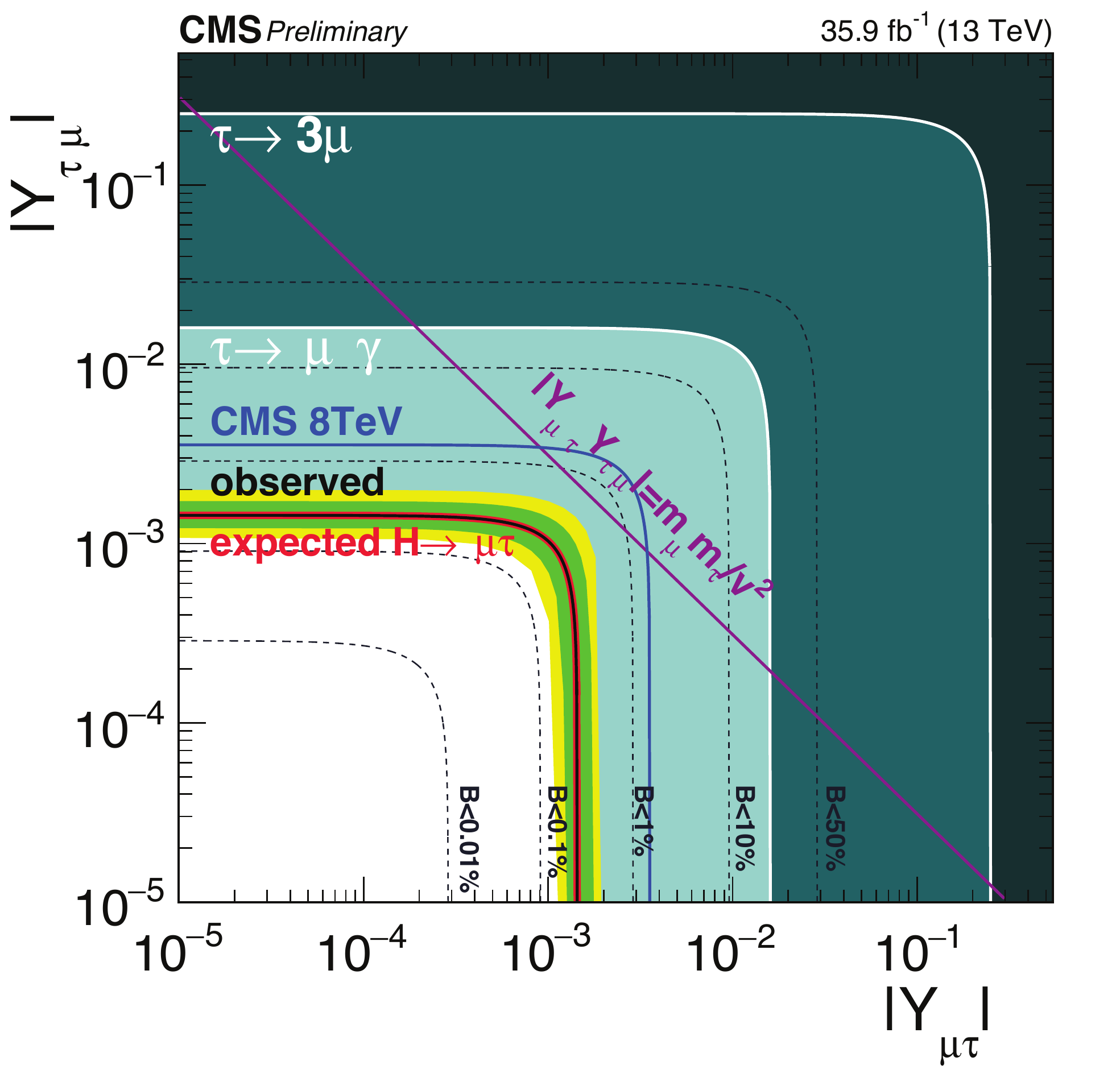}
\caption{
Comparison of CMS results
%~\cite{Khachatryan:2015kon} of Eq.~(\ref{eq:hmutau-CMS})
for $h^0 \to \mu\tau$ ($h^0$ marked as $H$ in plot)
with $\tau \to \mu\gamma$, $3\mu$ bounds
in the $|Y_{\mu\tau}|$--$|Y_{\tau\mu}|$ coupling plane.
The heavy (thin) solid contour is the observed (expected),
dashed contours are marked with ${\cal B}(h^0 \to \mu\tau)$ values,
while thin solid diagonal line corresponds to the ``Cheng--Sher'' value of
$|Y_{\mu\tau}Y_{\tau\mu}| = m_\mu m_\tau/v^2$.
[Left]:  8 TeV result, hinting at excess (observed above expected);
[Right] 13 TeV result (2016 data), which rules out 8 TeV hint
 (lower observed coinciding with expected).
 See text for further discussions.
 [Source: courtesy CMS Collaboration, from Refs.~\cite{Khachatryan:2015kon}
  and~\cite{CMS:2017onh}]
}
 \label{hmutau-CMS}
\end{center}
\end{figure}

The importance of $h \to \mu\tau$ direct search at LHC had been
emphasized~\cite{Blankenburg:2012ex, Harnik:2012pb} before the CMS
study~\cite{Khachatryan:2015kon} that gave the result of Eq.~(\ref{eq:hmutau-CMS}),
and in fact provided strong motivation for it.
Let us discuss the development of thoughts and studies.
After the $t\to ch$ suggestion~\cite{Hou:1991un} within 2HDM-III,
impact of FCNH in the lepton sector was explored~\cite{Chang:1993kw} for
$\mu \to e\gamma$ which involves $\mu\tau h$ and $\tau eh$ couplings,
and the importance of the two-loop diagram was emphasized.
This brings in a top loop with $\rho_{tt}$ coupling~\cite{Chen:2013qta}
to heavy Higgs boson $H^0$, which was mentioned below Eq.~(\ref{eq:rho-ct}).
That is, an effective $H\gamma^*\gamma$ correction is attached to
the $\mu \to \tau \to e$ line, which could compete with the
one loop diagram for $\mu \to e \gamma$, as the latter is suppressed by
three chirality flips, while the former has only one.
The fact that two-loop diagrams may in fact dominate in these transitions
were originally pointed out by Bjorken and Weinberg~\cite{Bjorken:1977vt},
but it is often called the Barr--Zee mechanism~\cite{Barr:1990vd},
from the independent but similar diagrammatic discussion for electron EDM.
With discovery of Higgs boson $h^0$ in 2012, and following the earlier
suggestion~\cite{Blankenburg:2012ex} that $h^0 \to \mu\tau$ is still allowed even
at ${\cal O}(10\%)$, Ref.~\cite{Harnik:2012pb} converted the detailed formulas for
$\mu\to e\gamma$~\cite{Chang:1993kw} to the case of $\tau \to \mu\gamma$
for a more thorough study.
While confirming the suggestion of Ref.~\cite{Blankenburg:2012ex},
the authors of Ref.~\cite{Harnik:2012pb} pointed out further that direct search
for $h^0 \to \mu\tau$ at the LHC would quickly give rise to a better bound,
if not discovery, of finite $\mu\tau h$ coupling, or
\be
\rho_{\mu\tau}\,\cos(\beta-\alpha)\,\bar \mu\tau h^0,
 \label{eq:rho-mutau}
\ee
that is more stringent than coming from $\tau\to\mu\gamma$.
In fact, the CMS study~\cite{Khachatryan:2015kon} took over the
banner plot of Ref.~\cite{Harnik:2012pb}, which we display in Fig.~\ref{hmutau-CMS}[left].
The plot compares $\tau \to \mu\mu\mu,\ \mu\gamma$ indirect bounds
with direct bound from $h \to \mu\tau$.
We note that $|Y_{\mu\tau}| = |\rho_{\mu\tau}\,\cos(\beta-\alpha)|$,
reflecting the $Y_{\mu\tau}$ notation used in Ref.~\cite{Harnik:2012pb},
which treated the FCNH of $h^0$ directly. This may be an oversimplification,
while we have not distinguished between $\rho_{\mu\tau}$ and $\rho_{\tau\mu}$
for simplicity of discussion.
We stress that exotic $h^0 \to \mu\tau$ decay is naturally suppressed
by a small $h^0$--$H^0$ mixing angle (called ``alignment'').
If a sizable rate at the level of Eq.~(\ref{eq:hmutau-CMS}) is confirmed,
it implies a sizable $\rho_{\mu\tau}$, but if $h^0 \to \mu\tau$ disappears,
it could just be due to a small $|\cos(\beta-\alpha)|$,
and should not diminish the possibility of FCNH $\mu\tau H^0$ coupling.

\vskip0.2cm
\noindent {\bf Note Added (Post LHCP 2017)}:
At the LHCP 2017 conference held in Shanghai, the CMS experiment
unveiled the result of 2016 dataset at 13 TeV, with a total of 35.9 fb$^{-1}$,
which does not support the 8 TeV result of Eq.~(\ref{eq:hmutau-CMS}).
The new bound (see Fig.~\ref{hmutau-CMS}[right]) is~\cite{CMS:2017onh}
\be
{\cal B}(h \to \mu\tau) < 0.25\%,
 \quad {\rm (CMS\ 13\ TeV,\ 35.9\ fb}^{-1})
 \label{eq:hmutau-CMS2016}
\ee
at 95\% C.L. We have kept our discussion to give a sense of
excitement/disappointment and development.

\

\noindent \underline{\bf 2HDM-III, or SM2?} \vskip0.1cm

We have by now introduced the flavor changing couplings
$\rho_{ct}$ (Eq.~(\ref{eq:rho-ct})), $\rho_{\mu\tau}$ (Eq.~(\ref{eq:rho-mutau})),
as well as the flavor conserving couplings $\rho_{cc}$, $\rho_{tt}$,
of the heavy Higgs boson $H^0$. These couplings mix into the coupling of $h^0$
by the small mixing parameter $\cos(\beta - \alpha)$,
which means that the $\rho_{ij}$ couplings of the physical $H^0$ boson
should be modulated by a $\sin(\beta - \alpha)$ factor.
Likewise, the SM couplings of $h^0$ would pick up a $\sin(\beta - \alpha)$ factor,
while these mix into $H^0$ with a $-\cos(\beta - \alpha)$ factor.
The $h^0$--$H^0$ mixing arises from the Higgs potential,
where data indicates that $|\cos(\beta - \alpha)|$ is quite small.
Our stated context is the 2HDM, but we have debunked the myth of NFC.
The observed light $h^0$ boson (with small $H^0$ admixture) gives rise to VEV,
and hence the three other components of the mass-giving doublet become Goldstone bosons
that are absorbed into the longitudinal parts of the massive $W^\pm$, $W^0$ gauge bosons.
The apparent decoupling/alignment indicates that
the second Higgs doublet is considerably heavier, with a scalar $H^0$,
a pseudoscalar $A^0$, and charged scalar $H^\pm$ components
that are not too far apart in mass.
We shall assume $CP$ is conserved within the Higgs sector.\footnote{
Otherwise it could have easily been detected by neutron EDM.
}
The key point is that,
with the simultaneous diagonalization of the fermion mass matrices and
Yukawa matrices of the mass-giving doublet,
the $\sin(\beta-\alpha)$-dependent couplings of $h^0$ are flavor diagonal.
For the second Higgs doublet, however, the Yukawa matrices \emph{cannot} be
simultaneously diagonalized with the mass matrices,
hence are naturally flavor changing.
This general phenomenon of FCNH is now being probed at the LHC, as described above,
although Nature seems to have been hiding it by a small $|\cos(\beta - \alpha)|$.

Interest in 2HDM-III, which we now prefer calling SM2, in fact grew already
after BaBar announced the $B \to D^{(*)}\tau\nu$ anomaly~\cite{Lees:2012xj},
claiming that it cannot be explained by 2HDM-II.
It was then pointed out~\cite{Crivellin:2012ye, Fajfer:2012jt} that
the BaBar anomaly could possibly be explained by the charged Higgs boson
$H^\pm$ of 2HDM-III,\footnote{
As well as other NP such as leptoquarks, see Sec.~\ref{sec:babar-A}.
}
i.e. 2HDM that allows for FCNH at tree-level.
It thus became quite an active field.
But a charged scalar, or scalar operator, explanation for the BaBar anomaly
has been challenged~\cite{Li:2016vvp, Alonso:2016oyd} recently by argument of the
$B_c$ lifetime, i.e. the $B_c \to \tau\nu$ fraction.
Thus, in this backdrop, it is the direct searches such as
$t \to ch$ or $h \to \mu\tau$ that would be the unequivocal tests for
the presence of an extra Higgs doublet, attesting to the power of direct search.
Discovering these modes, of course, does not directly
prove the existence of a second doublet with FCNH couplings,
as other models are possible (see e.g. Refs.~\cite{Altmannshofer:2016oaq}
and \cite{Galon:2017qes}).

Why would we like to call 2HDM-III the ``standard'' 2HDM, or SM2?
Yukawa couplings are provided, but not predicted in SM,
and we just learn their values from Nature by experimentation.
Now that a mass-giving Higgs doublet seems affirmed,
given the generation repetition of fermions,
it seems rather plausible that Nature would provide a second Higgs doublet.
In contrast, an extra scalar singlet could be conveniently added
for model building, but it would be relatively arbitrary
as it is not related to electroweak symmetry breaking,
while triplets have issue with electroweak precision measurables.
Although 2HDM traditionally comes with NFC that eliminates FCNH couplings,
but in comparing with SM, we should
``learn their values from Nature by experimentation''.
It is in this sense that it should be called SM2,
since FCNH is general, while Nature has apparently
opted for decoupling/alignment, such that the exotic component of
$h(125)$ is quite suppressed.
There is a second reason why we would like to call it SM2.
With a myriad of new Yukawa couplings, including FCNH couplings
such as $\rho_{ct}$, $\rho_{\mu\tau}$ that we have discussed,
it was pointed out recently~\cite{Fuyuto:2017ewj} that the extra $\rho_{tt}$ coupling,
which can carry CPV phase (since phase freedom has already been used
in establishing the CKM matrix), could be sufficient for electroweak baryogenesis,
i.e. generating baryon asymmetry of the Universe (BAU)
by weak scale physics.
Besides $\rho_{tt}$, the second Higgs doublet also supplies
 the needed first order phase transition by Higgs sector couplings.
If so, given the known deficiency of SM towards this
very important effect for our own existence,
it may well be Nature's design to achieve matter--antimatter asymmetry of the Universe
through a second Higgs doublet. Calling it SM2 would only be fitting.

With the myriad of new Yukawa couplings and associated CPV phases,
SM2 would greatly enrich the future FPCP program.
For example, on the direct search front, given $\rho_{tt} \neq 0$,
even if one is in the decoupling/alignment limit,
there is still $gg \to H^0$ production via a top triangle loop.
Assuming $m_{H^0} > 2m_t$, although establishing a resonance in $t\bar t$ may be challenging
because of interference with $gg \to t\bar t$
background~\cite{Dicus:1994bm, Frederix:2007gi, Carena:2016npr},
one could search for, and perhaps discover the $H^0$ (and $A^0$)
boson through FCNH $H^0 \to t\bar c$~\cite{Altunkaynak:2015twa} search.
The Higgs--Top flavor physics may be just starting.

%\hskip0.65cm  tCNC; \quad FCH

\begin{figure}[ht]
\begin{center}
\includegraphics[width=7cm]{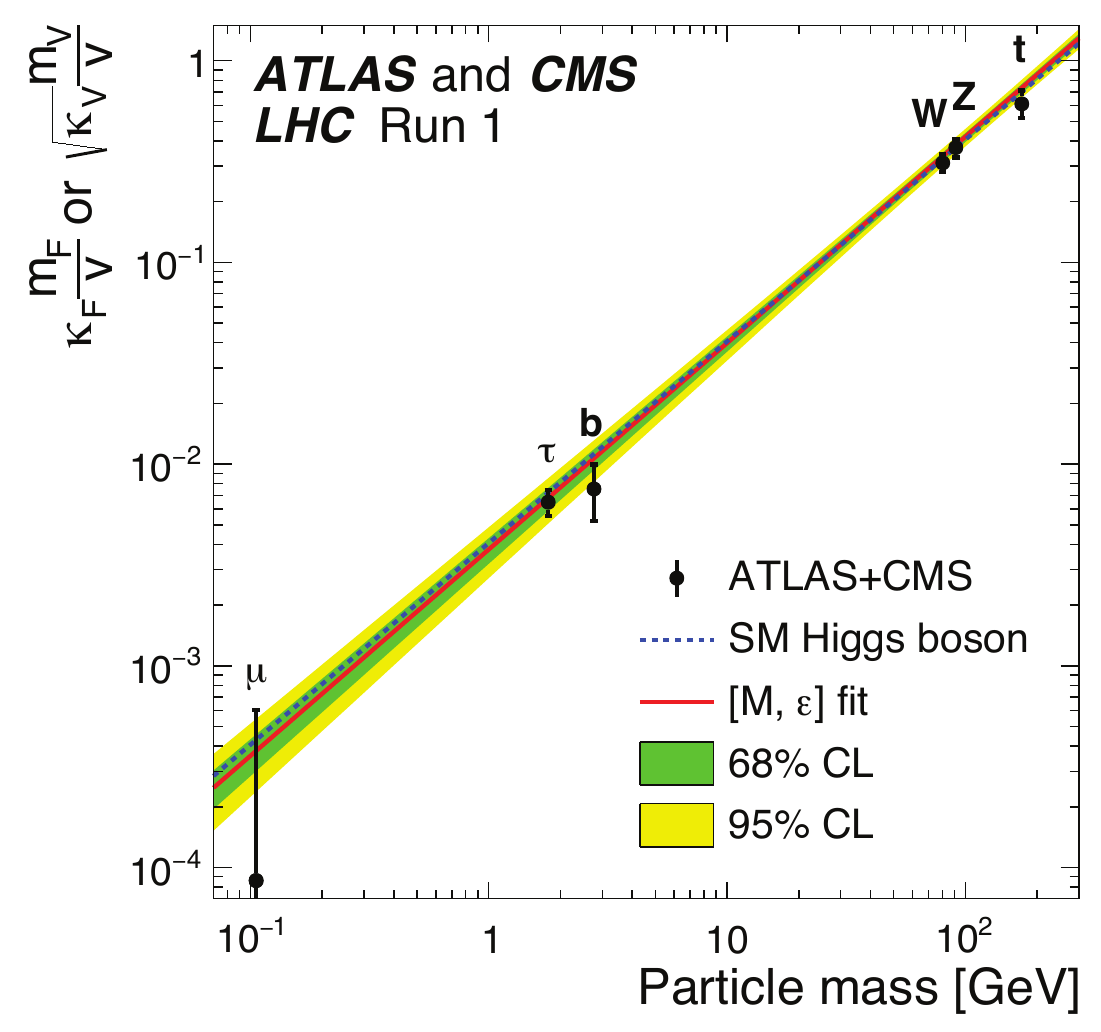} \hskip0.3cm
\includegraphics[width=7.3cm]{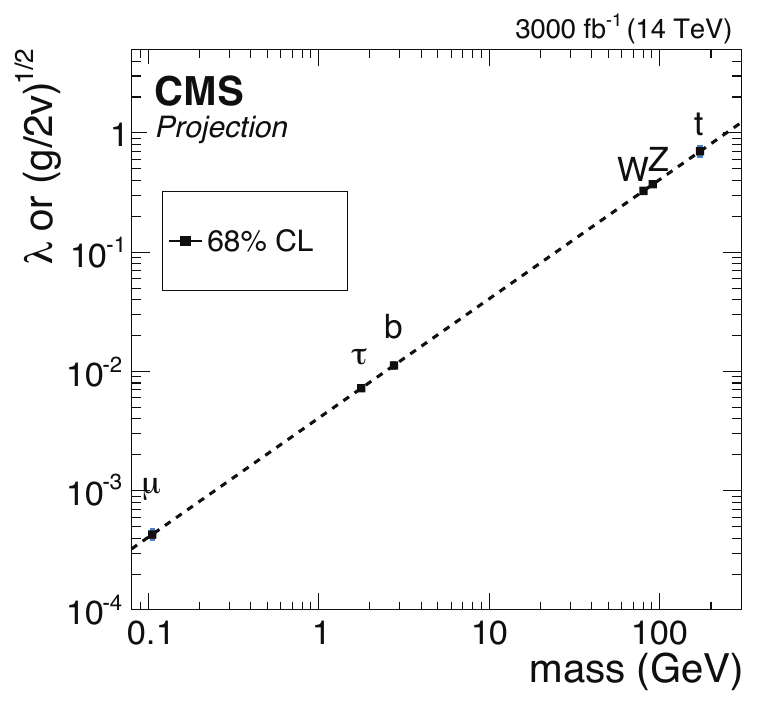}
\caption{
Mass vs Yukawa coupling ``linear plot'':
 [left] combined Run 1 measurement, and
 [right] CMS projection for HL-LHC.
%The linear relation between Yukawa coupling and mass is striking.
Note the absence of charm.
 [Source: left figure, courtesy ATLAS and CMS Collaborations, from Ref.~\cite{Khachatryan:2016vau};
  right figure, courtesy CMS Collaboration, from Ref.~\cite{LHCC-P-008}]
}
\label{fig:linear_plot}
\end{center}
\end{figure}

\section{Conclusion}

Taking on the topic of Flavor Physics and $CP$ violation,
we have covered a large swath of ground.
The CKM edifice that governs charged current weak interactions,
including through interesting loop effects such as box and penguin
diagrams, shows no sign of fraying. As it stands so well through
the B factory era, and by now the first round of LHC, it largely
convinces the FPCP practitioner of the ``reality'' (and, well,
complexity) behind, or underneath the edifice: Yukawa couplings
and their structure are very real! The CKM matrix elements,
including the CPV phase(s), are due to Yukawa couplings, which
are dynamic. These are the same couplings that generate fermion
masses through the Higgs mechanism, which is nicely illustrated
in the mass--coupling linear plot, Fig.~\ref{fig:linear_plot}
from the LHC.
The lefthand side of Fig.~\ref{fig:linear_plot} summarizes our
current knowledge of direct measurements~\cite{Khachatryan:2016vau}
of Yukawa couplings from Higgs boson $h$ decays or interactions.
While the muon case is not yet really measured, and the charm quark
is also absent (difficulty to charm-tag and measure $h \to c\bar c$),
the apparent linear relation between mass and coupling is truly
remarkable. But it is just the more recent echo from hadron
collider experiments that discovered the $h$ boson, where
FPCP physicists have known since observation of $B^0$ mixing,
that the dynamical, nondecoupled top mass effect is there.

The righthand side of Fig.~\ref{fig:linear_plot} is the
projection~\cite{LHCC-P-008} of the mass--Yukawa linear plot to
the end of the HL-LHC running, ca. 2035, where the absence of
charm is still conspicuous. Methods would supposedly be developed
to cover this.
Impressive as Fig.~\ref{fig:linear_plot} is, the current CKM
measurement as reflected in Fig.\ref{fig:CKM2016} is no less
impressive, and probably more beautiful. And, with advent of
$B$ physics agenda at Belle~II coming soon, and with the
excellent performance of LHCb (and also CMS and ATLAS), even
before HL-LHC starts, the improvement of Fig.\ref{fig:CKM2016}
would already be comparable to Fig.~\ref{fig:linear_plot}[right].
What is more, just like the pursuit of New Physics by direct search
at the LHC, FPCP has a broad program to indirectly probe New Physics.
The aim of this review is not just for the current status,
but to give the prospect for the early 2020s, i.e. by time of
LHC Run~3.

Perhaps the CKM fit would show some cracks? The CPV phase of
$\phi_s$ may still be more promising, and LHCb is the
experiment to watch, which is luckily crosschecked by CMS and ATLAS.
$B_s \to \mu^+\mu^-$ would become better measured, for sure.
But would we discover $B^0 \to \mu^+\mu^-$ with Run 2 data?
If so, it would necessarily mean New Physics beyond SM.
The world, however, would need to be convinced that LHCb
and CMS have backgrounds under control.

The $P_5'$ and BaBar anomalies look quite enticing, and would
certainly improve soon, although the BaBar anomaly would really
need Belle~II for a more convincing probe with high statistics.
Concurrently, the $B\to \tau\nu$ and $\mu\nu$ modes should be scrutinized
to give us a better picture of what type of New Physics, if any,
could it be.
The recent update by CMS show consistency of $P_5'$ with SM.
Though with larger errors than LHCb, it is a cautionary note to be taken,
but it also makes Run 2 exciting on this front.
On the other hand, the very recent LHCb result based on Run 1 data
finds $R_{K^*}$ to be consistent with $R_K$ (despite the issue
of lower bin fr $R_{K^*}$), which together
offer important and theoretically clean probes of New Physics.
We not only await Run 2 results to unfold, but also the imminent turn-on of Belle II.

$B$ physics is not just composed of these highlights. There is
a large number of more traditional B factory subjects, some of
which can be studied by LHCb, but some would need Belle~II
data to scrutinize, such as semileptonic, leptonic (with neutrino)
and radiative decays. We would soon start to enjoy the influx of
such information. For instance, is there anomaly in the $B \to \tau\nu$,
$\mu\nu$ rate or not? Is the mild hint of $B\to K^{(*)}\nu\nu$ above
SM expectation anywhere real? How about $\Delta S$, i.e. deviation
of indirect CPV for $b\to s\bar qq$ modes from $b\to c\bar cs$ modes?
What would be the outcome of the $B\to K\pi$ direct CPV sum rule test?

We are in a lucky period where major progress is expected in the
next few years on $K^+ \to \pi^+\nu\nu$ and $K_L \to \pi^0\nu\nu$.
NA62 should measure the $K^+$ mode with some precision by 2020,
while KOTO should start to see the evidence for the $K_L$ mode.
Perhaps New Physics would unfold? KOTO has some chance to uncover
a Dark particle with mass of $\pi^0$.
In the same time frame, diligent work by lattice theorists may
expose a new problem, or anomaly, in $\varepsilon'/\varepsilon$,
and at the same time demonstrate the coming of age of lattice
itself in FPCP.

Turning to low energy, we will see major development in both
$\mu\to e\gamma$ (MEG-II) and $\mu\to e$ conversion (COMET and Mu2e),
with prospect for new measurements, if not discovery.
This program has some parallel in $\tau \to \mu\gamma$, $\mu\mu\mu$,
$eee$ with Belle~II, while LHCb can also pursue the all charged track
processes.
Back to muons, finally we expect to see the muon $g-2$ anomaly
scrutinized experimentally again, and there is great anticipation
for the outcome from the new Muon g-2 experiment at Fermilab.
At even lower energy, there is good promise for new measurements
of electron and neutron EDM, and new developments maybe even for
the proton. This is the far, but truly important reach, of the
CPV quest under FPCP banner.

Returning to rather high energies, ATLAS and CMS are vigorously
pursuing $t \to cZ$ (and hopefully start on $t \to cZ'$)
and $t \to ch$ search, where there is better prospect for
improvement in the latter with Run 2 data.
Despite the Run 1 hint for $h \to \mu\tau$ was not confirmed
at 13 TeV, $h \to \mu\tau$ could still reemerge with full Run 2 data.
If so, or if $t \to ch$ is discovered, it would exclude 2HDM-II,
the favorite two Higgs doublet model from SUSY, as it would call for
tree level flavor changing neutral Higgs couplings.
But if these are not seen, it would add another support to alignment,
that $h$ is an SM-like Higgs particle.
However, from the BaBar anomaly, interest in $\tau\to\mu\gamma$,
top and Higgs FCNH couplings, to electroweak baryogenesis, we
find this General 2HDM, which we would prefer calling SM2 if
any genuine evidence emerges, appealing.
It would provide many new observables and parameters to test and check,
and usher in a new era for FPCP.

We have put forth many questions on many fronts, many of which
could soon make progress. We can only hope that
some genuine New Physics would emerge from at least one of these
probes.

\vskip0.75cm
\noindent{\bf Acknowledgements}\vskip0.1cm

We have enjoyed discussions and communications with
F.~Ambrosino, T.~Browder, M.~Ciuchini, M.~Corvino, G.~Cowan, P.~Crivelli, A.~Crivellin,
G.~D'Ambrosio, U.~Egede, M.~Gersabeck, T.~Gershon, Y.~Grossman, Y.B.~Hsiung, O.~Iorio, 
M.~Kohda, J.~Kubo, A.~Lenz, P.~Massarotti, M.~Moulson, F.~Muheim, S.~Nussinov, M.~Pospelov,
C.~Schwanda, E.~Senaha, A.~Soni, S.~Stone, S.~Suzuki, B.~Svetitsky, T.~Yamanaka and R.~Zwicky.
WSH thanks the Universities of Edinburgh, Hawaii, Kanazawa, Roma, Saga and Tel Aviv,
as well as INFN Napoli, for pleasant and fruitful visits.


\begin{thebibliography}{99}
\itemsep -2pt
%
\bibitem{Cabibbo:1963yz}
  N.~Cabibbo,
  %``Unitary Symmetry and Leptonic Decays,''
  \Journal{\em Phys.\ Rev.\ Lett.}{10}{531.}{1963}
 % doi:10.1103/PhysRevLett.10.531
  %%CITATION = doi:10.1103/PhysRevLett.10.531;%%
  %5372 citations counted in INSPIRE as of 18 Dec 2016
%
\bibitem{Glashow:1970gm}
  S.L.~Glashow, J.~Iliopoulos and L.~Maiani,
  %``Weak Interactions with Lepton-Hadron Symmetry,''
  \Journal{\em Phys.\ Rev.}{D 2}{1285.}{1970}
 % doi:10.1103/PhysRevD.2.1285
  %%CITATION = doi:10.1103/PhysRevD.2.1285;%%
  %5331 citations counted in INSPIRE as of 18 Dec 2016
%
\bibitem{Aubert:1974js}
  J.J.~Aubert {\it et al.} [E598 Collaboration],
  %``Experimental Observation of a Heavy Particle J,''
  \Journal{\em Phys.\ Rev.\ Lett.}{33}{1404.}{1974}
 % doi:10.1103/PhysRevLett.33.1404
  %%CITATION = doi:10.1103/PhysRevLett.33.1404;%%
  %2304 citations counted in INSPIRE as of 18 Dec 2016
%
\bibitem{Augustin:1974xw}
  J.E.~Augustin {\it et al.} [SLAC-SP-017 Collaboration],
  %``Discovery of a Narrow Resonance in e+ e- Annihilation,''
  \Journal{\em Phys.\ Rev.\ Lett.}{33}{1406.}{1974}
 % [Adv.\ Exp.\ Phys.\  {\bf 5}, 141 (1976)].
 % doi:10.1103/PhysRevLett.33.1406
  %%CITATION = doi:10.1103/PhysRevLett.33.1406;%%
  %2201 citations counted in INSPIRE as of 18 Dec 2016
%
\bibitem{Christenson:1964fg}
  J.H.~Christenson, J.W.~Cronin, V.L.~Fitch and R.~Turlay,
  %``Evidence for the 2 pi Decay of the k(2)0 Meson,''
  \Journal{\em Phys.\ Rev.\ Lett.}{13}{138.}{1964}
 % doi:10.1103/PhysRevLett.13.138
  %%CITATION = doi:10.1103/PhysRevLett.13.138;%%
  %2858 citations counted in INSPIRE as of 18 Dec 2016
%
\bibitem{Kobayashi:1973fv}
  M.~Kobayashi and T.~Maskawa,
  %``CP Violation in the Renormalizable Theory of Weak Interaction,''
  \Journal{\em Prog.\ Theor.\ Phys.}{49}{652.}{1973}
 % doi:10.1143/PTP.49.652
  %%CITATION = doi:10.1143/PTP.49.652;%%
  %8903 citations counted in INSPIRE as of 18 Dec 2016
%
\bibitem{Perl:1975bf}
  M.L.~Perl {\it et al.},
  %``Evidence for Anomalous Lepton Production in e+ - e- Annihilation,''
  \Journal{\em Phys.\ Rev.\ Lett.}{35}{1489.}{1975}
 % doi:10.1103/PhysRevLett.35.1489
  %%CITATION = doi:10.1103/PhysRevLett.35.1489;%%
  %1372 citations counted in INSPIRE as of 18 Dec 2016
%
\bibitem{Herb:1977ek}
  S.W.~Herb {\it et al.},
  %``Observation of a Dimuon Resonance at 9.5-GeV in 400-GeV Proton-Nucleus Collisions,''
  \Journal{\em Phys.\ Rev.\ Lett.}{39}{252.}{1977}
 % doi:10.1103/PhysRevLett.39.252
  %%CITATION = doi:10.1103/PhysRevLett.39.252;%%
  %1805 citations counted in INSPIRE as of 19 Dec 2016
%
\bibitem{Abe:1995hr}
  F.~Abe {\it et al.} [CDF Collaboration],
  %``Observation of top quark production in $\bar{p}p$ collisions,''
  \Journal{\em Phys.\ Rev.\ Lett.}{74}{2626.}{1995}
 % doi:10.1103/PhysRevLett.74.2626
 % [hep-ex/9503002].
  %%CITATION = doi:10.1103/PhysRevLett.74.2626;%%
  %2949 citations counted in INSPIRE as of 18 Dec 2016
%
\bibitem{Abachi:1995iq}
  S.~Abachi {\it et al.} [D\O\ Collaboration],
  %``Observation of the top quark,''
  \Journal{\em Phys.\ Rev.\ Lett.}{74}{2632.}{1995}
 % doi:10.1103/PhysRevLett.74.2632
 % [hep-ex/9503003].
  %%CITATION = doi:10.1103/PhysRevLett.74.2632;%%
  %2634 citations counted in INSPIRE as of 18 Dec 2016
%
\bibitem{Albrecht:1987ap}
  H.~Albrecht {\it et al.} [ARGUS Collaboration],
  %``Observation of B0 - anti-B0 Mixing,''
  \Journal{\em Phys.\ Lett.}{B 192}{245.}{1987}
 % doi:10.1016/0370-2693(87)91177-4
  %%CITATION = doi:10.1016/0370-2693(87)91177-4;%%
  %1338 citations counted in INSPIRE as of 18 Dec 2016
%
\bibitem{Lockyer:1983ev}
  N.~Lockyer {\it et al.},
  %``Measurement of the Lifetime of Bottom Hadrons,''
  \Journal{\em Phys.\ Rev.\ Lett.}{51}{1316.}{1983}
 % doi:10.1103/PhysRevLett.51.1316
  %%CITATION = doi:10.1103/PhysRevLett.51.1316;%%
  %324 citations counted in INSPIRE as of 18 Dec 2016
%
\bibitem{Fernandez:1983az}
  E.~Fernandez {\it et al.},
  %``Lifetime of Particles Containing B Quarks,''
  \Journal{\em Phys.\ Rev.\ Lett.}{51}{1022.}{1983}
 % doi:10.1103/PhysRevLett.51.1022
  %%CITATION = doi:10.1103/PhysRevLett.51.1022;%%
  %317 citations counted in INSPIRE as of 18 Dec 2016
%
\bibitem{Klopfenstein:1983nz}
  C.~Klopfenstein {\it et al.} [CUSB Collaboration],
  %``Semileptonic Decay of the B Meson,''
  \Journal{\em Phys.\ Lett.}{130 B}{444.}{1983}
 % doi:10.1016/0370-2693(83)91540-X
  %%CITATION = doi:10.1016/0370-2693(83)91540-X;%%
  %184 citations counted in INSPIRE as of 18 Dec 2016
%
\bibitem{Chen:1984ci}
  A.~Chen {\it et al.} [CLEO Collaboration],
  %``Limit on the B ---> u Coupling from Semileptonic B Decay,''
  \Journal{\em Phys.\ Rev.\ Lett.}{52}{1084.}{1984}
 % doi:10.1103/PhysRevLett.52.1084
  %%CITATION = doi:10.1103/PhysRevLett.52.1084;%%
  %195 citations counted in INSPIRE as of 18 Dec 2016
%
\bibitem{Wolfenstein}
  L.~Wolfenstein,
  %``Parametrization of the Kobayashi-Maskawa Matrix,''
  \Journal{\em Phys.\ Rev.\ Lett.}{51}{1945.}{1983}
 % doi:10.1103/PhysRevLett.51.1945
  %%CITATION = doi:10.1103/PhysRevLett.51.1945;%%
  %2733 citations counted in INSPIRE as of 18 Dec 2016
%
\bibitem{PDG}
  C.~Patrignani {\it et al.} [Particle Data Group],
  %``Review of Particle Physics,''
  \Journal{\em Chin.\ Phys.}{C 40}{100001.}{2016}
 % doi:10.1088/1674-1137/40/10/100001
  %%CITATION = doi:10.1088/1674-1137/40/10/100001;%%
  %173 citations counted in INSPIRE as of 18 Dec 2016
%
\bibitem{Carter:1980tk}
  A.B.~Carter and A.I.~Sanda,
  %``CP Violation in B Meson Decays,''
  \Journal{\em Phys.\ Rev.}{D 23}{1567.}{1981}
 % doi:10.1103/PhysRevD.23.1567
  %%CITATION = doi:10.1103/PhysRevD.23.1567;%%
  %729 citations counted in INSPIRE as of 18 Dec 2016
%
\bibitem{Bigi:1981qs}
  I.I.Y.~Bigi and A.I.~Sanda,
  %``Notes on the Observability of CP Violations in B Decays,''
  \Journal{\em Nucl.\ Phys.}{B 193}{85.}{1981}
 % doi:10.1016/0550-3213(81)90519-8
  %%CITATION = doi:10.1016/0550-3213(81)90519-8;%%
  %948 citations counted in INSPIRE as of 18 Dec 2016
%
\bibitem{Aubert:2001nu}
  B.~Aubert {\it et al.} [BaBar Collaboration],
  %``Observation of CP violation in the $B^0$ meson system,''
  \Journal{\em Phys.\ Rev.\ Lett.}{87}{091801.}{2001}
 % doi:10.1103/PhysRevLett.87.091801
 % [hep-ex/0107013].
  %%CITATION = doi:10.1103/PhysRevLett.87.091801;%%
  %799 citations counted in INSPIRE as of 18 Dec 2016
%
\bibitem{Abe:2001xe}
  K.~Abe {\it et al.} [Belle Collaboration],
  %``Observation of large CP violation in the neutral $B$ meson system,''
  \Journal{\em Phys.\ Rev.\ Lett.}{87}{091802.}{2001}
 % doi:10.1103/PhysRevLett.87.091802
 % [hep-ex/0107061].
  %%CITATION = doi:10.1103/PhysRevLett.87.091802;%%
  %831 citations counted in INSPIRE as of 18 Dec 2016
%
\bibitem{FlavTeV}
  G.W.-S. Hou, {\it Flavor Physics and the TeV Scale}
  (Springer-Verlag Berlin Heidelberg, 2009), Appendix A.
%
\bibitem{Peskin:2008zz}
  See e.g. M.E.~Peskin,
  %``Particle physics: Song of the electroweak penguin,''
  \Journal{\em Nature}{452}{293.}{2008}
 % doi:10.1038/452293a
  %%CITATION = doi:10.1038/452293a;%%
  %20 citations counted in INSPIRE as of 07 Jan 2017
%
\bibitem{Lin:2008zzaa}
  S.-W.~Lin, Y.~Unno, W.-S.~Hou, P.~Chang {\it et al.} [Belle Collaboration],
  %``Difference in direct charge-parity violation between charged and neutral $B$ meson decays,''
  \Journal{\em Nature}{452}{332.}{2008}
 % doi:10.1038/nature06827
  %%CITATION = doi:10.1038/nature06827;%%
  %141 citations counted in INSPIRE as of 18 Dec 2016
%
\bibitem{Hou:1986ug}
  W.-S.~Hou, R.S.~Willey and A.~Soni,
  %``Implications of a Heavy Top Quark and a Fourth Generation on the Decays B ---> K Lepton+ Lepton-, K Neutrino anti-neutrino,''
  \Journal{\em Phys.\ Rev.\ Lett.}{58}{1608.}{1987}
 % Erratum: [Phys.\ Rev.\ Lett.\  {\bf 60}, 2337 (1988)].
 % doi:10.1103/PhysRevLett.58.1608
  %%CITATION = doi:10.1103/PhysRevLett.58.1608;%%
  %248 citations counted in INSPIRE as of 19 Dec 2016
%
\bibitem{Hou:1999st}
  W.-S.~Hou and K.-C.~Yang,
  %``Possibility of large final state interaction phases in light of B ---> K pi and pi pi data,''
  \Journal{\em Phys.\ Rev.\ Lett.}{84}{4806.}{2000}
%  Phys.\ Rev.\ Lett.\  {\bf 84}, 4806 (2000)
%  Erratum: [Phys.\ Rev.\ Lett.\  {\bf 90}, 039901 (2003)]
%  doi:10.1103/PhysRevLett.84.4806, 10.1103/PhysRevLett.90.039901  [hep-ph/9911528].
  %%CITATION = doi:10.1103/PhysRevLett.84.4806, 10.1103/PhysRevLett.90.039901;%%
  %43 citations counted in INSPIRE as of 31 Mar 2017
%
\bibitem{Chua:2002wk}
  C.-K.~Chua, W.-S.~Hou and K.-C.~Yang,
  %``Indication for large rescatterings in charmless rare B decays,''
  \Journal{\em Mod.\ Phys.\ Lett.}{A 18}{1763.}{2003}
%  Mod.\ Phys.\ Lett.\ A {\bf 18}, 1763 (2003)
%  doi:10.1142/S0217732303011551  [hep-ph/0210002].
  %%CITATION = doi:10.1142/S0217732303011551;%%
  %41 citations counted in INSPIRE as of 31 Mar 2017
%
\bibitem{Beneke:1999br}
  M.~Beneke, G.~Buchalla, M.~Neubert and C.T.~Sachrajda,
  %``QCD factorization for B ---> pi pi decays: Strong phases and CP violation in the heavy quark limit,''
  \Journal{\em Phys.\ Rev.\ Lett.}{83}{1914.}{1999}
%  Phys.\ Rev.\ Lett.\  {\bf 83}, 1914 (1999)
%  doi:10.1103/PhysRevLett.83.1914  [hep-ph/9905312].
  %%CITATION = doi:10.1103/PhysRevLett.83.1914;%%
  %1160 citations counted in INSPIRE as of 31 Mar 2017
%
\bibitem{Keum:2000wi}
  Y.-Y.~Keum, H.-n.~Li and A.I.~Sanda,
  %``Penguin enhancement and $B \to K \pi$ decays in perturbative QCD,''
  \Journal{\em Phys.\ Rev.}{D 63}{054008.}{2001}
%  Phys.\ Rev.\ D {\bf 63}, 054008 (2001)
%  doi:10.1103/PhysRevD.63.054008  [hep-ph/0004173].
  %%CITATION = doi:10.1103/PhysRevD.63.054008;%%
  %710 citations counted in INSPIRE as of 31 Mar 2017
%
\bibitem{Aaij:2011in}
  R.~Aaij {\it et al.} [LHCb Collaboration],
  %``Evidence for CP violation in time-integrated $D^0 \to h^-h^+$ decay rates,''
  \Journal{\em Phys.\ Rev.\ Lett.}{108}{111602.}{2012}
%  doi:10.1103/PhysRevLett.108.129903, 10.1103/PhysRevLett.108.111602
 % [arXiv:1112.0938 [hep-ex]].
  %%CITATION = doi:10.1103/PhysRevLett.108.129903, 10.1103/PhysRevLett.108.111602;%%
  %310 citations counted in INSPIRE as of 13 Feb 2017
%
\bibitem{Aaij:2014gsa}
  R.~Aaij {\it et al.} [LHCb Collaboration],
  %``Measurement of $CP$ asymmetry in $D^0 \rightarrow K^- K^+$ and $D^0 \rightarrow \pi^- \pi^+$ decays,''
  \Journal{\em JHEP}{1407}{041.}{2014}
%  doi:10.1007/JHEP07(2014)041  [arXiv:1405.2797 [hep-ex]].
  %%CITATION = doi:10.1007/JHEP07(2014)041;%%
  %80 citations counted in INSPIRE as of 13 Feb 2017
%
\bibitem{Stone}
  S.L. Stone, this volume.
%
\bibitem{SFKing}
  For a review, see e.g. S.F. King, previous volume.

\bibitem{vudfor}
  W.J. Marciano and A. Sirlin,
  \Journal{\em Phys.\ Rev.\ Lett.}{71}{3629;}{1993}
%Phys. Rev. Lett. 71 (1993) 3629;
  J.C. Harfy and I.S. Towner,
  \Journal{\em Phys.\ Rev.}{C 79}{055502.}{2009}
%Phys. Rev. C79 (2009), 055502.
%
\bibitem{vud}
  J.C. Harfy and I.S. Towner,
  \Journal{\em Phys.\ Rev.}{C 91}{025501.}{2015}
%Phys. Rev. C91 (2015) 025501.
% arXiv:1411.5987].%

\bibitem{flavia} M. Antonelli {\it et al.} [FlaviaNet Working Group on Kaon Decays],
 arXiv:0801.1817; see also http://www.lnf.infn.it/wg/vus.

\bibitem{three-flavor-lattice}
  S. Aoki {\it et al.} [FLAG Working Group],
  \Journal{\em Eur.\ Phys.\ J.}{C 74}{2890;}{2014}
%  Eur. Phys. J., C74 (2014) 2890;
  http://itpwiki.unibe.ch/flag.

\bibitem{kloe}
  F. Ambrosino {\it et al.} [KLOE Collaboration],
  \Journal{\em Phys.\ Lett.}{B 632}{76.}{2006}
%  Phys. Lett. B632 (2006) 76.

\bibitem{vushyp}
  N. Cabibbo, E.C.~Swallow and R.~Winston,,
%  \Journal{\em Ann.\ Rev.\ Nucl.\ Part.\ Sci.}{53}{39.}{2009}
%  Ann. Rev. Nucl. Part. Sci. 53 (2009) 39.
  \Journal{\em Phys.\ Rev.\ Lett.}{92}{251803.}{2004}
%  doi:10.1103/PhysRevLett.92.251803  [hep-ph/0307214].
  %%CITATION = doi:10.1103/PhysRevLett.92.251803;%%
  %76 citations counted in INSPIRE as of 09 Apr 2017

%\bibitem{mateu} V. Mateu and A. Pich, JHEP 0510 (2005) 041.

\bibitem{hfag} Y.~Amhis {\it et al.} [Heavy Flavor Averaging Group], arXiv:1412.7515,
  and on-line updates at http://www.slac.stanford.edu/xorg/hfag/ prior to PDG 2016.



\bibitem{sirlin}
  A. Sirlin,
  \Journal{\em Rev.\ Mod.\ Phys.}{50}{573.}{1978}
%  Rev. Mod. Phys. 50 (1978)573.
%
\bibitem{Cirigliano:2009wk}
  See e.g. V.~Cirigliano, J.~Jenkins and M.~Gonzalez-Alonso,
  %``Semileptonic decays of light quarks beyond the Standard Model,''
  \Journal{\em Nucl.\ Phys.}{B 830}{95.}{2010}
%  doi:10.1016/j.nuclphysb.2009.12.020
%  [arXiv:0908.1754 [hep-ph]].
  %%CITATION = doi:10.1016/j.nuclphysb.2009.12.020;%%
  %86 citations counted in INSPIRE as of 25 Jun 2017

%%\bibitem{pdg-rev/} See the review of the CKM quark-mixing matrix in PDG 2016 \cite{PDG}.

%\bibitem{bvcd1}  % <== PDG2004!
%  F.J. Gilman {\it et al.},
%  \Journal{\em Phys.\ Lett.}{B 592}{793.}{2004}
%  Phys. Lett. B592 (2004) 793.

\bibitem{bvcd2}
  G. De Lellis, P.~Migliozzi and P.~Santorelli,
  \Journal{\em Phys.\ Rept.}{399}{277.}{2004}
%  Phys. Rept. 399 (2004) 227
%(Erratum {\it ibid.} 411 (2005) 323).

\bibitem{bvcd3}
  A. Kayis-Topaksu {\it et al.} [CHORUS Collaboration],
  \Journal{\em Phys.\ Lett.}{B 626}{24.}{2005}
%  Phys. Lett. B626 (2005)24.

%%\bibitem{pdg-vcb-vub} See the review of $|V_{cb}|$ and $|V_{ub}|$ in PDG 2016 \cite{PDG}.
%
\bibitem{lep2w}
  Contribution of LEP W branching fraction results, LEPEWWG/XSEC/2005-01
  to PDG2005,
  http://lepewwg.web.cern.ch/LEPEWWG/lepww/4f/Winter05/.
%
\bibitem{Bailey:2014tva}
  J.A.~Bailey {\it et al.} [Fermilab Lattice and MILC Collaborations],
  %``Update of $|V_{cb}|$ from the $\bar{B}\to D^*\ell\bar{\nu}$ form factor at zero recoil with three-flavor lattice QCD,''
  \Journal{\em Phys.\ Rev.}{D 89}{114504.}{2014}
%  Phys.\ Rev.\ D {\bf 89}, 114504 (2014)
%  doi:10.1103/PhysRevD.89.114504
%  [arXiv:1403.0635 [hep-lat]].
  %%CITATION = doi:10.1103/PhysRevD.89.114504;%%
  %77 citations counted in INSPIRE as of 25 Jun 2017

\bibitem{Lattice:2015rga}
  J.A.~Bailey {\it et al.} [MILC Collaboration],
  %``B→Dℓν form factors at nonzero recoil and |V$_{cb}$| from 2+1-flavor lattice QCD,''
  \Journal{\em Phys.\ Rev.}{D 92}{034506.}{2015}
%  Phys.\ Rev.\ D {\bf 92}, no. 3, 034506 (2015)
%  doi:10.1103/PhysRevD.92.034506
%  [arXiv:1503.07237 [hep-lat]].
  %%CITATION = doi:10.1103/PhysRevD.92.034506;%%
  %83 citations counted in INSPIRE as of 25 Jun 2017
%
\bibitem{Na:2015kha}
  H.~Na {\it et al.} [HPQCD Collaboration],
  %``$B \rightarrow D l \nu$ form factors at nonzero recoil and extraction of $|V_{cb}|$,''
  \Journal{\em Phys.\ Rev.}{D 92}{054510.}{2015}
%  Phys.\ Rev.\ D {\bf 92}, no. 5, 054510 (2015)
%  Erratum: [Phys.\ Rev.\ D {\bf 93}, no. 11, 119906 (2016)]
%  doi:10.1103/PhysRevD.93.119906, 10.1103/PhysRevD.92.054510
%  [arXiv:1505.03925 [hep-lat]].
  %%CITATION = doi:10.1103/PhysRevD.93.119906, 10.1103/PhysRevD.92.054510;%%
  %94 citations counted in INSPIRE as of 25 Jun 2017
%
\bibitem{ope1}
  I.I. Bigi, M.A.~Shifman, N.G.~Uraltsev and A.I.~Vainshtein,
  \Journal{\em Phys.\ Rev.\ Lett.}{71}{496.}{1993}

\bibitem{ope2}
  A.V. Manohar and M.B. Wise,
  \Journal{\em Phys.\ Rev.}{D 49}{1310.}{1994}
%  Phys. Rev. D49 (1994) 1310.

\bibitem{shape1}
  M. Neubert,
  \Journal{\em Phys.\ Rev.}{D 49}{3392;}{1994}
  \Journal{\em Phys.\ Rev.}{D 49}{4623.}{1994}
%  Phys. Rev. D49, (1994) 4623;  {\it ibid.} D 49 (1994) 3392.

\bibitem{shape2}
  I.I. Bigi, M.A.~Shifman, N.G.~Uraltsev and A.I.~Vainshtein,
  \Journal{\em Int.\ J.\ Mod.\ Phys.}{A 9}{2467.}{1994}
%  Int.  J. Mod. Phys. A9 (1994) 2467.

\bibitem{lqcd1}
  E. Dalgic {\it et al.,}
  \Journal{\em Phys.\ Rev.}{D 73}{074502.}{2006}
%  Phys. Rev. D73 (2006) 074502
  %; Erratum {\it ibid.} D75 (2007) 119906.

\bibitem{lqcd2}
  J.A. Bailey {\it et al.} [Fermilab Lattice and MILC Collaborations],
  \Journal{\em Phys.\ Rev.}{D 92}{014024.}{2015}
%  Phys. Rev. D92 (2015) 014024.
%
\bibitem{Detmold:2015aaa}
  W.~Detmold, C.~Lehner and S.~Meinel,
  %``$\Lambda_b \to p \ell^- \bar{\nu}_\ell$ and $\Lambda_b \to \Lambda_c \ell^- \bar{\nu}_\ell$ form factors from lattice QCD with relativistic heavy quarks,''
  \Journal{\em Phys.\ Rev.}{D 92}{034503.}{2015}
%  Phys.\ Rev.\ D {\bf 92}, no. 3, 034503 (2015)
%  doi:10.1103/PhysRevD.92.034503
%  [arXiv:1503.01421 [hep-lat]].
  %%CITATION = doi:10.1103/PhysRevD.92.034503;%%
  %67 citations counted in INSPIRE as of 25 Jun 2017

\bibitem{ubcb_lhcb}
  R. Aaij {\it et al.} [LHCb Collaboration],
  \Journal{\em Nature\ Phys.}{11}{743.}{2015}
%  Nature Phys. 11 (2015) 743.

%\bibitem{pdg-mixing} See the PDG review for $B^0-\overline{B}^0$ mixing.

\bibitem{tev-vtb}
  T.A. Aaltonen {\it et al.} [CDF and D\O\ Collaborations],
  \Journal{\em Phys.\ Rev.\ Lett.}{115}{152003.}{2015}
%  Phys. Rev. Lett. 115 (2015) 152003.

\bibitem{lhctop} LHC Top Working Group, https://lpcc.web.cern.ch/lpcc/index.php?page=top\_wg.
     %,           WG Plots: Single Top Quark Production, November 2015 in addition, a new
     %            t-channel results at 13 TeV, ATLAS-CONF-2015-079 is included.


\bibitem{vtb-d0}
  V.M. Abazov {\it et al.} [D\O\ Collaboration],
  \Journal{\em Phys.\ Rev.\ Lett.}{107}{121802.}{2011}
%  Phys. Rev. Lett. 107 (2011) 121802.

\bibitem{vtb-cdf}
  D. Acosta {\it et al.} [CDF Collaboration],
  \Journal{\em Phys.\ Rev.\ Lett.}{95}{102002.}{2005}
%  Phys. Rev. Lett. 95 (2005) 102002.

\bibitem{vtb-cms} %CMS-PAS-TOP-11-029, CMS Collaboration  (2011).
  V.~Khachatryan {\it et al.} [CMS Collaboration],
  %``Measurement of the ratio $\mathcal B(t \to Wb)/\mathcal B(t \to Wq)$ in pp collisions at $\sqrt{s}$ = 8 TeV,''
  \Journal{\em Phys.\ Lett.}{B 736}{33.}{2014}
%  Phys.\ Lett.\ B {\bf 736}, 33 (2014)
%  doi:10.1016/j.physletb.2014.06.076  [arXiv:1404.2292 [hep-ex]].
  %%CITATION = doi:10.1016/j.physletb.2014.06.076;%%
  %50 citations counted in INSPIRE as of 07 Apr 2017
%
\bibitem{Flynn:1989iu}
  J.M.~Flynn and L.~Randall,
  %``The Electromagnetic Penguin Contribution to Epsilon-prime / Epsilon for Large Top Quark Mass,''
  \Journal{\em Phys.\ Lett.}{B 224}{221.}{1989}
 % [Phys.\ Lett.\ B {\bf 235}, 412 (1990)]
 % Erratum: [Phys.\ Lett.\ B {\bf 235}, 412 (1990)].
%  doi:10.1016/0370-2693(89)91078-2, 10.1016/0370-2693(90)91986-L
  %%CITATION = doi:10.1016/0370-2693(89)91078-2, 10.1016/0370-2693(90)91986-L;%%
  %186 citations counted in INSPIRE as of 28 Feb 2017

%\bibitem{eekt} J.M. Flynn and L. Randall. Phys. Lett. B224 (1989) 221;
%G. Buchalla, A. J. Buras, and M. K. Harlander, Nucl. Phys. B408 (1990) 313.

%\bibitem{hfag-triangle} Heavy Flavor Averaging Group \cite{hfag}, Summer 2016
% updates for Unitarity Triangle Parameters:http://www.slac.stanford.edu/xorg/hfag/triangle/summer2016/.

\bibitem{pipi-phi2}
  M. Gronau and D. London,
  \Journal{\em Phys.\ Rev.\ Lett.}{65}{3381.}{1990}
%  Phys. Rev. Lett. 65, (1990) 3381.

\bibitem{rhorho-babar}
  B. Aubert {\it et al.} [BaBar Collaboration],
  \Journal{\em Phys.\ Rev.}{D 78}{071104.}{2008}
%  Phys. Rev. D78 (2008) 071104.

\bibitem{dalitz}
  H.R. Quinn and A.E. Synder,
  \Journal{\em Phys.\ Rev.}{D 48}{2139.}{1993}
%  Phys. Rev. D48 (1993) 2139.

\bibitem{bellerhopi}
  A.~Kusaka, C.-C.~Wang, H.~Ishino {\it et al.} [Belle Collaboration],
  \Journal{\em Phys.\ Rev.\ Lett.}{98}{221602.}{2007}
%  Phys. Rev. Lett. 98 (2007) 221602.

\bibitem{babarrhopi}
  B. Aubert {\it et al.} [BaBar Collaboration],
  \Journal{\em Phys.\ Rev.}{D 88}{121003.}{2013}
%  Phys. Rev. D88 (2013) 121003.

\bibitem{glw1}
  M. Gronau and D. London,
  \Journal{\em Phys.\ Lett.}{B 253}{483.}{1991}
%  Phys. Lett. B253 (1991) 483.

\bibitem{glw2}
  M. Gronau and D. Wyler,
  \Journal{\em Phys.\ Lett.}{B 265}{172.}{1991}
%  Phys. Lett. B265 (1991) 172.

\bibitem{ads}
  D. Atwood, I.~Dunietz and A.~Soni,
  \Journal{\em Phys.\ Rev.\ Lett.}{78}{3257;}{1997}
%  Phys. Rev. Lett. 78 (1997) 3257;
  \Journal{\em Phys.\ Rev.}{D 63}{036005.}{2001}
%  Phys. Rev. D63 (2001) 036005.

\bibitem{dalitz1-phi3}
  A. Bondar, talk at Belle analysis workshop, Novosibirsk, September 2002;
  A. Poluektov {\it et al.} [Belle Collaboration],
  \Journal{\em Phys.\ Rev.}{D 70}{072003.}{2004}
%  Phys. Rev. D70 (2004) 072003.

\bibitem{dalitz2-phi3}
  A. Giri, Y.~Grossman, A.~Soffer and J.~Zupan,
  \Journal{\em Phys.\ Rev.}{D 68}{054018.}{2003}
%   Phys. Rev. D68 (2003) 054018.

\bibitem{belle-phi3}
  A.~Poluektov, A.~Bondar, B.D.~Yabsley {\it et al.} [Belle Collaboration],
  \Journal{\em Phys.\ Rev.}{D 81}{112002.}{2010}
%  Phys. Rev. D81 (2010) 112002.

\bibitem{babar-phi3}
  B. Aubert {\it et al.} [BaBar Collaboration],
  \Journal{\em Phys.\ Rev.\ Lett.}{105}{121801.}{2010}
%  Phys. Rev. Lett. 105 (2010) 121801.
%
\bibitem{lhcb-phi3}
  R. Aaij {\it et al.} [LHCb Collaboration],
  \Journal{\em JHEP}{1410}{097.}{2014}
%  JHEP 1410 (2014) 097.
%
\bibitem{lhcb-gamma-EPS}
  The LHCb Collaboration,
  LHCb-CONF-2017-004.

\bibitem{hocker}
  A. H\"ocker, H. Lacker, S. Laplace and F. Le Diberder,
  \Journal{\em Eur.\ Phys.\ J.}{C 21}{225.}{2001}
%  Eur. Phys. J. C21 (2001) 225.

\bibitem{ckmfitter}
  J. Charles {\it et al.} [CKMfitter Group],
  \Journal{\em Eur.\ Phys.\ J.}{C 41}{1.}{2005}
%  Eur. Phys. J. C41 (2005) 1.

%\bibitem{frequent}
%  G. Eigen, G. Dubois-Felsmann, D.G. Hitlin and F.C. Porter,
%  \Journal{\em Phys.\ Rev.}{D 89}{033004.}{2014}
%  Phys. Rev. D89 (2014) 033004.

\bibitem{utfit1}
  M. Bona {\it et al.} [UTfit Collaboration],
  \Journal{\em JHEP}{0507}{028.}{2005}
%  JHEP 0507 (2005) 028.

\bibitem{utfit2}
  M. Bona {\it et al.} [UTfit Collaboration],
  \Journal{\em JHEP}{0803}{049.}{2008}
%  JHEP 0803 (2008) 049.

\bibitem{jarlskog}
  C. Jarlskog,
  \Journal{\em Phys.\ Rev.\ Lett.}{55}{1039.}{1985}
%  Phys. Rev. Lett. 55 (1985) 1039.
%
\bibitem{Amhis:2016xyh}
  Y.~Amhis {\it et al.} [Heavy Flavor Averaging Group],
  %``Averages of $b$-hadron, $c$-hadron, and $\tau$-lepton properties as of summer 2016,''
  arXiv:1612.07233 [hep-ex];
  and online updates at {http://www.slac.stanford.edu/xorg/hfag}.
  %%CITATION = ARXIV:1612.07233;%%
  %16 citations counted in INSPIRE as of 01 Mar 2017
%
\bibitem{Abulencia:2006ze}
  A.~Abulencia {\it et al.} [CDF Collaboration],
  %``Observation of $B^0_s - \bar{B}^0_s$ Oscillations,''
  \Journal{\em Phys.\ Rev.\ Lett.}{97}{242003.}{2006}
%  Phys.\ Rev.\ Lett.\  {\bf 97}, 242003 (2006)
%  doi:10.1103/PhysRevLett.97.242003  [hep-ex/0609040].
  %%CITATION = doi:10.1103/PhysRevLett.97.242003;%%
  %649 citations counted in INSPIRE as of 11 Apr 2017
%
\bibitem{Artuso:2015swg}
  M.~Artuso, G.~Borissov and A.~Lenz,
  %``CP violation in the $B_s^0$ system,''
  \Journal{\em Rev.\ Mod.\ Phys.}{88}{045002.}{2016}
%  Rev.\ Mod.\ Phys.\ {\bf 88}, 045002 (2016)
%  doi:10.1103/RevModPhys.88.045002  [arXiv:1511.09466 [hep-ph]].
  %%CITATION = doi:10.1103/RevModPhys.88.045002;%%
  %26 citations counted in INSPIRE as of 19 Mar 2017

%\cite{Charles:2004jd}
%\bibitem{Charles:2004jd}
%  J.~Charles {\it et al.} [CKMfitter Group],
  %``CP violation and the CKM matrix: Assessing the impact of the asymmetric $B$ factories,''
%  Eur.\ Phys.\ J.\ C {\bf 41}, no. 1, 1 (2005)
%  doi:10.1140/epjc/s2005-02169-1
%  [hep-ph/0406184], updated results and plots available at: http://ckmfitter.in2p3.fr.
  %%CITATION = doi:10.1140/epjc/s2005-02169-1;%%
  %1503 citations counted in INSPIRE as of 19 Mar 2017

%
\bibitem{Charles:2011va}
  J.~Charles {\it et al.},
  %``Predictions of selected flavour observables within the Standard Model,''
  \Journal{\em Phys.\ Rev.}{D 84}{033005.}{2001}
%  Phys.\ Rev.\ D {\bf 84}, 033005 (2011)
%  doi:10.1103/PhysRevD.84.033005  [arXiv:1106.4041 [hep-ph]].
  %%CITATION = doi:10.1103/PhysRevD.84.033005;%%
  %186 citations counted in INSPIRE as of 13 Mar 2017

%
\bibitem{Dunietz:2000cr}
  I.~Dunietz, R.~Fleischer and U.~Nierste,
  %``In pursuit of new physics with $B_s$ decays,''
  \Journal{\em Phys.\ Rev.}{D 63}{114015.}{2001}
%  Phys.\ Rev.\ D {\bf 63}, 114015 (2001)
%  doi:10.1103/PhysRevD.63.114015  [hep-ph/0012219].
  %%CITATION = doi:10.1103/PhysRevD.63.114015;%%
  %327 citations counted in INSPIRE as of 10 Apr 2017

%\cite{Buras:2009if}
%\bibitem{Buras:2009if}
%  A.~J.~Buras,
  %``Flavour Theory: 2009,''
%  PoS EPS {\bf -HEP2009}, 024 (2009)
%  [arXiv:0910.1032 [hep-ph]].
  %%CITATION = ARXIV:0910.1032;%%
  %66 citations counted in INSPIRE as of 13 Mar 2017

%\cite{Chiang:2009ev}
%\bibitem{Chiang:2009ev}
%  C.~W.~Chiang, A.~Datta, M.~Duraisamy, D.~London, M.~Nagashima and A.~Szynkman,
  %``New Physics in B0(s) ---> J/psi phi: A General Analysis,''
 % JHEP {\bf 1004}, 031 (2010)
%  doi:10.1007/JHEP04(2010)031
%  [arXiv:0910.2929 [hep-ph]].
  %%CITATION = doi:10.1007/JHEP04(2010)031;%%
  %52 citations counted in INSPIRE as of 13 Mar 2017
%
\bibitem{Dighe:1995pd}
  A.S.~Dighe, I.~Dunietz, H.J.~Lipkin and J.L.~Rosner,
%  ``Angular distributions and lifetime differences in $B_s \to J/\psi \phi$ decays,''
  \Journal{\em Phys.\ Lett.}{B 369}{144.}{1996}
%  Phys.\ Lett.\ B {\bf 369}, 144 (1996)
%  [hep-ph/9511363].
  %%CITATION = HEP-PH/9511363;%%
  %262 citations counted in INSPIRE as of 16 Sep 2014
%
\bibitem{Hou:2005hd}
  W.-S.~Hou, M.~Nagashima and A.~Soddu,
  %``Difference in B+ and B0 direct CP asymmetry as effect of a fourth generation,''
  \Journal{\em Phys.\ Rev.\ Lett.}{95}{141601;}{2005}
%  Phys.\ Rev.\ Lett.\  {\bf 95}, 141601 (2005)
%  doi:10.1103/PhysRevLett.95.141601  [hep-ph/0503072].
  %%CITATION = doi:10.1103/PhysRevLett.95.141601;%%
  %116 citations counted in INSPIRE as of 10 Apr 2017
%\cite{Hou:2006mx}
%\bibitem{Hou:2006mx}
%  W.-S.~Hou, M.~Nagashima and A.~Soddu,
  %``Large time-dependent CP violation in B/s0 system and finite $D^0$ - $\bar{D}^0$ mass difference in four generation standard model,''
  \Journal{\em Phys.\ Rev.}{D 76}{016004.}{2007}
%  Phys.\ Rev.\ D {\bf 76}, 016004 (2007)
%  doi:10.1103/PhysRevD.76.016004  [hep-ph/0610385].
  %%CITATION = doi:10.1103/PhysRevD.76.016004;%%
  %134 citations counted in INSPIRE as of 10 Apr 2017
%
\bibitem{Bona:2008jn}
  M.~Bona {\it et al.} [UTfit Collaboration],
  %``First Evidence of New Physics in $b \longleftrightarrow s$ Transitions,''
  \Journal{\em PMC\ Phys.}{A 3}{6.}{2009}
%  PMC Phys.\ A {\bf 3}, 6 (2009)
%  doi:10.1186/1754-0410-3-6  [arXiv:0803.0659 [hep-ph]].
  %%CITATION = doi:10.1186/1754-0410-3-6;%%
  %205 citations counted in INSPIRE as of 19 Mar 2017
%
\bibitem{Golutvin11}
  See p. 23 of the talk by A. Golutvin presented at the CERN LHC RRB held in April 2011.
%
\bibitem{Abazov:2013uma}
  V.M.~Abazov {\it et al.} [D\O\ Collaboration],
  %``Study of CP-violating charge asymmetries of single muons and like-sign dimuons in collisions,''
  \Journal{\em Phys.\ Rev.}{D 89}{012002.}{2014}
%  Phys.\ Rev.\ D {\bf 89}, 012002 (2014)
%  doi:10.1103/PhysRevD.89.012002  [arXiv:1310.0447 [hep-ex]].
  %%CITATION = doi:10.1103/PhysRevD.89.012002;%%
  %63 citations counted in INSPIRE as of 05 Apr 2017
%
\bibitem{Chua:2011er}
  C.-K.~Chua, W.-S.~Hou and C.-H.~Shen,
  %``Long-Distance Contribution to {\Delta}{\Gamma}s/{\Gamma}s of the $B_s-\bar{B}_s$ System,''
  \Journal{\em Phys.\ Rev.}{D 84}{074037.}{2011}
%  Phys.\ Rev.\ D {\bf 84}, 074037 (2011)
%  doi:10.1103/PhysRevD.84.074037  [arXiv:1107.4325 [hep-ph]].
  %%CITATION = doi:10.1103/PhysRevD.84.074037;%%
  %19 citations counted in INSPIRE as of 05 Apr 2017
%
\bibitem{CMS:2014xfa}
  V.~Khachatryan {\it et al.} [CMS and LHCb Collaborations],
  %``Observation of the rare $B^0_s\to\mu^+\mu^-$ decay from the combined analysis of CMS and LHCb data,''
  \Journal{\em Nature}{522}{68.}{2015}
%  Nature {\bf 522}, 68 (2015)
%  doi:10.1038/nature14474  [arXiv:1411.4413 [hep-ex]].
  %%CITATION = doi:10.1038/nature14474;%%
  %263 citations counted in INSPIRE as of 25 Mar 2017
%
\bibitem{Bobeth:2013uxa}
  C.~Bobeth {\it et al.}, %M.~Gorbahn, T.~Hermann, M.~Misiak, E.~Stamou and M.~Steinhauser,
  %``$B_{s,d} \to l^+ l^-$ in the Standard Model with Reduced Theoretical Uncertainty,''
  \Journal{\em Phys.\ Rev.\ Lett.}{112}{101801.}{2014}
%  Phys.\ Rev.\ Lett.\  {\bf 112}, 101801 (2014)
%  doi:10.1103/PhysRevLett.112.101801  [arXiv:1311.0903 [hep-ph]].
  %%CITATION = doi:10.1103/PhysRevLett.112.101801;%%
  %210 citations counted in INSPIRE as of 24 Mar 2017
%
\bibitem{Huang:1998vb}
  C.-S.~Huang, W.~Liao and Q.-S.~Yan,
  %``The Promising process to distinguish supersymmetric models with large tan Beta from the standard model: B ---> X(s) mu+ mu-,''
  \Journal{\em Phys.\ Rev.}{D 59}{011701.}{1999}
%  Phys.\ Rev.\ D {\bf 59}, 011701 (1999)
%  doi:10.1103/PhysRevD.59.011701  [hep-ph/9803460].
  %%CITATION = doi:10.1103/PhysRevD.59.011701;%%
  %130 citations counted in INSPIRE as of 10 Apr 2017
%
\bibitem{Hamzaoui:1998nu}
  C.~Hamzaoui, M.~Pospelov and M.~Toharia,
  %``Higgs mediated FCNC in supersymmetric models with large tan Beta,''
  \Journal{\em Phys.\ Rev.}{D 59}{095005.}{1999}
%  Phys.\ Rev.\ D {\bf 59}, 095005 (1999)
%  doi:10.1103/PhysRevD.59.095005  [hep-ph/9807350].
  %%CITATION = doi:10.1103/PhysRevD.59.095005;%%
  %184 citations counted in INSPIRE as of 19 Jun 2017
%
\bibitem{Choudhury:1998ze}
  S.R.~Choudhury and N.~Gaur,
  %``Dileptonic decay of B(s) meson in SUSY models with large tan Beta,''
  \Journal{\em Phys.\ Lett.}{B 451}{86.}{1999}
%  Phys.\ Lett.\ B {\bf 451}, 86 (1999)
%  doi:10.1016/S0370-2693(99)00203-8  [hep-ph/9810307].
  %%CITATION = doi:10.1016/S0370-2693(99)00203-8;%%
  %256 citations counted in INSPIRE as of 10 Apr 2017
%
\bibitem{Babu:1999hn}
  K.S.~Babu and C.F.~Kolda,
  %``Higgs mediated $B^0 \to \mu^{+} \mu^{-}$ in minimal supersymmetry,''
  \Journal{\em Phys.\ Rev.\ Lett.}{84}{228.}{2000}
%  Phys.\ Rev.\ Lett.\  {\bf 84}, 228 (2000)
%  doi:10.1103/PhysRevLett.84.228  [hep-ph/9909476].
  %%CITATION = doi:10.1103/PhysRevLett.84.228;%%
  %485 citations counted in INSPIRE as of 10 Apr 2017
%
\bibitem{Bobeth:2001sq}
  C.~Bobeth, T.~Ewerth, F.~Kruger and J.~Urban,
  %``Analysis of neutral Higgs boson contributions to the decays $\bar{B}$( $s^{)} \to \ell^{+} \ell^{-}$ and $\bar{B} \to K \ell^{+} \ell^{-}$,''
  \Journal{\em Phys.\ Rev.}{D 64}{074014.}{2001}
%  Phys.\ Rev.\ D {\bf 64}, 074014 (2001)
%  doi:10.1103/PhysRevD.64.074014  [hep-ph/0104284].
  %%CITATION = doi:10.1103/PhysRevD.64.074014;%%
  %329 citations counted in INSPIRE as of 10 Apr 2017
%
\bibitem{Aaltonen:2013as}
  T.~Aaltonen {\it et al.} [CDF Collaboration],
  %``Search for $B_s^0 \rightarrow \mu^{+}\mu^{-}$ and $B^0 \rightarrow \mu^{+}\mu^{-}$ decays with the full CDF Run II data set,''
  \Journal{\em Phys.\ Rev.}{D 87}{072003.}{2013}
%  Phys.\ Rev.\ D {\bf 87}, 072003 (2013)
%  doi:10.1103/PhysRevD.87.072003  [arXiv:1301.7048 [hep-ex]].
  %%CITATION = doi:10.1103/PhysRevD.87.072003;%%
  %27 citations counted in INSPIRE as of 10 Apr 2017

%\cite{Ellis:2006jy}
%\bibitem{Ellis:2006jy}
%  J.~R.~Ellis, K.~A.~Olive, Y.~Santoso and V.~C.~Spanos,
  %``On $B_s \to \mu^{+} \mu^{-}$ and cold dark matter scattering in the MSSM with non-universal Higgs masses,''
%  JHEP {\bf 0605}, 063 (2006)
%  doi:10.1088/1126-6708/2006/05/063
%  [hep-ph/0603136].
  %%CITATION = doi:10.1088/1126-6708/2006/05/063;%%
  %50 citations counted in INSPIRE as of 24 Mar 2017

 %\cite{Davidson:2010uu}
%\bibitem{Davidson:2010uu}
%  S.~Davidson and S.~Descotes-Genon,
  %``Minimal Flavour Violation for Leptoquarks,''
 % JHEP {\bf 1011}, 073 (2010)
%  doi:10.1007/JHEP11(2010)073
%  [arXiv:1009.1998 [hep-ph]].
  %%CITATION = doi:10.1007/JHEP11(2010)073;%%
  %41 citations counted in INSPIRE as of 24 Mar 2017

%\cite{Ellis:2007kb}
%\bibitem{Ellis:2007kb}
%  J.~R.~Ellis, J.~S.~Lee and A.~Pilaftsis,
  %``B-Meson Observables in the Maximally CP-Violating MSSM with Minimal Flavour Violation,''
%  Phys.\ Rev.\ D {\bf 76}, 115011 (2007)
%  doi:10.1103/PhysRevD.76.115011
%  [arXiv:0708.2079 [hep-ph]].
  %%CITATION = doi:10.1103/PhysRevD.76.115011;%%
  %104 citations counted in INSPIRE as of 24 Mar 2017
%
\bibitem{Aaboud:2016ire}
  M.~Aaboud {\it et al.} [ATLAS Collaboration],
  %``Study of the rare decays of $B^0_s$ and $B^0$ into muon pairs from data collected during the LHC Run 1 with the ATLAS detector,''
  \Journal{\em Eur.\ Phys.\ J.}{C 76}{513.}{2016}
%  Eur.\ Phys.\ J.\ C {\bf 76}, 513 (2016)
%  doi:10.1140/epjc/s10052-016-4338-8  [arXiv:1604.04263 [hep-ex]].
  %%CITATION = doi:10.1140/epjc/s10052-016-4338-8;%%
  %28 citations counted in INSPIRE as of 25 Mar 2017
%
\bibitem{Buras:2003td}
  A.J.~Buras,
  %``Relations between $\Delta$ M($s$, $d^{)}$ and B($s$, $d^{)} \to \mu \bar{\mu}$ in models with minimal flavor violation,''
  \Journal{\em Phys.\ Lett.}{B 566}{115.}{2003}
%  Phys.\ Lett.\ B {\bf 566}, 115 (2003)
%  doi:10.1016/S0370-2693(03)00561-6  [hep-ph/0303060].
  %%CITATION = doi:10.1016/S0370-2693(03)00561-6;%%
  %274 citations counted in INSPIRE as of 24 Mar 2017
%
\bibitem{Straub:2010ih}
  D.M.~Straub,
  %``New physics correlations in rare decays,''
  arXiv:1012.3893 [hep-ph], and its further evolutions.
  %%CITATION = ARXIV:1012.3893;%%
  %46 citations counted in INSPIRE as of 05 Apr 2017
%
\bibitem{Hou:2013btm}
  W.-S.~Hou, M.~Kohda and F.~Xu,
  %``Implication of possible observation of enhanced $B_d^0→μ^+μ^-$ decay,''
  \Journal{\em Phys.\ Rev.}{D 87}{094005.}{2013}
%  Phys.\ Rev.\ D {\bf 87}, 094005 (2013)
%  doi:10.1103/PhysRevD.87.094005  [arXiv:1302.1471 [hep-ph]].
  %%CITATION = doi:10.1103/PhysRevD.87.094005;%%
  %13 citations counted in INSPIRE as of 11 Apr 2017
%
\bibitem{Dutta:2015dla}
  B.~Dutta and Y.~Mimura,
  %``Enhancement of $Br(B_d \to \mu^+\mu^-)/Br(B_s \to \mu^+\mu^-)$ in supersymmetric unified models,''
  \Journal{\em Phys.\ Rev.}{D 91}{095011.}{2015}
%  Phys.\ Rev.\ D {\bf 91}, 095011 (2015)
%  doi:10.1103/PhysRevD.91.095011  [arXiv:1501.02044 [hep-ph]].
  %%CITATION = doi:10.1103/PhysRevD.91.095011;%%
  %3 citations counted in INSPIRE as of 11 Apr 2017
%
\bibitem{Aaij:2017vad}
  R.~Aaij {\it et al.} [LHCb Collaboration],
  %``Measurement of the $B^0_s\to\mu^+\mu^-$ branching fraction and effective lifetime and search for $B^0\to\mu^+\mu^-$ decays,''
  \Journal{\em Phys.\ Rev.\ Lett.}{D 118}{191801.}{2017}
%  Phys.\ Rev.\ Lett.\  {\bf 118}, 191801 (2017)
%  doi:10.1103/PhysRevLett.118.191801  [arXiv:1703.05747 [hep-ex]].
  %%CITATION = doi:10.1103/PhysRevLett.118.191801;%%
  %26 citations counted in INSPIRE as of 03 Jul 2017
%\cite{Yan:2000dc}
%\bibitem{Yan:2000dc}
%  Q.~S.~Yan, C.~S.~Huang, W.~Liao and S.~H.~Zhu,
  %``Exclusive semileptonic rare decays $B \to$ ($K$, $K^{*)} \ell^+ \ell^-$ in supersymmetric theories,''
%  Phys.\ Rev.\ D {\bf 62}, 094023 (2000)
%  doi:10.1103/PhysRevD.62.094023
%  [hep-ph/0004262].
  %%CITATION = doi:10.1103/PhysRevD.62.094023;%%
  %90 citations counted in INSPIRE as of 26 Mar 2017
%
\bibitem{Ali:1999mm}
  A.~Ali, P.~Ball, L.T.~Handoko and G.~Hiller,
  %``A Comparative study of the decays $B \to$ ($K$, $K^{*)} \ell^+ \ell^-$ in standard model and supersymmetric theories,''
  \Journal{\em Phys.\ Rev.}{D 61}{074024.}{2000}
%  Phys.\ Rev.\ D {\bf 61}, 074024 (2000)
%  doi:10.1103/PhysRevD.61.074024  [hep-ph/9910221].
  %%CITATION = doi:10.1103/PhysRevD.61.074024;%%
  %520 citations counted in INSPIRE as of 26 Mar 2017
%
\bibitem{Buchalla:2000sk}
  G.~Buchalla, G.~Hiller and G.~Isidori,
  %``Phenomenology of nonstandard $Z$ couplings in exclusive semileptonic $b \to s$ transitions,''
  \Journal{\em Phys.\ Rev.}{D 63}{014015.}{2000}
%  Phys.\ Rev.\ D {\bf 63}, 014015 (2000)
%  doi:10.1103/PhysRevD.63.014015  [hep-ph/0006136].
  %%CITATION = doi:10.1103/PhysRevD.63.014015;%%
  %267 citations counted in INSPIRE as of 26 Mar 2017
%
\bibitem{Altmannshofer:2008dz}
  W.~Altmannshofer {\it et al.}, %P.~Ball, A.~Bharucha, A.J.~Buras, D.M.~Straub and M.~Wick,
  %``Symmetries and Asymmetries of $B \to K^{*} \mu^{+} \mu^{-}$ Decays in the Standard Model and Beyond,''
  \Journal{\em JHEP}{0901}{019.}{2009}
%  JHEP {\bf 0901}, 019 (2009)
%  doi:10.1088/1126-6708/2009/01/019  [arXiv:0811.1214 [hep-ph]].
  %%CITATION = doi:10.1088/1126-6708/2009/01/019;%%
  %380 citations counted in INSPIRE as of 26 Mar 2017
%
\bibitem{Ali:1991is}
  A.~Ali, T.~Mannel and T.~Morozumi,
  %``Forward backward asymmetry of dilepton angular distribution in the decay b ---> s l+ l-,''
  \Journal{\em Phys.\ Lett.}{B 273}{505.}{1991}
%  Phys.\ Lett.\ B {\bf 273}, 505 (1991).
%  doi:10.1016/0370-2693(91)90306-B
  %%CITATION = doi:10.1016/0370-2693(91)90306-B;%%
  %329 citations counted in INSPIRE as of 12 Apr 2017
%
\bibitem{Prescott:1978tm}
  C.Y.~Prescott {\it et al.},
  %``Parity Nonconservation in Inelastic Electron Scattering,''
  \Journal{\em Phys.\ Lett.}{B 77}{347.}{1978}
%  Phys.\ Lett.\  {\bf 77B}, 347 (1978).
%  doi:10.1016/0370-2693(78)90722-0
  %%CITATION = doi:10.1016/0370-2693(78)90722-0;%%
  %768 citations counted in INSPIRE as of 12 Apr 2017
%
\bibitem{Wei:2009zv}
  J.-T.~Wei, P.~Chang {\it et al.} [Belle Collaboration],
  %``Measurement of the Differential Branching Fraction and Forward-Backword Asymmetry for $B \to K^{(*)}\ell^+\ell^-$,''
  \Journal{\em Phys.\ Rev.\ Lett.}{103}{171801.}{2009}
% Phys.\ Rev.\ Lett.\  {\bf 103}, 171801 (2009)
%  doi:10.1103/PhysRevLett.103.171801  [arXiv:0904.0770 [hep-ex]].
  %%CITATION = doi:10.1103/PhysRevLett.103.171801;%%
  %351 citations counted in INSPIRE as of 12 Apr 2017
%
%\bibitem{Feldmann:2002iw}
%  T.~Feldmann and J.~Matias,
  %``Forward backward and isospin asymmetry for $B \to K^* l^+ l^-$ decay in the standard model and in supersymmetry,''
%  JHEP {\bf 0301}, 074 (2003)
%  doi:10.1088/1126-6708/2003/01/074
%  [hep-ph/0212158].
  %%CITATION = doi:10.1088/1126-6708/2003/01/074;%%
  %132 citations counted in INSPIRE as of 26 Mar 2017
%
\bibitem{Descotes-Genon:2013vna}
  S.~Descotes-Genon, T.~Hurth, J.~Matias and J.~Virto,
  %``Optimizing the basis of $B\to K^*ll$ observables in the full kinematic range,''
  \Journal{\em JHEP}{1305}{137.}{2013}
%  JHEP {\bf 1305}, 137 (2013)
%  doi:10.1007/JHEP05(2013)137  [arXiv:1303.5794 [hep-ph]].
  %%CITATION = doi:10.1007/JHEP05(2013)137;%%
  %161 citations counted in INSPIRE as of 26 Mar 2017
%
\bibitem{Aaij:2015oid}
  R.~Aaij {\it et al.} [LHCb Collaboration],
  %``Angular analysis of the $B^{0} \to K^{*0} \mu^{+} \mu^{-}$ decay using 3 fb$^{-1}$ of integrated luminosity,''
  \Journal{\em JHEP}{1602}{104.}{2016}
%  JHEP {\bf 1602}, 104 (2016)
%  doi:10.1007/JHEP02(2016)104  [arXiv:1512.04442 [hep-ex]].
  %%CITATION = doi:10.1007/JHEP02(2016)104;%%
  %120 citations counted in INSPIRE as of 12 Apr 2017
%
\bibitem{Descotes-Genon:2014uoa}
  S.~Descotes-Genon, L.~Hofer, J.~Matias and J.~Virto,
  %``On the impact of power corrections in the prediction of $B \to K^*\mu^+\mu^-$ observables,''
  \Journal{\em JHEP}{1412}{125.}{2014}
%  JHEP {\bf 1412}, 125 (2014)
%  doi:10.1007/JHEP12(2014)125  [arXiv:1407.8526 [hep-ph]].
  %%CITATION = doi:10.1007/JHEP12(2014)125;%%
  %114 citations counted in INSPIRE as of 26 Mar 2017
%
\bibitem{Lyon:2014hpa}
  J.~Lyon and R.~Zwicky,
  %``Resonances gone topsy turvy - the charm of QCD or new physics in $b \to s \ell^+ \ell^-$?,''
  arXiv:1406.0566 [hep-ph].
  %%CITATION = ARXIV:1406.0566;%%
  %122 citations counted in INSPIRE as of 19 Jun 2017
%
\bibitem{Jager:2014rwa}
  S.~J\" ager and J.~Martin Camalich,
  %``Reassessing the discovery potential of the $B \to K^{*} \ell^+\ell^-$ decays in the large-recoil region: SM challenges and BSM opportunities,''
  \Journal{\em Phys.\ Rev.}{D 93}{014028.}{2016}
%  Phys.\ Rev.\ D {\bf 93}, 014028 (2016).
%  doi:10.1103/PhysRevD.93.014028
%  [arXiv:1412.3183 [hep-ph]].
  %%CITATION = doi:10.1103/PhysRevD.93.014028;%%
  %110 citations counted in INSPIRE as of 19 Jun 2017
%
\bibitem{Ciuchini:2015qxb}
  M.~Ciuchini {\it et al.},
   %, M.~Fedele, E.~Franco, S.~Mishima, A.~Paul, L.~Silvestrini and M.~Valli,
  %``$B\to K^* \ell^+ \ell^-$ decays at large recoil in the Standard Model: a theoretical reappraisal,''
  \Journal{\em JHEP}{1606}{116.}{2016}
%  JHEP {\bf 1606}, 116 (2016)
%  doi:10.1007/JHEP06(2016)116  [arXiv:1512.07157 [hep-ph]].
  %%CITATION = doi:10.1007/JHEP06(2016)116;%%
  %45 citations counted in INSPIRE as of 26 Mar 2017
%
\bibitem{Descotes-Genon:2015uva}
  S.~Descotes-Genon, L.~Hofer, J.~Matias and J.~Virto,
  %``Global analysis of $b\to s\ell\ell$ anomalies,''
  \Journal{\em JHEP}{1606}{092.}{2016}
%  JHEP {\bf 1606}, 092 (2016)
%  doi:10.1007/JHEP06(2016)092  [arXiv:1510.04239 [hep-ph]].
  %%CITATION = doi:10.1007/JHEP06(2016)092;%%
  %134 citations counted in INSPIRE as of 13 Apr 2017
%
\bibitem{Descotes-Genon:2013wba}
  S.~Descotes-Genon, J.~Matias and J.~Virto,
  %``Understanding the $B\to K^*\mu^+\mu^-$ Anomaly,''
  \Journal{\em Phys.\ Rev.}{D 88}{074002.}{2013}
%  Phys.\ Rev.\ D {\bf 88}, 074002 (2013)
%  doi:10.1103/PhysRevD.88.074002  [arXiv:1307.5683 [hep-ph]].
  %%CITATION = doi:10.1103/PhysRevD.88.074002;%%
  %242 citations counted in INSPIRE as of 13 Apr 2017
%
\bibitem{Altmannshofer:2013foa}
  W.~Altmannshofer and D.M.~Straub,
  %``New Physics in $B \to K^*\mu\mu$?,''
  \Journal{\em Eur.\ Phys.\ J.}{C 73}{2646.}{2013}
%  Eur.\ Phys.\ J.\ C {\bf 73}, 2646 (2013)
%  doi:10.1140/epjc/s10052-013-2646-9  [arXiv:1308.1501 [hep-ph]].
  %%CITATION = doi:10.1140/epjc/s10052-013-2646-9;%%
  %182 citations counted in INSPIRE as of 13 Apr 2017
%
\bibitem{Altmannshofer:2017fio}
  W.~Altmannshofer, C.~Niehoff, P.~Stangl and D.M.~Straub,
  %``Status of the $B\rightarrow K^*\mu ^+\mu ^-$ anomaly after Moriond 2017,''
  \Journal{\em Eur.\ Phys.\ J.}{C 77}{377.}{2017}
%  Eur.\ Phys.\ J.\ C {\bf 77}, 377 (2017).
%  doi:10.1140/epjc/s10052-017-4952-0
%  [arXiv:1703.09189 [hep-ph]].
  %%CITATION = doi:10.1140/epjc/s10052-017-4952-0;%%
  %18 citations counted in INSPIRE as of 19 Jun 2017
%
\bibitem{Wehle:2016yoi}
  S.~Wehle, C.~Niebuhr, S.~Yashchenko {\it et al.} [Belle Collaboration],
  %``Lepton-Flavor-Dependent Angular Analysis of $B\to K^\ast \ell^+\ell^-$,''
  \Journal{\em Phys.\ Rev.\ Lett.}{118}{111801.}{2017}
%  Phys.\ Rev.\ Lett.\  {\bf 118}, 111801 (2017)
%  doi:10.1103/PhysRevLett.118.111801  [arXiv:1612.05014 [hep-ex]].
  %%CITATION = doi:10.1103/PhysRevLett.118.111801;%%
  %12 citations counted in INSPIRE as of 12 Apr 2017

%
\bibitem{CMS:2017ivg}
  CMS Collaboration,
  %``Measurement of the $P_1$ and $P_5'$ angular parameters of the decay $\mathrm{B}^0 \to \mathrm{K}^{*0} \mu^+ \mu^-$ in proton-proton collisions at $\sqrt{s}=8~\mathrm{TeV}$,''
  CMS-PAS-BPH-15-008.
  %%CITATION = CMS-PAS-BPH-15-008;%%

\bibitem{ATLASp5prime}
  ATLAS Collaboration,
  %``Angular analysis of $B_d^0\toK^{*0}\mu^+\mu^-$ decays in pp collisions at $\sqrt{s}=8$~TeV with the ATLAS detector,''
  ATLAS-CONF-2017-023.
%
\bibitem{Aaij:2014ora}
  R.~Aaij {\it et al.} [LHCb Collaboration],
  %``Test of lepton universality using $B^{+}\rightarrow K^{+}\ell^{+}\ell^{-}$ decays,''
  \Journal{\em Phys.\ Rev.\ Lett.}{113}{151601.}{2014}
%  Phys.\ Rev.\ Lett.\  {\bf 113}, 151601 (2014)
%  doi:10.1103/PhysRevLett.113.151601  [arXiv:1406.6482 [hep-ex]].
  %%CITATION = doi:10.1103/PhysRevLett.113.151601;%%
  %306 citations counted in INSPIRE as of 02 Apr 2017
%
\bibitem{Aaij:2017vbb}
  R.~Aaij {\it et al.} [LHCb Collaboration],
  %``Test of lepton universality with $B^{0} \rightarrow K^{*0}\ell^{+}\ell^{-}$ decays,''
  arXiv:1705.05802 [hep-ex].
  %%CITATION = ARXIV:1705.05802;%%
  %15 citations counted in INSPIRE as of 23 Jun 2017
%
\bibitem{Altmannshofer:2017yso}
  W.~Altmannshofer, P.~Stangl and D.~M.~Straub,
  %``Interpreting Hints for Lepton Flavor Universality Violation,''
  arXiv:1704.05435 [hep-ph].
  %%CITATION = ARXIV:1704.05435;%%
  %31 citations counted in INSPIRE as of 23 Jun 2017
%
\bibitem{Lees:2012xj}
  J.P.~Lees {\it et al.} [BaBar Collaboration],
  %``Evidence for an excess of $\bar{B} \to D^{(*)} \tau^-\bar{\nu}_\tau$ decays,''
  \Journal{\em Phys.\ Rev.\ Lett.}{109}{101802.}{2012}
%  Phys.\ Rev.\ Lett.\  {\bf 109}, 101802 (2012)
%  doi:10.1103/PhysRevLett.109.101802  [arXiv:1205.5442 [hep-ex]].
  %%CITATION = doi:10.1103/PhysRevLett.109.101802;%%
  %387 citations counted in INSPIRE as of 14 Apr 2017
%
\bibitem{Fajfer:2012vx}
  S.~Fajfer, J.F.~Kamenik and I.~Nisandzic,
  %``On the $B \to D^* \tau \bar \nu_{\tau}$ Sensitivity to New Physics,''
  \Journal{\em Phys.\ Rev.}{D 85}{094025.}{2012}
%  Phys.\ Rev.\ D {\bf 85}, 094025 (2012)
%  doi:10.1103/PhysRevD.85.094025  [arXiv:1203.2654 [hep-ph]].
  %%CITATION = doi:10.1103/PhysRevD.85.094025;%%
  %232 citations counted in INSPIRE as of 02 Apr 2017
%
\bibitem{Bigi:2016mdz}
  D.~Bigi and P.~Gambino,
  %``Revisiting $B\to D \ell \nu$,''
  \Journal{\em Phys.\ Rev.}{D 94}{094008.}{2016}
%  Phys.\ Rev.\ D {\bf 94}, 094008 (2016)
%  doi:10.1103/PhysRevD.94.094008  [arXiv:1606.08030 [hep-ph]].
  %%CITATION = doi:10.1103/PhysRevD.94.094008;%%
  %16 citations counted in INSPIRE as of 02 Apr 2017
%
\bibitem{Bernlochner:2017jka}
  F.U.~Bernlochner, Z.~Ligeti, M.~Papucci and D.J.~Robinson,
  %``Combined analysis of semileptonic $B$ decays to $D$ and $D^*$: $R(D^{(*)})$, $|V_{cb}|$, and new physics,''
  \Journal{\em Phys.\ Rev.}{D 95}{115008.}{2017}
%  Phys.\ Rev.\ D {\bf 95}, 115008 (2017).
%  doi:10.1103/PhysRevD.95.115008
%  [arXiv:1703.05330 [hep-ph]].
  %%CITATION = doi:10.1103/PhysRevD.95.115008;%%
  %14 citations counted in INSPIRE as of 23 Jun 2017
%
\bibitem{Lees:2013uzd}
  J.P.~Lees {\it et al.} [BaBar Collaboration],
  %``Measurement of an Excess of $\bar{B} \to D^{(*)}\tau^- \bar{\nu}_\tau$ Decays and Implications for Charged Higgs Bosons,''
  \Journal{\em Phys.\ Rev.}{D 88}{072012.}{2013}
%  Phys.\ Rev.\ D {\bf 88}, 072012 (2013)
%  doi:10.1103/PhysRevD.88.072012  [arXiv:1303.0571 [hep-ex]].
  %%CITATION = doi:10.1103/PhysRevD.88.072012;%%
  %206 citations counted in INSPIRE as of 02 Apr 2017
%
\bibitem{Huschle:2015rga}
  M.~Huschle, T.~Kuhr, M.~Heck, P.~Goldenzweig {\it et al.} [Belle Collaboration],
  %``Measurement of the branching ratio of $\bar{B} \to D^{(\ast)} \tau^- \bar{\nu}_\tau$ relative to $\bar{B} \to D^{(\ast)} \ell^- \bar{\nu}_\ell$ decays with hadronic tagging at Belle,''
  \Journal{\em Phys.\ Rev.}{D 92}{072014.}{2015}
%  Phys.\ Rev.\ D {\bf 92}, 072014 (2015)
%  doi:10.1103/PhysRevD.92.072014  [arXiv:1507.03233 [hep-ex]].
  %%CITATION = doi:10.1103/PhysRevD.92.072014;%%
  %165 citations counted in INSPIRE as of 02 Apr 2017
%
\bibitem{Sato:2016svk}
  Y.~Sato, T.~Iijima, K.~Adamczyk {\it et al.} [Belle Collaboration],
  %``Measurement of the branching ratio of $\bar{B}^0 \rightarrow D^{*+} \tau^- \bar{\nu}_{\tau}$ relative to $\bar{B}^0 \rightarrow D^{*+} \ell^- \bar{\nu}_{\ell}$ decays with a semileptonic tagging method,''
  \Journal{\em Phys.\ Rev.}{D 94}{072007.}{2016}
%  Phys.\ Rev.\ D {\bf 94}, 072007 (2016)
%  doi:10.1103/PhysRevD.94.072007  [arXiv:1607.07923 [hep-ex]].
  %%CITATION = doi:10.1103/PhysRevD.94.072007;%%
  %23 citations counted in INSPIRE as of 02 Apr 2017
%
\bibitem{Hirose:2016wfn}
  S.~Hirose, T.~Iijima {\it et al.} [Belle Collaboration],
  %``Measurement of the $\tau$ lepton polarization and $R(D^*)$ in the decay $\bar{B} \to D^* \tau^- \bar{\nu}_\tau$,''
  \Journal{\em Phys.\ Rev.\ Lett.}{118}{211801.}{2017}
%  Phys.\ Rev.\ Lett.\  {\bf 118}, 211801 (2017)
%  doi:10.1103/PhysRevLett.118.211801  [arXiv:1612.00529 [hep-ex]].
  %%CITATION = doi:10.1103/PhysRevLett.118.211801;%%
  %33 citations counted in INSPIRE as of 03 Jul 2017%
\bibitem{Aaij:2015yra}
  R.~Aaij {\it et al.} [LHCb Collaboration],
  %``Measurement of the ratio of branching fractions $\mathcal{B}(\bar{B}^0 \to D^{*+}\tau^{-}\bar{\nu}_{\tau})/\mathcal{B}(\bar{B}^0 \to D^{*+}\mu^{-}\bar{\nu}_{\mu})$,''
  \Journal{\em Phys.\ Rev.\ Lett.}{115}{111803.}{2015}
%  Phys.\ Rev.\ Lett.\  {\bf 115}, 111803 (2015)
%  Addendum: [Phys.\ Rev.\ Lett.\  {\bf 115}, no. 15, 159901 (2015)]
%  doi:10.1103/PhysRevLett.115.159901, 10.1103/PhysRevLett.115.111803  [arXiv:1506.08614 [hep-ex]].
  %%CITATION = doi:10.1103/PhysRevLett.115.159901, 10.1103/PhysRevLett.115.111803;%%
  %193 citations counted in INSPIRE as of 02 Apr 2017
%
\bibitem{Crivellin:2012ye}
  A.~Crivellin, C.~Greub and A.~Kokulu,
  %``Explaining $B\to D\tau\nu$, $B\to D^*\tau\nu$ and $B\to \tau\nu$ in a 2HDM of type III,''
   \Journal{\em Phys.\ Rev.}{D 86}{054014.}{2012}
%  Phys.\ Rev.\ D {\bf 86}, 054014 (2012)
%  doi:10.1103/PhysRevD.86.054014  [arXiv:1206.2634 [hep-ph]].
  %%CITATION = doi:10.1103/PhysRevD.86.054014;%%
  %160 citations counted in INSPIRE as of 19 Mar 2017
%
\bibitem{Fajfer:2012jt}
  S.~Fajfer, J~F.~Kamenik, I.~Nisandzic and J.~Zupan,
  %``Implications of Lepton Flavor Universality Violations in B Decays,''
   \Journal{\em Phys.\ Rev.\ Lett.}{109}{161801.}{2012}
%  Phys.\ Rev.\ Lett.\  {\bf 109}, 161801 (2012)
%  doi:10.1103/PhysRevLett.109.161801  [arXiv:1206.1872 [hep-ph]].
  %%CITATION = doi:10.1103/PhysRevLett.109.161801;%%
  %136 citations counted in INSPIRE as of 19 Mar 2017
%
\bibitem{Sakaki:2013bfa}
  Y.~Sakaki, M.~Tanaka, A.~Tayduganov and R.~Watanabe,
  %``Testing leptoquark models in $\bar B \to D^{(*)} \tau \bar\nu$,''
  \Journal{\em Phys.\ Rev.}{D 88}{094012.}{2013}
%  Phys.\ Rev.\ D {\bf 88}, 094012 (2013)
%  doi:10.1103/PhysRevD.88.094012  [arXiv:1309.0301 [hep-ph]].
  %%CITATION = doi:10.1103/PhysRevD.88.094012;%%
  %93 citations counted in INSPIRE as of 02 Apr 2017
%
\bibitem{Hiller:2014ula}
  G.~Hiller and M.~Schmaltz,
  %``Diagnosing lepton-nonuniversality in $b \to s \ell \ell$,''
  \Journal{\em JHEP}{1502}{055.}{2015}
%  JHEP {\bf 1502}, 055 (2015)
%  doi:10.1007/JHEP02(2015)055  [arXiv:1411.4773 [hep-ph]].
  %%CITATION = doi:10.1007/JHEP02(2015)055;%%
  %52 citations counted in INSPIRE as of 14 Apr 2017
%
\bibitem{Gripaios:2014tna}
  B.~Gripaios, M.~Nardecchia and S.A.~Renner,
  %``Composite leptoquarks and anomalies in $B$-meson decays,''
  \Journal{\em JHEP}{1505}{006.}{2015}
%  JHEP {\bf 1505}, 006 (2015)
%  doi:10.1007/JHEP05(2015)006  [arXiv:1412.1791 [hep-ph]].
  %%CITATION = doi:10.1007/JHEP05(2015)006;%%
  %86 citations counted in INSPIRE as of 14 Apr 2017
%
\bibitem{Faroughy:2016osc}
  D.A.~Faroughy, A.~Greljo and J.F.~Kamenik,
  %``Confronting lepton flavor universality violation in B decays with high-$p_T$ tau lepton searches at LHC,''
  \Journal{\em Phys.\ Let.}{B 764}{126.}{2017}
%  Phys.\ Lett.\ B {\bf 764}, 126 (2017)
%  doi:10.1016/j.physletb.2016.11.011  [arXiv:1609.07138 [hep-ph]].
  %%CITATION = doi:10.1016/j.physletb.2016.11.011;%%
  %16 citations counted in INSPIRE as of 14 Apr 2017
%
\bibitem{Li:2016vvp}
  X.-Q.~Li, Y.-D.~Yang and X.~Zhang,
  %``Revisiting the one leptoquark solution to the R(D$^{(∗)}$) anomalies and its phenomenological implications,''
   \Journal{\em JHEP}{1608}{054.}{2016}
%  JHEP {\bf 1608}, 054 (2016).
%  doi:10.1007/JHEP08(2016)054
%  [arXiv:1605.09308 [hep-ph]].
  %%CITATION = doi:10.1007/JHEP08(2016)054;%%
  %32 citations counted in INSPIRE as of 23 Jun 2017
%
\bibitem{Alonso:2016oyd}
  R.~Alonso, B.~Grinstein and J.~Martin Camalich,
  %``The lifetime of the $B_c^-$ meson and the anomalies in $B\to D^{(*)}\tau\nu$,''
   \Journal{\em Phys.\ Rev.\ Lett.}{118}{081802.}{2017}
%  Phys.\ Rev.\ Lett.\  {\bf 118}, 081802 (2017)
%  doi:10.1103/PhysRevLett.118.081802  [arXiv:1611.06676 [hep-ph]].
  %%CITATION = doi:10.1103/PhysRevLett.118.081802;%%
  %11 citations counted in INSPIRE as of 18 Mar 2017
%
\bibitem{LHCb-had-tau}
   The LHCb Collaboration, LHCb-PAPER-2017-017; Federico Betti for the LHCb Collaboration, arXiv:1705.10651.
%
\bibitem{Feruglio:2016gvd}
  F.~Feruglio, P.~Paradisi and A.~Pattori,
  %``Revisiting Lepton Flavor Universality in B Decays,''
   \Journal{\em Phys.\ Rev.\ Lett.}{118}{011801.}{2017}
%  Phys.\ Rev.\ Lett.\  {\bf 118}, 011801 (2017).
%  doi:10.1103/PhysRevLett.118.011801
%  [arXiv:1606.00524 [hep-ph]].
  %%CITATION = doi:10.1103/PhysRevLett.118.011801;%%
  %40 citations counted in INSPIRE as of 23 Jun 2017
%
\bibitem{Aaij:2014iva}
  R.~Aaij {\it et al.} [LHCb Collaboration],
  %``Measurements of $CP$ violation in the three-body phase space of charmless $B^{\pm}$ decays,''
  \Journal{\em Phys.\ Rev.}{D 90}{112004.}{2014}
%  Phys.\ Rev.\ D {\bf 90}, 112004 (2014)
%  doi:10.1103/PhysRevD.90.112004  [arXiv:1408.5373 [hep-ex]].
  %%CITATION = doi:10.1103/PhysRevD.90.112004;%%
  %46 citations counted in INSPIRE as of 03 Apr 2017
%
\bibitem{Hou:2000bz}
  W.-S.~Hou and A.~Soni,
  %``Pathways to rare baryonic B decays,''
  \Journal{\em Phys.\ Rev.\ Lett.}{86}{4247.}{2001}
%  Phys.\ Rev.\ Lett.\  {\bf 86}, 4247 (2001)
%  doi:10.1103/PhysRevLett.86.4247  [hep-ph/0008079].
  %%CITATION = doi:10.1103/PhysRevLett.86.4247;%%
  %98 citations counted in INSPIRE as of 15 Apr 2017
%
\bibitem{Abe:2002ds}
  K.~Abe {\it et al.} [Belle Collaboration],
  %``Observation of B+- ---> p anti-p K+-,''
  \Journal{\em Phys.\ Rev.\ Lett.}{88}{181803.}{2002}
%  Phys.\ Rev.\ Lett.\  {\bf 88}, 181803 (2002)
%  doi:10.1103/PhysRevLett.88.181803  [hep-ex/0202017].
  %%CITATION = doi:10.1103/PhysRevLett.88.181803;%%
  %130 citations counted in INSPIRE as of 15 Apr 2017
%
\bibitem{Aaij:2013fta}
  R.~Aaij {\it et al.} [LHCb Collaboration],
  %``First evidence for the two-body charmless baryonic decay $B^0 \to p \bar{p}$,''
  \Journal{\em JHEP}{1310}{005.}{2013}
%  JHEP {\bf 1310}, 005 (2013)
%  doi:10.1007/JHEP10(2013)005  [arXiv:1308.0961 [hep-ex]].
  %%CITATION = doi:10.1007/JHEP10(2013)005;%%
  %12 citations counted in INSPIRE as of 15 Apr 2017
%
\bibitem{Aaij:2016xfa}
  R.~Aaij {\it et al.} [LHCb Collaboration],
  %``Evidence for the two-body charmless baryonic decay $B^+ \to p \bar\Lambda$,''
  \Journal{\em JHEP}{1704}{162.}{2017}
%  JHEP {\bf 1704}, 162 (2017)
%  doi:10.1007/JHEP04(2017)162  arXiv:1611.07805 [hep-ex].
  %%CITATION = ARXIV:1611.07805;%%
  %2 citations counted in INSPIRE as of 15 Apr 2017
%
\bibitem{Hou:2005iu}
  W.-S.~Hou, M.~Nagashima and A.~Soddu,
  %``Baryon number violation involving higher generations,''
  \Journal{\em Phys.\ Rev.}{D 72}{095001.}{2005}
%  Phys.\ Rev.\ D {\bf 72}, 095001 (2005)
%  doi:10.1103/PhysRevD.72.095001  [hep-ph/0509006].
  %%CITATION = doi:10.1103/PhysRevD.72.095001;%%
  %21 citations counted in INSPIRE as of 15 Apr 2017
%
\bibitem{Chatrchyan:2013bba}
  S.~Chatrchyan {\it et al.} [CMS Collaboration],
  %``Search for baryon number violation in top-quark decays,''
  \Journal{\em Phys.\ Lett.}{B 731}{173.}{2014}
%  Phys.\ Lett.\ B {\bf 731}, 173 (2014)
%  doi:10.1016/j.physletb.2014.02.033  [arXiv:1310.1618 [hep-ex]].
  %%CITATION = doi:10.1016/j.physletb.2014.02.033;%%
  %10 citations counted in INSPIRE as of 15 Apr 2017

\bibitem{sumrule_gronau}
  M.~Gronau,
  %``A Precise sum rule among four B ---> K pi CP asymmetries,''
  \Journal{\em Phys.\ Lett.}{B 627}{82.}{2005}
%  Phys.\ Lett.\ B {\bf 627}, 82 (2005)
%  doi:10.1016/j.physletb.2005.09.014  [hep-ph/0508047].
  %%CITATION = doi:10.1016/j.physletb.2005.09.014;%%
  %81 citations counted in INSPIRE as of 15 Apr 2017

\bibitem{sumrule_soni}
  D. Atwood and A. Soni,
  \Journal{\em Phys.\ Rev.}{D 58}{036005.}{1998}
%  Phys. Rev. D58 (1998) 036005.
%
\bibitem{Gerard:1990ni}
  J.-M.~G\'erard and W.-S.~Hou,
  %``CP violation in inclusive and exclusive charmless B decays,''
  \Journal{\em Phys.\ Rev.}{D 43}{2909.}{1991}
%  Phys.\ Rev.\ D {\bf 43}, 2909 (1991).
%  doi:10.1103/PhysRevD.43.2909
  %%CITATION = doi:10.1103/PhysRevD.43.2909;%%
  %209 citations counted in INSPIRE as of 20 Apr 2017
%
\bibitem{Hou:1988wt}
  W.-S.~Hou,
  %``Qcd Induced Charmless B Decays,''
  \Journal{\em Nucl.\ Phys.}{B 308}{561.}{1988}
%  Nucl.\ Phys.\ B {\bf 308}, 561 (1988).
%  doi:10.1016/0550-3213(88)90578-0
  %%CITATION = doi:10.1016/0550-3213(88)90578-0;%%
  %90 citations counted in INSPIRE as of 20 Apr 2017
%
\bibitem{Grzadkowski:1991kb}
  B.~Grzadkowski and W.-S.~Hou,
  %``Solutions to the B meson semileptonic branching ratio puzzle within two Higgs doublet models,''
  \Journal{\em Phys.\ Lett.}{B 272}{383.}{1991}
%  Phys.\ Lett.\ B {\bf 272}, 383 (1991).
%  doi:10.1016/0370-2693(91)91847-O
  %%CITATION = doi:10.1016/0370-2693(91)91847-O;%%
  %80 citations counted in INSPIRE as of 20 Apr 2017
%
\bibitem{Lees:2013kla}
  J.P.~Lees {\it et al.} [BaBar Collaboration],
  %``Search for $B \to K^{(*)} \nu \overline \nu$ and invisible quarkonium decays,''
  \Journal{\em Phys.\ Rev.}{D 87}{112005.}{2013}
%  doi:10.1103/PhysRevD.87.112005  [arXiv:1303.7465 [hep-ex]].
  %%CITATION = doi:10.1103/PhysRevD.87.112005;%%
  %49 citations counted in INSPIRE as of 17 Feb 2017

\bibitem{tautau-babar}
  B. Aubert {\it et al.} [BaBar Collaboration],
  \Journal{\em Phys.\ Rev.\ Lett.}{96}{241801.}{2006}
%  Phys.Rev.Lett. 96 (2006) 241802.

\bibitem{tautau-lhcb}
  R. Aaij {\it et al.} [LHCb Collaboration],
  \Journal{\em Phys.\ Rev.\ Lett.}{118}{251802.}{2017}
%  Phys.\ Rev.\ Lett.\  {\bf 118}, 251802 (2017)
%  doi:10.1103/PhysRevLett.118.251802  arXiv:1703.02508.
%
\bibitem{Aubert:2008ac}
  B.~Aubert {\it et al.} [BaBar Collaboration],
  %``A Search for $B^+ \to \ell^+ \nu_{\ell}$ Recoiling Against $B^{-}\to D^{0} \ell^{-}\bar{\nu} X$,''
  \Journal{\em Phys.\ Rev.}{D 81}{051101.}{2010}
%  arXiv:0809.4027 [hep-ex].
  %%CITATION = ARXIV:0809.4027;%%
  %55 citations counted in INSPIRE as of 02 Apr 2017
%
\bibitem{Lees:2012ju}
  J.P.~Lees {\it et al.} [BaBar Collaboration],
  %``Evidence of $B^+ \to \tau^+\nu$ decays with hadronic B tags,''
  \Journal{\em Phys.\ Rev.}{D 88}{031102.}{2013}
%  Phys.\ Rev.\ D {\bf 88}, 031102 (2013)
%  doi:10.1103/PhysRevD.88.031102  [arXiv:1207.0698 [hep-ex]].
  %%CITATION = doi:10.1103/PhysRevD.88.031102;%%
  %149 citations counted in INSPIRE as of 02 Apr 2017
%
\bibitem{Hara:2010dk}
  K.~Hara, T.~Iijima {\it et al.} [Belle Collaboration],
  %``Evidence for $B^- -> \tau^- \bar{\nu}$ with a Semileptonic Tagging Method,''
  \Journal{\em Phys.\ Rev.}{D 82}{071101.}{2010}
%  Phys.\ Rev.\ D {\bf 82}, 071101 (2010)
%  doi:10.1103/PhysRevD.82.071101  [arXiv:1006.4201 [hep-ex]].
  %%CITATION = doi:10.1103/PhysRevD.82.071101;%%
  %131 citations counted in INSPIRE as of 02 Apr 2017
%
\bibitem{Adachi:2012mm}
  K.~Hara, Y.~Horii, T.~Iijima {\it et al.} [Belle Collaboration],
  %``Evidence for $B^- \to \tau^- \bar{\nu}_\tau$  with a Hadronic Tagging Method Using the Full Data Sample of Belle,''
  \Journal{\em Phys.\ Rev.\ Lett.}{110}{131801.}{2013}
%  Phys.\ Rev.\ Lett.\  {\bf 110}, 131801 (2013)
%  doi:10.1103/PhysRevLett.110.131801  [arXiv:1208.4678 [hep-ex]].
  %%CITATION = doi:10.1103/PhysRevLett.110.131801;%%
  %201 citations counted in INSPIRE as of 02 Apr 2017
%
\bibitem{Kronenbitter:2015kls}
  B.~Kronenbitter, M.~Heck, P.~Goldenzweig, T.~Kuhr {\it et al.} [Belle Collaboration],
  %``Measurement of the branching fraction of B^+ -> tau^+ nu_tau decays with the semileptonic tagging method,''
  \Journal{\em Phys.\ Rev.}{D 92}{051102.}{2015}
%  Phys.\ Rev.\ D {\bf 92}, 051102 (2015)
%  doi:10.1103/PhysRevD.92.051102  [arXiv:1503.05613 [hep-ex]].
  %%CITATION = doi:10.1103/PhysRevD.92.051102;%%
  %41 citations counted in INSPIRE as of 02 Apr 2017
%
\bibitem{Hou:1992sy}
  W.-S.~Hou,
  %``Enhanced charged Higgs boson effects in B- ---> tau anti-neutrino, mu anti-neutrino and b ---> tau anti-neutrino + X,''
  \Journal{\em Phys.\ Rev.}{D 48}{2342.}{1993}
%  Phys.\ Rev.\ D {\bf 48}, 2342 (1993).
%  doi:10.1103/PhysRevD.48.2342
  %%CITATION = doi:10.1103/PhysRevD.48.2342;%%
  %427 citations counted in INSPIRE as of 13 Jul 2017

\bibitem{munu-babar}
  B. Aubert {\it et al.} [BaBar Collaboration],
  \Journal{\em Phys.\ Rev.}{D 79}{091101.}{2009}

\bibitem{munu-belle}
  N. Satoyama {\it et al.} [Belle Collaboration],
  \Journal{\em Phys.\ Lett.}{B 647}{67.}{2007}
%
\bibitem{Yook:2014kga}
  Y.~Yook, Y.-J. Kwon {\it et al.} [Belle Collaboration],
  %``Search for $B^+ \to e^+ \nu$ and $B^+ \to \mu^+ \nu$ decays using hadronic tagging,''
  \Journal{\em Phys.\ Rev.}{D 91}{052016.}{2015}
%  Phys.\ Rev.\ D {\bf 91}, 052016 (2015)
%  doi:10.1103/PhysRevD.91.052016  [arXiv:1406.6356 [hep-ex]].
  %%CITATION = doi:10.1103/PhysRevD.91.052016;%%
  %6 citations counted in INSPIRE as of 20 Apr 2017
%
\bibitem{Hou:2012hh}
  W.-S.~Hou,
  %``Theory Summary (a Perspective,''
  arXiv:1207.7275 [hep-ph].
  %%CITATION = ARXIV:1207.7275;%%
  %2 citations counted in INSPIRE as of 20 Apr 2017
%
\bibitem{Belle-munu}
   Talk by S. Falke at EPS 2017, and paper under preparation by the Belle Collaboration.

\bibitem{K*gamma}
  R. Ammar {\it et al.} [CLEO Collaboration],
  \Journal{\em Phys.\ Rev.\ Lett.}{71}{674.}{1993}

\bibitem{Buchmuller}
  O.L. Buchm\"uller and H.U. Fl\"{a}cher,
  \Journal{\em Phys.\ Rev.}{D 73}{073008.}{2006}

\bibitem{misiak}
%  T. Hermann, M. Misiak and M. Steinhauser,
%  \Journal{\em JHEP}{1211}{036.}{2012}
  M.~Misiak {\it et al.},
  %``Updated NNLO QCD predictions for the weak radiative B-meson decays,''
  \Journal{\em Phys.\ Rev.\ Lett.}{114}{221801.}{2015}
%  Phys.\ Rev.\ Lett.\  {\bf 114}, 221801 (2015)
%  doi:10.1103/PhysRevLett.114.221801  [arXiv:1503.01789 [hep-ph]].
  %%CITATION = doi:10.1103/PhysRevLett.114.221801;%%
  %156 citations counted in INSPIRE as of 20 Apr 2017
%
\bibitem{Bertolini:1986th}
  S.~Bertolini, F.~Borzumati and A.~Masiero,
  %``QCD Enhancement of Radiative b Decays,''
  \Journal{\em Phys.\ Rev.\ Lett.}{59}{180.}{1987}
%  Phys.\ Rev.\ Lett.\ {\bf 59}, 180 (1987).
%  doi:10.1103/PhysRevLett.59.180
  %%CITATION = doi:10.1103/PhysRevLett.59.180;%%
  %364 citations counted in INSPIRE as of 20 Apr 2017
%
\bibitem{Deshpande:1987nr}
  N.G.~Deshpande, P.~Lo, J.~Trampetic, G.~Eilam and P.~Singer,
  %``B ---> K* gamma and the Top Quark Mass,''
  \Journal{\em Phys.\ Rev.\ Lett.}{59}{183.}{1987}
%  Phys.\ Rev.\ Lett.\  {\bf 59}, 183 (1987).
%  doi:10.1103/PhysRevLett.59.183
  %%CITATION = doi:10.1103/PhysRevLett.59.183;%%
  %368 citations counted in INSPIRE as of 20 Apr 2017
%
\bibitem{Grinstein:1987pu}
  B.~Grinstein and M.B.~Wise,
  %``Weak Radiative B Meson Decay as a Probe of the Higgs Sector,''
  \Journal{\em Phys.\ Lett.}{B 201}{274.}{1988}
%  Phys.\ Lett.\ B {\bf 201}, 274 (1988).
%  doi:10.1016/0370-2693(88)90227-4
  %%CITATION = doi:10.1016/0370-2693(88)90227-4;%%
  %142 citations counted in INSPIRE as of 20 Apr 2017
%
\bibitem{Hou:1987kf}
  W.-S.~Hou and R.S.~Willey,
  %``Effects of Charged Higgs Bosons on the Processes b ---> s Gamma, b ---> s g* and b ---> s Lepton+ Lepton-,''
  \Journal{\em Phys.\ Lett.}{B 202}{591.}{1988}
%  Phys.\ Lett.\ B {\bf 202}, 591 (1988).
%  doi:10.1016/0370-2693(88)91870-9
  %%CITATION = doi:10.1016/0370-2693(88)91870-9;%%
  %207 citations counted in INSPIRE as of 20 Apr 2017
%
\bibitem{Belle:2016ufb}
  A.~Abdesselam {\it et~al.}~[Belle~Collaboration],
  %``Measurement of the inclusive $B\to X_{s+d} \gamma$ branching fraction, photon energy spectrum and HQE parameters,''
  arXiv:1608.02344 [hep-ex].
  %%CITATION = ARXIV:1608.02344;%%
  %15 citations counted in INSPIRE as of 18 Jun 2017
%
\bibitem{Misiak:2017bgg}
  M.~Misiak and M.~Steinhauser,
  %``Weak Radiative Decays of the B Meson and Bounds on $M_{H^\pm}$ in the Two-Higgs-Doublet Model,''
  \Journal{\em Eur.\ Phys.\ J.}{C 77}{201.}{2017}
%  Eur.\ Phys.\ J.\ C {\bf 77}, 201 (2017).
%  doi:10.1140/epjc/s10052-017-4776-y
%  [arXiv:1702.04571 [hep-ph]].
  %%CITATION = doi:10.1140/epjc/s10052-017-4776-y;%%
  %23 citations counted in INSPIRE as of 18 Jun 2017
%
\bibitem{darkt1}
  P. Fayet,
  \Journal{\em Phys.\ Rev.}{D 75}{115017.}{2007}
%  Phys. Rev. D75 (2007) 115017.  	

\bibitem{darkt2}
  M. Pospelov, A. Ritz, and M.B. Voloshin,
  \Journal{\em Phys.\ Lett.}{B 662}{53.}{2008}
%  Phys. Lett. B 662 (2008) 53.

\bibitem{darkt3}
  C. Cheung, J.T. Ruderman, L. T. Wang, and I. Yavin,
  \Journal{\em Phys.\ Rev.}{D 80}{035008.}{2009}
%  Phys. Rev. D80, (2009) 035008.
%
\bibitem{Holdom:1985ag}
  B.~Holdom,
  %``Two U(1)'s and Epsilon Charge Shifts,''
  \Journal{\em Phys.\ Lett.}{B 166}{196.}{1986}
%  Phys.\ Lett.\  {\bf 166B}, 196 (1986).
%  doi:10.1016/0370-2693(86)91377-8
  %%CITATION = doi:10.1016/0370-2693(86)91377-8;%%
  %1100 citations counted in INSPIRE as of 06 Apr 2017

\bibitem{darkpho1}
  J.P. Lees {\it et al.} [BaBar Collaboration],
  \Journal{\em Phys.\ Rev.\ Lett.}{113}{201801.}{2014}
%  Phys. Rev. Lett. 113 (2014) 201801.

\bibitem{darkpho2}
  J.P. Lees {\it et al.} [BaBar Collaboration],
  arXiv:1702.03327.

\bibitem{darkhigg1}
  J.P. Lees {\it et al.} [BaBar Collaboration],
  \Journal{\em Phys.\ Rev.\ Lett.}{108}{211801.}{2012}
%  Phys. Rev. Lett. 108 (2012) 211801.

\bibitem{darkhigg2}
  I. Jaegle {\it et al.} [Belle Collaboration],
  \Journal{\em Phys.\ Rev.\ Lett.}{114}{211801.}{2015}
%  Phys. Rev. Lett. 114 (2015) 211801.

\bibitem{darkforce}
  J.P. Lees {\it et al.} [BaBar Collaboration],
  \Journal{\em Phys.\ Rev.}{D 94}{011102.}{2016}
%  Phys.Rev. D94 (2016) 011102.

\bibitem{darkbos}
  E. Won {\it et al.} [Belle Collaboration],
  \Journal{\em Phys.\ Rev.}{D 94}{092006.}{2016}
%  Phys.Rev. D94 (2016) 092006
%
\bibitem{Buras:2016egb}
  See A.J.~Buras,
  %``Kaon Flavour Physics Strikes Back,''
  arXiv:1611.06206 [hep-ph], and references therein.
  %%CITATION = ARXIV:1611.06206;%%
%
\bibitem{Buras:2015qea}
  A.J.~Buras, D.~Buttazzo, J.~Girrbach-Noe and R.~Knegjens,
  %``$ {K}^{+}\to {\pi}^{+}\nu \overline{\nu} $ and $ {K}_L\to {\pi}^0\nu \overline{\nu} $ in the Standard Model: status and perspectives,''
  \Journal{\em JHEP}{1511}{033.}{2015}
%  doi:10.1007/JHEP11(2015)033  [arXiv:1503.02693 [hep-ph]].
  %%CITATION = doi:10.1007/JHEP11(2015)033;%%
  %70 citations counted in INSPIRE as of 14 Feb 2017
%
\bibitem{Artamonov:2008qb}
  A.V.~Artamonov {\it et al.} [E949 Collaboration],
  %``New measurement of the $K^{+} \to \pi^{+} \nu \bar{\nu}$ branching ratio,''
  \Journal{\em Phys.\ Rev.\ Lett.}{101}{191802.}{2008}
%  doi:10.1103/PhysRevLett.101.191802  [arXiv:0808.2459 [hep-ex]].
  %%CITATION = doi:10.1103/PhysRevLett.101.191802;%%
  %169 citations counted in INSPIRE as of 14 Feb 2017
%
\bibitem{Artamonov:2009sz}
  A.V.~Artamonov {\it et al.} [E949 Collaboration],
  %``Study of the decay $K^+\to\pi^+\nu \bar\nu$ in the momentum region $140 < P_\pi < 199$ MeV/c,''
  \Journal{\em Phys.\ Rev.}{D 79}{092004.}{2009}
%  doi:10.1103/PhysRevD.79.092004  [arXiv:0903.0030 [hep-ex]].
  %%CITATION = doi:10.1103/PhysRevD.79.092004;%%
  %168 citations counted in INSPIRE as of 14 Feb 2017
%
\bibitem{CERNNA48/2:2016tdo}
  J.R.~Batley {\it et al.} [NA48/2 Collaboration],
  %``Searches for Lepton Number Violation and Resonances in $K^{\pm}\to\pi\mu\mu$ Decays,''
  \Journal{\em Phys.\ Let.}{B 769}{67.}{2017}
%  Phys.\ Lett.\ B {\bf 769}, 67 (2017)
%  doi:10.1016/j.physletb.2017.03.029  [arXiv:1612.04723 [hep-ex]].
  %%CITATION = doi:10.1016/j.physletb.2017.03.029;%%
%
\bibitem{Atre:2009rg}
  A.~Atre, T.~Han, S.~Pascoli and B.~Zhang,
  %``The Search for Heavy Majorana Neutrinos,''
  \Journal{\em JHEP}{0905}{030.}{2009}
%  doi:10.1088/1126-6708/2009/05/030  [arXiv:0901.3589 [hep-ph]].
  %%CITATION = doi:10.1088/1126-6708/2009/05/030;%%
  %358 citations counted in INSPIRE as of 15 Feb 2017
%
\bibitem{Crivellin:2016vjc}
  A.~Crivellin, G.~D'Ambrosio, M.~Hoferichter and L.C.~Tunstall,
  %``Violation of lepton flavor and lepton flavor universality in rare kaon decays,''
  \Journal{\em Phys.\ Rev.}{D 93}{074038.}{2016}
%  doi:10.1103/PhysRevD.93.074038  [arXiv:1601.00970 [hep-ph]].
  %%CITATION = doi:10.1103/PhysRevD.93.074038;%%
  %15 citations counted in INSPIRE as of 15 Feb 2017
%
\bibitem{Batley:2015lha}
  J.R.~Batley {\it et al.} [NA48/2 Collaboration],
  %``Search for the dark photon in $\pi^0$ decays,''
  \Journal{\em Phys.\ Lett.}{B 746}{178.}{2015}
%  doi:10.1016/j.physletb.2015.04.068  [arXiv:1504.00607 [hep-ex]].
  %%CITATION = doi:10.1016/j.physletb.2015.04.068;%%
  %75 citations counted in INSPIRE as of 15 Feb 2017
%
\bibitem{Ahn:2009gb}
  J.K.~Ahn {\it et al.} [E391a Collaboration],
  %``Experimental study of the decay K0(L) ---> pi0 nu nu-bar,''
  \Journal{\em Phys.\ Rev.}{D 81}{072004.}{2010}
%  doi:10.1103/PhysRevD.81.072004  [arXiv:0911.4789 [hep-ex]].
  %%CITATION = doi:10.1103/PhysRevD.81.072004;%%
  %107 citations counted in INSPIRE as of 15 Feb 2017
%
\bibitem{Grossman:1997sk}
  Y.~Grossman and Y.~Nir,
  %``K(L) ---> pi0 neutrino anti-neutrino beyond the standard model,''
  \Journal{\em Phys.\ Lett.}{B 398}{163.}{1997}
%  doi:10.1016/S0370-2693(97)00210-4  [hep-ph/9701313].
  %%CITATION = doi:10.1016/S0370-2693(97)00210-4;%%
  %263 citations counted in INSPIRE as of 15 Feb 2017
%
\bibitem{Ahn:2016kja}
  J.K.~Ahn {\it et al.} [KOTO Collaboration],
  %``A new search for the $K_{L} \to \pi^0 \nu \overline{\nu}$ and $K_{L} \to \pi^{0} X^{0}$ decays,''
  \Journal{\em PTEP}{2017}{021C01.}{2017}
%  PTEP {\bf 2017}, 021C01 (2017)
%  doi:10.1093/ptep/ptx001  [arXiv:1609.03637 [hep-ex]].
  %%CITATION = doi:10.1093/ptep/ptx001;%%
  %8 citations counted in INSPIRE as of 02 Apr 2017%
\bibitem{Fuyuto:2014cya}
  K.~Fuyuto, W.-S.~Hou and M.~Kohda,
  %``Loophole in $K \to \pi\nu\bar{\nu}$ Search and New Weak Leptonic Forces,''
  \Journal{\em Phys.\ Rev.\ Lett.}{114}{171802.}{2015}
%  doi:10.1103/PhysRevLett.114.171802  [arXiv:1412.4397 [hep-ph]].
  %%CITATION = doi:10.1103/PhysRevLett.114.171802;%%
  %7 citations counted in INSPIRE as of 16 Feb 2017
%
\bibitem{Artamonov:2005cu}
  A.V.~Artamonov {\it et al.} [E949 Collaboration],
  %``Upper Limit on the Branching Ratio for the Decay $\pi^0 \to \nu \overline {\nu}$,''
  \Journal{\em Phys.\ Rev.}{D 72}{091102.}{2005}
%  doi:10.1103/PhysRevD.72.091102  [hep-ex/0506028].
  %%CITATION = doi:10.1103/PhysRevD.72.091102;%%
  %30 citations counted in INSPIRE as of 16 Feb 2017
%
\bibitem{Altmannshofer:2014cfa}
  W.~Altmannshofer, S.~Gori, M.~Pospelov and I.~Yavin,
  %``Quark flavor transitions in $L_\mu-L_\tau$ models,''
  \Journal{\em Phys.\ Rev.}{D 89}{095033.}{2014}
%  doi:10.1103/PhysRevD.89.095033  [arXiv:1403.1269 [hep-ph]].
  %%CITATION = doi:10.1103/PhysRevD.89.095033;%%
  %140 citations counted in INSPIRE as of 17 Feb 2017
%
\bibitem{Altmannshofer:2014pba}
  W.~Altmannshofer, S.~Gori, M.~Pospelov and I.~Yavin,
  %``Neutrino Trident Production: A Powerful Probe of New Physics with Neutrino Beams,''
  \Journal{\em Phys.\ Rev.\ Lett.}{113}{091801.}{2014}
%  doi:10.1103/PhysRevLett.113.091801  [arXiv:1406.2332 [hep-ph]].
  %%CITATION = doi:10.1103/PhysRevLett.113.091801;%%
  %72 citations counted in INSPIRE as of 17 Feb 2017
%
\bibitem{Fuyuto:2015gmk}
  K.~Fuyuto, W.-S.~Hou and M.~Kohda,
  %``Z′ -induced FCNC decays of top, beauty, and strange quarks,''
  \Journal{\em Phys.\ Rev.}{D 93}{054021.}{2016}
%  doi:10.1103/PhysRevD.93.054021  [arXiv:1512.09026 [hep-ph]].
  %%CITATION = doi:10.1103/PhysRevD.93.054021;%%
  %8 citations counted in INSPIRE as of 17 Feb 2017
%
\bibitem{Aaij:2015tna}
  R.~Aaij {\it et al.} [LHCb Collaboration],
  %``Search for hidden-sector bosons in $B^0 \!\to K^{*0}\mu^+\mu^-$ decays,''
  \Journal{\em Phys.\ Rev.\ Lett.}{115}{161802.}{2015}
%  doi:10.1103/PhysRevLett.115.161802  [arXiv:1508.04094 [hep-ex]].
  %%CITATION = doi:10.1103/PhysRevLett.115.161802;%%
  %25 citations counted in INSPIRE as of 17 Feb 2017
%
\bibitem{Yamanaka:2016wmx}
  T.~Yamanaka,
  %``Rare kaon decay experiments,''
  \Journal{\em PoS\ FPCP}{2016}{018.}{2017}
  %%CITATION = POSCI,FPCP2016,018;%%
%
\bibitem{NA62-KLEVER}
  A. Bradley {\it et al.}, NA62-16-03.
%
\bibitem{Davoudiasl:2014kua}
  H.~Davoudiasl, H.-S.~Lee and W.J.~Marciano,
  %``Muon $g−2$, rare kaon decays, and parity violation from dark bosons,''
  \Journal{\em Phys.\ Rev.}{D 89}{095006.}{2014}
%  Phys.\ Rev.\ D {\bf 89}, 095006 (2014)
%  doi:10.1103/PhysRevD.89.095006  [arXiv:1402.3620 [hep-ph]].
  %%CITATION = doi:10.1103/PhysRevD.89.095006;%%
  %58 citations counted in INSPIRE as of 06 Apr 2017
%
\bibitem{Banerjee:2016tad}
  D.~Banerjee {\it et al.} [NA64 Collaboration],
  %``Search for invisible decays of sub-GeV dark photons in missing-energy events at the CERN SPS,''
  \Journal{\em Phys.\ Rev.\ Lett.}{118}{011802.}{2017}
%  Phys.\ Rev.\ Lett.\  {\bf 118}, 011802 (2017)
%  doi:10.1103/PhysRevLett.118.011802  [arXiv:1610.02988 [hep-ex]].
  %%CITATION = doi:10.1103/PhysRevLett.118.011802;%%
  %13 citations counted in INSPIRE as of 06 Apr 2017
%
\bibitem{Gninenko:2014pea}
  S.N.~Gninenko, N.V.~Krasnikov and V.A.~Matveev,
  %``Muon g-2 and searches for a new leptophobic sub-GeV dark boson in a missing-energy experiment at CERN,''
  \Journal{\em Phys.\ Rev.}{D 91}{095015.}{2015}
%  Phys.\ Rev.\ D {\bf 91}, 095015 (2015)
%  doi:10.1103/PhysRevD.91.095015  [arXiv:1412.1400 [hep-ph]].
  %%CITATION = doi:10.1103/PhysRevD.91.095015;%%
  %16 citations counted in INSPIRE as of 06 Apr 2017
%
\bibitem{AlaviHarati:1999xp}
  A.~Alavi-Harati {\it et al.} [KTeV Collaboration],
  %``Observation of direct CP violation in $K_{S,L} \to \pi \pi$ decays,''
  \Journal{\em Phys.\ Rev.\ Lett.}{83}{22.}{1999}
%  doi:10.1103/PhysRevLett.83.22  [hep-ex/9905060].
  %%CITATION = doi:10.1103/PhysRevLett.83.22;%%
  %677 citations counted in INSPIRE as of 28 Feb 2017
%
\bibitem{Fanti:1999nm}
  V.~Fanti {\it et al.} [NA48 Collaboration],
  %``A New measurement of direct CP violation in two pion decays of the neutral kaon,''
  \Journal{\em Phys.\ Lett.}{B 465}{335.}{1999}
%  doi:10.1016/S0370-2693(99)01030-8  [hep-ex/9909022].
  %%CITATION = doi:10.1016/S0370-2693(99)01030-8;%%
  %520 citations counted in INSPIRE as of 28 Feb 2017
%
\bibitem{Aubert:2004qm}
  B.~Aubert {\it et al.} [BaBar Collaboration],
  %``Observation of direct CP violation in $B^0 \to K^+ \pi^-$ decays,''
  \Journal{\em Phys.\ Rev.\ Lett.}{93}{131801.}{2004}
%  Phys.\ Rev.\ Lett.\  {\bf 93}, 131801 (2004)
%  doi:10.1103/PhysRevLett.93.131801  [hep-ex/0407057].
  %%CITATION = doi:10.1103/PhysRevLett.93.131801;%%
  %323 citations counted in INSPIRE as of 02 Mar 2017
%
\bibitem{Chao:2004mn}
  Y.~Chao, P.~Chang {\it et al.} [Belle Collaboration],
  %``Evidence for direct CP violation in B0 ---> K+ pi- decays,''
  \Journal{\em Phys.\ Rev.\ Lett.}{93}{191802.}{2004}
%  doi:10.1103/PhysRevLett.93.191802  [hep-ex/0408100].
  %%CITATION = doi:10.1103/PhysRevLett.93.191802;%%
  %213 citations counted in INSPIRE as of 02 Mar 2017
%
\bibitem{Wu:1964qx}
  T.T.~Wu and C.-N.~Yang,
  %``Phenomenological Analysis of Violation of CP Invariance in Decay of K0 and anti-K0,''
  \Journal{\em Phys.\ Rev.\ Lett.}{13}{380.}{1964}
%  doi:10.1103/PhysRevLett.13.380
  %%CITATION = doi:10.1103/PhysRevLett.13.380;%%
  %340 citations counted in INSPIRE as of 28 Feb 2017
%
\bibitem{Winstein:1991gz}
  B.~Winstein,
  %``CP violation in neutral B meson decays in the standard and superweak models,''
   \Journal{\em Phys.\ Rev.\ Lett.}{68}{1271.}{1992}
%  doi:10.1103/PhysRevLett.68.1271
  %%CITATION = doi:10.1103/PhysRevLett.68.1271;%%
  %44 citations counted in INSPIRE as of 28 Feb 2017
%
\bibitem{Wolfenstein:1964ks}
  L.~Wolfenstein,
  %``Violation of CP Invariance and the Possibility of Very Weak Interactions,''
   \Journal{\em Phys.\ Rev.\ Lett.}{13}{562.}{1964}
%  doi:10.1103/PhysRevLett.13.562
  %%CITATION = doi:10.1103/PhysRevLett.13.562;%%
  %818 citations counted in INSPIRE as of 28 Feb 2017
%
\bibitem{Bai:2015nea}
  Z.~Bai {\it et al.} [RBC+UKQCD Collaboration],
  %``Standard Model Prediction for Direct CP Violation in K→ππ Decay,''
   \Journal{\em Phys.\ Rev.\ Lett.}{115}{212001.}{2015}
%  doi:10.1103/PhysRevLett.115.212001  [arXiv:1505.07863 [hep-lat]].
  %%CITATION = doi:10.1103/PhysRevLett.115.212001;%%
  %57 citations counted in INSPIRE as of 28 Feb 2017
%
\bibitem{Blum:2015ywa}
  T.~Blum {\it et al.} [RBC+UKQCD Collaboration],
  %``$K \rightarrow \pi\pi$ $\Delta I=3/2$ decay amplitude in the continuum limit,''
   \Journal{\em Phys.\ Rev.}{D 91}{074502.}{2015}
%  doi:10.1103/PhysRevD.91.074502  [arXiv:1502.00263 [hep-lat]].
  %%CITATION = doi:10.1103/PhysRevD.91.074502;%%
  %45 citations counted in INSPIRE as of 28 Feb 2017
%
\bibitem{Hill:2017wzi}
  For a recent brief review, see R.J.~Hill,
  %``Review of experimental and theoretical status of the proton radius puzzle,''
  arXiv:1702.01189 [hep-ph], and references therein.
  %%CITATION = ARXIV:1702.01189;%%
%
\bibitem{Antognini:1900ns}
  A.~Antognini {\it et al.},
  %``Proton Structure from the Measurement of $2S-2P$ Transition Frequencies of Muonic Hydrogen,''
   \Journal{\em Science}{339}{417.}{2013}
%  doi:10.1126/science.1230016
  %%CITATION = doi:10.1126/science.1230016;%%
  %263 citations counted in INSPIRE as of 02 Mar 2017
%
\bibitem{Gorringe:2015cma}
  T.P.~Gorringe and D.W.~Hertzog,
  %``Precision Muon Physics,''
   \Journal{\em Prog.\ Part.\ Nucl.\ Phys.}{84}{73.}{2015}
%  doi:10.1016/j.ppnp.2015.06.001  [arXiv:1506.01465 [hep-ex]].
  %%CITATION = doi:10.1016/j.ppnp.2015.06.001;%%
  %13 citations counted in INSPIRE as of 02 Mar 2017
%
\bibitem{MEG:2016wtm}
  A.M.~Baldini {\it et al.} [MEG Collaboration],
  %``Search for the lepton flavour violating decay $\mu ^+ \rightarrow \mathrm {e}^+ \gamma $ with the full dataset of the MEG experiment,''
   \Journal{\em Eur.\ Phys.\ J.}{C 76}{434.}{2016}
%  doi:10.1140/epjc/s10052-016-4271-x  [arXiv:1605.05081 [hep-ex]].
  %%CITATION = doi:10.1140/epjc/s10052-016-4271-x;%%
  %67 citations counted in INSPIRE as of 02 Mar 2017
%
\bibitem{Brooks:1999pu}
  M.L.~Brooks {\it et al.} [MEGA Collaboration],
  %``New limit for the family number nonconserving decay mu+ ---> e+ gamma,''
   \Journal{\em Phys.\ Rev.\ Lett.}{83}{1521.}{1999}
%  doi:10.1103/PhysRevLett.83.1521  [hep-ex/9905013].
  %%CITATION = doi:10.1103/PhysRevLett.83.1521;%%
  %645 citations counted in INSPIRE as of 04 Mar 2017
%
\bibitem{Kersten:2014xaa}
  J.~Kersten, J.h.~Park, D.~St\" ockinger and L.~Velasco-Sevilla,
  %``Understanding the correlation between $(g-2)_\mu$ and $\mu \rightarrow e \gamma$ in the MSSM,''
   \Journal{\em JHEP}{1408}{118.}{2014}
%  doi:10.1007/JHEP08(2014)118  [arXiv:1405.2972 [hep-ph]].
  %%CITATION = doi:10.1007/JHEP08(2014)118;%%
  %15 citations counted in INSPIRE as of 04 Mar 2017
%
\bibitem{Blankenburg:2012nx}
  G.~Blankenburg, G.~Isidori and J.~Jones-Perez,
  %``Neutrino Masses and LFV from Minimal Breaking of U(3)^5 and U(2)^5 flavor Symmetries,''
   \Journal{\em Eur.\ Phys.\ J.}{C 72}{2126.}{2012}
%  doi:10.1140/epjc/s10052-012-2126-7  [arXiv:1204.0688 [hep-ph]].
  %%CITATION = doi:10.1140/epjc/s10052-012-2126-7;%%
  %42 citations counted in INSPIRE as of 04 Mar 2017
%
\bibitem{Antusch:2006vw}
  S.~Antusch, E.~Arganda, M.~J.~Herrero and A.~M.~Teixeira,
  %``Impact of theta(13) on lepton flavour violating processes within SUSY seesaw,''
   \Journal{\em JHEP}{0611}{090.}{2006}
%  doi:10.1088/1126-6708/2006/11/090  [hep-ph/0607263].
  %%CITATION = doi:10.1088/1126-6708/2006/11/090;%%
  %131 citations counted in INSPIRE as of 04 Mar 2017
%
\bibitem{Bertl:2006up}
  W.H.~Bertl {\it et al.} [SINDRUM II Collaboration],
  %``A Search for muon to electron conversion in muonic gold,''
   \Journal{\em Eur.\ Phys.\ J.}{C 47}{337.}{2006}
%  doi:10.1140/epjc/s2006-02582-x
  %%CITATION = doi:10.1140/epjc/s2006-02582-x;%%
  %295 citations counted in INSPIRE as of 04 Mar 2017
%
\bibitem{Dzhilkibaev:1989zb}
  R.M.~Dzhilkibaev and V.M.~Lobashev,
  %``On the Search for $\mu \to e$ Conversion on Nuclei. (In Russian),''
   \Journal{\em Sov.\ J.\ Nucl.\ Phys.}{49}{384.}{1989}
%  [Yad.\ Fiz.\  {\bf 49}, 622 (1989)].
  %%CITATION = SJNCA,49,384;%%
  %50 citations counted in INSPIRE as of 04 Mar 2017
%
\bibitem{Bartoszek:2014mya}
  L.~Bartoszek {\it et al.} [Mu2e Collaboration],
  %``Mu2e Technical Design Report,''
  arXiv:1501.05241 [physics.ins-det].
  %%CITATION = ARXIV:1501.05241;%%
  %66 citations counted in INSPIRE as of 04 Mar 2017
%
\bibitem{Kuno:2013mha}
  Y.~Kuno [for the COMET Collaboration],
  %``A search for muon-to-electron conversion at J-PARC: The COMET experiment,''
   \Journal{\em PTEP}{2013}{022C01.}{2013}
%  doi:10.1093/ptep/pts089
  %%CITATION = doi:10.1093/ptep/pts089;%%
  %73 citations counted in INSPIRE as of 04 Mar 2017
%
\bibitem{DeeMe}
  See webpage http://nasubi.hep.sci.osaka-u.ac.jp/plone.
%
\bibitem{deGouvea:2013zba}
  A.~de Gouv\^ea and P.~Vogel,
  %``Lepton Flavor and Number Conservation, and Physics Beyond the Standard Model,''
   \Journal{\em Prog.\ Part.\ Nucl.\ Phys.}{71}{75.}{2013}
%  doi:10.1016/j.ppnp.2013.03.006  [arXiv:1303.4097 [hep-ph]].
  %%CITATION = doi:10.1016/j.ppnp.2013.03.006;%%
  %77 citations counted in INSPIRE as of 04 Mar 2017
%
\bibitem{Mu3e}
  See webpage https://www.psi.ch/mu3e/.
%
\bibitem{Beneke:2015lba}
  M.~Beneke, P.~Moch and J.~Rohrwild,
  %``Lepton flavour violation in RS models with a brane- or nearly brane-localized Higgs,''
   \Journal{\em Nucl.\ Phys.}{B 906}{561.}{2016}
%  doi:10.1016/j.nuclphysb.2016.02.037  [arXiv:1508.01705 [hep-ph]].
  %%CITATION = doi:10.1016/j.nuclphysb.2016.02.037;%%
  %11 citations counted in INSPIRE as of 04 Mar 2017
%
\bibitem{Bennett:2006fi}
  G.W.~Bennett {\it et al.} [Muon g-2 Collaboration],
  %``Final Report of the Muon E821 Anomalous Magnetic Moment Measurement at BNL,''
   \Journal{\em Phys.\ Rev.}{D 73}{072003.}{2006}
%  doi:10.1103/PhysRevD.73.072003  [hep-ex/0602035].
  %%CITATION = doi:10.1103/PhysRevD.73.072003;%%
  %1339 citations counted in INSPIRE as of 05 Mar 2017
%
\bibitem{Davier:2016iru}
  M.~Davier,
  %``Update of the Hadronic Vacuum Polarisation Contribution to the muon g-2,''
  arXiv:1612.02743 [hep-ph].
  %%CITATION = ARXIV:1612.02743;%%
  %3 citations counted in INSPIRE as of 06 Mar 2017
%
\bibitem{Stockinger:2006zn}
  See e.g. the review by D.~St\"ockinger,
  %``The Muon Magnetic Moment and Supersymmetry,''
   \Journal{\em J.\ Phys.}{G 34}{R45.}{2007}
  J.\ Phys.\ G {\bf 34}, R45 (2007)
%  doi:10.1088/0954-3899/34/2/R01  [hep-ph/0609168].
  %%CITATION = doi:10.1088/0954-3899/34/2/R01;%%
  %194 citations counted in INSPIRE as of 19 Apr 2017
%
\bibitem{Grange:2015fou}
  J.~Grange {\it et al.} [Muon g-2 Collaboration],
  %``Muon (g-2) Technical Design Report,''
  arXiv:1501.06858 [physics.ins-det].
  %%CITATION = ARXIV:1501.06858;%%
  %76 citations counted in INSPIRE as of 06 Mar 2017
%
\bibitem{Benayoun:2014tra}
  M.~Benayoun {\it et al.},
  %``Hadronic contributions to the muon anomalous magnetic moment Workshop. $(g-2)_{\mu}$: Quo vadis? Workshop. Mini proceedings,''
  arXiv:1407.4021 [hep-ph] (workshop mini-proceedings).
  %%CITATION = ARXIV:1407.4021;%%
  %51 citations counted in INSPIRE as of 06 Mar 2017
%
\bibitem{Blum:2016lnc}
  T.~Blum {\it et al.}, %N.~Christ, M.~Hayakawa, T.~Izubuchi, L.~Jin, C.~Jung and C.~Lehner,
  %``Connected and Leading Disconnected Hadronic Light-by-Light Contribution to the Muon Anomalous Magnetic Moment with a Physical Pion Mass,''
   \Journal{\em Phys.\ Rev.\ Lett.}{118}{022005.}{2017}
%  doi:10.1103/PhysRevLett.118.022005  [arXiv:1610.04603 [hep-lat]].
  %%CITATION = doi:10.1103/PhysRevLett.118.022005;%%
  %4 citations counted in INSPIRE as of 06 Mar 2017
%
\bibitem{Prades:2009tw}
  A standard quote is
  J.~Prades, E.~de Rafael and A.~Vainshtein,
  %``The Hadronic Light-by-Light Scattering Contribution to the Muon and Electron Anomalous Magnetic Moments,''
   \Journal{\em Adv.\ Ser.\ Direct.\ High Energy Phys.}{20}{303.}{2009}
%  doi:10.1142/9789814271844_0009  [arXiv:0901.0306 [hep-ph]].
  %%CITATION = doi:10.1142/9789814271844_0009;%%
  %207 citations counted in INSPIRE as of 06 Mar 2017
%
\bibitem{Blum:2015you}
  T.~Blum {\it et al.},
  %``Calculation of the hadronic vacuum polarization disconnected contribution to the muon anomalous magnetic moment,''
   \Journal{\em Phys.\ Rev.\ Lett.}{116}{232002.}{2016}
%  Phys.\ Rev.\ Lett.\  {\bf 116}, 232002 (2016)
%  doi:10.1103/PhysRevLett.116.232002  [arXiv:1512.09054 [hep-lat]].
  %%CITATION = doi:10.1103/PhysRevLett.116.232002;%%
  %21 citations counted in INSPIRE as of 03 Jul 2017
%
\bibitem{FLAG}
   Flavor Lattice Averaging Group, http://itpwiki.unibe.ch/flag/.
%
\bibitem{Engel:2013lsa}
  J.~Engel, M.J.~Ramsey-Musolf and U.~van Kolck,
  %``Electric Dipole Moments of Nucleons, Nuclei, and Atoms: The Standard Model and Beyond,''
   \Journal{\em Prog.\ Part.\ Nucl.\ Phys.}{71}{21.}{2013}
%  doi:10.1016/j.ppnp.2013.03.003  [arXiv:1303.2371 [nucl-th]].
  %%CITATION = doi:10.1016/j.ppnp.2013.03.003;%%
  %145 citations counted in INSPIRE as of 02 Mar 2017
%
\bibitem{Afach:2015sja}
  J.M.~Pendlebury {\it et al.},
  %``Revised experimental upper limit on the electric dipole moment of the neutron,''
   \Journal{\em Phys.\ Rev.}{D 92}{092003.}{2015}
%  doi:10.1103/PhysRevD.92.092003  [arXiv:1509.04411 [hep-ex]].
  %%CITATION = doi:10.1103/PhysRevD.92.092003;%%
  %52 citations counted in INSPIRE as of 07 Mar 2017
%
\bibitem{nEDM-PSI}
  See webpage https://www.psi.ch/nedm/.
%
\bibitem{nEDM-SNS}
  See webpage http://www.phy.ornl.gov/nedm/.
 %
\bibitem{Golub:1994cg}
  R.~Golub and S.K.~Lamoreaux,
  %``Neutron electric dipole moment, ultracold neutrons and polarized He-3,''
  \Journal{\em Phys.\ Rept.}{237}{1.}{1994}
%  Phys.\ Rept.\  {\bf 237}, 1 (1994).
%  doi:10.1016/0370-1573(94)90084-1
  %%CITATION = doi:10.1016/0370-1573(94)90084-1;%%
  %97 citations counted in INSPIRE as of 09 Mar 2017
%
\bibitem{Baron:2013eja}
  J.~Baron {\it et al.} [ACME Collaboration],
  %``Order of Magnitude Smaller Limit on the Electric Dipole Moment of the Electron,''
   \Journal{\em Science}{343}{269.}{2014}
%  doi:10.1126/science.1248213  [arXiv:1310.7534 [physics.atom-ph]].
  %%CITATION = doi:10.1126/science.1248213;%%
  %313 citations counted in INSPIRE as of 07 Mar 2017
%
\bibitem{Inoue:2014nva}
  S.~Inoue, M.J.~Ramsey-Musolf and Y.~Zhang,
  %``CP-violating phenomenology of flavor conserving two Higgs doublet models,''
   \Journal{\em Phys.\ Rev.}{D 89}{115023.}{2014}
%  doi:10.1103/PhysRevD.89.115023  [arXiv:1403.4257 [hep-ph]].
  %%CITATION = doi:10.1103/PhysRevD.89.115023;%%
  %53 citations counted in INSPIRE as of 08 Mar 2017
%
\bibitem{Brod:2011re}
  J.~Brod, A.L.~Kagan and J.~Zupan,
  %``Size of direct CP violation in singly Cabibbo-suppressed D decays,''
   \Journal{\em Phys.\ Rev.}{D 86}{014023.}{2012}
%  Phys.\ Rev.\ D {\bf 86}, 014023 (2012)
%  doi:10.1103/PhysRevD.86.014023  [arXiv:1111.5000 [hep-ph]].
  %%CITATION = doi:10.1103/PhysRevD.86.014023;%%
  %113 citations counted in INSPIRE as of 18 Feb 2017
%
\bibitem{Bhattacharya:2012ah}
  B.~Bhattacharya, M.~Gronau and J.L.~Rosner,
  %``CP asymmetries in singly-Cabibbo-suppressed $D$ decays to two pseudoscalar mesons,''
   \Journal{\em Phys.\ Rev.}{D 85}{054014.}{2012}
%  Phys.\ Rev.\ D {\bf 85}, 054014 (2012)
%  [Phys.\ Rev.\ D {\bf 85}, no. 7, 079901 (2012)]
%  doi:10.1103/PhysRevD.85.079901, 10.1103/PhysRevD.85.054014  [arXiv:1201.2351 [hep-ph]].
  %%CITATION = doi:10.1103/PhysRevD.85.079901, 10.1103/PhysRevD.85.054014;%%
  %117 citations counted in INSPIRE as of 18 Feb 2017
%
\bibitem{MarSir}
  W. Marciano and A. Sirlin,
  \Journal{\em Phys.\ Rev.\ Lett.}{61}{1815.}{1988}
%
\bibitem{tau_tau-Belle}
  K. Belous \emph{et al.} (Belle collaboration),
  \Journal{\em Phys.\ Rev.\ Lett.}{112}{031801.}{2014}
%
\bibitem{Dedes:2002rh}
  A.~Dedes, J.R.~Ellis and M.~Raidal,
  %``Higgs mediated $B^0_{s,d} \to \mu \tau$, $e \tau$ and $\tau \to 3 \mu$, $e \mu \mu$ decays in supersymmetric seesaw models,''
   \Journal{\em Phys.\ Lett.}{B 549}{159.}{2002}
%  Phys.\ Lett.\ B {\bf 549}, 159 (2002)
%  doi:10.1016/S0370-2693(02)02900-3  [hep-ph/0209207].
  %%CITATION = doi:10.1016/S0370-2693(02)02900-3;%%
  %121 citations counted in INSPIRE as of 13 Mar 2017
%
\bibitem{Goto:2014vga}
  T.~Goto, Y.~Okada, T.~Shindou, M.~Tanaka and R.~Watanabe,
  %``Lepton flavor violation in the supersymmetric seesaw model after the LHC 8 TeV run,''
   \Journal{\em Phys.\ Rev.}{D 91}{033007.}{2015}
%  Phys.\ Rev.\ D {\bf 91}, 033007 (2015)
%  doi:10.1103/PhysRevD.91.033007  [arXiv:1412.2530 [hep-ph]].
  %%CITATION = doi:10.1103/PhysRevD.91.033007;%%
  %3 citations counted in INSPIRE as of 21 Mar 2017
%
\bibitem{Cvetic:2002jy}
  G.~Cvetic, C.~Dib, C.S.~Kim and J.D.~Kim,
  %``On lepton flavor violation in tau decays,''
   \Journal{\em Phys.\ Rev.}{D 66}{034008.}{2002}
%  Phys.\ Rev.\ D {\bf 66}, 034008 (2002)
%  Erratum: [Phys.\ Rev.\ D {\bf 68}, 059901 (2003)]
%  doi:10.1103/PhysRevD.66.034008, 10.1103/PhysRevD.68.059901  [hep-ph/0202212].
  %%CITATION = doi:10.1103/PhysRevD.66.034008, 10.1103/PhysRevD.68.059901;%%
  %75 citations counted in INSPIRE as of 13 Mar 2017
%
\bibitem{Yue:2002ja}
  C.-x.~Yue, Y.-m.~Zhang and L.-j.~Liu,
  %``Nonuniversal gauge bosons Z-prime and lepton flavor violation tau decays,''
  \Journal{\em Phys.\ Lett.}{B 547}{252.}{2002}
%  Phys.\ Lett.\ B {\bf 547}, 252 (2002)
%  doi:10.1016/S0370-2693(02)02781-8  [hep-ph/0209291].
  %%CITATION = doi:10.1016/S0370-2693(02)02781-8;%%
  %50 citations counted in INSPIRE as of 13 Mar 2017
%
\bibitem{Chiang:2016vgf}
  For some discussion, see C.-W.~Chiang, K.~Fuyuto and E.~Senaha,
  %``Electroweak Baryogenesis with Lepton Flavor Violation,''
  \Journal{\em Phys.\ Lett.}{B 762}{315.}{2016}
%  Phys.\ Lett.\ B {\bf 762}, 315 (2016)
%  doi:10.1016/j.physletb.2016.09.052  [arXiv:1607.07316 [hep-ph]].
  %%CITATION = doi:10.1016/j.physletb.2016.09.052;%%
  %5 citations counted in INSPIRE as of 21 Mar 2017
%
\bibitem{Hayasaka:2010np}
  K.~Hayasaka, K.~Inami, Y.~Miyazaki {\it et al.} [Belle Collaboration],
  %``Search for Lepton Flavor Violating Tau Decays into Three Leptons with 719 Million Produced Tau+Tau- Pairs,''
   \Journal{\em Phys.\ Lett.}{B 687}{139.}{2010}
%  Phys.\ Lett.\ B {\bf 687}, 139 (2010)
%  doi:10.1016/j.physletb.2010.03.037  [arXiv:1001.3221 [hep-ex]].
  %%CITATION = doi:10.1016/j.physletb.2010.03.037;%%
  %190 citations counted in INSPIRE as of 13 Mar 2017
%
\bibitem{Aaij:2016cfh}
  R.~Aaij {\it et al.} [LHCb Collaboration],
  %``Measurement of the difference of time-integrated CP asymmetries in $D^0 \rightarrow K^{-} K^{+} $ and $D^0 \rightarrow \pi^{-} \pi^{+} $ decays,''
   \Journal{\em Phys.\ Rev.\ Lett.}{116}{191601.}{2016}
%  Phys.\ Rev.\ Lett.\  {\bf 116}, 191601 (2016)
%  doi:10.1103/PhysRevLett.116.191601  [arXiv:1602.03160 [hep-ex]].
  %%CITATION = doi:10.1103/PhysRevLett.116.191601;%%
  %23 citations counted in INSPIRE as of 01 Mar 2017
%
\bibitem{Gersabeck:2011xj}
  M.~Gersabeck, M.~Alexander, S.~Borghi, V.V.~Gligorov and C.~Parkes,
  %``On the interplay of direct and indirect CP violation in the charm sector,''
   \Journal{\em J.\ Phys.}{G 39}{045005.}{2012}
%  J.\ Phys.\ G {\bf 39}, 045005 (2012)
%  doi:10.1088/0954-3899/39/4/045005  [arXiv:1111.6515 [hep-ex]].
  %%CITATION = doi:10.1088/0954-3899/39/4/045005;%%
  %42 citations counted in INSPIRE as of 01 Mar 2017
%
\bibitem{Nierste:2015zra}
  U.~Nierste and S.~Schacht,
  %``CP Violation in $D^0\rightarrow K_SK_S$,''
   \Journal{\em Phys.\ Rev.}{D 92}{054036.}{2015}
%  Phys.\ Rev.\ D {\bf 92}, 054036 (2015)
%  doi:10.1103/PhysRevD.92.054036  [arXiv:1508.00074 [hep-ph]].
  %%CITATION = doi:10.1103/PhysRevD.92.054036;%%
  %5 citations counted in INSPIRE as of 06 Mar 2017
%
\bibitem{Abdesselam:2016gqq}
  A.~Abdesselam {\it et al.} [Belle Collaboration],
  %``Measurement of $CP$ asymmetry in the $D^{0} \to K^0_S K^0_S$ decay at Belle,''
  arXiv:1609.06393 [hep-ex].
  %%CITATION = ARXIV:1609.06393;%%
  %2 citations counted in INSPIRE as of 06 Mar 2017
%
\bibitem{Aaij:2015lsa}
  R.~Aaij {\it et al.} [LHCb Collaboration],
  %``Studies of the resonance structure in $D^0\to K^0_S K^{\pm}\pi^{\mp}$ decays,''
   \Journal{\em Phys.\ Rev.}{D 93}{052018.}{2016}
%  Phys.\ Rev.\ D {\bf 93}, 052018 (2016)
%  doi:10.1103/PhysRevD.93.052018  [arXiv:1509.06628 [hep-ex]].
  %%CITATION = doi:10.1103/PhysRevD.93.052018;%%
  %9 citations counted in INSPIRE as of 06 Mar 2017
%
\bibitem{Aaij:2016nki}
  R.~Aaij {\it et al.} [LHCb Collaboration],
  %``Search for $CP$ violation in the phase space of $D^0\rightarrow\pi^+\pi^-\pi^+\pi^-$ decays,''
   \Journal{\em Phys.\ Lett.}{B 769}{345.}{2017}
%  Phys.\ Lett.\ B {\bf 769}, 345 (2017)
%  doi:10.1016/j.physletb.2017.03.062  arXiv:1612.03207 [hep-ex].
  %%CITATION = ARXIV:1612.03207;%%
  %1 citations counted in INSPIRE as of 06 Mar 2017
%
\bibitem{Aaij:2016roz}
  R.~Aaij {\it et al.} [LHCb Collaboration],
  %``Measurements of charm mixing and $C\!P$ violation using $D^0 \to K^\pm \pi^\mp$ decays,''
   \Journal{\em Phys.\ Rev.}{D 95}{052004.}{2017}
%  Phys.\ Rev.\ D {\bf 95}, 052004 (2017)
%  doi:10.1103/PhysRevD.95.052004  [arXiv:1611.06143 [hep-ex]].
  %%CITATION = doi:10.1103/PhysRevD.95.052004;%%
  %2 citations counted in INSPIRE as of 02 Apr 2017%
\bibitem{Aaij:2016rhq}
  R.~Aaij {\it et al.} [LHCb Collaboration],
  %``First observation of $D^0-\bar D^0$ oscillations in $D^0\to K^+\pi^-\pi^+\pi^-$ decays and measurement of the associated coherence parameters,''
   \Journal{\em Phys.\ Rev.\ Lett.}{116}{241801.}{2016}
%  Phys.\ Rev.\ Lett.\  {\bf 116}, 241801 (2016)
%  doi:10.1103/PhysRevLett.116.241801  [arXiv:1602.07224 [hep-ex]].
  %%CITATION = doi:10.1103/PhysRevLett.116.241801;%%
  %15 citations counted in INSPIRE as of 11 Mar 2017
%
\bibitem{Staric:2015sta}
  M.~Stari\v c {\it et al.} [Belle Collaboration],
  %``Measurement of $D^0 ? \bar {D^0}$ mixing and search for CP violation in $D^0 \to K^+ K^?, \pi^+ \pi^?$ decays with the full Belle data set,''
   \Journal{\em Phys.\ Lett.}{B 753}{412.}{2016}
%  Phys.\ Lett.\ B {\bf 753}, 412 (2016)
%  doi:10.1016/j.physletb.2015.12.025  [arXiv:1509.08266 [hep-ex]].
  %%CITATION = doi:10.1016/j.physletb.2015.12.025;%%
  %4 citations counted in INSPIRE as of 11 Mar 2017
%
\bibitem{LHCb:2016ifp}
  LHCb Collaboration,
  %``$CP$-violating asymmetries from the decay-time distribution of prompt $D^0 \to K^+ K^-$ and $D^0 \to \pi^+\pi^-$ decays in the full $\mbox{LHCb}$ Run 1 data sample. Measurement using unbinned, acceptance corrected decay-time.,''
  LHCb-CONF-2016-010. %, CERN-LHCb-CONF-2016-010.
  %%CITATION = LHCB-CONF-2016-010, CERN-LHCB-CONF-2016-010;%%
%
\bibitem{LHCb:2016lbh}
  LHCb Collaboration,
  %``$CP$-violating asymmetries from the decay-time distribution of  prompt $D^0 \to K^+K^-$  and  $D^0 \to \pi^+\pi^-$ decays in the  full LHCb Run~1 data sample. Measurement using yield asymmetries in bins of decay time.,''
  LHCb-CONF-2016-009. %, CERN-LHCb-CONF-2016-009.
  %%CITATION = LHCB-CONF-2016-009, CERN-LHCB-CONF-2016-009;%%
%
\bibitem{Glashow:1976nt}
  S.L.~Glashow and S.~Weinberg,
  %``Natural Conservation Laws for Neutral Currents,''
   \Journal{\em Phys.\ Rev.}{D 15}{1958.}{1977}
%  doi:10.1103/PhysRevD.15.1958
  %%CITATION = doi:10.1103/PhysRevD.15.1958;%%
  %1497 citations counted in INSPIRE as of 12 Mar 2017
%
\bibitem{Fritzsch:1977vd}
  H.~Fritzsch,
  %``Weak Interaction Mixing in the Six - Quark Theory,''
   \Journal{\em Phys.\ Lett.}{73B}{317.}{1978}
%  doi:10.1016/0370-2693(78)90524-5
  %%CITATION = doi:10.1016/0370-2693(78)90524-5;%%
  %794 citations counted in INSPIRE as of 12 Mar 2017
%
\bibitem{Cheng:1987rs}
  T.-P.~Cheng and M.~Sher,
  %``Mass Matrix Ansatz and Flavor Nonconservation in Models with Multiple Higgs Doublets,''
   \Journal{\em Phys.\ Rev.}{D 35}{3484.}{1987}
%  doi:10.1103/PhysRevD.35.3484
  %%CITATION = doi:10.1103/PhysRevD.35.3484;%%
  %496 citations counted in INSPIRE as of 12 Mar 2017
%
\bibitem{Hou:1991un}
  W.-S.~Hou,
  %``Tree level t ---> c h or h ---> t anti-c decays,''
   \Journal{\em Phys.\ Lett.}{B 296}{179.}{1992}
%  doi:10.1016/0370-2693(92)90823-M
  %%CITATION = doi:10.1016/0370-2693(92)90823-M;%%
  %206 citations counted in INSPIRE as of 12 Mar 2017
%
\bibitem{Chatrchyan:2013nwa}
  S.~Chatrchyan {\it et al.} [CMS Collaboration],
  %``Search for Flavor-Changing Neutral Currents in Top-Quark Decays $t \to Zq$ in $pp$ Collisions at $\sqrt{s}=8$  TeV,''
   \Journal{\em Phys.\ Rev.\ Lett.}{112}{171802.}{2014}
%  doi:10.1103/PhysRevLett.112.171802  [arXiv:1312.4194 [hep-ex]].
  %%CITATION = doi:10.1103/PhysRevLett.112.171802;%%
  %76 citations counted in INSPIRE as of 12 Mar 2017
%
\bibitem{Aad:2015uza}
  G.~Aad {\it et al.} [ATLAS Collaboration],
  %``Search for flavour-changing neutral current top-quark decays to $qZ$ in $pp$ collision data collected with the ATLAS detector at $\sqrt s =8$  TeV,''
   \Journal{\em Eur.\ Phys.\ J.}{C 76}{12.}{2016}
%  Eur.\ Phys.\ J.\ C {\bf 76}, 12 (2016)
%  doi:10.1140/epjc/s10052-015-3851-5  [arXiv:1508.05796 [hep-ex]].
  %%CITATION = doi:10.1140/epjc/s10052-015-3851-5;%%
  %23 citations counted in INSPIRE as of 19 Apr 2017
%
\bibitem{Azatov:2014lha}
  A.~Azatov, G.~Panico, G.~Perez and Y.~Soreq,
  %``On the Flavor Structure of Natural Composite Higgs Models & Top Flavor Violation,''
   \Journal{\em JHEP}{1412}{082.}{2014}
%  doi:10.1007/JHEP12(2014)082  [arXiv:1408.4525 [hep-ph]].
  %%CITATION = doi:10.1007/JHEP12(2014)082;%%
  %20 citations counted in INSPIRE as of 12 Mar 2017
%
\bibitem{Bardhan:2016txk}
  See e.g. D.~Bardhan, G.~Bhattacharyya, D.~Ghosh, M.~Patra and S.~Raychaudhuri,
  %``Detailed analysis of flavor-changing decays of top quarks as a probe of new physics at the LHC,''
   \Journal{\em Phys.\ Rev.}{D 94}{015026.}{2016}
%  doi:10.1103/PhysRevD.94.015026  [arXiv:1601.04165 [hep-ph]].
  %%CITATION = doi:10.1103/PhysRevD.94.015026;%%
  %9 citations counted in INSPIRE as of 12 Mar 2017
%
\bibitem{CMS:2013xfa}
  CMS Collaboration,
  %``Projected Performance of an Upgraded CMS Detector at the LHC and HL-LHC: Contribution to the Snowmass Process,''
  arXiv:1307.7135 [hep-ex] (contribution to Snowmass 2013).
  %%CITATION = ARXIV:1307.7135;%%
  %229 citations counted in INSPIRE as of 12 Mar 2017
%
\bibitem{ATLAS:2013hta}
  ATLAS Collaboration,
  %``Physics at a High-Luminosity LHC with ATLAS,''
  arXiv:1307.7292 [hep-ex] (contribution to Snowmass 2013).
  %%CITATION = ARXIV:1307.7292;%%
  %151 citations counted in INSPIRE as of 12 Mar 2017
%
\bibitem{Aad:2015gea}
  See e.g. G.~Aad {\it et al.} [ATLAS Collaboration],
  %``Search for single top-quark production via flavour-changing neutral currents at 8 TeV with the ATLAS detector,''
   \Journal{\em Eur.\ Phys.\ J.}{C 76}{55.}{2016}
%  doi:10.1140/epjc/s10052-016-3876-4  [arXiv:1509.00294 [hep-ex]].
  %%CITATION = doi:10.1140/epjc/s10052-016-3876-4;%%
  %25 citations counted in INSPIRE as of 12 Mar 2017
%
\bibitem{Sirunyan:2017kkr}
  A.M.~Sirunyan {\it et al.} [CMS Collaboration],
  %``Search for associated production of a Z boson with a single top quark and for tZ flavour-changing interactions in pp collisions at $\sqrt{s}$ = 8 TeV,''
  \Journal{\em JHEP}{1707}{003.}{2017}.
%  arXiv:1702.01404 [hep-ex].
  %%CITATION = ARXIV:1702.01404;%%
  %1 citations counted in INSPIRE as of 12 Mar 2017
%
\bibitem{Hou:2017ozb}
  W.-S.~Hou, M.~Kohda and T.~Modak,
  %``Search for $tZ'$ associated production induced by $tcZ'$ couplings at the LHC,''
  %arXiv:1702.07275 [hep-ph].
  \Journal{\em Phys.\ Rev.}{D 96}{015037}{2017}.
  %%CITATION = ARXIV:1702.07275;%%
%
\bibitem{Chen:2013qta}
  K.-F.~Chen, W.-S.~Hou, C.~Kao and M.~Kohda,
  %``When the Higgs meets the Top: Search for t --> ch^0 at the LHC,''
   \Journal{\em Phys.\ Lett.}{B 725}{378.}{2013}
%  doi:10.1016/j.physletb.2013.07.060  [arXiv:1304.8037 [hep-ph]].
  %%CITATION = doi:10.1016/j.physletb.2013.07.060;%%
  %38 citations counted in INSPIRE as of 13 Mar 2017
%
\bibitem{TheATLAScollaboration:2013nia}
  ATLAS collaboration,
  %``Search for flavor-changing neutral currents in $t\rightarrow cH$, with $H\to\gamma\gamma$, and limit on the tcH coupling,''
  ATLAS-CONF-2013-081.
  %%CITATION = ATLAS-CONF-2013-081;%%
  %23 citations counted in INSPIRE as of 13 Mar 2017
%
\bibitem{Craig:2012vj}
  N.~Craig {\it et al.}, %J.A.~Evans, R.~Gray, M.~Park, S.~Somalwar, S.~Thomas and M.~Walker,
  %``Searching for $t \to c h$ with Multi-Leptons,''
   \Journal{\em Phys.\ Rev.}{D 86}{075002.}{2012}
%  Phys.\ Rev.\ D {\bf 86}, 075002 (2012)
%  doi:10.1103/PhysRevD.86.075002  [arXiv:1207.6794 [hep-ph]].
  %%CITATION = doi:10.1103/PhysRevD.86.075002;%%
  %44 citations counted in INSPIRE as of 13 Mar 2017
%
\bibitem{Aad:2014dya}
  G.~Aad {\it et al.} [ATLAS Collaboration],
  %``Search for top quark decays $t \to qH$ with $H \to \gamma\gamma$ using the ATLAS detector,''
   \Journal{\em JHEP}{1406}{008.}{2014}
%  JHEP {\bf 1406}, 008 (2014)
%  doi:10.1007/JHEP06(2014)008  [arXiv:1403.6293 [hep-ex]].
  %%CITATION = doi:10.1007/JHEP06(2014)008;%%
  %82 citations counted in INSPIRE as of 03 Apr 2017
%
\bibitem{Khachatryan:2014jya}
  V.~Khachatryan {\it et al.} [CMS Collaboration],
  %``Searches for heavy Higgs bosons in two-Higgs-doublet models and for $t→ch$ decay using multilepton and diphoton final states in $pp$ collisions at 8 TeV,''
   \Journal{\em Phys.\ Rev.}{D 90}{112013.}{2014}
%  Phys.\ Rev.\ D {\bf 90}, 112013 (2014)
%  doi:10.1103/PhysRevD.90.112013  [arXiv:1410.2751 [hep-ex]].
  %%CITATION = doi:10.1103/PhysRevD.90.112013;%%
  %66 citations counted in INSPIRE as of 03 Apr 2017
%
\bibitem{Aad:2015pja}
  G.~Aad {\it et al.} [ATLAS Collaboration],
  %``Search for flavour-changing neutral current top quark decays $t\to Hq$ in $pp$ collisions at $\sqrt{s}=8$ TeV with the ATLAS detector,''
   \Journal{\em JHEP}{1512}{061.}{2015}
%  JHEP {\bf 1512}, 061 (2015)
%  doi:10.1007/JHEP12(2015)061  [arXiv:1509.06047 [hep-ex]].
  %%CITATION = doi:10.1007/JHEP12(2015)061;%%
  %34 citations counted in INSPIRE as of 13 Mar 2017
%
\bibitem{Khachatryan:2016atv}
  V.~Khachatryan {\it et al.} [CMS Collaboration],
  %``Search for top quark decays via Higgs-boson-mediated flavor-changing neutral currents in pp collisions at $ \sqrt{s}=8 $ TeV,''
   \Journal{\em JHEP}{1702}{079.}{2017}
%  JHEP {\bf 1702}, 079 (2017)
%  doi:10.1007/JHEP02(2017)079  [arXiv:1610.04857 [hep-ex]].
  %%CITATION = doi:10.1007/JHEP02(2017)079;%%
  %5 citations counted in INSPIRE as of 13 Mar 2017
%
\bibitem{Aaboud:2017mfd}
  M.~Aaboud {\it et al.} [ATLAS Collaboration],
  %``Search for top quark decays $t\rightarrow qH$, with $H\to\gamma\gamma$, in $\sqrt{s}=13$ TeV $pp$ collisions using the ATLAS detector,''
  arXiv:1707.01404 [hep-ex].
  %%CITATION = ARXIV:1707.01404;%%
%
\bibitem{Kao:2011aa}
  C.~Kao, H.-Y.~Cheng, W.-S.~Hou and J.~Sayre,
  %``Top Decays with Flavor Changing Neutral Higgs Interactions at the LHC,''
   \Journal{\em Phys.\ Lett.}{B 716}{225.}{2012}
%  Phys.\ Lett.\ B {\bf 716}, 225 (2012)
%  doi:10.1016/j.physletb.2012.08.032  [arXiv:1112.1707 [hep-ph]].
  %%CITATION = doi:10.1016/j.physletb.2012.08.032;%%
  %17 citations counted in INSPIRE as of 03 Apr 2017
%
\bibitem{ATL-tch-gaga}
   ATLAS Collaboration, ATL-PHYS-PUB-2013-012.
%
\bibitem{Altunkaynak:2015twa}
  B.~Altunkaynak, W.-S.~Hou, C.~Kao, M.~Kohda and B.~McCoy,
  %``Flavor Changing Heavy Higgs Interactions at the LHC,''
   \Journal{\em Phys.\ Lett.}{B 751}{135.}{2015}
%  Phys.\ Lett.\ B {\bf 751}, 135 (2015)
%  doi:10.1016/j.physletb.2015.10.024  [arXiv:1506.00651 [hep-ph]].
  %%CITATION = doi:10.1016/j.physletb.2015.10.024;%%
  %14 citations counted in INSPIRE as of 13 Mar 2017
%
\bibitem{DiazCruz:1999xe}
  J.L.~Diaz-Cruz and J.J.~Toscano,
  %``Lepton flavor violating decays of Higgs bosons beyond the standard model,''
   \Journal{\em Phys.\ Rev.}{D 62}{116005.}{2000}
%  Phys.\ Rev.\ D {\bf 62}, 116005 (2000)
%  doi:10.1103/PhysRevD.62.116005  [hep-ph/9910233].
  %%CITATION = doi:10.1103/PhysRevD.62.116005;%%
  %106 citations counted in INSPIRE as of 14 Mar 2017
%
\bibitem{Han:2000jz}
  T.~Han and D.~Marfatia,
  %``h ---> mu tau at hadron colliders,''
   \Journal{\em Phys.\ Rev.\ Lett.}{86}{1442.}{2001}
%  Phys.\ Rev.\ Lett.\  {\bf 86}, 1442 (2001)
%  doi:10.1103/PhysRevLett.86.1442  [hep-ph/0008141].
  %%CITATION = doi:10.1103/PhysRevLett.86.1442;%%
  %66 citations counted in INSPIRE as of 12 Mar 2017
%
\bibitem{Khachatryan:2015kon}
  V.~Khachatryan {\it et al.} [CMS Collaboration],
  %``Search for Lepton-Flavour-Violating Decays of the Higgs Boson,''
   \Journal{\em Phys.\ Lett.}{B 749}{337.}{2015}
%  Phys.\ Lett.\ B {\bf 749}, 337 (2015)
%  doi:10.1016/j.physletb.2015.07.053  [arXiv:1502.07400 [hep-ex]].
  %%CITATION = doi:10.1016/j.physletb.2015.07.053;%%
  %192 citations counted in INSPIRE as of 13 Mar 2017
%
\bibitem{Aad:2015gha}
  G.~Aad {\it et al.} [ATLAS Collaboration],
  %``Search for lepton-flavour-violating H → μτ decays of the Higgs boson with the ATLAS detector,''
   \Journal{\em JHEP}{1511}{211.}{2015}
%  JHEP {\bf 1511}, 211 (2015)
%  doi:10.1007/JHEP11(2015)211  [arXiv:1508.03372 [hep-ex]].
  %%CITATION = doi:10.1007/JHEP11(2015)211;%%
  %98 citations counted in INSPIRE as of 13 Mar 2017
%
\bibitem{CMS:2016qvi}
  CMS Collaboration,
  %``Search for Lepton Flavour Violating Decays of the Higgs Boson in the mu-tau final state at 13 TeV,''
  CMS-PAS-HIG-16-005.
  %%CITATION = CMS-PAS-HIG-16-005;%%
  %17 citations counted in INSPIRE as of 13 Mar 2017
%
\bibitem{Blankenburg:2012ex}
  G.~Blankenburg, J.~Ellis and G.~Isidori,
  %``Flavour-Changing Decays of a 125 GeV Higgs-like Particle,''
   \Journal{\em Phys.\ Lett.}{B 712}{386.}{2012}
%  Phys.\ Lett.\ B {\bf 712}, 386 (2012)
%  doi:10.1016/j.physletb.2012.05.007  [arXiv:1202.5704 [hep-ph]].
  %%CITATION = doi:10.1016/j.physletb.2012.05.007;%%
  %139 citations counted in INSPIRE as of 15 Mar 2017
%
\bibitem{Harnik:2012pb}
  R.~Harnik, J.~Kopp and J.~Zupan,
  %``Flavor Violating Higgs Decays,''
   \Journal{\em JHEP}{1303}{026.}{2013}
%  JHEP {\bf 1303}, 026 (2013)
%  doi:10.1007/JHEP03(2013)026  [arXiv:1209.1397 [hep-ph]].
  %%CITATION = doi:10.1007/JHEP03(2013)026;%%
  %188 citations counted in INSPIRE as of 12 Mar 2017
%
\bibitem{Chang:1993kw}
  D.~Chang, W.-S.~Hou and W.-Y.~Keung,
  %``Two loop contributions of flavor changing neutral Higgs bosons to mu ---> e gamma,''
   \Journal{\em Phys.\ Rev.}{D 48}{217.}{1993}
%  Phys.\ Rev.\ D {\bf 48}, 217 (1993)
%  doi:10.1103/PhysRevD.48.217  [hep-ph/9302267].
  %%CITATION = doi:10.1103/PhysRevD.48.217;%%
  %147 citations counted in INSPIRE as of 12 Mar 2017
%
\bibitem{Bjorken:1977vt}
  J.D.~Bjorken and S.~Weinberg,
  %``A Mechanism for Nonconservation of Muon Number,''
   \Journal{\em Phys.\ Rev.\ Lett.}{38}{622.}{1977}
%  Phys.\ Rev.\ Lett.\  {\bf 38}, 622 (1977).
%  doi:10.1103/PhysRevLett.38.622
  %%CITATION = doi:10.1103/PhysRevLett.38.622;%%
  %174 citations counted in INSPIRE as of 15 Mar 2017
%
\bibitem{Barr:1990vd}
  S.M.~Barr and A.~Zee,
  %``Electric Dipole Moment of the Electron and of the Neutron,''
   \Journal{\em Phys.\ Rev.\ Lett.}{65}{21.}{1990}
%  Phys.\ Rev.\ Lett.\  {\bf 65}, 21 (1990)
%  Erratum: [Phys.\ Rev.\ Lett.\  {\bf 65}, 2920 (1990)].
%  doi:10.1103/PhysRevLett.65.21
  %%CITATION = doi:10.1103/PhysRevLett.65.21;%%
  %382 citations counted in INSPIRE as of 15 Mar 2017
%
\bibitem{CMS:2017onh}
  The CMS Collaboration, %[CMS Collaboration],
  %``Search for lepton flavour violating  decays of the Higgs boson to $\mu\tau$  and  $\textrm{e}\tau$ in proton-proton collisions at $\sqrt{s}=13~\mathrm{TeV}$,''
  CMS-PAS-HIG-17-001.
  %%CITATION = CMS-PAS-HIG-17-001;%%
  %1 citations counted in INSPIRE as of 25 Jun 2017
%
\bibitem{Altmannshofer:2016oaq}
  W.~Altmannshofer, M.~Carena and A.~Crivellin,
  %``$L_\mu - L_\tau$ theory of Higgs flavor violation and $(g-2)_\mu$,''
   \Journal{\em Phys.\ Rev.}{D 94}{095026.}{2016}
%  Phys.\ Rev.\ D {\bf 94}, 095026 (2016).
%  doi:10.1103/PhysRevD.94.095026
%  [arXiv:1604.08221 [hep-ph]].
  %%CITATION = doi:10.1103/PhysRevD.94.095026;%%
  %22 citations counted in INSPIRE as of 23 Jun 2017
%
\bibitem{Galon:2017qes}
  I.~Galon and J.~Zupan,
  %``Dark sectors and enhanced $h\to \tau \mu$ transitions,''
   \Journal{\em JHEP}{1705}{083.}{2017}
%  JHEP {\bf 1705}, 083 (2017).
%  doi:10.1007/JHEP05(2017)083
%  [arXiv:1701.08767 [hep-ph]].
  %%CITATION = doi:10.1007/JHEP05(2017)083;%%
  %1 citations counted in INSPIRE as of 23 Jun 2017
%
\bibitem{Fuyuto:2017ewj}
  K.~Fuyuto, W.-S.~Hou and E.~Senaha,
  %``Electroweak baryogenesis via top transport,''
  arXiv:1705.05034 [hep-ph].
  %%CITATION = ARXIV:1705.05034;%%
%
\bibitem{Dicus:1994bm}
  D.~Dicus, A.~Stange and S.~Willenbrock,
  %``Higgs decay to top quarks at hadron colliders,''
   \Journal{\em Phys.\ Lett.}{B 333}{126.}{1994}
%  Phys.\ Lett.\ B {\bf 333}, 126 (1994).
%  doi:10.1016/0370-2693(94)91017-0
%  [hep-ph/9404359].
  %%CITATION = doi:10.1016/0370-2693(94)91017-0;%%
  %99 citations counted in INSPIRE as of 23 Jun 2017
%
\bibitem{Frederix:2007gi}
  R.~Frederix and F.~Maltoni,
  %``Top pair invariant mass distribution: A Window on new physics,''
   \Journal{\em JHEP}{0901}{047.}{2009}
%  JHEP {\bf 0901}, 047 (2009).
%  doi:10.1088/1126-6708/2009/01/047
%  [arXiv:0712.2355 [hep-ph]].
  %%CITATION = doi:10.1088/1126-6708/2009/01/047;%%
  %191 citations counted in INSPIRE as of 21 Jun 2017
%
\bibitem{Carena:2016npr}
  For a recent reference, see M.~Carena and Z.~Liu,
  %``Challenges and opportunities for heavy scalar searches in the $ t\overline{t} $ channel at the LHC,''
   \Journal{\em JHEP}{1611}{159.}{2016}
%  JHEP {\bf 1611}, 159 (2016), and references therein.
%  doi:10.1007/JHEP11(2016)159
%  [arXiv:1608.07282 [hep-ph]].
  %%CITATION = doi:10.1007/JHEP11(2016)159;%%
  %16 citations counted in INSPIRE as of 18 Jun 2017
%
\bibitem{Khachatryan:2016vau}
  G.~Aad {\it et al.} [ATLAS and CMS Collaborations],
  %``Measurements of the Higgs boson production and decay rates and constraints on its couplings from a combined ATLAS and CMS analysis of the LHC pp collision data at $ \sqrt{s}=7 $ and 8 TeV,''
   \Journal{\em JHEP}{1608}{045.}{2016}
%  JHEP {\bf 1608}, 045 (2016)
%  doi:10.1007/JHEP08(2016)045  [arXiv:1606.02266 [hep-ex]].
  %%CITATION = doi:10.1007/JHEP08(2016)045;%%
  %213 citations counted in INSPIRE as of 03 Apr 2017
%
\bibitem{LHCC-P-008}
  CMS Collaboration, Technical Proposal for Phase II Upgrade, LHCC-P-008.
%
%\bibitem{rupr0} R. Machleidt, \Journal{\em Adv. Nucl. Phys.}{19}{189}{1989}

\end{thebibliography}
\end{document}